\pdfoutput=1
\RequirePackage{latex/atlaslatexpath}

\documentclass[cernpreprint, atlasdraft=true, texlive=2016, coverpage=false, UKenglish, ]{atlasdoc}

\usepackage[backend=bibtex]{atlaspackage}
\usepackage{atlasbiblatex}
\usepackage{atlasphysics}
\usepackage{comment}

\usepackage{scrlayer-scrpage}
\usepackage{xcolor}
\usepackage[T1]{fontenc}

\graphicspath{{figures/}} 

\AtlasTitle{
Probe of  soft-QCD in minimum bias  events  of   \(pp\) collisions   with   the  ATLAS at the LHC
\\[10mm]
{\Large   
Yuri~A.~Kulchitsky\footnote{E-mail: Yuri.Koultchitski@cern.ch,Kulchitsky@jinr.ru}
}
\\
{\large   Joint Institute for Nuclear Research,  Dubna,  Russia}
\\[-3mm]
{\large    B.I.~Stepanov Institute of Physics, National Academy of Sciences, Minsk, Belarus }
}

\AtlasAbstract{
The study of the minimum-bias charged-particle distributions is reviewed.
The data are obtained using the ATLAS detector at the LHC in proton-proton collisions  at 
centre-of-mass energies from \(0.9\) to \(13\)~\TeV.
The particles are required to have  an absolute pseudorapidity of  less than \(2.5\).
For charged-particle distributions, two transverse momentum threshold cases,  greater than \(100\)~\MeV\ 
and \(500\)~\MeV,  were taken.
The charged-particle multiplicity,   its dependence on  the transverse momentum and pseudorapidity,  
the dependence  of the  average transverse momentum on the  charged-particle multiplicity, 
and  the KNO-scaling study  are presented. 
The measured distributions are compared with the predictions of various  tunings of   the
Monte Carlo generator, which implement different minimum-bias models.
The Monte Carlo model predictions qualitatively describe the data well, but with some significant discrepancies.
Measurements of minimum-bias  events by the ALICE and CMS  collaborations  are presented. \\[6mm]
}

 \AtlasDate{\today}

 \arXivId{2307.03925 }
  
\PreprintIdNumber{arXiv:2307.03925 [hep-ex]}  

\begin{document}

\maketitle    


{
\tableofcontents
}

\newpage
\section{Introduction}
\label{intro}  

The study of soft Quantum Chromodynamics (QCD)  charged-particle distributions in 
proton--proton,  \(pp\), and   proton--antiproton,  \(p\bar{p}\),   collisions    
probes the strong interaction in the low transverse momentum, \(p_{\mathrm{T}}\),
regime or non-perturbative QCD (non-pQCD).
A theoretical description of low-\(p_{\mathrm{T}}\) processes  within pQCD is not possible.
Predictions can be made with  phenomenological models  inspired by QCD 
(see  reviews in  \cite{ParticleDataGroup:2022pth,Grosse-Oetringhaus:2018wkk}). 
In  the low-\(p_{\mathrm{T}}\) region,  charged-particle interactions are typically described by  
quantum QCD-inspired  models implemented in Monte Carlo (MC)  event generators. 
Data are used to constrain such MC models and gain further insight into the particle 
dynamics of the low-\(p_{\mathrm{T}}\) regime. 
Measurements are used to constrain the free parameters of these models. 

Low-\(p_{\mathrm{T}}\)  processes arising from pile-up events\footnote{Pile-up events are 
\(pp\) interactions  in the same bunch crossing at higher instantaneous luminosities in
addition to the triggered collision  between two protons.} 
may also affect the topologies of events involving an interaction with a high-\(p_{\mathrm{T}}\)  scale. 
An understanding of soft-QCD processes is therefore important both   on its own and as a means of 
reducing systematic uncertainties in measurements of  high-\(p_{\mathrm{T}}\)  phenomena.
An accurate description of low-\(p_{\mathrm{T}}\) strong interaction processes  is 
essential for simulating single   \( p p \) and \( p \bar{p} \) interactions    and  the pile-up effects. 

Understanding of soft-QCD interactions has  a direct impact  on precision measurements of  
high-\(p_{\mathrm{T}}\)  phenomena and  searches for new physics;
it provides insight into strong interactions in    the non-pQCD  regime: 
soft-QCD results are used in  MC generator tuning, 
soft-QCD description is essential for simulating   an  underlying event (UE)  with    
multiple parton interactions (MPI),   and initial and final state gluon radiation (ISR, FSR).
An important example of a process  that is entirely governed by soft-QCD physics  is  hadronization. 
Since there is no uniform description of the phenomena that occur at low \(p_{\mathrm{T}}\), 
there  are a variety of models trying to explain them through comparisons with extracted data. 
There is a wealth of  CERN's  Large Hadron Collider (LHC)  \cite{Evans:2008zzb}
measurements that probe the soft-QCD region,  and basically all LHC experiments   measure soft-QCD phenomena. 

Minimum bias  (MB)  events were used  for soft-QCD studies. 
MB are inelastic events selected by  an MB trigger  with as little bias as possible  or   with low-\(p_{\mathrm{T}}\) events. 
MB events include non-diffractive (ND),   single-diffractive  (SD),   double-diffractive  (DD),   and 
central-diffractive (CD) processes.
In order to make a more complete study of particle properties in  MB events, 
results are given for different multiplicity and kinematic selections termed  ``phase spaces'' (PS). 

Measurements of charged-particle distributions by the  ATLAS  \cite{PERF-2007-01}  detector
\cite{STDM-2010-01,STDM-2010-06,STDM-2014-19,STDM-2015-02,STDM-2015-17}
at the centre-of-mass (CM)  energies  \(\sqrt{s} = 0.9\),  \(2.36\), \(7\),  \(8\) and \(13\)~\TeV\  
were performed for  the  pseudorapidity (\(\eta\) ) region   \(\mid\eta\mid <2.5\)  
and  for   the samples of events with  the primary charged-particle multiplicity  (\( n_{\mathrm{ch}} \)) 
more than  or equal to 2 with  the  charged-particle  transverse momentum  \(p_{\mathrm{T}}>100\)~\MeV\ 
and  with  the primary charged-particle multiplicity   \(n_{\mathrm{ch}} \ge 1,\ 6,\ 20,\ 50\) 
with  the  charged-particle  transverse momentum   \(p_{\mathrm{T}} > 500\)~\MeV.
Charged-particle transverse momentum results for  \(pp\) and  \(Pb + Pb\) interactions  at \(2.76\) \TeV\
\cite{ATLAS:2015qmb}, for  \(pp\) and  \(p + Pb\) interactions  at \(5.02\) \TeV\
\cite{ATLAS:2016kvp,ATLAS:2014cza}
in the pseudorapidity range   \(\mid\eta\mid <2\)  of  particles with    \(p_{\mathrm{T}} > 500\)~\MeV\ 
and  \(p_{\mathrm{T}} > 4000\)~\MeV, respectively,   and with     \(p_{\mathrm{T}}  \lessapprox 200\)~\GeV\  
\cite{ATLAS:2015qmb} were studied by ATLAS.

Charged-particle distributions  were  measured  
by the ALICE  \cite{Crochet:2008zz}  collaboration
\cite{ALICE:2009wpl,ALICE:2010cin,ALICE:2010syw,ALICE:2010mty,ALICE:2013rdo,ALICE:2013bva,ALICE:2013txf,ALICE:2015olq,ALICE:2015qqj,ALICE:2017pcy,ALICE:2019dfi,ALICE:2022xip,ALICE:2023csm},
the CMS \cite{CMS:2008xjf}   collaboration 
\cite{CMS:2010wcx,CMS:2010tjh,CMS:2010qvf,CMS:2011dsa,CMS:2011mry,CMS:2011xjg,CMS:2015zrm,CMS:2016gde,CMS:2018nhd}, 
the CMS and TOTEM   \cite{TOTEM:2008lue}   collaborations  \cite{CMS:2014kix}, 
the LHCb  \cite{LHCb:2008vvz}  collaboration \cite{LHCb:2011jir,LHCb:2014wmv}, 
the LHCf  \cite{LHCf:2008lfy}  collaboration,  and
the TOTEM  \cite{TOTEM:2008lue}  collaboration \cite{TOTEM:2014zyx}.

Similar measurements aimed at probing  strong interactions at  low \(p_{\mathrm{T}}\) 
have been made in lower-energy  from \( \sqrt{s} = 0.03\) to  \(0.9\)~\TeV\ 
for \( e^{+} e^{-} \), \( e p \) and  \(  p \bar{p}\) collisions. 
The low \(p_{\mathrm{T}}\)  studies  were carried out in    \(pp\)  collisions  at the ISR  (CERN) by the 
ACHM and ABCDHW    collaborations  at \( \sqrt{s} = 0.0304,\ 0.0445,\ 0.0526\) and \(0.0622\)~\TeV\  
\cite{Aachen:1977izz,Ames:1983cqw}.
Similar studies  were  also carried out in \(p \bar{p} \)  collisions   at the SPS  (CERN) by the  
NA22 \cite{EHSNA22:1987syd}, UA1 \cite{UA1:1982fux,UA1:1989bou},  UA4  \cite{UA4:1985idi}   and  UA5 \cite{UA5:1981ccg,UA5:1982ygd,UA5:1984zxi,UA5:1985hzd,UA5:1985fid,UA5:1986yef,UA5:1986xrd,UA5:1987rzq,UA5:1988osh,UA5:1988gup}
collaborations   at  \( \sqrt{s} = 0.022\), \(0.2\), \(0.54\) and  \(0.9\)~\TeV.

Important results on this subject were   also obtained  in  \(  p \bar{p}\)  collisions  
at Tevatron (Fermilab)   by the CDF  \cite{CDF:1988lbl,CDF:1996yrf} 
collaboration   at  \( \sqrt{s} = 0.63,\ 1.8\) and \(1.96\)~\TeV\  
\cite{CDF:1988evs,CDF:1989nkn,CDF:2001hmt,CDF:2009cxa,Moggi:2009zza} 
and  by the E735  collaboration   at  \( \sqrt{s} = 0.3,\ 0.54,\ 0.9\) and \(1.8\)~\TeV\    \cite{E735:1994joq}.

The hypothesis that at very high  energies  the probability distributions   \( P (n, \sqrt{s}) \)  
of producing  \( n \)   particles in a certain collision process should exhibit  a scaling relation 
was  proposed  in   \cite{Polyakov:1970lyy,Koba:1972ng,Proceedings:1973mza}.
This scaling behaviour is a property of particle multiplicity distributions known  as the KNO scaling hypothesis.
The  main assumption of  the KNO scaling is  the Feynman scaling   \cite{Feynman:1969ej},
where it was  concluded that for asymptotically large energies, 
the mean total number of any kind of particle rises logarithmically  with  the CM energy as 
\(  \langle n \rangle  \propto \ln{\sqrt{s}}\).

The results of  the KNO scaling study using the ATLAS experiment data are presented in 
\cite{Kulchitsky:2022gkm}. 
The KNO scaling was also studied at the LHC energies by   the  CMS   \cite{CMS:2010qvf}  
and  ALICE    \cite{ALICE:2010cin,ALICE:2015olq,ALICE:2022xip,Fan:2022bbp,ALICE:2023csm}.

Charged-particle multiplicity and transverse momentum distributions in  \(pp\) collisions  at   CM energies    
\(\sqrt{s} = 0.2\) -- \(14\)~\TeV\    within the MC Quark-Gluon String Model (QGSM) 
\cite{Kaidalov:1982xg,Kaidalov:1982xe}   based on Gribov’s Reggeon Field Theory  (RFT)
\cite{Gribov:1967vfb,Gribov:1983ivg} were  studied  in   \cite{Bleibel:2010ar,Bravina:2016cme}, 
where  special attention was given to the origin of violation of the KNO scaling.
A  detailed theoretical description of the KNO scaling was  done in 
\cite{Kittel:2005fu,Dremin:2000ep,Grosse-Oetringhaus:2009eis}.
The  novel,  physically well-motivated scaling rules for high-energy data  were  introduced  in 
\cite{Hegyi:2000sp}.

The MB events  were also used by the  LHC experiments  to study UE,  Bose-Einstein correlations (BEC),
an inelastic cross section,  track jets,  particle correlations,  hadronization,  and  colour reconnection.
To perform precise Standard Model measurements or to search for new physics phenomena at  Hadron Colliders,  
it is important to have a good understanding not only of the primary short-distance hard scattering process
but also of the accompanying interactions of the rest of the \( pp \) collision, 
collectively termed the UE.
It is impossible to uniquely  separate  the UE from the hard scattering process on an event-by-event basis,
but observables can be defined  that are particularly sensitive to the properties of the UE. 
Such observables have been studied using the MB events measurements 
performed by the ATLAS detector in \(pp\) collisions  at  \(\sqrt{s} = 0.9\) and \(7\)~\TeV\  
\cite{ATLAS:2010kmf,ATLAS:2011wqb} and  at \(\sqrt{s} = 13 \)~\TeV\   \cite{ATLAS:2017blj}. 
Using the  MB events,  the BEC effect  with one size parameter, the source radius, has been studied
by the ATLAS detector  in \(pp\) collisions  at  \(\sqrt{s} = 0.9\) and \(7\)~\TeV\   \cite{ATLAS:2015dqi}  
and  at \(\sqrt{s} = 13 \)~\TeV\   \cite{ATLAS:2022wvk}. 
Fiducial inelastic cross-sections were measured by  ATLAS  at \(\sqrt{s} = 7\)~\TeV\   
\cite{ATLAS:2011zrx}  and   at \(\sqrt{s} = 13 \)~\TeV\    \cite{ATLAS:2016ygv}. 
The recent soft-QCD measurement results of the  LHC  experiments  are reported, for example,  in 
\cite{Tasevsky:2018efz,Kulchitsky:2020sdv,Sarkisyan-Grinbaum:2013dca}.

This paper is organised as follows:
A short description of the  soft-QCD physics  is presented in  Sec.~\ref{soft_qcd}. 
The  ATLAS detector for  the study of MB events is described in  Sec.~\ref{atlasdetector}.
The  MC model tunes are presented in Sec.~\ref{MC}.
The charged-particle  analysis is performed in Sec.~\ref{CP}.
A study of the KNO scaling is presented in Sec.~\ref{KNO_scaling}.
The summary and conclusions are given in Sec.~\ref{summary}.

\clearpage
\section{Soft QCD}
\label{soft_qcd}

\begin{figure*}[t!]
\centering
\begin{minipage}[h]{0.85\textwidth} 
\center{\includegraphics[width=1.0\linewidth]{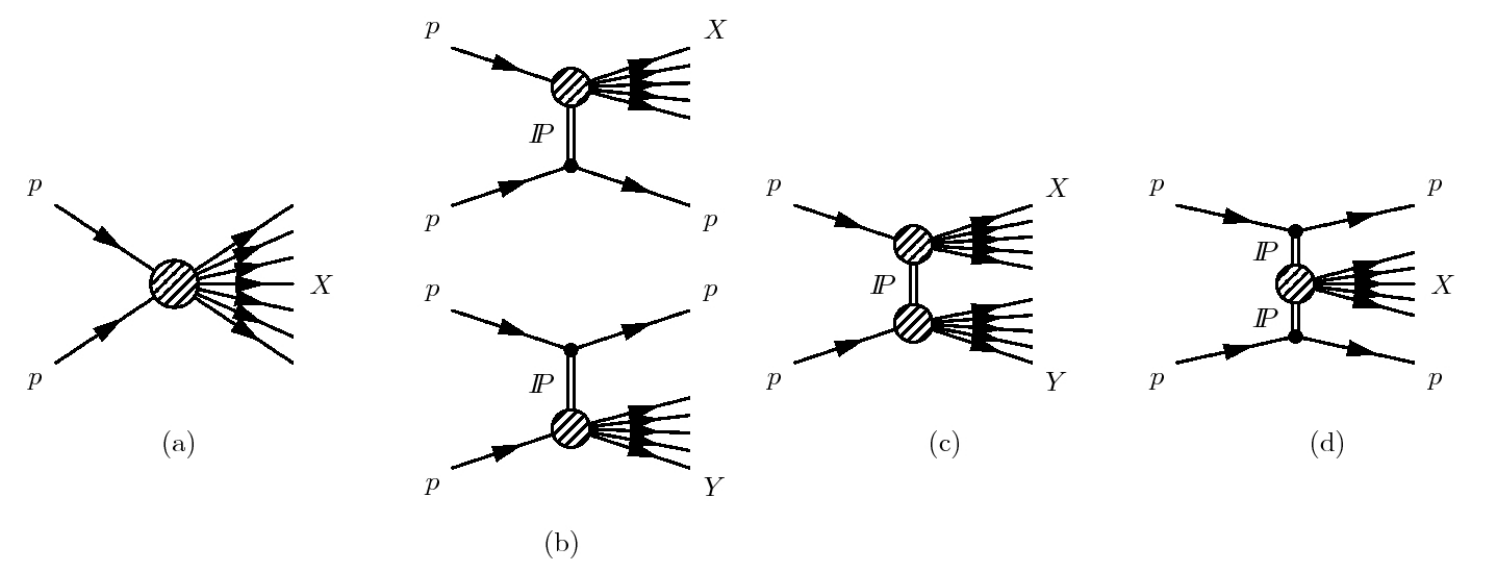}}
\\
\end{minipage}
\caption{
Schematic diagrams of 
(a) non-diffractive (ND),  \(pp \rightarrow X\),  and  diffractive processes  with 
(b) single-diffraction (SD),  \(pp \rightarrow Xp\) or  \(pp \rightarrow pY\),
(c) double-diffraction (DD),  \(pp \rightarrow XY\), and 
(d)  central-diffraction (CD),  \(pp \rightarrow pXp\);  \(X\) (\(Y\)) 
represents a dissociated proton or a centrally produced hadronic system.
The double line \(\mathbb{P}\)  corresponds to the Pomeron exchange and \(p\)  for proton.
Taken from Ref.~\cite{CMS:2015inp}.
}
\label{fig_diagram_01}
\end{figure*}

Understanding of soft-QCD interactions has  a direct impact on   precision measurements in 
high energy physics and   searches for new physics  that
provides insight into strong interactions in   the non-pQCD regime:  the soft-QCD results are used
\begin{itemize}
\item
in MC generator tuning,
\item
for  description of UE simulation,  
\item
for description of multiple parton interactions, MPI,
\item
for description of  initial and  final state gluon radiation, ISR and FSR.
\end{itemize}
Schematic diagrams of non-diffractive 
and  diffractive processes  with  single dissociation, 
double dissociation, 
and central diffraction 
are  shown in Fig.\ \ref{fig_diagram_01}.

As discussed in   Ref.~\cite{ParticleDataGroup:2020ssz}, the Ryskin-Martin-Khoze (RMK) model introduced in 
\cite{Ryskin:2011qe} 
based on a modification of the classic Gribov's Reggeon Field Theory  (RFT)
\cite{Gribov:1967vfb,Gribov:1983ivg}
allows one to trace the smooth transition from the pure perturbative  
region  with large parton transverse momentum  (\(k_{\mathrm{T}}\))  into the soft domain. 
Strong absorption of  low-\(k_{\mathrm{T}}\)  partons 
plays a crucial role here since it produces  an effective infrared cut-off 
and provides a possibility of extending  the parton approach  
used for hard processes to also describe high-energy soft and  semi-hard interactions. 
This approach combines a description of 
soft physics and diffraction with  jet physics in  a coherent self-consistent way.
The soft and hard components,  independently included
\cite{Frankfurt:2013ria,Werner:2005jf,Wang:1996yf,Ostapchenko:2019hxu}
are also possible.
In this approach,  the soft part is described in terms of RFT with the phenomenological 
soft Pomeron pole,  while the hard part is calculated in terms of the 
Parton model for mini-jet production with the energy-dependent  cut-off  
\(k_{\mathrm{T}} > k_{\mathrm{0}} (s)\).
A combined description of soft and hard processes in hadronic collisions is reached within the 
\textsc{QGSJET-II}  MC model  \cite{Ostapchenko:2010vb}  using 
the   semi-hard Pomeron approach  \cite{Liu:2001hz}.
In  Ref.~\cite{Gotsman:2015acv}  a model was constructed that incorporated  attractive features of  
two successful theoretical approaches to  high-energy  QCD: 
Balitsky-Fadin-Kuraev-Lipatov (BFKL) Pomeron calculus 
\cite{Fadin:1975cb,Kuraev:1976ge,Kuraev:1977fs,Balitsky:1978ic,Ioffe:2010zz} and 
the Colour Glass Condensate  approach  (leads to a saturation of parton density with \(s\))
\cite{Kovchegov:2012mbw}.

In  Refs.~\cite{CDF:2001hmt,Rimondi:2000jc,Rimondi:2001by,Rimondi:2004se} 
an analysis was done  for the data set  divided into two classes corresponding  to soft and hard interactions. 
The term hard interactions is typically understood to mean   high-\(p_{\mathrm{T}}\)  parton-parton 
interactions associated with such phenomena as  jets,   while the soft component consists of everything else.
A comparison of  the results shows distinct differences in the behaviour 
of the two samples  as a function of the CM energy. 
Evidence  was found  that the properties of the soft sample are invariant  as a function of  the CM energy.
The separation of  hard   and soft interactions  in the LHC experiments  can be done  using   the event shape observables  \cite{ATLAS:2016hjr},  for example, spherocity or transverse trust.

\section{ATLAS detector}
\label{atlasdetector}

\begin{figure*}[t!]
\centering
\begin{minipage}[h]{0.96\textwidth} 
\center{\includegraphics[width=1.0\linewidth]{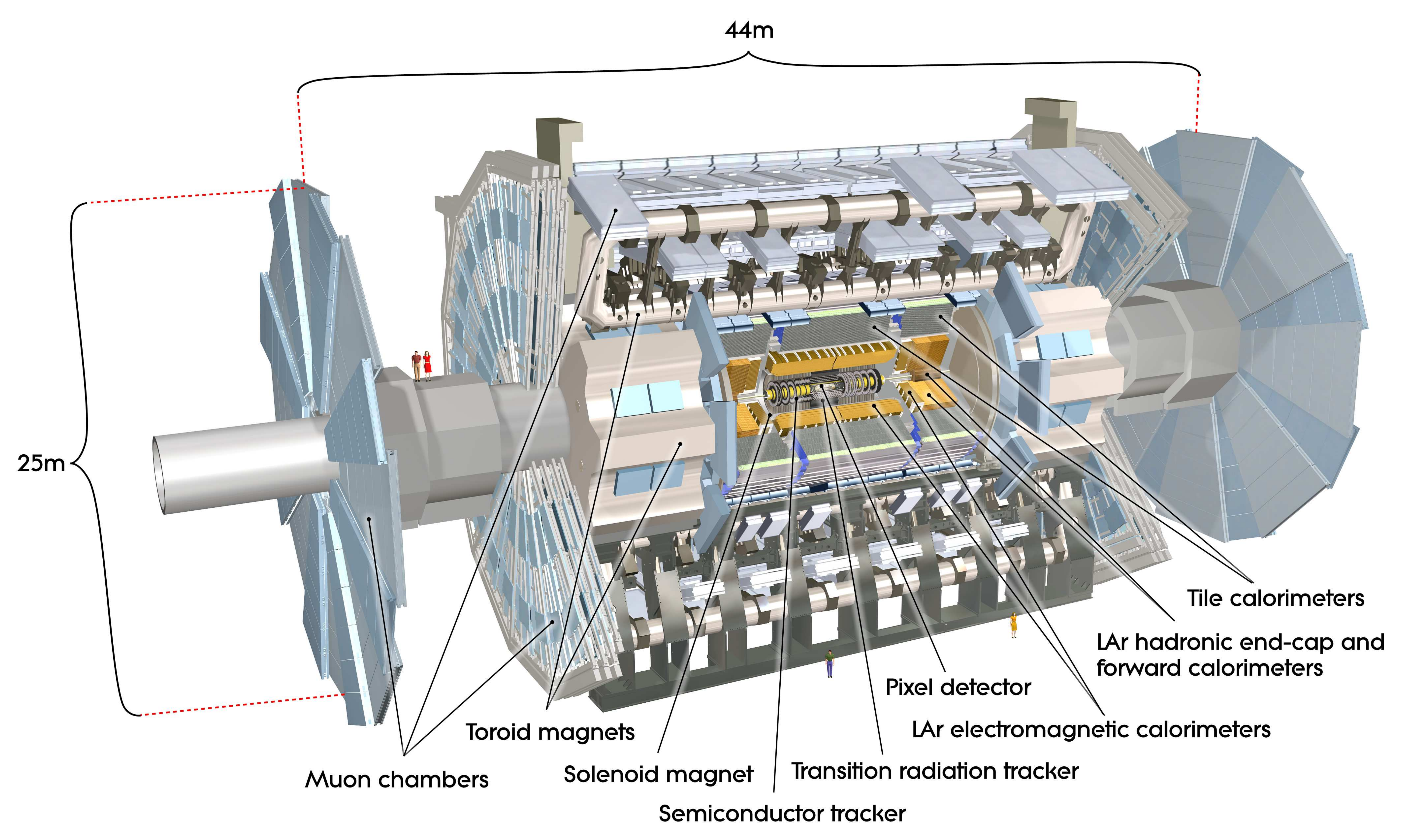}}
\\
\end{minipage}
\caption{
 Cut-away view of 
the 
 ATLAS detector. 
The dimensions of the detector are \(25\)~m in  diameter and \(44\)~m in length. 
%
%
The overall weight of the detector is approximately \(7000\) tonnes.
The number of electronic channels in the detector is about \(100\) million. 
Taken from Ref.~\cite{PERF-2007-01}.
}
\label{fig_ATLAS}
\end{figure*}

ATLAS  is a multipurpose particle physics experiment  
\cite{PERF-2007-01}
operating at  one of the beam  interaction points at the LHC 
\cite{Evans:2008zzb}.
The cut-away view of 
ATLAS detector\footnote{ATLAS uses a right-handed coordinate system 
with its origin at the nominal interaction point (IP) 
in the centre of the detector and the \(z\)-axis along the beam pipe. 
The \(x\)-axis points from the IP to the centre of the LHC ring,  and the \(y\)-axis points upward.
Cylindrical coordinates \((r, \phi)\) are used in the transverse plane, \(\phi\) being 
the azimuthal angle around the beam pipe. 
The pseudorapidity is defined in terms of the polar angle  \(\theta\) as  \(\eta=-\ln\tan(\theta/2)\).
The angular distance is measured in units of  \(\Delta R = \sqrt{ (\Delta \eta)^{2} + (\Delta\phi)^{2} }\).
}   
is shown  in Fig.~\ref{fig_ATLAS}.
The ATLAS detector covers almost the whole solid angle around 
the collision point with layers of  tracking detectors,  calorimeters, 
 and  muon chambers. 
It is designed to study a wide range of physics topics at LHC energies. 
The tracking devices and the  trigger system 
\cite{ATLAS:2012nks,ATLAS:2016wtr} are of particular importance  for the study of MB events.

\begin{figure*}[t!]
\centering
\begin{minipage}[h]{0.64\textwidth} 
\center{\includegraphics[width=1.0\linewidth]{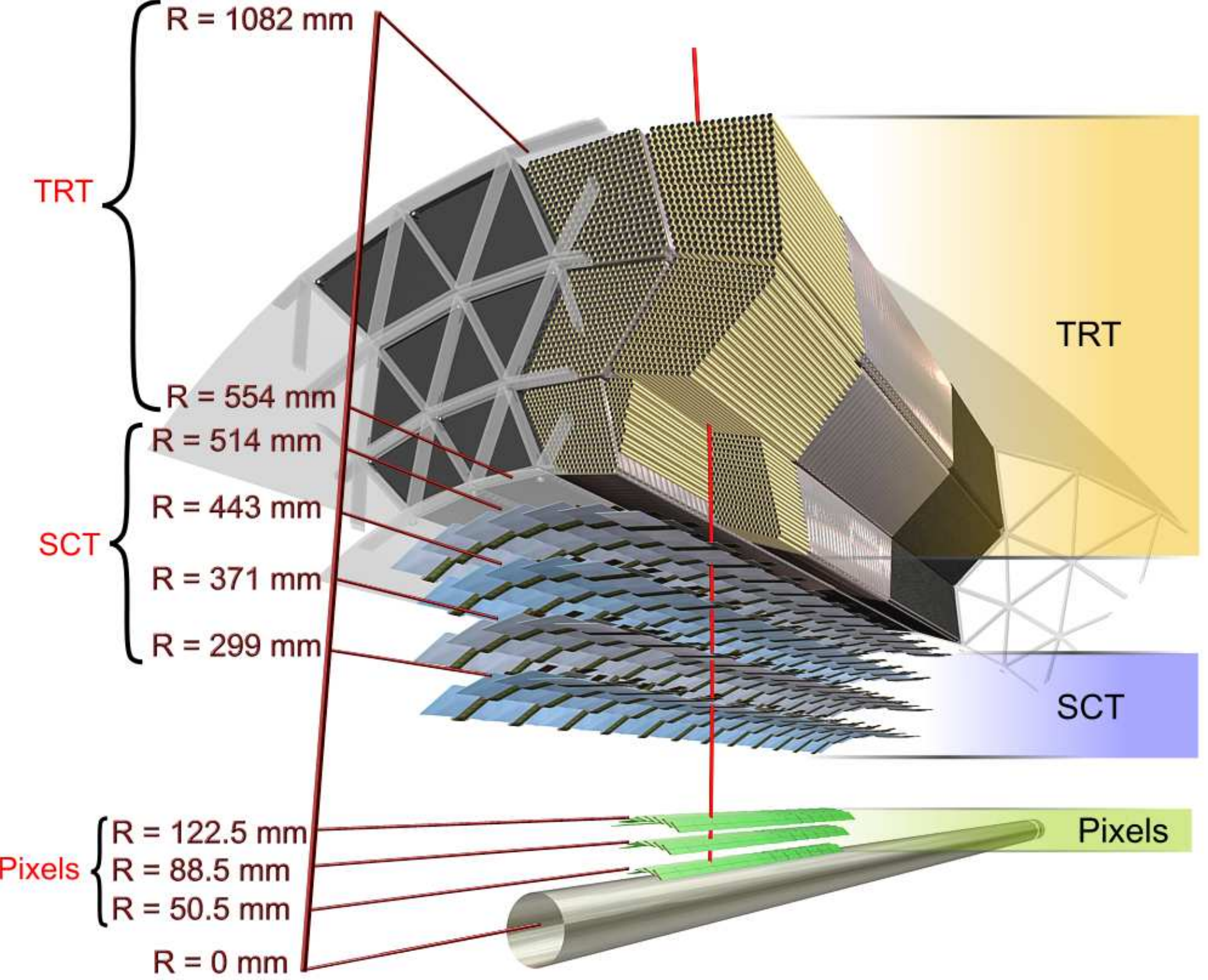}}
\\
\end{minipage}
\caption{
The cross-section of the  ATLAS Inner Detector tracker (ID), showing the
detection layers provided by  three different detector technologies.
The ATLAS ID comprises three detector types dedicated to tracking (from inside out): 
the Silicon Pixel Detector (Pixels),  
the Semi Conductor Tracker (SCT),  
and
the Transition Radiation Tracker (TRT).
During the first long shutdown of the LHC,
the Insertable B-Layer (IBL) was constructed, inserted,  
and commissioned to
become an additional (innermost) layer of the existing Pixel Detector. 
All these detectors allow precision measurement of charged particle
trajectories in the environment of numerous tracks.
The IBL and Pixels detectors mainly contribute to the accurate measurement of 
vertices;  
the SCT is to measure precisely the particle momenta; 
and  the TRT  is to ease 
pattern recognition with its very large number of close hits 
(while also contributing to electron identification).
Taken from Ref.~\cite{ATLAS:2010ojh}.
}
\label{fig_ATLAS_ID}
\end{figure*}

The  innermost part of the   ATLAS  detector is the  
Inner Detector  tracker (ID),
which  has full coverage in \(\phi\) and  covers the pseudorapidity range \(\mid\eta\mid<2.5\).
The cut-away view of the ATLAS ID  is shown  in Fig.~\ref{fig_ATLAS_ID}.
The ID is immersed in the 
\(\mathrm{2\, T}\)  
axial magnetic field of a superconducting solenoid 
and measures  
the
trajectories of charged particles.
It consists of a silicon pixel detector (Pixels), 
a silicon microstrip detector (SCT), 
and
a straw-tube transition radiation tracker (TRT), 
each of which is split into a barrel and two endcap components. 
The Pixels, 
SCT, 
and TRT  are located around the interaction point, 
spanning radial distances of 
\(33\)--\(150\)~mm, \(299\)--\(560\)~mm and \(563\)--\(1066\)~mm, respectively. 
The barrel (each endcap)  consists of four (three) pixel layers, four (nine) double layers 
of silicon microstrips,  
and 73 (160) layers of TRT straws. 
The Pixels, 
SCT, 
and TRT have \((r, \phi)\)-position resolutions of  
\(10\)~\(\mu\)m,  \(17\)~\(\mu\)m,  and  \(130\)~\(\mu\)m,   respectively.
 
During the first long shutdown of the LHC,
the Insertable B-Layer (IBL) \cite{ATLAS:2010ojh}
was constructed, inserted, 
and commissioned to
become an additional (innermost) layer of the existing Pixel Detector. 
The IBL is composed of \(14\) lightweight staves arranged in a cylindrical geometry, 
each made of \(12\) silicon planar sensors in its central region and \(2\times 4\) 
three-dimensional sensors at the ends. 
The IBL pixel dimensions are  \(50\)~\(\mu\)m in the \(\phi\)-direction and  \(250\)~\(\mu\)m 
in the \(z\)-direction (compared with \(50\)~\(\mu\)m by \(400\)~\(\mu\)m for  the other pixel layers).
The intrinsic spatial resolution of the IBL readout is \(10\)~\(\mu\)m in the \((r, \phi)\)-position 
and \(75\)~\(\mu\)m in  the \(z\)-position  \cite{ATLAS:2012rar}.
The smaller radius and the reduced pixel size result in improvements
in both the transverse and longitudinal impact parameter resolutions
\cite{STDM-2015-02,STDM-2015-17}.
The services for the existing pixel detector were upgraded, 
significantly reducing the amount of material in  
the region \(\mid\eta\mid > 1.5\), in particular at the boundaries of the active tracking volume. 

A track from a charged particle traversing the barrel detector typically has \(12\) silicon 
measurement points (hits), of which  
\(4\) at the Pixels 
and  \(8\) at the  SCT, 
and 
more than \(30\) TRT  straw hits.
Requirements on an IBL hit and on impact parameters 
strongly suppress the number of tracks from secondary particles. 

The ATLAS detector has a two-level trigger system:   the first-level (L1) trigger and the 
high-level trigger (HLT)   \cite{ATLAS:2012nks,ATLAS:2016wtr}.
MB events  were required to satisfy L1 triggers using the  MB trigger scintillators (MBTS). 
These are mounted at each end of the detector in front of the liquid-argon 
endcap-calorimeter cryostats  at \(z = \pm3.56\)~m,
and are segmented into  two rings in pseudorapidity 
(\(2.07 < \mid\eta\mid < 2.76\) and \(2.76 < \mid\eta\mid < 3.86\)).
The inner (outer) ring consists of eight (four) azimuthal sectors,  giving a total of \(12\) sectors on each side. 
The MB events were selected on the basis of the MBTS alone.
The trigger used in this measurement requires at least one signal in a scintillator on one side to 
be above threshold.
The MB ATLAS trigger collects  
inelastic  events (INEL) in the definition of 
ALICE or the CMS.

The methods developed for the measurement of the properties of MB events during 
low luminosity  runs using the ATLAS detector 
are
described in 
Ref.~\cite{ATLAS:2009cch}.
An extensive software suite \cite{ATL-SOFT-PUB-2021-001}  is used in the reconstruction and analysis of 
real and simulated data, in detector operations, 
and in the  trigger and data acquisition  systems of the experiment.

\section{Monte Carlo models}
\label{MC}

Inclusive  MB data are modelled in  MC event generators, 
assuming three different diffractive processes: 
non-diffractive, single-diffractive,  and  double-diffractive. 
Low-\(p_{\mathrm{T}}\)  scattering processes may be described by the  lowest-order (LO) pQCD
two-to-two parton scatters, where the divergence of the cross section at 
\( p_{\mathrm{T}} = 0 \)~\MeV\ is regulated by phenomenological models. 
A summary of MC generator tunes   used for comparison with  the  MB results based on the  ATLAS measurements
\cite{STDM-2010-01,STDM-2010-06,STDM-2014-19,STDM-2015-02,STDM-2015-17} 
is presented in  Table~\ref{tab:MC_enegy}. 

\begin{table*}[t!]
\centering 
\caption{
Summary of Monte Carlo 
generators used for comparison with 
the minimum-bias   results of 
ATLAS 
\cite{STDM-2010-01,STDM-2010-06,STDM-2014-19,STDM-2015-02,STDM-2015-17}. 
The version number, the corresponding tune name, 
and 
the parton distribution function (PDF) are presented for each MC generator.
}
\label{tab:MC_enegy}
\medskip
 \begin{tabular}{lllcllclc}
\hline
\hline

\(\sqrt{s}\)	
&
&Monte Carlo 
& 
&Version 			
&Tune 	
& 
&PDF 					
& 
\\

\( \mathrm{[TeV]}\)
&		
&Generator		
& Ref.\
& 					
& 					
& Ref.\						
& 
& Ref.\
\\

\hline
13
	&\cite{ATLAS:2016puo}
  	&\textsc{Pythia\,8}	
     &\cite{Sjostrand:2007gs}						
     &\textsc{8.186} 	
     &\textsc{A3} 	
     &\cite{ATLAS:2016puo}	
     &\textsc{NNPDF23LO}
     &\cite{Ball:2012cx}
\\[2mm]
     
\hline
13
	&\cite{STDM-2015-02,STDM-2015-17} 
	&\textsc{Pythia\,8}	
	&\cite{Sjostrand:2007gs} 						
	&\textsc{8.186} 	
	&\textsc{MONASH} 
	&\cite{Skands:2014pea} 
	&\textsc{NNPDF23LO}   
	&\cite{Ball:2012cx}  
	\\

8  	
	&\cite{STDM-2014-19}	
	&\textsc{Pythia\,8}	
     & \cite{Sjostrand:2007gs}						
     &\textsc{8.185}
     &\textsc{A2} 	
     &\cite{ATLAS:2011krm}	
     &\textsc{MSTW2008LO}
     &\cite{Martin:2009iq}
     \\
	
	&	
	&\textsc{EPOS} 	
	&\cite{Werner:2010ny}						
	&\textsc{LHCv3400}
	&\textsc{LHC} 	
	&\cite{Pierog:2013ria}	
	&\textsc{--} 			 
	&					   
	\\
	
	&	
	&\textsc{QGSJET-II} 
	&\cite{Ostapchenko:2010vb}					
	&\textsc{04} 
	&\textsc{Default} 
	&	
	&\textsc{--}  			 
	&					   
\\[2mm]
	
\hline
8	
	&\cite{STDM-2014-19}	
	&\textsc{Pythia\,6}
	&\cite{Sjostrand:2006za}
     &
     &\textsc{AMBT2B} 
     &\cite{ATLAS:2011zja}			
     &\textsc{CTEQ6L1}		
     &\cite{Pumplin:2002vw}
\\[2mm]

\hline
7 	
	&\cite{STDM-2010-06}	
  	&\textsc{Pythia\,8}	
     &\cite{Sjostrand:2007gs}						
     &\textsc{8.130}
     &\textsc{Default} 
     &	
     &\textsc{MRST LO*}
     &\cite{Sherstnev:2007nd}
     \\
2.36     
	&\cite{STDM-2010-06}		
	&\textsc{Pythia\,6}
	&\cite{Sjostrand:2006za}
     &\textsc{6.4.21} 
     &\textsc{AMBT1} 
     &\cite{STDM-2010-06}	
     &\textsc{MRST LO*}		
     &\cite{Sherstnev:2007nd}
     \\
0.9
	&\cite{STDM-2010-06,STDM-2010-01}		
	&\textsc{Pythia\,6}
	&\cite{Sjostrand:2006za}
     &\textsc{6.4.21} 
     &\textsc{MC09} 
     &\cite{ATLAS:2010zyu}		
     &\textsc{MRST LO*}		
     &\cite{Sherstnev:2007nd}
     \\

	&	
	&\textsc{Pythia\,6}
	&\cite{Sjostrand:2006za}
     &\textsc{6.4.21} 
     &\textsc{DW} 
     &\cite{TeV4LHCQCDWorkingGroup:2006fht}
     &\textsc{MRST LO*}		
     &\cite{Sherstnev:2007nd}
     \\

	&	
  	&\textsc{PHOJET}	
     &\cite{Engel:1994vs}						
     &\textsc{1.12} 
     &\textsc{Default} 	
     &
     &\textsc{MRST LO*}
     &\cite{Sherstnev:2007nd}
\\[2mm]

\hline
 0.9
	&\cite{STDM-2010-01}		
	&\textsc{Pythia\,6}
	&\cite{Sjostrand:2006za}
     &\textsc{6.4.21} 
     &\textsc{MC09c} 
     &\cite{ATLAS:2010zyu}		
     &\textsc{MRST LO*}		
     &\cite{Sherstnev:2007nd}
     \\
    
	&
	&\textsc{Pythia\,6}
	&\cite{Sjostrand:2006za}
     &\textsc{6.4} 
     &\textsc{Perugia\,0} 
     &\cite{Skands:2009zm}
     &
     &
     \\
    
\hline
\hline
\end{tabular}
\end{table*}


The 
\textsc{Pythia\,6} \cite{Sjostrand:2000wi,Sjostrand:2006za}, 
\textsc{Pythia\,8} \cite{Sjostrand:2007gs,Sjostrand:2014zea,Bierlich:2022pfr}, 
\textsc{PHOJET} \cite{Engel:1994vs}, 	
\textsc{EPOS} 	\cite{Werner:2010ny},	 
and 
\textsc{QGSJET-II} 
\cite{Ostapchenko:2010vb}		 
MC generators are used to correct the data for detector effects and to compare 
with particle-level corrected data. 
For the purpose of comparing the present measurements to different phenomenological
models describing  MB events, the following particle-level MC samples were generated.

\textsc{Pythia\,8}  \cite{Sjostrand:2007gs}   and  \textsc{EPOS}  \cite{Werner:2010ny}  models use
the effects of colour coherence, which is important in dense parton environments and effectively
reduces the number of particles produced in multiple parton--parton interactions. 
In \textsc{Pythia\,8} 
the simulation is split into non-diffractive and diffractive processes, the former dominated by 
\(t\)-channel gluon exchange and amounting  to approximately 80\% of the selected events, 
and the latter described by a Pomeron-based approach 
\cite{Corke:2010yf}. 

Different parameter settings in the models are used in 
simulation to reproduce  the existing experimental data and are referred to as tunes. 
A tune is a particular configuration or set of values 
for
the parameters of a particular MC model. 

The  \textsc{Pythia\,8}  MC generator \cite{Sjostrand:2007gs}   was used with  the 
parameter values set to the  \textsc{A2} tune \cite{ATLAS:2011krm} 
and with the \textsc{MSTW2008LO}  PDF set  \cite{Martin:2009iq}.   
The contributions from ND, SD, 
and DD processes were included in proportion to the 
cross sections predicted by  \textsc{Pythia\,8} with the  \textsc{A2} tune. 
The ATLAS MB tune  \textsc{Pythia\,8} \textsc{A2}  was used for  
the
determination of detector corrections. 
This was tuned using ATLAS MB  data at 7~\TeV\ for the MPI parameters.

The \textsc{Pythia\,8} \textsc{Monash} \cite{Skands:2014pea} 
is used the tune 
using 
MB and UE results. 
It was constructed using Drell--Yan 
and UE data from ATLAS  
and also data from the CMS,  SPS, 
and Tevatron in order to constrain energy scaling. 
The  Monash  UE tune  is based on  the   \textsc{NNPDF2.3LO} PDF    \cite{Ball:2012cx}  
and incorporates updated fragmentation parameters
as well as SPS and 
Tevatron data to constrain the energy scaling.  

The \textsc{Pythia\,8} version \textsc{8.130} MC generator \cite{Sjostrand:2007gs}  uses 
a
diffraction model that  produces much harder \(p_{\mathrm{T}}\)   and \(n_{\mathrm{cn}}\) 
spectra  for the SD and DD contributions  than \textsc{Pythia\,6}. 
The default parton shower model is similar to that used in  \textsc{Pythia\,6} \textsc{MC09}.

The new \textsc{Pythia\,8} \textsc{A3} tune  \cite{ATLAS:2016puo} is suitable for inclusive QCD 
modelling for LHC Run 3. 
The \textsc{Pythia\,8} \textsc{A3}  uses the ATLAS Run 2 charged particle 
distribution and inelastic cross section results  in addition to the Run 1  results
used previously to construct MB tunes. 
The \textsc{A3} uses the same \textsc{NNPDF 2.3LO} PDF and  demonstrates that an acceptable description 
of data can be achieved by using the  Donnachie--Landshoff  (DL) model for diffraction.  

%

The ATLAS \textsc{Pythia\,6} \cite{Sjostrand:2006za} \textsc{MC09} tune \cite{ATLAS:2010zyu} 
uses a specific set of 
optimised 
parameters; it employs the \textsc{MRST LO*} PDF \cite{Sherstnev:2007nd} 
and the \(p_{\mathrm{T}}\)-ordered parton shower \cite{Sjostrand:2004qj}. 
These parameters were derived by tuning to the UE and MB  Tevatron results  from
the
energy region \(\sqrt{s} = 0.63\) -- \(1.96\)~\TeV.

The ATLAS \textsc{Pythia\,6} \textsc{MC09c} tune \cite{ATLAS:2010zyu}
is an extension of the ATLAS \textsc{MC09} tune  where the  strength of the colour reconnection (CR) 
was tuned to describe the 
\(\langle p_{\mathrm{T}}\rangle\) distributions as a function of \(n_{\mathrm{ch}}\) 
measured by CDF in \(p \bar{p}\) collisions at the Tevatron \cite{CDF:2009cxa}.

The CR phenomenon is a pure soft-QCD effect. 
The point is that after a number of coloured secondary partons are produced,  there are different possibilities 
of forming the colour flow between these partons and  grouping
the partons into colourless clusters. 
In the process of reconnection,  one rearranges the colour flow in such a way as to 
minimise 
the size of the clusters. 
This is especially important when dealing with 
the
contributions of  MPI.  
The reconnection between the different 
cuts 
of Pomeron diagrams diminishes
the final multiplicity and can change the form of the  \(n_{\mathrm{ch}}\) distributions 
\cite{Campbell:2017hsr,Buckley:2011ms,Kundu:2019ajc}.

The \textsc{Pythia\,6} \textsc{AMBT1} tune   (ATLAS Minimum Bias Tune 1)  
\cite{STDM-2010-06}
was developed in order to adapt the free parameters of the  ND models  to the  experimental data
at \(\sqrt{s} = 0.9\) and \(7\)~\TeV\ in a diffraction-reduced  PS with 
\(n_{\mathrm{cn}} \ge 6\), \(p_{\mathrm{T}} > 500\)~\MeV, \(\mid\eta\mid <2.5\). 
The starting point for this tune is the ATLAS \textsc{Pythia\,6} \textsc{MC09c}  
\cite{ATLAS:2010zyu}.

The \textsc{Pythia\,6} \textsc{DW} tune \cite{TeV4LHCQCDWorkingGroup:2006fht} 
uses virtuality-ordered showers and was derived to describe the CDF Run II  UE and Drell--Yan data.

The \textsc{Pythia\,6} \textsc{AMBT2B} tune \cite{ATLAS:2011zja}	
with the \textsc{CTEQ6L1}	PDF  \cite{Pumplin:2002vw}  was 
evaluated  using  jet and  MB data.

\textsc{EPOS} \cite{Werner:2010ny}  provides  implementation of a parton-based  Gribov's Reggeon theory 
\cite{Gribov:1967vfb}
which  is an effective QCD-inspired field theory describing hard and soft scattering simultaneously.  
The \textsc{EPOS} generator, version \textsc{LHCv3400}, was used with   the \textsc{LHC}  tune 
\cite{Pierog:2013ria}.
The \textsc{EPOS} generator does not rely on PDF. 

The  \textsc{QGSJET-II} model  version \textsc{04}
\cite{Ostapchenko:2010vb} 
provides  a 
phenomenological  
treatment of hadronic and nuclear interactions  in 
the framework of  the Reggeon field theory.
The soft and semihard parton processes are included within the “semihard Pomeron” approach. 
For  \textsc{QGSJET-II} the default settings of the generator are applied. 
The   \textsc{QGSJET-II} generator does not rely on PDF. 

The \textsc{PHOJET} MC generator \cite{Engel:1994vs} version \textsc{1.12.1.35} 
is used as an alternative model to \textsc{Pythia}-based generators. 
It describes low-\(p_{\mathrm{T}}\)  physics using the two-component  Dual Parton Model (DPM)
\cite{Capella:1978ig,Capella:1992yb}  which includes  soft hadronic processes described by Pomeron 
exchange  and  semi-hard processes described by perturbative parton scattering. 
The \textsc{PHOJET} relies on \textsc{Pythia\,6} version \textsc{6.1.15}  for the fragmentation of partons.

The \textsc{Pythia\,6}  MC generator \textsc{Perugia\,0} tune  \cite{Skands:2009zm} 
with the soft-QCD part is tuned using only MB data from the  \(p \bar{p}\) Tevatron and CERN colliders. 

All large MC samples of MB  events were generated  and passed through the   ATLAS simulation  
programme
\cite{Aad:2010ah},   which is based on  \textsc{Geant4} \cite{GEANT4:2002zbu},  
and the reconstruction chain, which is exactly  the same as used for 
the 
collision dataset.


%
ATLAS used  \(13\) MC generators and 
their 
tunes
to correct the data for detector effects and to compare  with particle-level corrected MB results, 
which are presented in Table~\ref{tab:MC_enegy}. 
The comparisons of  the MC predictions with  the ATLAS MB results are presented in Sec.~\ref{CP}.

\section{Analysis of minimum-bias events }
\label{CP}

Measurements of inclusive particle spectra belong to 
the
basic items in the physics 
programmes 
of LHC experiments, 
and they are usually measured regularly at each collision energy. 
The charged-particle  multiplicity  is one of the key characteristics of high-energy hadron collisions 
and has been the subject of many experimental and  theoretical studies because, 
although quite simple to measure,  it is quite difficult to describe it in the full measured range. 
Measurements of charged-particle distributions  probe the non-pQCD regime  where
QCD-inspired models implemented in MC event generators  are used to 
describe  the data  and to constrain 
the
free parameters of MC models.
Accurate description of low-\(p_{\mathrm{T}}\) strong interaction processes is essential for 
simulating single \(pp\) and pile-up multiple \(pp\) interactions.
Such \(pp\) measurements are also used as input in many models  trying to describe heavy-ion results.

The results used in this review are based on the \(pp\)  data collected at
\( \sqrt{s} = 0.9\) -- \(13\)~\TeV\  recorded by the  ATLAS experiment 
\cite{PERF-2007-01}    at the LHC  \cite{Evans:2008zzb}  in 2010 -- 2015
\cite{STDM-2010-01,STDM-2010-06,STDM-2014-19,STDM-2015-02,STDM-2015-17}. 
The data were taken in  a special  configuration of the LHC  with low beam currents and  
reduced beam focusing,  producing 
a
low mean number of interactions per bunch-crossing  
in the range of
\(0.003\) -- \(0.007\).

The corrected distributions for primary charged particles  in five separate PS regions  for events with  
\(n_{\mathrm{ch}} \ge 2\), \(p_{\mathrm{T}} >100\)~\MeV, 
\(n_{\mathrm{ch}} \ge 1\),  \(p_{\mathrm{T}} >500\)~\MeV\  and
\(n_{\mathrm{ch}} \ge 6,\ 20,\ 50\),  \(p_{\mathrm{T}} >500\)~\MeV\  are used.
The results are compared to predictions of models tuned to a wide range of measurements. 
The measured distributions are presented as  inclusive-inelastic distributions  within a given 
PS region  with  minimal model-dependent corrections  to facilitate comparisons with models.

\subsection{Observables}
\label{obsrvables}
The following observables  were studied by ATLAS: 
\begin{equation}
\label{eq_eta}
\frac{1}{N_{\mathrm{ev}}}
\cdot
\frac{ \mathrm{d} N_{\mathrm{ch}}}{ \mathrm{d} \eta} ,
\end{equation}
%
\begin{equation}
\label{eq_eta_pT}
\frac{1}{N_{\mathrm{ev}}}
\cdot
\frac{1}{2 \pi p_{\mathrm{T}} }
\cdot
\frac{\mathrm{d}^2 N_{\mathrm{ch}}}{\mathrm{d} \eta \mathrm{d} p_{\mathrm{T}}} ,
\end{equation}
%
\begin{equation}
\label{eq_nch}
\frac{1}{N_{\mathrm{ev}}}
\cdot
\frac{\mathrm{d} N_{\mathrm{ev}}}{\mathrm{d} n_{\mathrm{ch}}} ,
\end{equation}
%
%
\begin{equation} 
\label{eq_pT}
\frac{\mathrm{d}\langle p_{\mathrm{T}}\rangle}{\mathrm{d} n_{\mathrm{ch}}} ,
\end{equation}
where, \( \eta\) is the particle pseudorapidity, \( p_{\mathrm{T}} \) is the 
charged-particle transverse  momentum,\footnote{The factor \( 2\pi p_{\mathrm{T}} \)  
in the \( p_{\mathrm{T}} \) spectrum comes from the Lorentz-invariant definition 
of the cross-section in terms of \( \mathrm{d}^{3} p\).  
The results could thus be interpreted as the massless approximation to \( \mathrm{d}^{3} p\).}
\(n_{\mathrm{ch}}\) is the number of primary charged particles in an event within the kinematic acceptance.
\(N_{\mathrm{ev}}\) is the event number yield for a given event selection,
\(N_{\mathrm{ch}}\) is the total number of primary charged particles in all selected events in the data sample,
\( \langle p_{\mathrm{T}}\rangle \) 
is  the average transverse momentum of primary charged particles within the kinematic acceptance. 

A primary charged particle is defined as a charged particle with a mean lifetime  \( \tau > 300\)~ps, 
which is either directly produced in \( p p \) interactions or from decays of directly produced particles with 
\( \tau  < 30\)~ps.
Charged particles produced from decays of particles with  \( \tau > 30\)~ps
are considered as secondary particles and are thus excluded. 

The usually used inclusive charged-particle spectra 
correspond to events with a minimum multiplicity 
\( n_{\mathrm{ch}} \ge 2 \) or \( n_{\mathrm{ch}} \ge 1 \) and contain 
primary charged particles possessing a minimum transverse momentum 
\( p_{\mathrm{T}} > 100 \)~\MeV\ or  \( p_{\mathrm{T}} > 500 \)~\MeV, respectively, 
for  the pseudorapidity region \( \mid\eta\mid < 2.5\). 
Primary charged-particle spectra are also shown for  higher-multiplicity events 
(\( n_{\mathrm{ch}} \ge 6,\ 20\ \mathrm{and}\ 50\),  \( p_{\mathrm{T}} > 500 \)~\MeV).

\clearpage
\subsection{Pseudorapidity dependence of charged-particle multiplicity }
\label{Nch_eta}
\subsubsection{ATLAS distributions of charged-particle multiplicity over \( \eta \)}

\begin{figure*}[t!]
\centering
\begin{minipage}[h]{0.45\textwidth} 
\center{\includegraphics[width=1.0\linewidth]{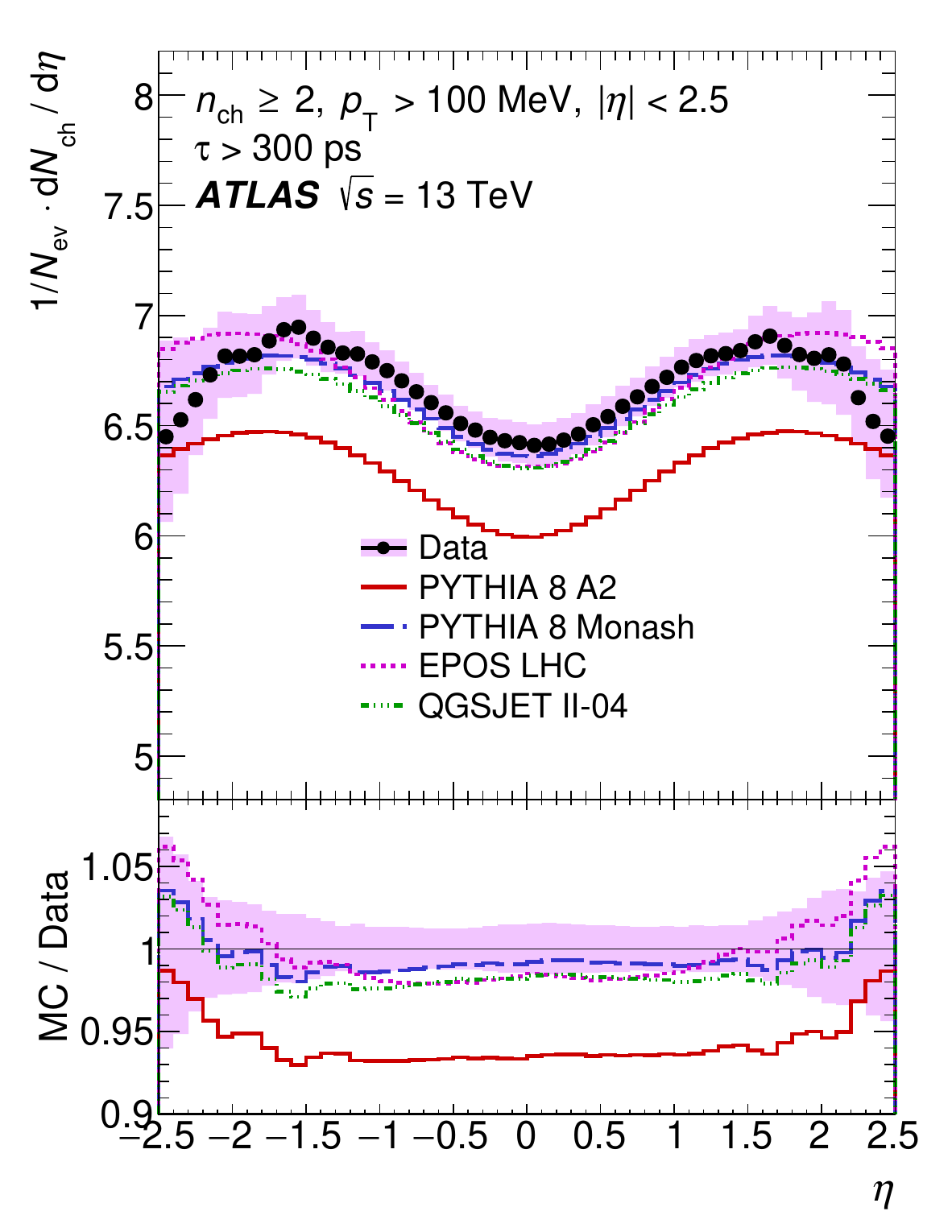}}
(a) 
\\
\end{minipage}
\hspace{2mm}
\begin{minipage}[h]{0.45\textwidth}
\center{\includegraphics[width=1.0\linewidth]{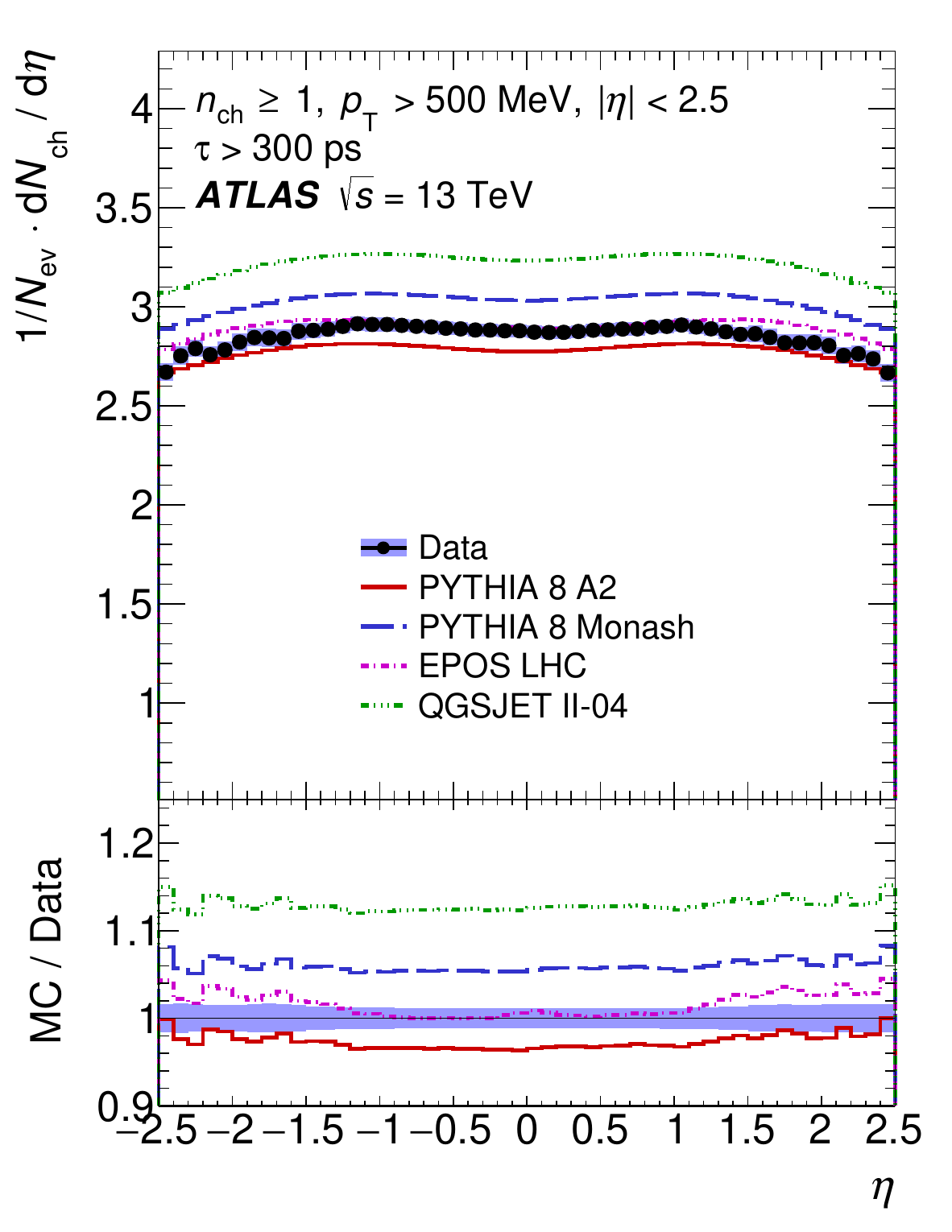}}
(b) 
\\
\end{minipage}
\caption{
Top panel: 
Primary charged-particle multiplicity  density  pseudorapidity distributions
for events for  \(\mid\eta\mid < 2.5\), each with a lifetime \(\tau >300\)~ps,
at  the centre-of-mass energy  \(\sqrt{s}= 13\)~\TeV\ 
with 
(a) \(n_{\mathrm{ch}} \ge 2\), \(p_{\mathrm{T}} >100\)~\MeV\  \cite{STDM-2015-17} 
and
(b) \(n_{\mathrm{ch}} \ge 1\), \(p_{\mathrm{T}} >500\)~\MeV\ \cite{STDM-2015-02}. 
The data represented by dots 
is
compared  to various particle-level MC predictions,  
which are shown by curves. 
The shaded areas around the data points represent the total statistical and systematic uncertainties added in quadrature.
Bottom panel: 
The ratios of the MC predictions to   the experimental results  are shown. 
Bands represent the uncertainties of the experimental results.
Taken from Ref.~\cite{STDM-2015-02,STDM-2015-17}.
}
\label{fig_13_eta}
\end{figure*}

\begin{figure*}[t!]
\centering
\begin{minipage}[h]{0.32\textwidth} 
\center{\includegraphics[width=1.0\linewidth]{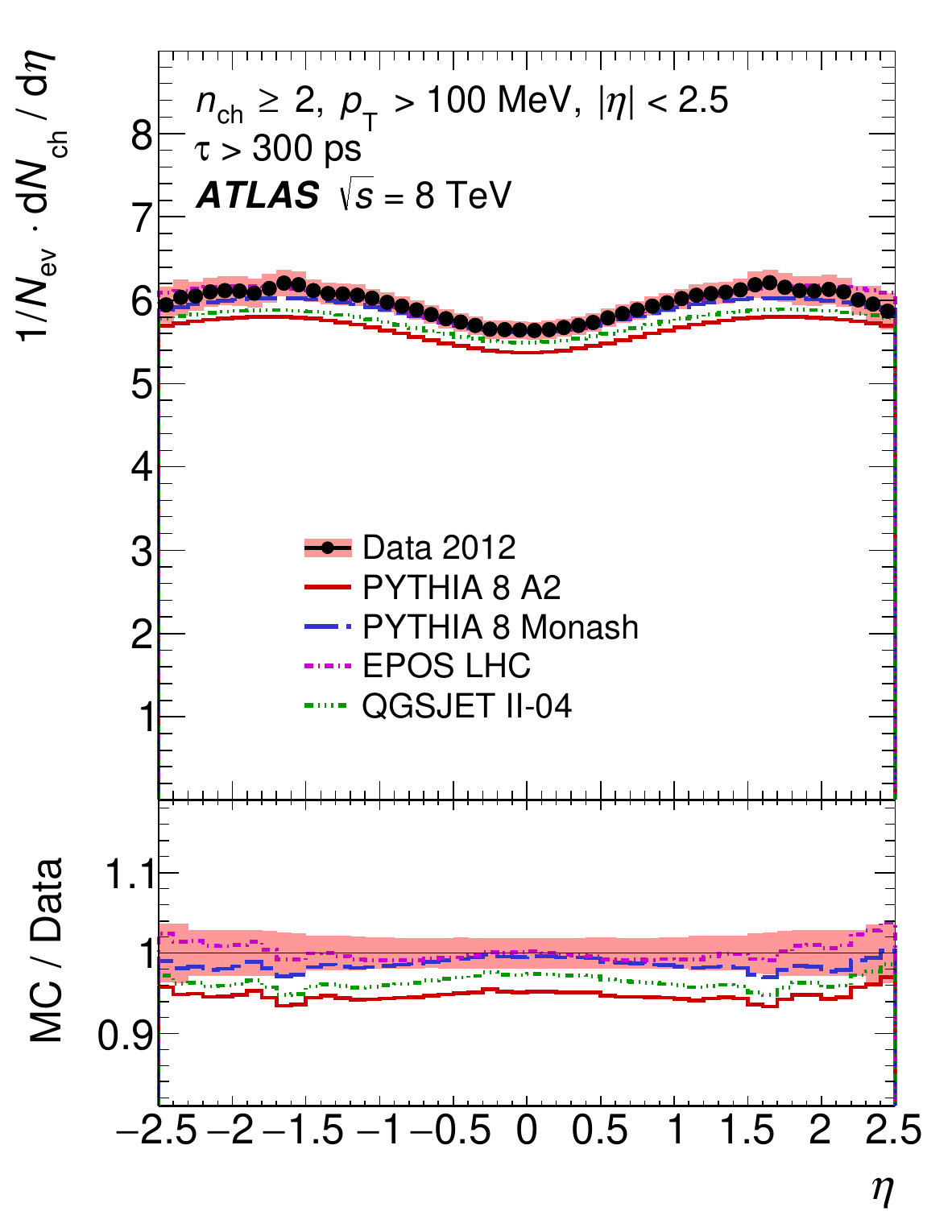}}
(a) 
\\
\end{minipage}
\hfill
\begin{minipage}[h]{0.32\textwidth}
\center{\includegraphics[width=1.0\linewidth]{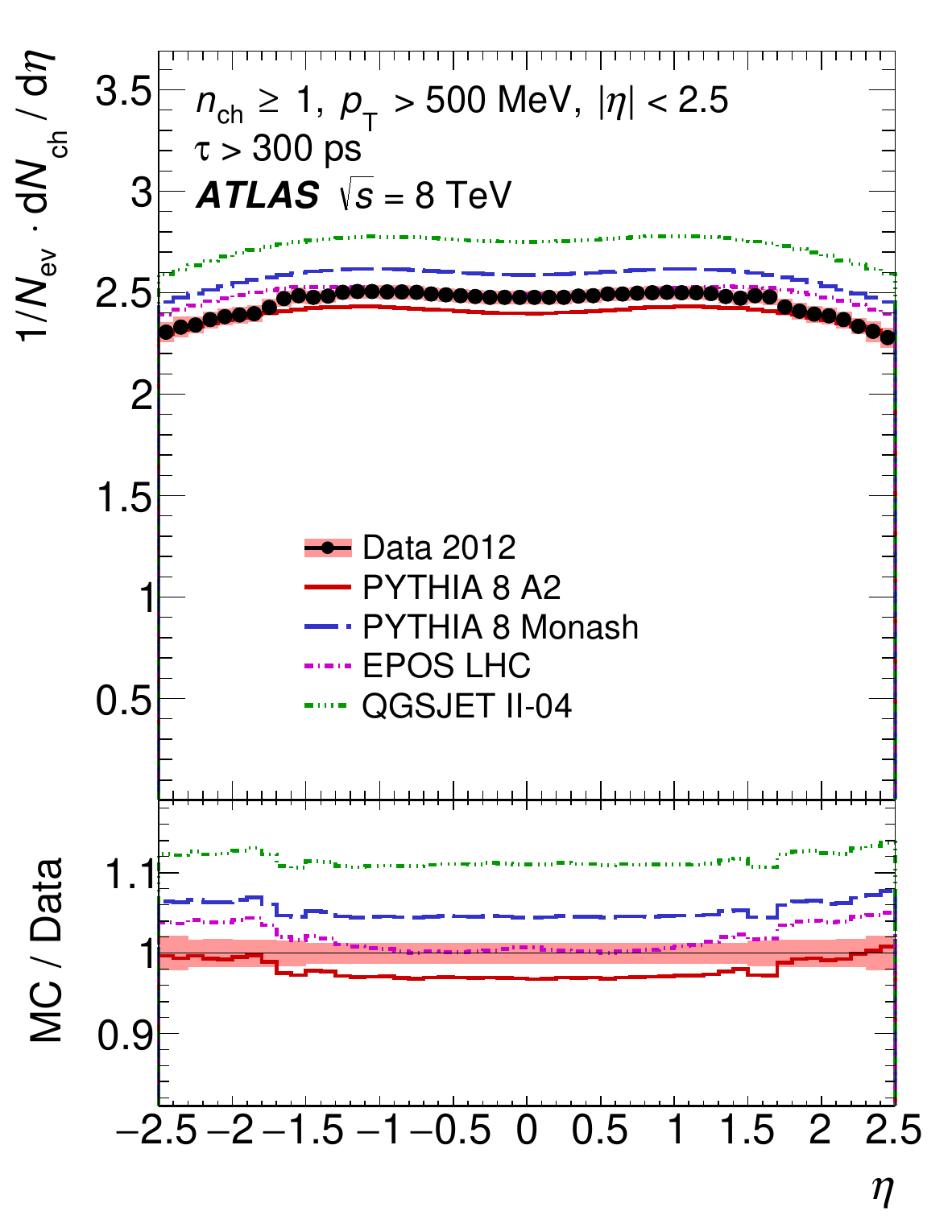}}
(b) 
\\
\end{minipage}
\hfill
\begin{minipage}[h]{0.32\textwidth} 
\center{\includegraphics[width=1.0\linewidth]{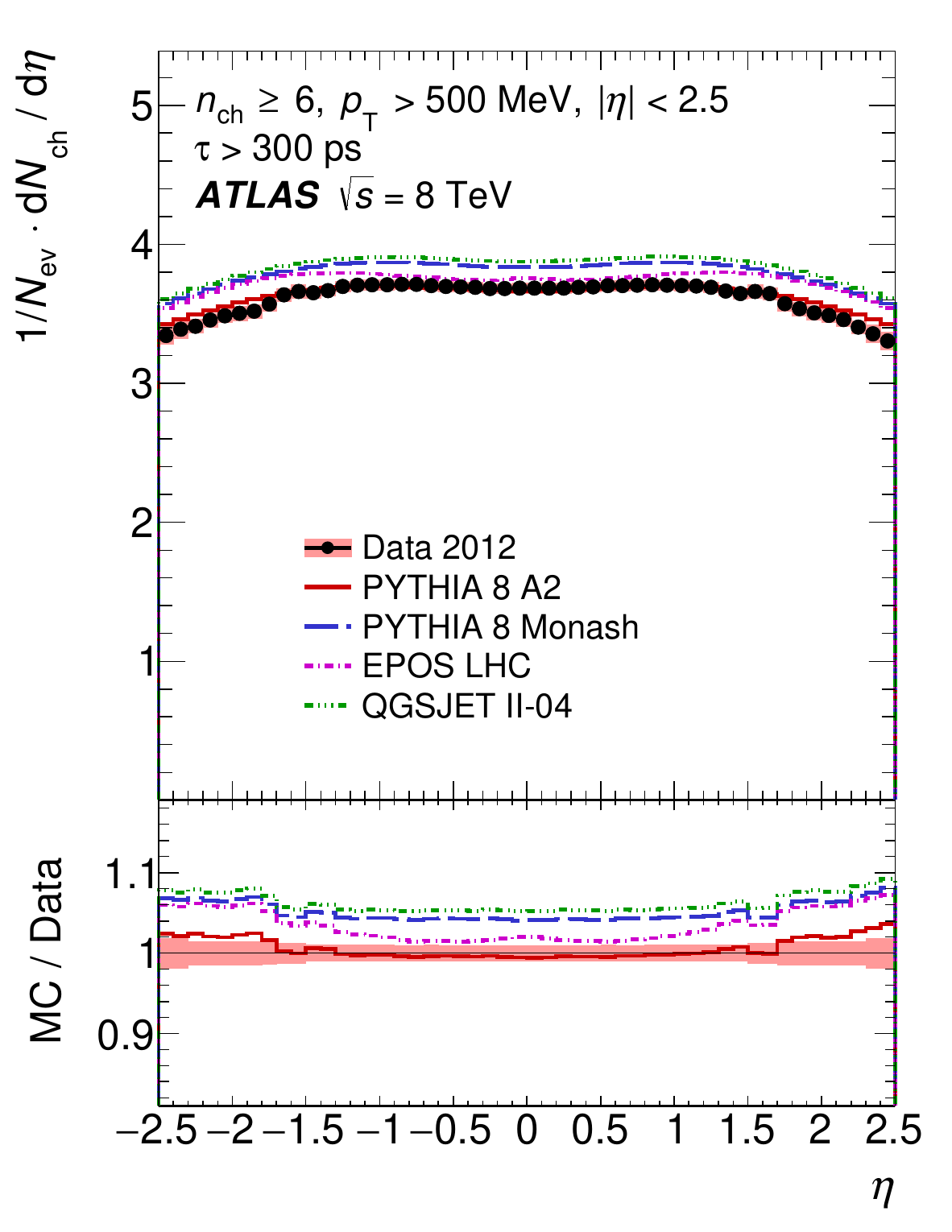}}
(c) 
\\
\end{minipage}
\vfill
\begin{minipage}[h]{0.32\textwidth}
\center{\includegraphics[width=1.0\linewidth]{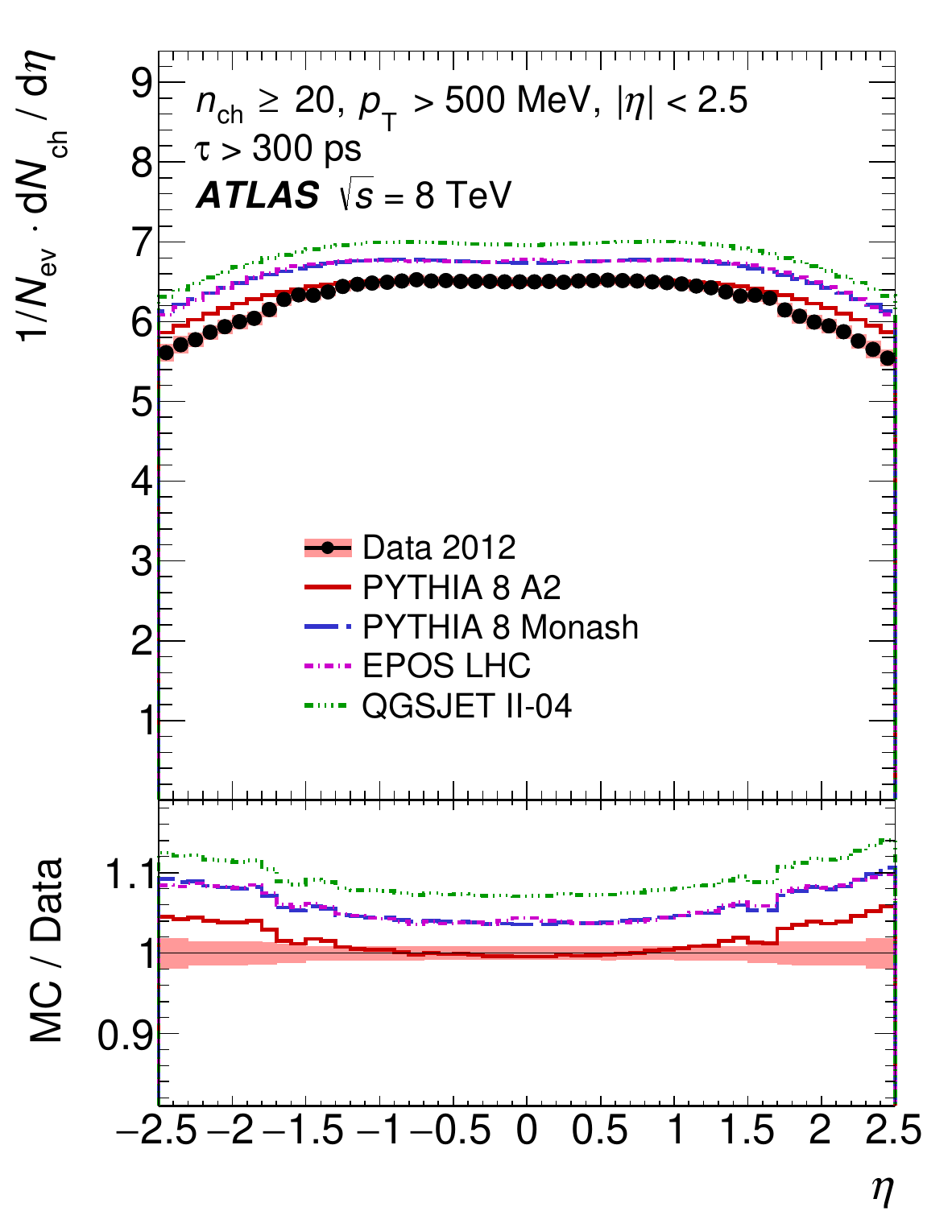}}
(d) 
\\
\end{minipage}
\hspace{2mm}
\begin{minipage}[h]{0.32\textwidth} 
\center{\includegraphics[width=1.0\linewidth]{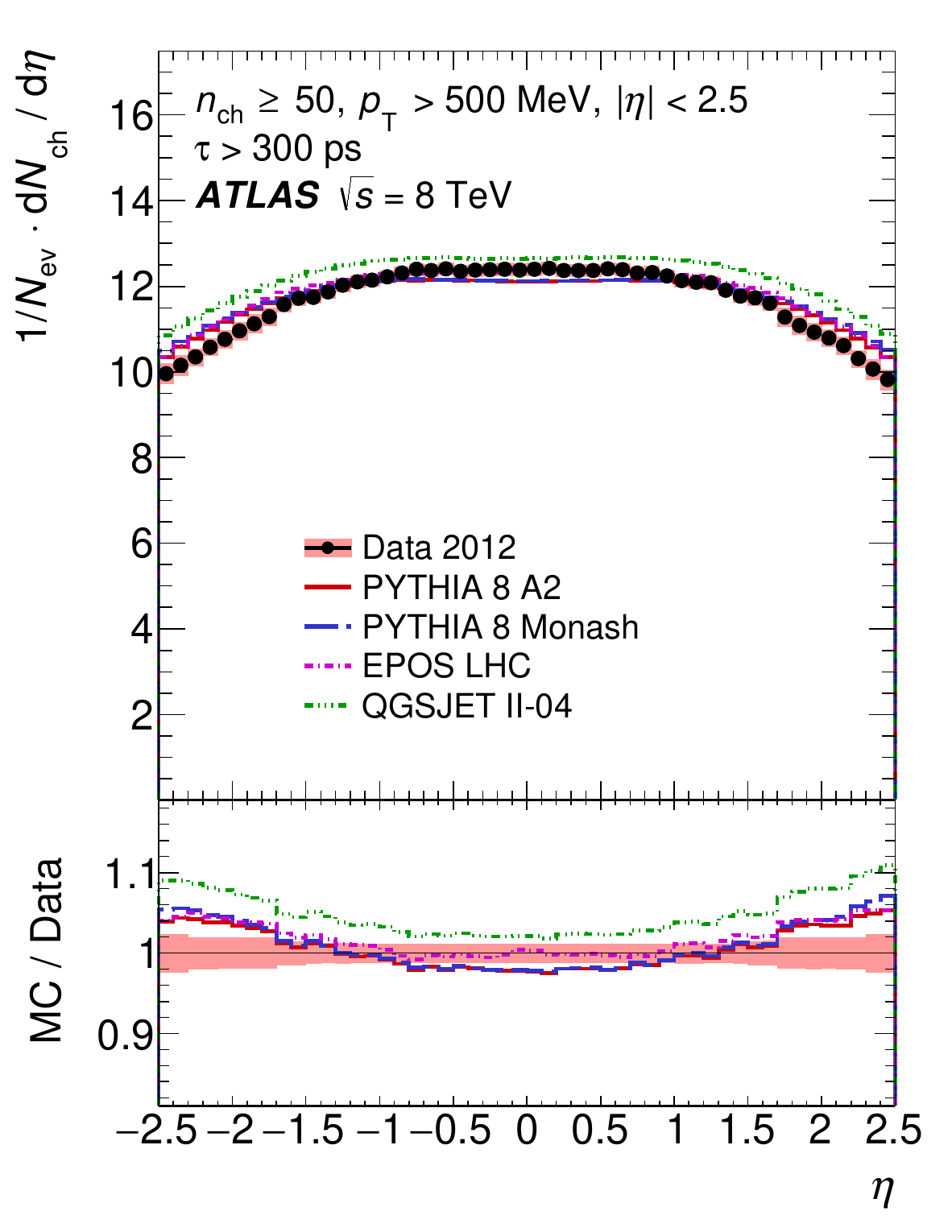}}
(e) 
\\
\end{minipage}
\caption{
Top panel: 
Primary charged-particle  multiplicity  density pseudorapidity distributions
for events for  \(\mid\eta\mid < 2.5\), each with a lifetime \(\tau >300\)~ps,
at the centre-of-mass energy  \(\sqrt{s}=  8\)~\TeV\  \cite{STDM-2014-19}  
with 
(a) \(n_{\mathrm{ch}} \ge 2\) and \(p_{\mathrm{T}} >100\)~\MeV\  and 
for \(p_{\mathrm{T}} >500\)~\MeV\ with 
(b) \(n_{\mathrm{ch}} \ge 1\),
(c)  \(n_{\mathrm{ch}} \ge 6\), 
(d)  \(n_{\mathrm{ch}} \ge 20\) and 
(e)  \(n_{\mathrm{ch}} \ge 50\).
The data represented by dots 
is
compared  to various particle-level MC predictions,  
which are shown by curves. 
The shaded areas around the data points represent the total statistical and systematic uncertainties added in quadrature.
Bottom panel: 
The ratios of the MC predictions to  the experimental results are shown. 
Bands represent the uncertainties of the experimental results.
Taken from Ref.~\cite{STDM-2014-19}.
}
\label{fig_8_eta}
\end{figure*}

\begin{figure*}[t!]
\centering
\begin{minipage}[h]{0.32\textwidth} 
\center{\includegraphics[width=1.0\linewidth]{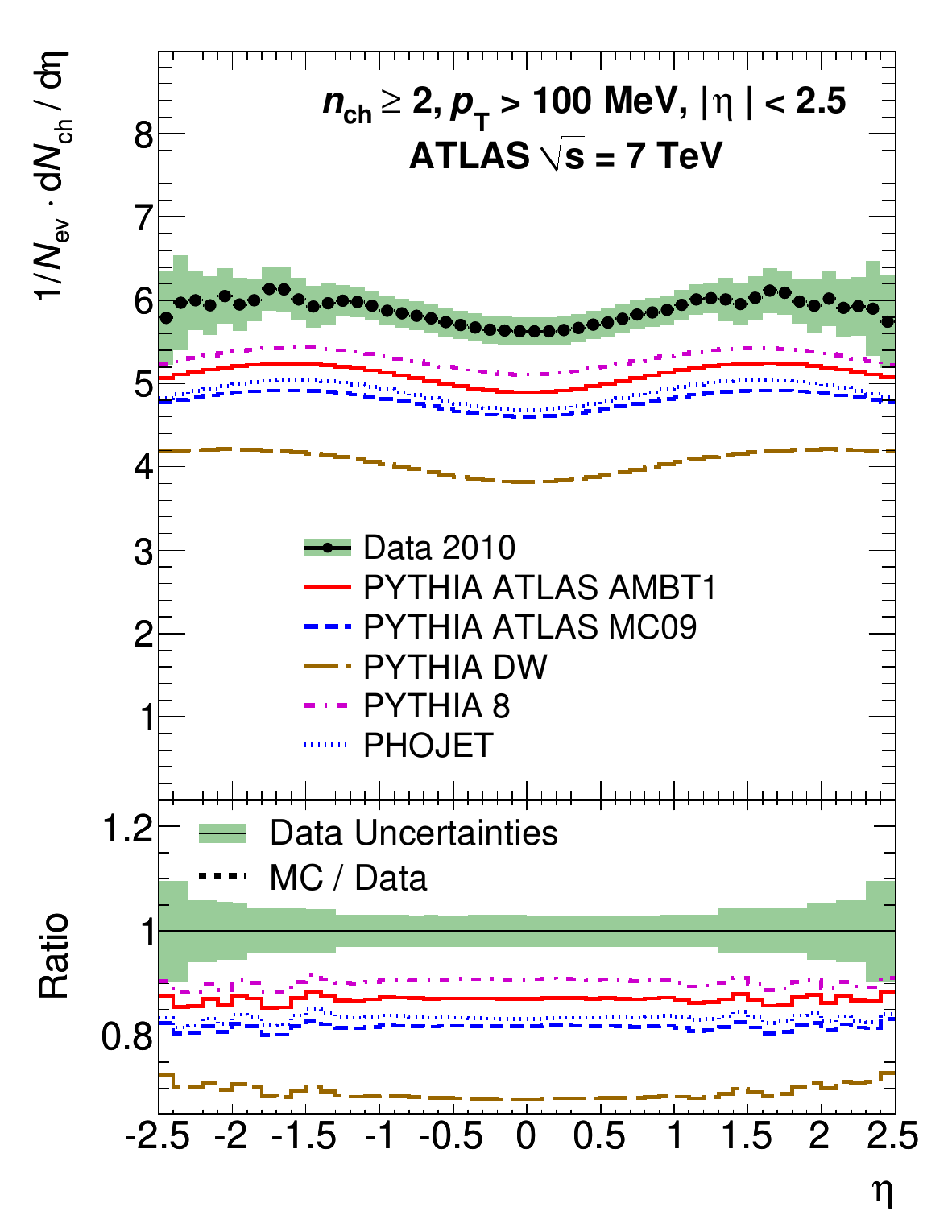}}
(a) 
\\
\end{minipage}
\hfill
\begin{minipage}[h]{0.32\textwidth}
\center{\includegraphics[width=1.0\linewidth]{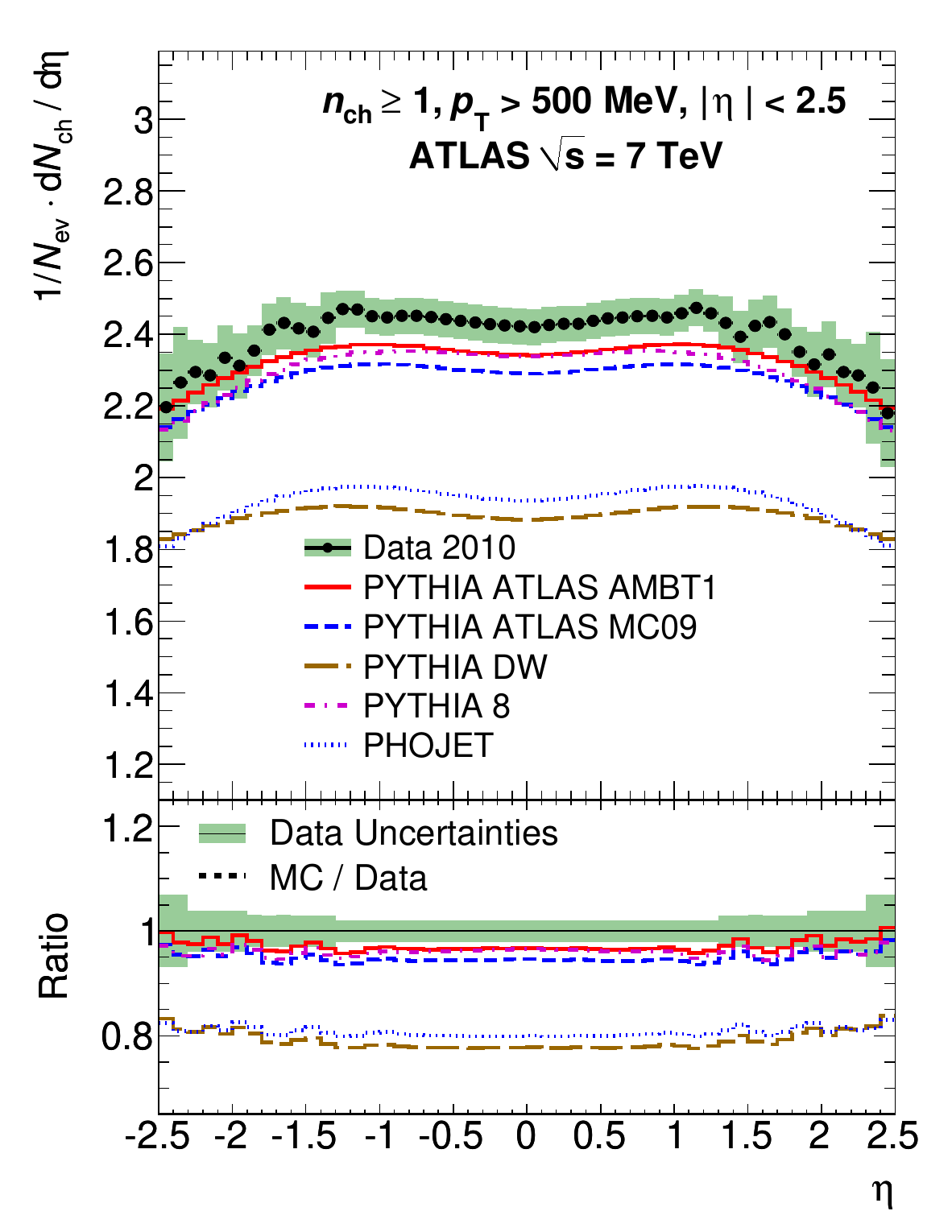}}
(b) 
\\
\end{minipage}
\hfill
\begin{minipage}[h]{0.32\textwidth} 
\center{\includegraphics[width=1.0\linewidth]{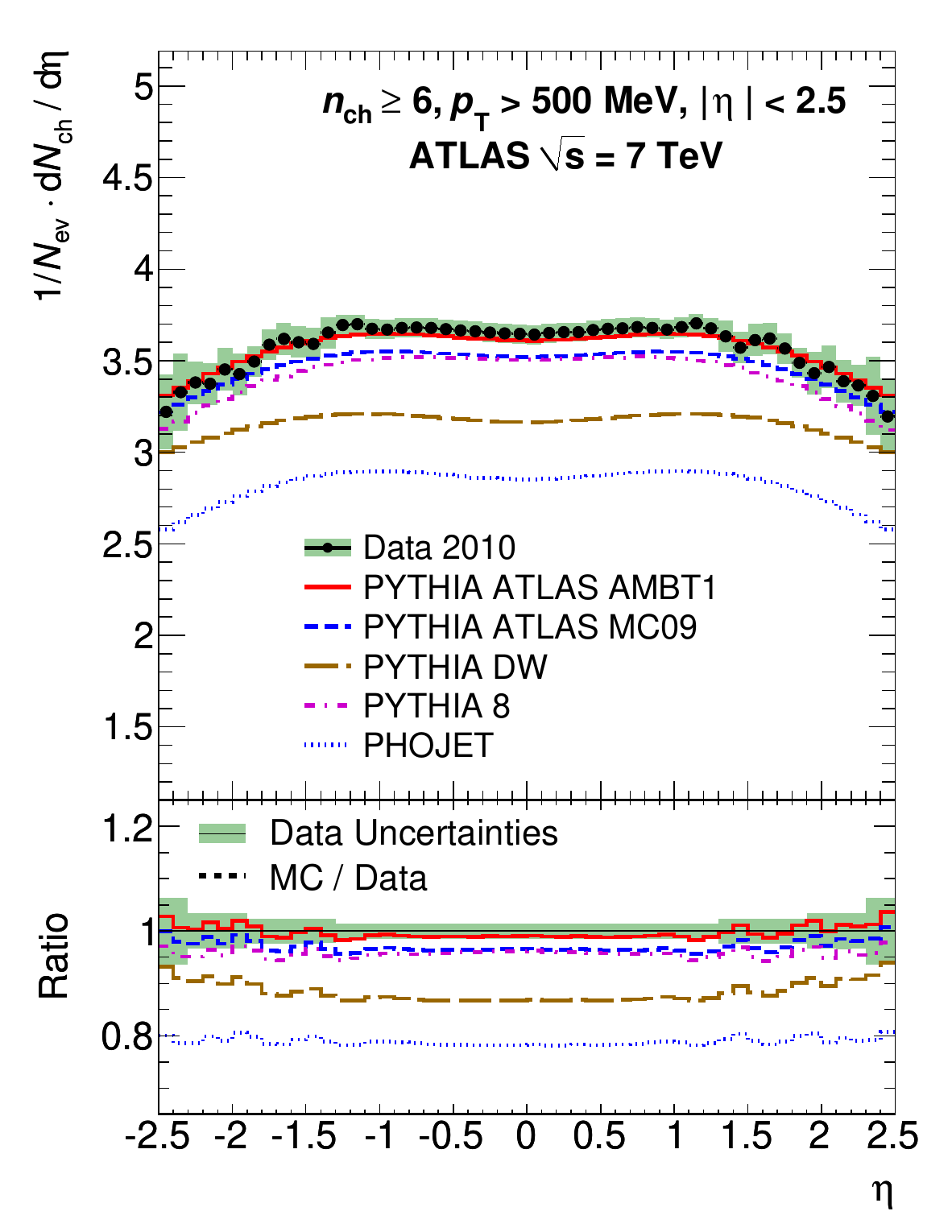}}
(c) 
\\
\end{minipage}
\caption{
Top panel: 
Primary charged-particle  multiplicity  density pseudorapidity distributions for events for  \(\mid\eta\mid < 2.5\)
at the centre-of-mass energy  \(\sqrt{s}= 7\)~\TeV\  \cite{STDM-2010-06}	  
with 
(a) \(n_{\mathrm{ch}} \ge 2\), \(p_{\mathrm{T}} >100\)~\MeV\  and 
for \(p_{\mathrm{T}} >500\)~\MeV\ with 
(b) \(n_{\mathrm{ch}} \ge 1\) and
(c)  \(n_{\mathrm{ch}} \ge 6\).
The data represented by dots 
is
compared 
to various particle-level MC predictions,  which are shown by curves. 
The shaded areas around the data points represent the total statistical and systematic uncertainties added in quadrature.
Bottom panel: 
The ratios of the MC predictions to  the experimental results are shown. 
Bands represent the uncertainties of the experimental results.
Taken from Ref.~\cite{STDM-2010-06}.
}
\label{fig_7_eta}
\end{figure*}

\begin{figure*}[t!]
\centering
\begin{minipage}[h]{0.45\textwidth} 
\center{\includegraphics[width=1.0\linewidth]{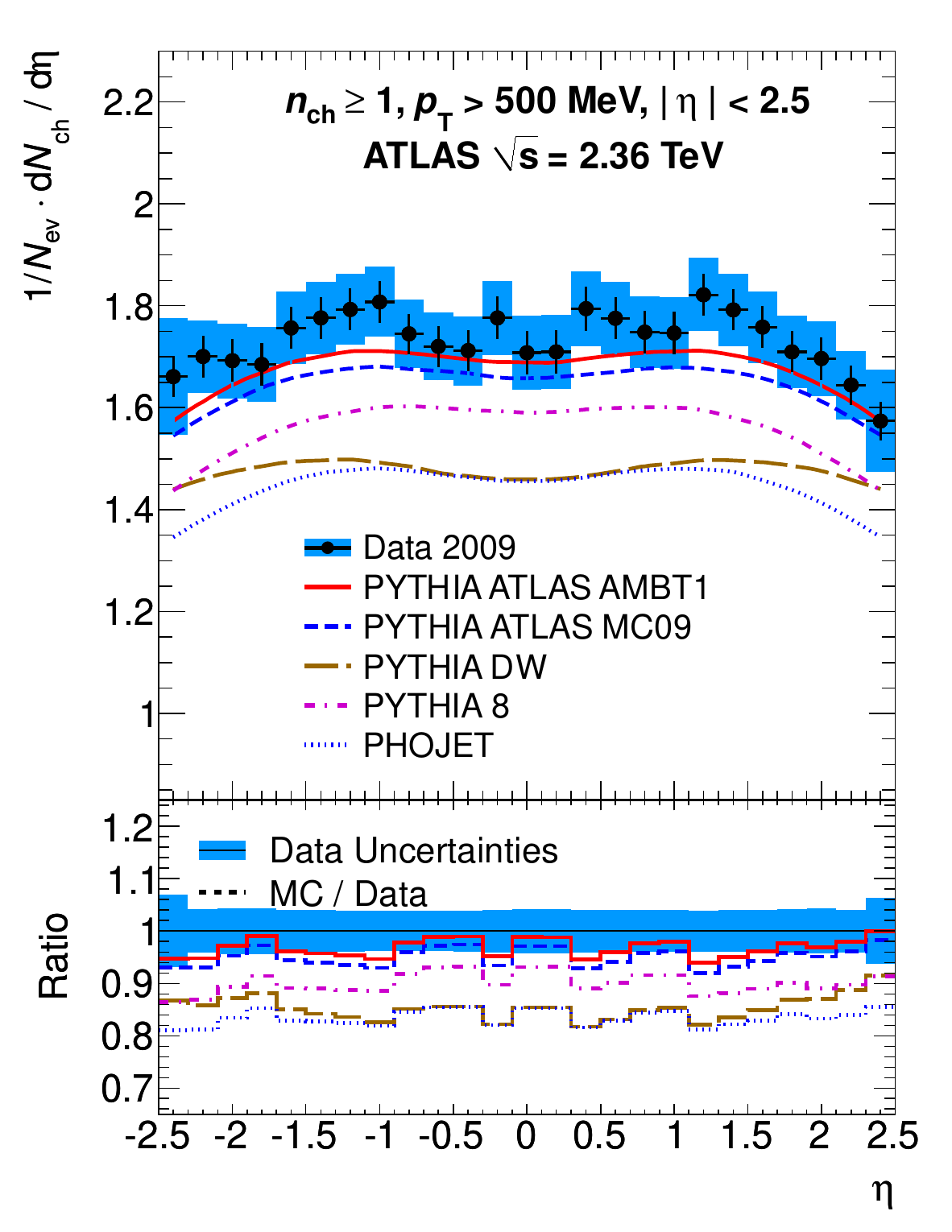}}
\\
\end{minipage}
\caption{
Top panel: 
Primary charged-particle  multiplicity  density pseudorapidity distribution
for events for  \(\mid\eta\mid < 2.5\) at the centre-of-mass energy  \(\sqrt{s}=  2.36\)~\TeV\ 
\cite{STDM-2010-06}	  
with  \(n_{\mathrm{ch}} \ge 1\) and \(p_{\mathrm{T}} >500\)~\MeV.
The data represented by dots 
is
compared 
to various particle-level MC predictions,  which are shown by curves. 
The shaded areas around the data points represent the total statistical and systematic uncertainties added in quadrature.
Bottom panel: 
The ratios of the MC predictions to  the experimental results are shown. 
Bands represent the uncertainties of the experimental results.
Taken from Ref.~\cite{STDM-2010-06}.
}
\label{fig_236_eta}
\end{figure*}

\begin{figure*}[t!]
\centering
\begin{minipage}[h]{0.32\textwidth} 
\center{\includegraphics[width=1.0\linewidth]{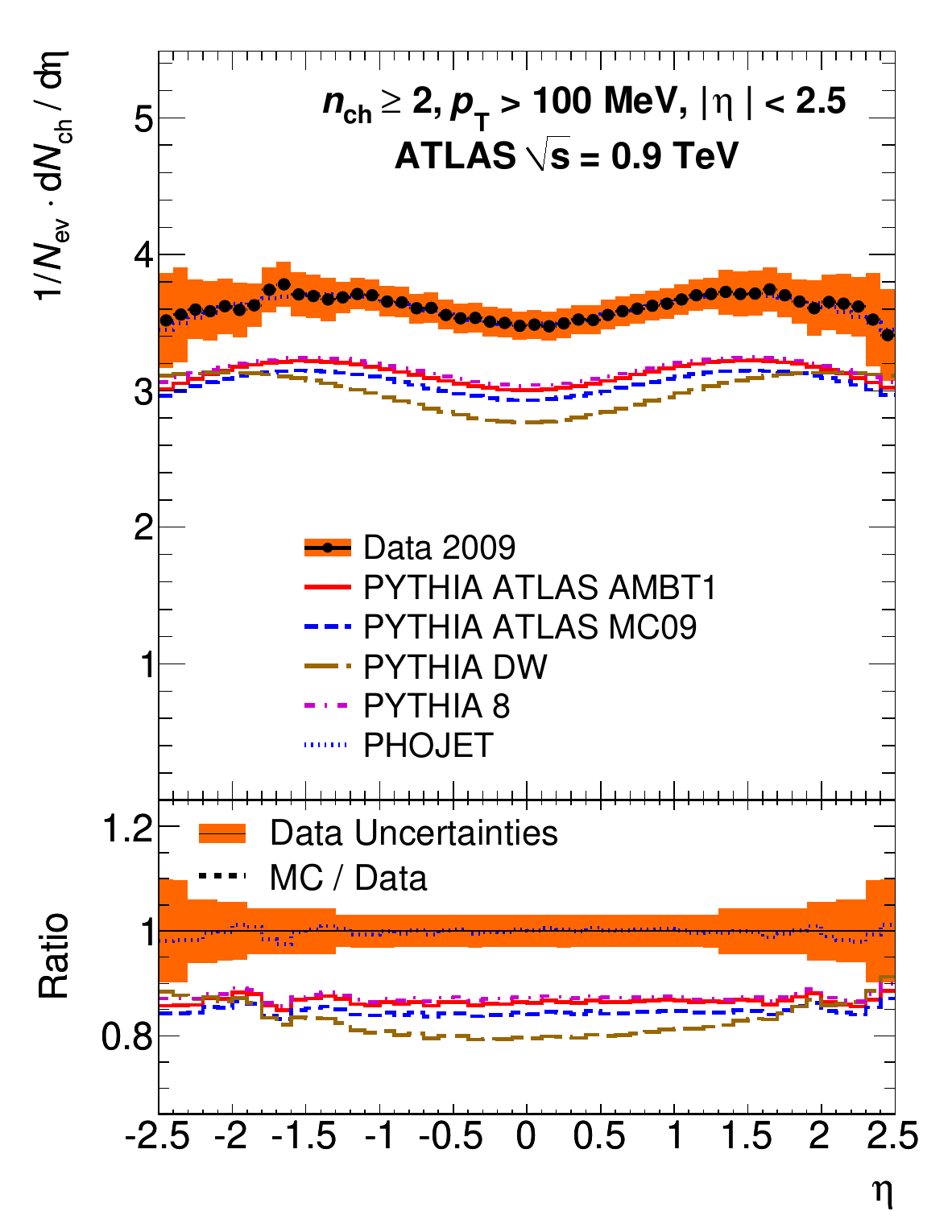}}
(a) 
\\
\end{minipage}
\hfill
\begin{minipage}[h]{0.32\textwidth}
\center{\includegraphics[width=1.0\linewidth]{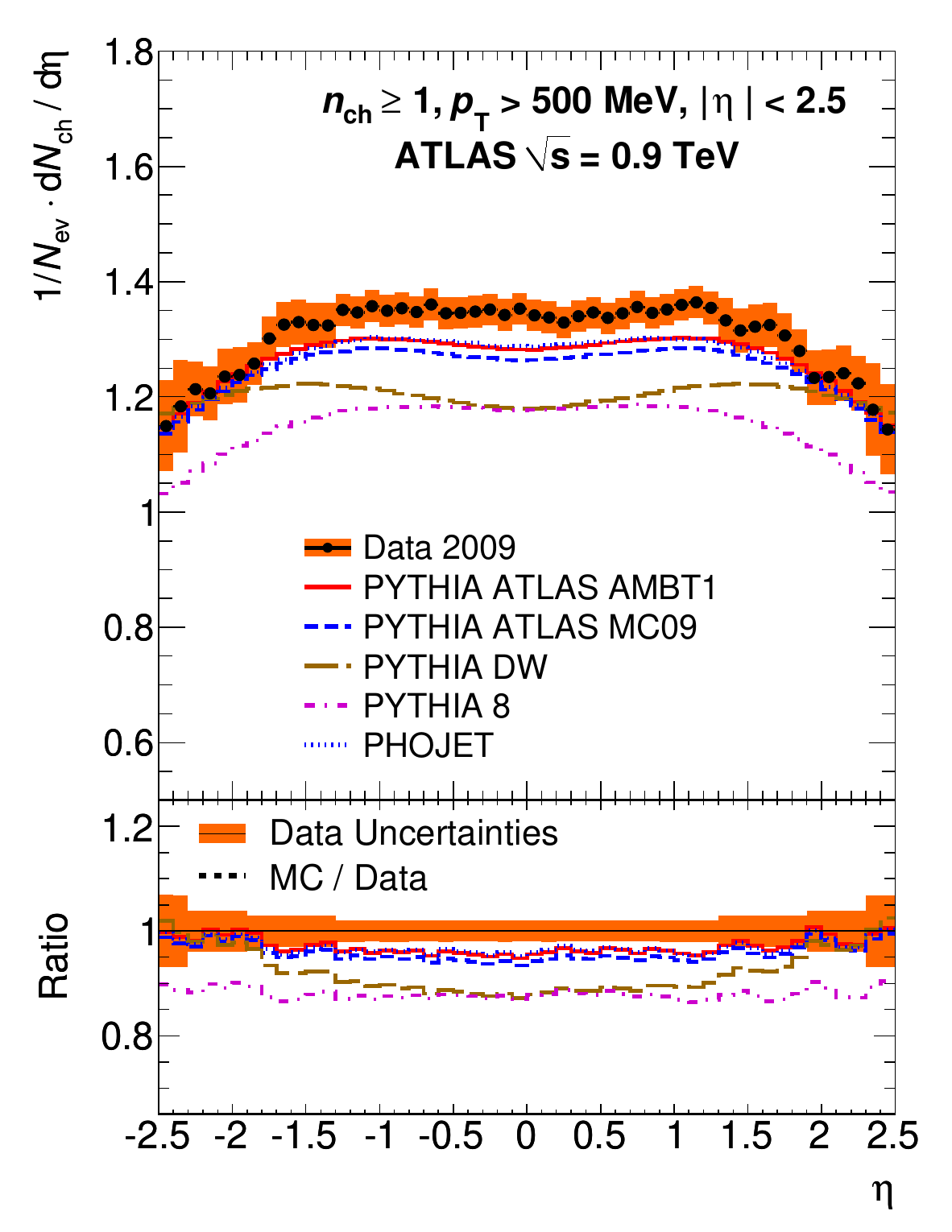}}
(b) 
\\
\end{minipage}
\hfill
\begin{minipage}[h]{0.32\textwidth} 
\center{\includegraphics[width=1.0\linewidth]{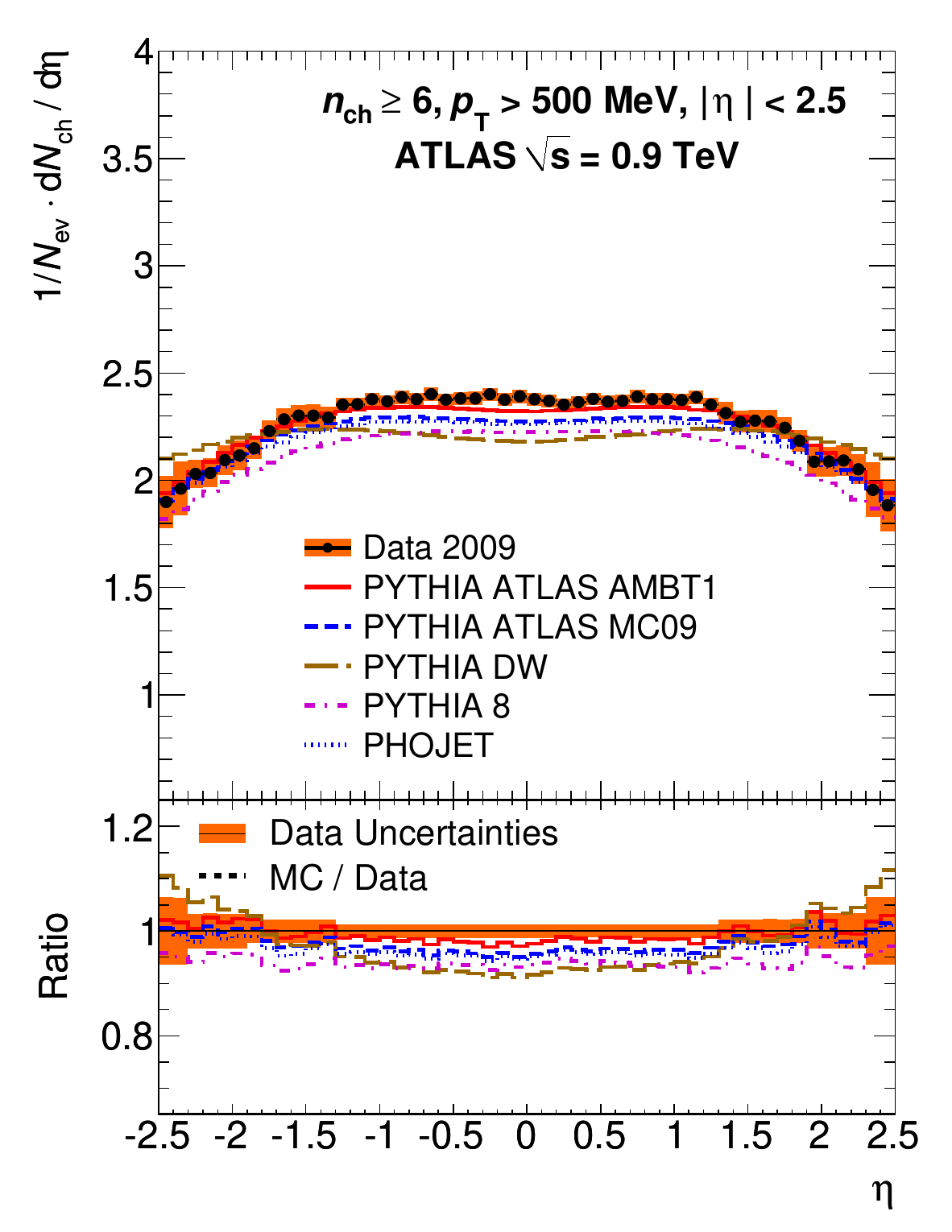}}
(c) 
\\
\end{minipage}
\caption{
Top panel: 
Primary charged-particle  multiplicity  density pseudorapidity distributions
for events for  \(\mid\eta\mid < 2.5\) at the centre-of-mass energy  \(\sqrt{s}=  0.9\)~\TeV\ 
\cite{STDM-2010-06}	   with 
(a) \(n_{\mathrm{ch}} \ge 2\) and \(p_{\mathrm{T}} >100\)~\MeV\ 
and 
for \(p_{\mathrm{T}} >500\)~\MeV\ with 
(b) \(n_{\mathrm{ch}} \ge 1\)
and
(c)  \(n_{\mathrm{ch}} \ge 6\).
The data represented by dots 
is
compared to various particle-level MC predictions,  
which are shown by curves. 
The shaded areas around the data points represent the total statistical and systematic uncertainties added in quadrature.
Bottom panel: 
The ratios of the MC predictions to  the experimental results are shown. 
Bands represent the uncertainties of the experimental results.
Taken from Ref.~\cite{STDM-2010-06}.
}
\label{fig_09_eta}
\end{figure*}

The primary charged-particle multiplicity  density  pseudorapidity distributions
(or ``pseudorapidity distribution'') 
for events with   \(n_{\mathrm{ch}} \ge 2\), \(p_{\mathrm{T}} >100\)~\MeV\  and 
\(n_{\mathrm{ch}} \ge 1\),  \(p_{\mathrm{T}} >500\)~\MeV\  for  \(\mid\eta\mid < 2.5\) 
studied by 
ATLAS 
\cite{STDM-2010-01,STDM-2010-06,STDM-2014-19,STDM-2015-02,STDM-2015-17}
at the CM energies   \(\sqrt{s}= 13\), \(8\), \(7\), \(2.36\) and \(0.9\)~\TeV\
are shown in 
Figs.~\ref{fig_13_eta},  \ref{fig_8_eta}(a) and (b),  \ref{fig_7_eta}(a) and (b),  \ref{fig_236_eta}
and \ref{fig_09_eta}, respectively.
The  pseudorapidity  distributions for particles with 
\(p_{\mathrm{T}} >500\)~\MeV\ and   higher minimum multiplicities per event
\(n_{\mathrm{ch}} \ge 6,\ 20,\ 50\)   at \(\sqrt{s}= 8\)~\TeV\  are shown  in   
Figs.~\ref{fig_8_eta}(c) -- (d),   and  for  \(n_{\mathrm{ch}} \ge 6\)   at \(\sqrt{s}= 7\) and \(0.9\)~\TeV\ 
in   Figs.~\ref{fig_7_eta}(c) and  \ref{fig_09_eta}(c),  respectively. 
The accuracy of measurement of   pseudorapidity distributions
increases   with  increasing  energy
because of the better  understanding of dead material 
values in the ATLAS ID in the data analysis for higher energies.

The ATLAS  experimental results  are compared to predictions of models tuned to a wide range 
of measurements described in Sec.~\ref{MC} and presented in Table~\ref{tab:MC_enegy}. 
The measured spectra are presented as inclusive distributions with  corrections that  minimally rely  
on the MC model used
in order to facilitate an accurate comparison with predictions.
In general, the systematic uncertainties are larger than the statistical uncertainties.
In most regions of all distributions, 
the dominant uncertainty comes  from 
track reconstruction efficiency.

Figure~\ref{fig_13_eta}   shows  the   pseudorapidity distributions at  \(\sqrt{s} = 13\)~\TeV.
%
The distribution corresponding to the   PS with  \(n_{\mathrm{ch}} \ge 2\),  \(p_{\mathrm{T}} >100\)~\MeV\ 
\cite{STDM-2015-17}
rises as \( \mid\eta\mid\) increases,  peaking at  \(\mid\eta\mid \approx  1.7  \)  before falling.
For the PS with  \(n_{\mathrm{ch}} \ge 1\), \(p_{\mathrm{T}} >500\)~\MeV\
\cite{STDM-2015-02}, 
the mean particle density is roughly constant at \(2.9\)  for \(\mid\eta\mid \lesssim 1.5 \) and falls at higher 
\(\eta\). 

For  pseudorapidity distributions at \(13\)~\TeV\ for  \(n_{\mathrm{ch}} \ge 2\) with 
\(p_{\mathrm{T}} >100\)~\MeV\  the   \textsc{Pythia\,8} \textsc{Monash} tune,   \textsc{EPOS} and  
\textsc{QGSJET-II}  give a good description for  \( \mid\eta\mid \lesssim 1.5 \) in   
Fig.~\ref{fig_13_eta}(a). 
The prediction from  the \textsc{Pythia\,8} \textsc{A2} tune
has the same shape as predictions from the other generators
but lies below the data.

In the case of   PS with  \(n_{\mathrm{ch}} \ge 1\), \(p_{\mathrm{T}} >500\)~\MeV,  
\textsc{EPOS} describes the data for  \( \mid\eta\mid \lesssim 1.0 \), 
and predicts a slightly larger multiplicity at larger \( \mid\eta\mid\) values. 
\textsc{QGSJET-II}  and  the  \textsc{Pythia\,8} \textsc{Monash} tune
predict multiplicities that are too large by approximately  \(15\)\%  and \(5\)\%,  respectively. 
The \textsc{Pythia\,8} \textsc{A2} tune  predicts a primary charged-particle multiplicity 
density  that is  \(3\)\%  too low in the central region but describes the data well in the forward region.

In Fig.~\ref{fig_8_eta}(a) at \(8\)~\TeV\  \cite{STDM-2014-19}
the distribution corresponding to the  PS with \(n_{\mathrm{ch}} \ge 2\), \(p_{\mathrm{T}} >100\)~\MeV\  
is well described by  \textsc{EPOS}  and   \textsc{Pythia\,8} \textsc{Monash} tune
but is underestimated by  the \textsc{Pythia\,8} \textsc{A2} tune and  \textsc{QGSJET-II}. 
In Fig.~\ref{fig_8_eta}(b) for the PS with  \(n_{\mathrm{ch}} \ge 1\), \(p_{\mathrm{T}} >500\)~\MeV\ 
\textsc{EPOS}   overestimates the distribution at  \( \mid\eta\mid > 1.7 \) 
and describes the data well for the rest of  the pseudorapidity range. 
The data are overestimated by the \textsc{QGSJET-II}   and
\textsc{Pythia\,8} \textsc{Monash} tune calculations and underestimated by the 
\textsc{Pythia\,8} \textsc{A2} tune prediction. 

A similar shape is seen for the PS  corresponding to higher multiplicities with 
\(n_{\mathrm{ch}} \ge 6,\ 20,\ 50\)  shown in  Fig.~\ref{fig_8_eta}(c) -- (e)  
with the extent of the plateau becoming shorter as the multiplicity threshold is raised. 
A small apparent structure in the distributions of the central values of the data points 
occurs at values of  \( \mid\eta\mid \sim 1.7 \). 
In these
figures, 
all models overestimate the overall yield for the PS with  \(n_{\mathrm{ch}} \ge 6,\ 20\) 
although  \textsc{Pythia\,8} \textsc{A2}  describes the plateau in the central region well. 
For the largest multiplicity threshold,  \(n_{\mathrm{ch}} \ge 50\),  
all of the models overestimate the data at \( \mid\eta\mid > 1.7 \) 
but provide a better description in the central region.

Figures \ref{fig_7_eta}(a)  and    \ref{fig_09_eta}(a)   show the \(\eta\) distributions for the 
most inclusive  PS region with  \(n_{\mathrm{ch}} \ge 2\), \(p_{\mathrm{T}} >100\)~\MeV.
In these cases, 
the distributions show weaker dependence on  \(\mid\eta\mid\)
than in the  other plots  at \(\sqrt{s}= 7\)~\TeV\   and \(\sqrt{s}= 0.9\)~\TeV. 
%
Figures~\ref{fig_7_eta}(b), \ref{fig_236_eta}  and \ref{fig_09_eta}(b)  show the 
pseudorapidity distributions in the  PS region with 
\(n_{\mathrm{ch}} \ge 1\), \(p_{\mathrm{T}} >500\)~\MeV\
at \(\sqrt{s}= 7\)~\TeV,  \(\sqrt{s}= 2.36\)~\TeV\ and \(\sqrt{s}= 0.9\)~\TeV,  respectively. 
The mean particle density is roughly constant for \(\mid\eta\mid < 1.0\) and
decreases at higher  \(\mid\eta\mid\). 
The distribution shapes of the models  are similar except for that of the  \textsc{Pythia\,6} \textsc{DW} tune, 
which has a flatter spectrum and a more pronounced dip at central \(\mid\eta\mid\), 
especially at low \(\sqrt{s}\). 
At energies  \(7\)~\TeV,  \(2.36\)~\TeV\ and \(0.9\)~\TeV\  the \textsc{Pythia\,6} \textsc{AMBT1} tune 
gives the best shape and normalisation description of the data,  although it was tuned for 
\(n_{\mathrm{ch}} \ge 6\)  in Figs.~\ref{fig_7_eta}(c) and \ref{fig_09_eta}(c).

At \(\sqrt{s}= 7\)~\TeV\   all the shapes seem to model the observed spectrum reasonably well, 
but at this energy, 
the difference in normalisation among the models varies more widely,
and no model reproduces the data.
At \(\sqrt{s}= 0.9\)~\TeV\  there is very little difference between the models, 
both in shape  and normalisation, 
with the exception of  \textsc{PHOJET},  which shows excellent agreement with the data. 
The other models show, 
on average, 
too few particles. 
The shape of the distribution is reasonably well described by all models.

\begin{table*}[t!]
\centering 
\caption{
Fiducial inelastic cross-section measured by ATLAS at  \( \sqrt{s} = 13\)~\TeV\   \cite{ATLAS:2016ygv}
and at  \( \sqrt{s} = 7\)~\TeV\ \cite{ATLAS:2011zrx} compared with  the ATLAS \textsc{Pythia\,8} \textsc{A3}  
\cite{ATLAS:2016puo}  and  
Schuler--Sj\(\ddot{\mathrm{o}}\)strand (SS)  model \cite{Schuler:1993wr} predictions. 
The SS model is used in both  the ATLAS \textsc{Pythia\,8} \textsc{A2}  and \textsc{Monash} tunes. 
\textsc{Pythia\,8} \textsc{A3} uses the   Donnachie--Landshoff  model \cite{Donnachie:1992ny}
with two tuned parameters. 
Taken from Ref.~\cite{ATLAS:2016puo}.
}
\label{tab:Cs_A3}
\medskip
\begin{tabular}{rlll}
\hline
\hline

\(\sqrt{s}\) [TeV]
& 
Experimental Results [mb] 
& 
SS model [mb]
& 
\textsc{Pythia\,8} \textsc{A3} [mb] 
\\
\hline
13		& 68.1\(\pm\)1.4& 74.4& 69.9\\
7		& 60.3\(\pm\)2.1& 66.1& 62.3\\
\hline
\hline
\end{tabular}
\end{table*}

\begin{figure*}[t!]
\centering
\begin{minipage}[h]{0.32\textwidth}
\center{\includegraphics[width=1.0\linewidth]{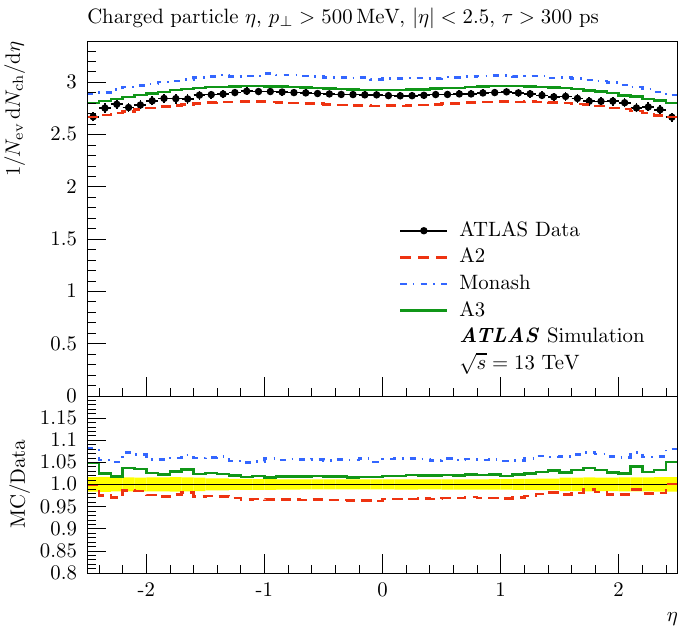}}
(a) 
\\
\end{minipage}
\hfill
\begin{minipage}[h]{0.32\textwidth}
\center{\includegraphics[width=1.0\linewidth]{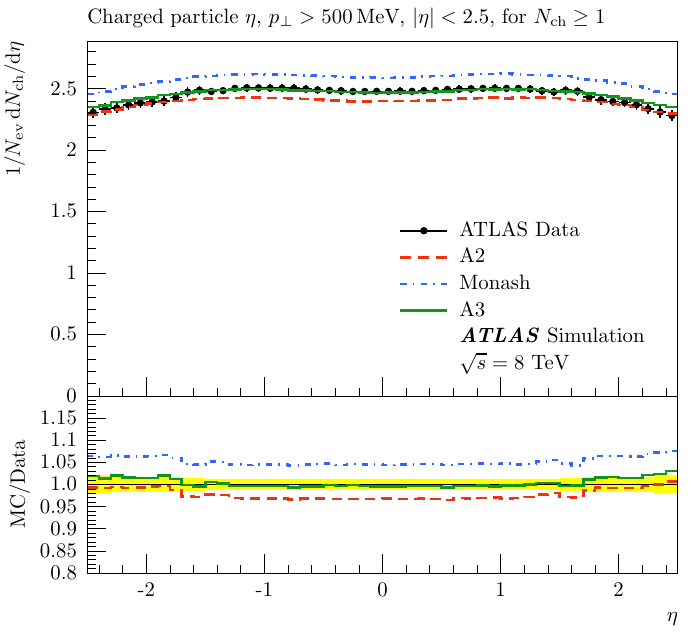}}
(b) 
\\
\end{minipage}
\hfill
\begin{minipage}[h]{0.32\textwidth} 
\center{\includegraphics[width=1.0\linewidth]{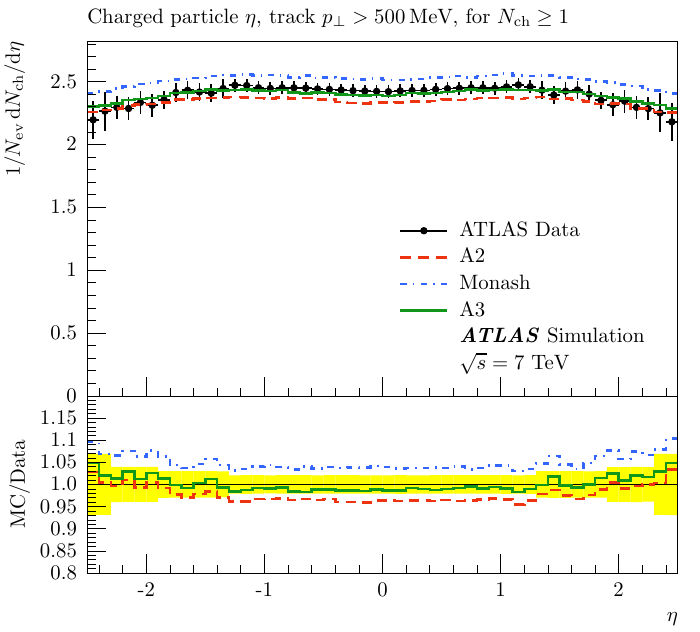}}
(c) 
\\
\end{minipage}
\vfill
\begin{minipage}[h]{0.32\textwidth} 
\center{\includegraphics[width=1.0\linewidth]{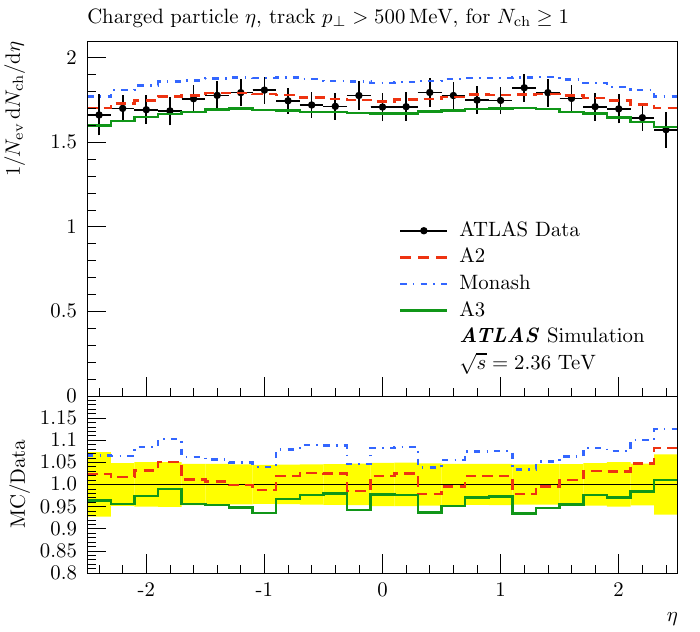}}
(d) 
\\
\end{minipage}
\hspace{2mm}
\begin{minipage}[h]{0.32\textwidth}
\center{\includegraphics[width=1.0\linewidth]{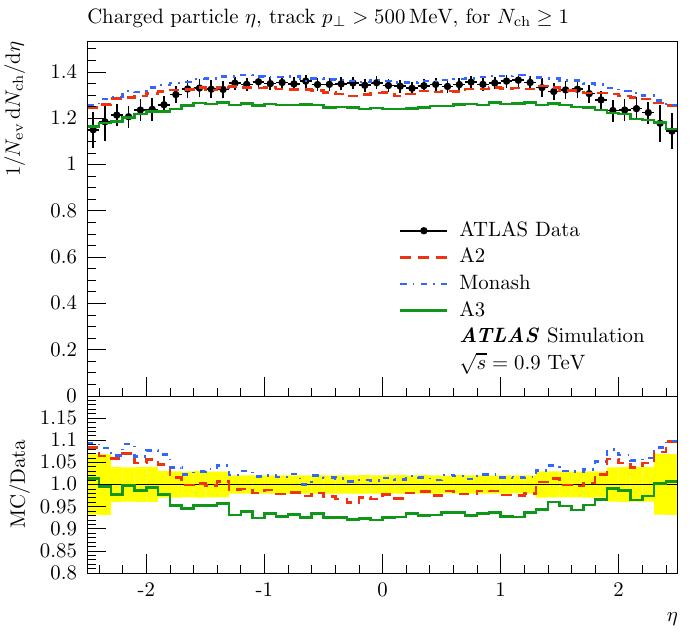}}
(e) 
\\
\end{minipage}
\caption{
Top panel: 
The \textsc{Pythia\,8} \textsc{A3}, \textsc{A2} and \textsc{Monash}   predictions 
\cite{ATLAS:2016puo} compared with ATLAS primarily charged-particle multiplicity 
density pseudorapidity distributions  for events with  \(n_{\mathrm{ch}} \ge 1\), 
\(p_{\mathrm{T}} >500\)~\MeV\  at  the CM energies 
(a) \(\sqrt{s}= 13\)~\TeV, 
(b) \(\sqrt{s}= 8\)~\TeV, 
(c) \(\sqrt{s}= 7\)~\TeV,
(d) \(\sqrt{s}= 2.36\)~\TeV\ 
and 
(e) \(\sqrt{s}= 0.9\)~\TeV\ 
\cite{STDM-2010-06,STDM-2014-19,STDM-2015-02,STDM-2015-17}.
The yellow-shaded areas represent the measurement uncertainty.
Bottom panel: 
The ratios of the MC predictions to  the experimental results at \(\sqrt{s}=13\)~\TeV\ are shown. 
Bands represent the uncertainties of the experimental results.
Taken from Ref.~\cite{ATLAS:2016puo}.
}
\label{fig_A3_eta}
\end{figure*}

In Ref.~\cite{ATLAS:2016puo}  the performance of the ATLAS \textsc{Pythia\,8} \textsc{A3} tune  was
presented  for   primary charged-particle multiplicity  density pseudorapidity  distributions,  
transverse momentum  distributions, and multiplicity distributions,
as well as
average transverse momentum multiplicity  distributions, 
compared to the  predictions of the previous ATLAS  \textsc{Pythia\,8} tunes   ---
\textsc{A2}  and  \textsc{Monash}.
Both 
of
these tunes use the default  Schuler--Sj\(\ddot{\mathrm{o}}\)strand  (SS)  diffraction model
\cite{Schuler:1993wr},  and predict the same value.
The SS model overestimates the inelastic cross-section measured by 
ATLAS at \(7\)~\TeV\ and  \(13\)~\TeV,  as can be seen  in Table \ref{tab:Cs_A3}; 
alternative models are therefore considered here. 
Changing the diffractive model affects the charged particle distributions not only at the 
low multiplicity  or in the  low \(p_{\mathrm{T}}\) region,  but also at intermediate values, and in each case, 
the MPI and  CR parameters  need retuning in order to preserve reasonable 
agreement
with data. 

The  DL model  \cite{Donnachie:1992ny}  is found to give the best description of the  MB observables
and the measured fiducial inelastic cross-section  \cite{ATLAS:2016ygv}.
The DL model  comes with two tunable parameters 
that
control the Pomeron Regge trajectory.

To understand the energy dependence of the parameters,  the tuning results at different   \(\sqrt{s}\)
individually using just MB distributions were initially determined. 
For each parameter at each \(\sqrt{s}\), a tuned value was determined and then compared 
to values of the same parameter  when a subset of sampling runs  is used.
The spread of these points was an indication of the statistical and extrapolation uncertainty  on the tune, 
as well as  how well  
the parameter was constrained by the observables used. 
The next step was to determine the sensitivity of each of these parameters to different observables
by successively adding distributions other than those from the MB analysis
and varying the relative weight.

The fiducial inelastic cross section predictions from  \textsc{Pythia\,8} \textsc{A3}  
are about \(5\)\% lower compared to SS,    which is somewhat closer to the values from  the data. 
This does not come at 
the
cost of sacrificing 
agreements 
with other distributions.

In  Figs.\ \ref{fig_A3_eta},  \ref{fig_A3_pT},  \ref{fig_A3_nch}, 
 and  \ref{fig_A3_pT_nch},
the performance of the  ATLAS \textsc{Pythia\,8} \textsc{A3}  tune  can be seen for  
primarily charged-particle multiplicity pseudorapidity  distributions, 
primary charged-particle multiplicity transverse momentum  distributions, 
primary charged-particle  multiplicity distributions,
and average transverse momentum multiplicity  distributions, 
compared to the previous   \textsc{Pythia\,8} \textsc{A2} and \textsc{Monash}  tunes. 

The predicted values of  the fiducial inelastic cross-section at  \( \sqrt{s} = 7\)~\TeV\ and \(13\)~\TeV\
for the tunes compared with 
the
data are shown in Table \ref{tab:Cs_A3}. 

Figure~\ref{fig_A3_eta} 
shows that  the \textsc{Pythia\,8} \textsc{A3} tune 
provides  a small improvement in the modelling of charged particle pseudorapidity distributions 
at  \(\sqrt{s}= 8\)~\TeV\ and, 
to a lesser extent,  
at \(\sqrt{s}= 13\)~\TeV,  at the expense of 
a
larger deterioration of the modelling of  \(\sqrt{s}= 0.9\)~\TeV\ data. 
Since the aim is to model soft collisions for pile-up at \(\sqrt{s} = 13\)~\TeV, 
the \textsc{Pythia\,8} \textsc{A3}  tune’s  mis-modelling of  the \(\sqrt{s}= 0.9\)~\TeV\ 
data is acceptable.


The  models  \textsc{EPOS} \textsc{LHC},  \textsc{PHOJET},  \textsc{QGSJET-II},
\textsc{Pythia\,6} and  \textsc{Pythia\,8}  show big troubles in describing the whole spectrum in the data, 
but the best agreement is achieved with  \textsc{EPOS}.
For \( p_{\mathrm{T}} > 100\)~\MeV\ at  the highest energies
\textsc{Pythia\,8}  \textsc{Monash}, \textsc{EPOS},    \textsc{QGSJET-II}  
give a good description for \(\mid\eta\mid< 1.5\).
The prediction from \textsc{Pythia\,8}  \textsc{A2}  has the same shape but lies below the data.
For \( p_{\mathrm{T}} > 500\)~\MeV\ at  the highest energies
the MCs have the same shape but different normalisation;
\textsc{EPOS}   and  \textsc{Pythia\,8}  \textsc{A2}  give remarkably good predictions.

\subsubsection{Distributions of charged-particle multiplicity over \( \eta \) of the LHC experiments}
\label{Comparison_Nch_eta_LHC}
\begin{figure*}[t!]
\centering  
\begin{minipage}[h]{0.40\textwidth} 
\center{\includegraphics[width=1.0\linewidth]{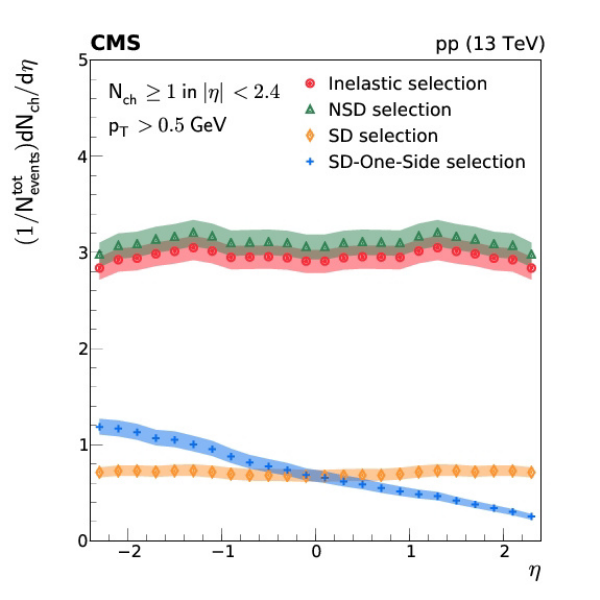}}
(a) 
\\
\end{minipage}
\vfill
\begin{minipage}[h]{0.32\textwidth}
\center{\includegraphics[width=1.0\linewidth]{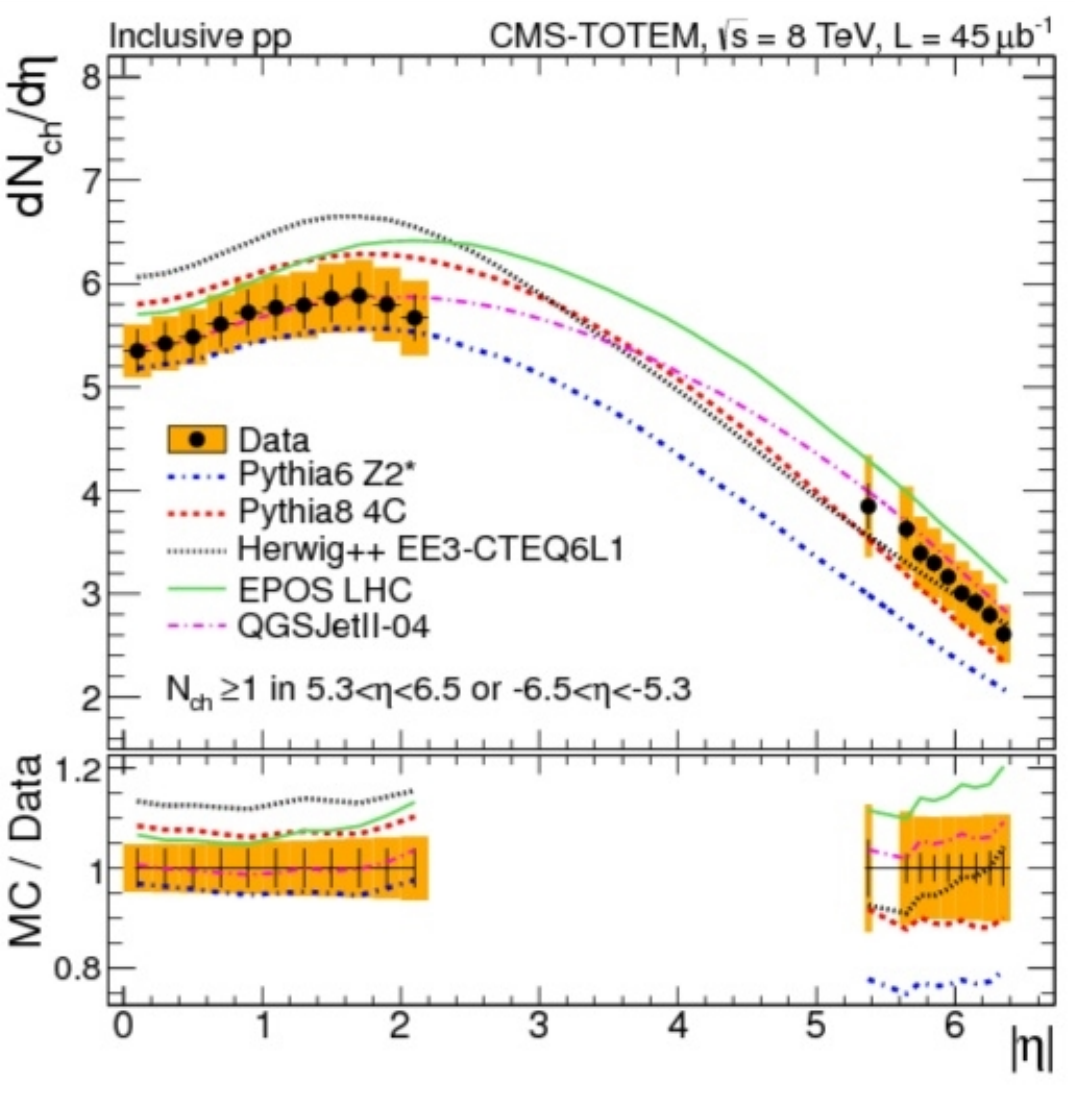}}
(b) 
\\
\end{minipage}
\hfill
\begin{minipage}[h]{0.32\textwidth}
\center{\includegraphics[width=1.0\linewidth]{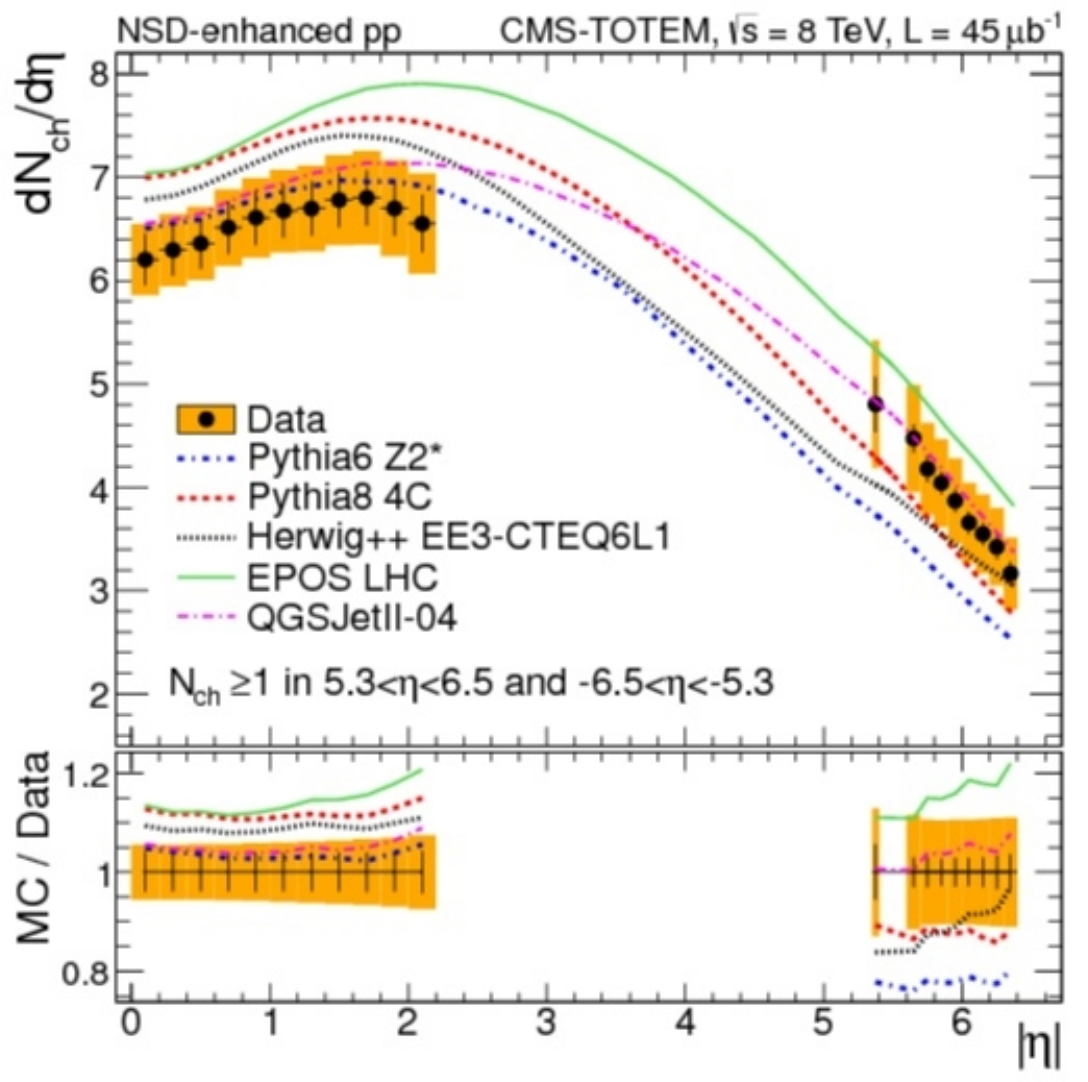}}
(c) 
\\
\end{minipage}
\hfill
\begin{minipage}[h]{0.32\textwidth}
\center{\includegraphics[width=1.0\linewidth]{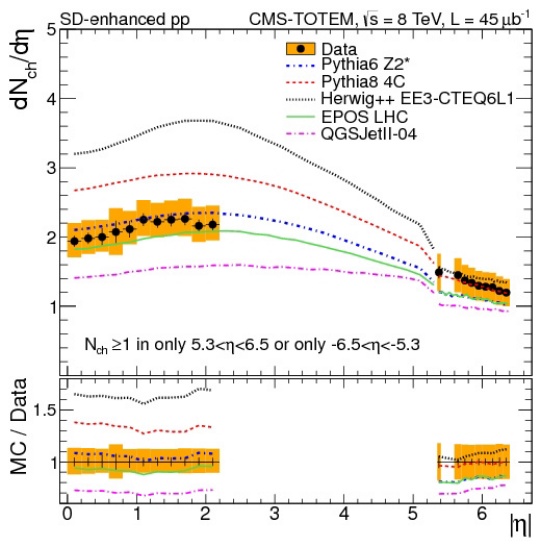}}
(d) 
\\
\end{minipage}
\caption{
(a)
Primary charged-particle multiplicity  density pseudorapidity distributions for events for  
\(\mid\eta\mid < 2.4\)  at the centre-of-mass energy \(\sqrt{s}= 13\)~\TeV\ 
with  \(n_{\mathrm{ch}} \ge 1\) and \(p_{\mathrm{T}} >500\)~\MeV. 
The multiplicity of charged particles per event for the  inelastic,
NSD-enhanced, 
SD-enhanced, 
 and  SD-One-Side enhanced  event samples are shown.
The band encompassing the data points represents 
the total systematic uncertainty, 
while the statistical uncertainty is included as a vertical bar for each data point.
Taken from Ref.~\cite{CMS:2018nhd}.
Primary charged-particle multiplicity  density pseudorapidity  distributions  from the
(b)  inelastic,
(c)  NSD-enhanced, 
and 
(d)  SD-enhanced  event samples
at the centre-of-mass energy  \(\sqrt{s}= 8\)~\TeV\  in \(\mid\eta\mid < 2.2\),  \(5.3 < \eta < 6.5\) 
and  \(-6.5 < \eta < -5.3\)  with   \(n_{\mathrm{ch}} \ge 1\) and  \(p_{\mathrm{T}} > 500\)~\MeV.
The error bars represent the statistical plus uncorrelated systematics between 
neighbouring bins and the bands show the combined systematic and statistical uncertainties. 
Taken from Ref.~\cite{CMS:2014kix}.
}
\label{fig_13_eta_CMS}
\end{figure*}

The CMS  results for  pseudorapidity distributions for events for  \(\mid\eta\mid < 2.4\) at 
the CM energies   \(\sqrt{s}= 13\)~\TeV\  with  \(n_{\mathrm{ch}} \ge 1\), \(p_{\mathrm{T}} >500\)~\MeV\
\cite{CMS:2018nhd} are shown in  Fig.\ \ref{fig_13_eta_CMS}(a). 
The measured distributions are presented for three different event data sets: 
\begin{enumerate}
\item
the most inclusive  sample  (inelastic), 
\item
the sample dominated by non-single diffractive dissociation events   (NSD-enhanced sample), 
\item
the sample enriched by single diffractive dissociation  events  (SD-enhanced sample). 
\end{enumerate}
The SD-minus and SD-plus samples are mutually exclusive, depending on the side of the
forward-detector that contains the hadronic activity. 
The  pseudorapidity distribution  of the SD-enhanced event sample is also presented 
as a symmetrized distribution constructed from the  SD-minus and SD-plus  enhanced samples 
and is referred to as the SD-One-Side enhanced event sample. 
The symmetrization is performed by reflecting the distribution with respect to
\( \mid\eta\mid  = 0\).
In general terms, the inelastic and NSD distributions are similar. 
The pseudorapidity density of the SD-enhanced event sample is about a factor of \(4\) 
lower than that of the most inclusive event samples. 

%
The combined CMS--TOTEM  pseudorapidity distributions are presented in
Figs.\ \ref{fig_13_eta_CMS}(b) -- (d) for  the inclusive  event selection sample, 
the NSD-enhanced event selection sample, 
 and the SD-enhanced event selection sample
\cite{CMS:2014kix}.
The measurements are compared to  the results from 
\textsc{Pythia\,6}  (version 6.426)  \cite{Sjostrand:2006za} tune \textsc{Z2*} \cite{CMS:2011xjg},
\textsc{Pythia\,8}  (version 8.153)  \cite{Sjostrand:2007gs} tune \textsc{4C} \cite{Corke:2010yf}, 
\textsc{HERWIG++}  (version 2.5.0) \cite{Bahr:2008pv} tune \textsc{UE-EE-3}  with \textsc{CTEQ6L1} 
\cite{Pumplin:2002vw} PDFs,  \textsc{EPOS} \textsc{LHCv3400} tune \textsc{LHC}  \cite{Pierog:2013ria}
and  \textsc{QGSJET-II}  version  \textsc{04}  \cite{Ostapchenko:2010vb}.

In Ref.~\cite{TOTEM:2014zyx},  
similar figures for the  pseudorapidity distributions
were presented  with additional \(\eta\) regions from TOTEM: \(3.7 < \eta < 4.8\)
and \(-7.0 < \eta < -6.0\). 
The results are derived in the central region by averaging the data points in the corresponding 
\( \pm \eta\) bins and in the forward region by averaging over the 
half-arms 
of
four  TOTEM T2 telescopes. 

The primarily charged-particle multiplicity density at   \( \eta = 0\)  is 
\( 5.35 \pm 0.36\) for the inclusive sample,
\( 6.20 \pm 0.46\) for the NSD-enhanced sample,  
and 
\( 1.94^{+ 0.26 }_{-0.23}\)  for the SD-enhanced sample,  with negligible statistical uncertainties. 
The CMS primarily charged-particle multiplicity density   at   \( \eta = 0\) 
for the NSD-enhanced sample  is in agreement  within error bars  with  the ATLAS  one 
presented in Table \ref{tab:average_nch_energy_13-8TeV} at
\(\sqrt{s}  =13\) \TeV\  for PS \(n_{\mathrm{ch}} \ge 2\), \(p_{\mathrm{T}} >100\)~\MeV. 

The predictions from various MC event generators differ from
the data by up to \(20\)\% for the inclusive and NSD-enhanced samples, 
with even larger discrepancies for the SD-enhanced sample. 
The data are well described by  \textsc{Pythia\,6}  and  \textsc{QGSJET-II} for the inclusive selection. 
For the NSD-enhanced sample, the predictions obtained from  \textsc{Pythia\,6}  and 
\textsc{QGSJET-II} agree with the data for most \(\eta\) bins. 
A good description of the measurement for the SD-enhanced sample is provided by both 
\textsc{EPOS} and  \textsc{Pythia\,6}. 

The forward  primarily charged-particle multiplicity density  over  pseudorapidity decreases  
with \(\mid\eta\mid\). 
In the inclusive sample, \( \mathrm{d}  N_{\mathrm{ch}} / \mathrm{d}\eta \) is 
\(3.85 \pm 0.49\)  at  \(\eta = 5.375\)  and 
\(2.61 \pm 0.28\)  at  \(\eta = 6.350\) with negligible statistical uncertainty. 
The pseudorapidity density of the NSD-enhanced sample  varies between 
\(4.80 \pm 0.62\)  and  \(3.17 \pm 0.35\),  while for the SD-enhanced sample it is in the range of 
\(1.49 \pm 0.27\)  to  \(1.20 \pm 0.20\).
The MC predictions for the three samples differ from the data by up to about \(\pm 30\)\%. 
For the inclusive and NSD-enhanced samples, the data in the forward region are in agreement with
the prediction from  \textsc{QGSJET-II} and are between the  \textsc{EPOS} and 
\textsc{Pythia\,8}  results. 
For the SD-enhanced selection, the TOTEM data points are close to the  \textsc{Pythia\,8}  and 
\textsc{HERWIG++}  predictions, while  \textsc{QGSJET-II} underestimates the data. 
The change in the slope of the MC curves close to  \(\eta = 5.3\), 
more visible for the NSD- and SD-enhanced distributions, is due to the event
selection requirement of at least one charged particle in the pseudorapidity region of 
the TOTEM T2 telescopes.

\subsection{Charged-particle multiplicity  density}
\label{nch_energy}
\subsubsection{Energy dependence of the multiplicity density at ATLAS}
\label{average_nch_energy}
\begin{figure*}[t!]
\centering
\begin{minipage}[h]{0.45\textwidth} 
\center{\includegraphics[width=1.0\linewidth]{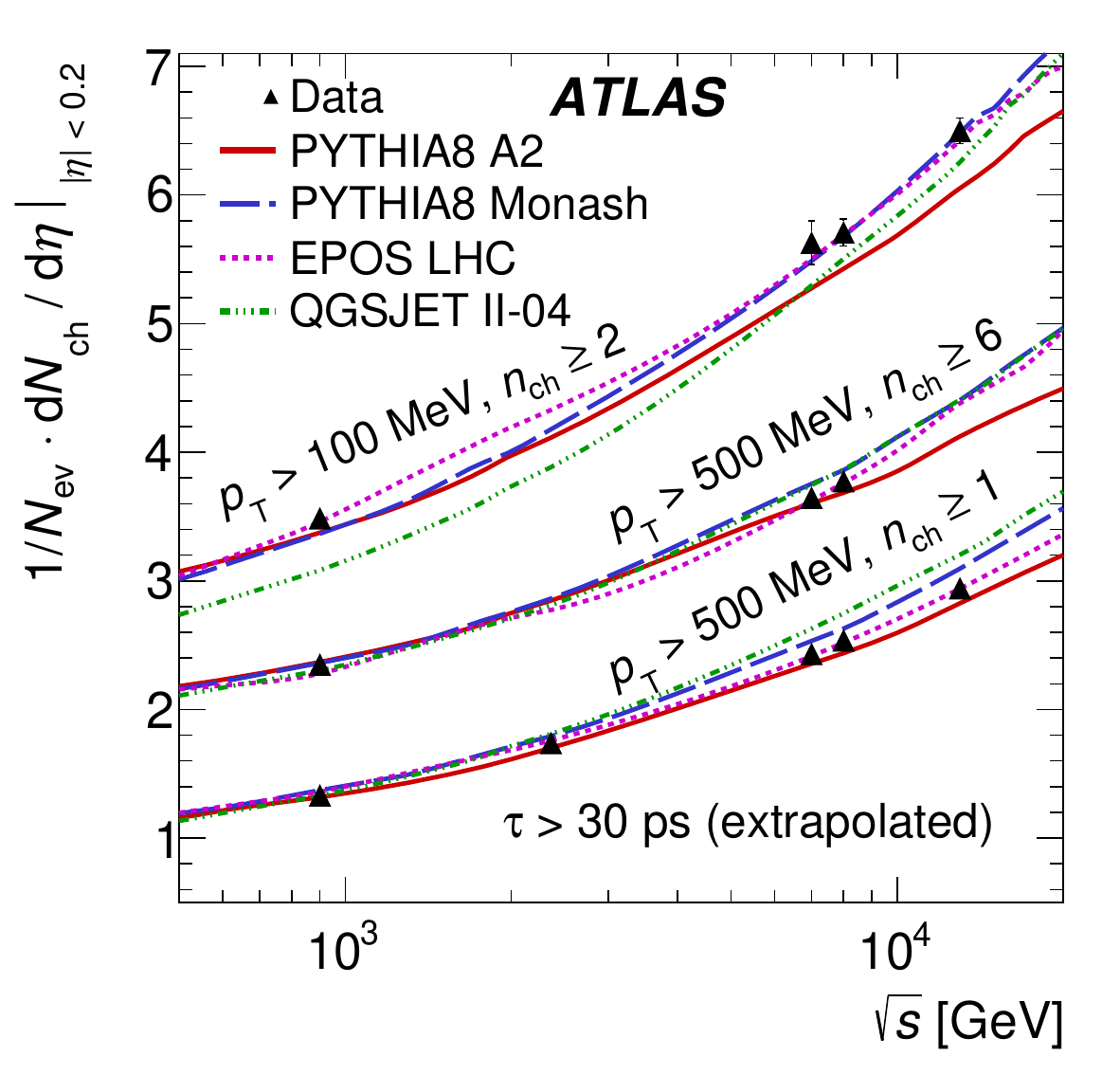}}
(a) 
\\
\end{minipage}
\hspace{2mm}
\begin{minipage}[h]{0.45\textwidth} 
\center{\includegraphics[width=1.0\linewidth]{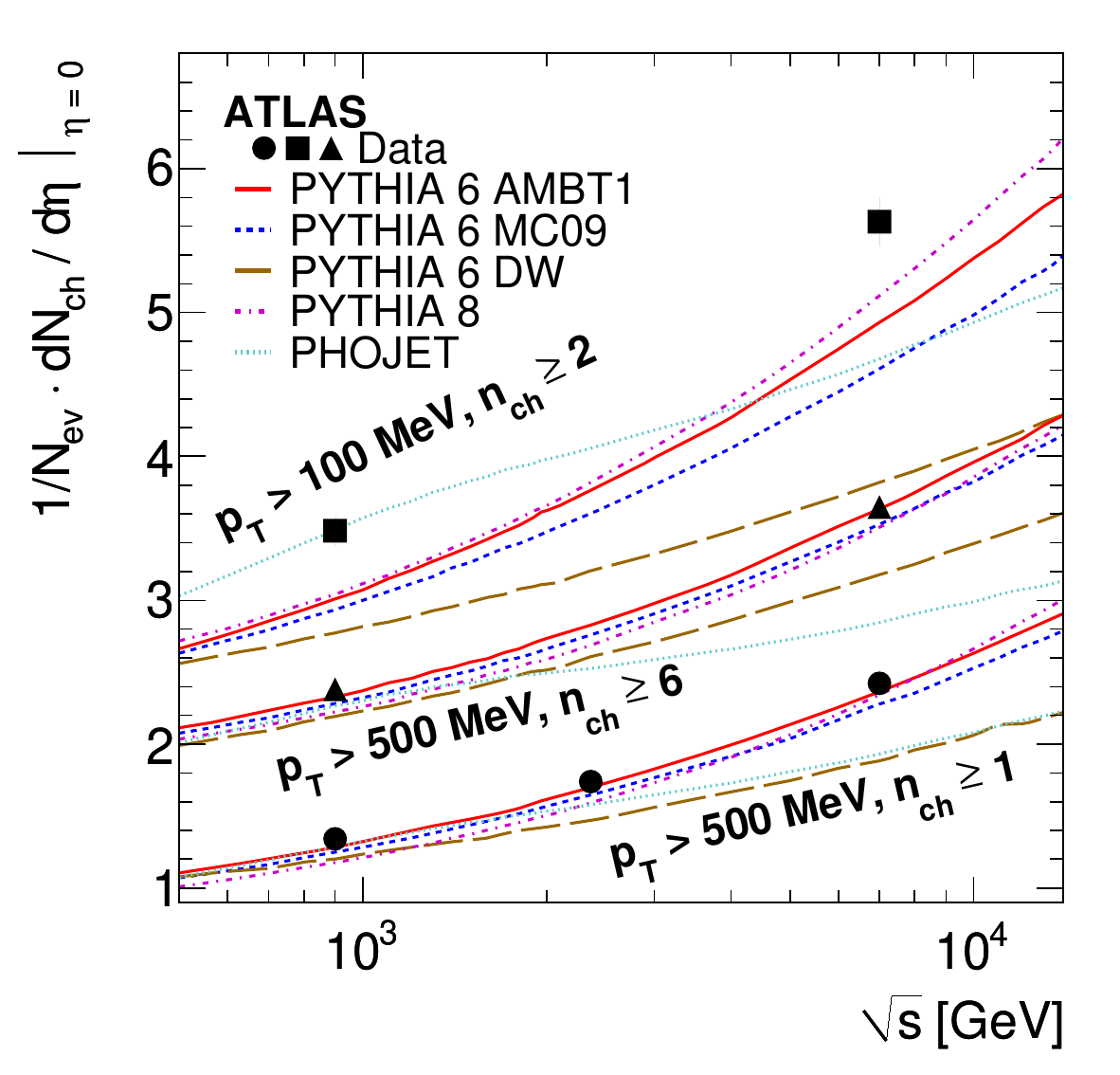}}
(b) 
\\
\end{minipage}
\caption{
The average primary charged-particle multiplicity density in \(pp\) interactions 
per unit of pseudorapidity for  \( \mid \eta \mid <0.2\) 
as a function of the centre-of-mass energy \(\sqrt{s}\)  for events with  
\(n_{\mathrm{ch}} \ge 2\), \(p_{\mathrm{T}} >100\)~\MeV, 
\(n_{\mathrm{ch}} \ge 1\),  \(p_{\mathrm{T}} >500\)~\MeV\ 
and
\(n_{\mathrm{ch}} \ge 6\),  \(p_{\mathrm{T}} >500\)~\MeV\ 
in comparison with predictions of Monte Carlo models
(a) 
\textsc{Pythia\,8} \textsc{A2}, \textsc{Pythia\,8} \textsc{Monash}, \textsc{EPOS} \textsc{LHC} and
\textsc{QGSJET-II}  for \(\sqrt{s}\) from \(0.9\) to \(13\)~\TeV\ \cite{STDM-2015-17} and 
(b) 
\textsc{Pythia\,6} \textsc{AMBT1}, \textsc{Pythia\,6} \textsc{MC09},  \textsc{Pythia\,6} \textsc{DW}, 
\textsc{Pythia\,8} and  \textsc{PHOJET} for \(\sqrt{s}\) from \(0.9\) to \(7\)~\TeV\
\cite{STDM-2010-06}.
The values for \(pp\) centre-of-mass energies are taken from  the ATLAS analyses 
\cite{STDM-2010-06,STDM-2014-19,STDM-2015-02,STDM-2015-17}. 
The results have been extrapolated to include charged strange baryons
(charged particles with a mean lifetime of \(30 < \tau < 300\)~ps). 
The data are shown as black triangles with vertical error bars representing the total uncertainty. 
They are compared to various MC predictions, 
which are shown as coloured lines.
Taken from (a) Ref.~\cite{STDM-2015-17}  and  (b)  Ref.~\cite{STDM-2010-06}.
}
\label{fig_energy}
\end{figure*}

The energy dependence of  primary charged-particle multiplicity density, 
\( 1/N_{\mathrm{ev}} \cdot \mathrm{d} N_{\mathrm{ch}}/ \mathrm{d} \eta \mid_{\eta =0} \),
is of interest  because it 
\begin{enumerate}
\item
provides information about the  basic properties of \(p p\) collisions, 
\item
is
related to the average energy density  achieved in the interaction of protons, 
\item
constitutes a reference for the comparison with heavy ion collisions. 
\end{enumerate}

The average primary charged-particle multiplicity  in \(pp\) interactions  per unit of pseudorapidity,
multiplicity density,   for  \(\mid\eta\mid <0.2\)  as a function of the  CM energy \(\sqrt{s}\) 
in three separate  PS regions for events with 
\(n_{\mathrm{ch}} \ge 2\), \(p_{\mathrm{T}} >100\)~\MeV, 
\(n_{\mathrm{ch}} \ge 1\),  \(p_{\mathrm{T}} >500\)~\MeV\ 
and
\(n_{\mathrm{ch}} \ge 6\),  \(p_{\mathrm{T}} >500\)~\MeV\ 
are shown in  Fig.~\ref{fig_energy}. 
The results are compared to predictions of MC models tuned to a wide range of measurements.
The comparison with  the MC models 
\textsc{Pythia\,8} \textsc{A2}, \textsc{Pythia\,8} \textsc{Monash}, \textsc{EPOS} \textsc{LHC},
\textsc{QGSJET-II}  for \(\sqrt{s}\) from \(0.9\) to \(13\)~\TeV\ 
\cite{STDM-2015-17}
and
\textsc{Pythia\,6} \textsc{AMBT1}, \textsc{Pythia\,6} \textsc{MC09},  \textsc{Pythia\,6} \textsc{DW}, 
\textsc{Pythia\,8}, \textsc{PHOJET} for \(\sqrt{s}\) from \(0.9\) to \(7\)~\TeV\ \cite{STDM-2010-06}
is shown  in Fig.~\ref{fig_energy}(a) and Fig.~\ref{fig_energy}(b),  respectively.

The  primary charged-particle multiplicity density in the central pseudorapidity 
region at \(\sqrt{s} = 13\)~\TeV\  for events with  \(n_{\mathrm{ch}} \ge 2\),
 \(p_{\mathrm{T}} >100\)~\MeV\  is measured for fiducial PS
to be  \(6.42 \pm 0.10\),   by averaging over \(\mid\eta\mid < 0.2\); 
the quoted error is the systematic uncertainty, the statistical uncertainty is negligible. 
In order to compare with other measurements, it is corrected for the contribution from 
strange baryons (and therefore extrapolated to primary charged particles with \(\tau > 30\)~ps) 
by a correction factor of  \(1.0121 \pm 0.0035\). 
The central value is taken from \textsc{EPOS};  the systematic uncertainty is taken from the difference  between 
\textsc{EPOS}  and  \textsc{Pythia\,8} \textsc{A2},  and the statistical uncertainty is negligible. 
The mean number of primary charged particles after the correction is 
\(6.50 \pm 0.10\)  at \(\sqrt{s} = 13\)~\TeV\   for events with 
\(n_{\mathrm{ch}} \ge 2\), \(p_{\mathrm{T}} >100\)~\MeV. 

The mean number of primary charged particles in the central region is computed 
by averaging over \(\mid\eta\mid < 0.2\)  and found to be 
\(2.874 \pm 0.001\  \mathrm{(stat)} \pm 0.033\ \mathrm{(syst)} \) 
at \(\sqrt{s} = 13\)~\TeV\  for events with 
\(n_{\mathrm{ch}} \ge 1\), \(p_{\mathrm{T}} >500\)~\MeV. 
This measurement is  corrected for the contribution from strange baryons. 
The prediction from \textsc{EPOS} is used to perform the extrapolation, 
and the deviation from the \textsc{Pythia\,8} \textsc{Monash}
prediction is taken as a systematic uncertainty and symmetrised to give \(1.024 \pm 0.009\).

\begin{table*}[t!]
\centering 
\caption{
Central primary charged-particle multiplicity density,
\(1/N_{\mathrm{ev}}\cdot d N_{\mathrm{ch}}/d \eta |_{\eta = 0} \),    
for five  phase spaces at \(\sqrt{s} = 13\)~\TeV\ \cite{STDM-2015-02,STDM-2015-17}
and \(\sqrt{s} = 8\)~\TeV\ \cite{STDM-2014-19}. 
The results are given for the fiducial definition \(\tau > 300\)~ps, 
as well as for the previously  used fiducial definition \(\tau > 30\)~ps. 
Taken from Refs.~\cite{STDM-2014-19,STDM-2015-02,STDM-2015-17}.
}
\label{tab:average_nch_energy_13-8TeV}
\medskip
 \begin{tabular}{lll@{~~~~}cll}
\hline
\hline
 \(\sqrt{s}\)
&\multicolumn{2}{l}{Phase Space}
&
&\multicolumn{2}{l}{Multiplicity Density 
}
\\
\cline{2-3}
\cline{5-6}

	[TeV]
&	\(n_{\mathrm{ch}}\ge\)
&  	\(p_{\mathrm{T}}^{\mathrm{min}}\) [MeV]
&
&  	\(\tau > 300\)~ps (Fiducial) 
&  	\(\tau > 30\)~ps  (Extrapolated)
\\

\hline
13
& 2 	& 100   &	&6.42\(\pm\)0.10	& 6.50\(\pm\)0.10	\\
& 1 	& 500 	&	&2.87\(\pm\)0.03 & 2.99\(\pm\)0.03 \\[1mm]

\hline
8
&2 	& 100 &	&5.64\(\pm\)0.10	 	& 5.71\(\pm\)0.11\\
&1 	& 500 &	&2.48\(\pm\)0.03 		& 2.54\(\pm\)0.04\\
&6 	& 500 &	&3.68\(\pm\)0.04 		& 3.78\(\pm\)0.05\\
&20 & 500&	&6.50\(\pm\)0.05 		& 6.66\(\pm\)0.07\\
&50 & 500&	&12.40\(\pm\)0.15 		& 12.71\(\pm\)0.18	\\
\hline
\hline
\end{tabular}
\end{table*}

A summary of central primary charged-particle multiplicity densities at \(\eta = 0 \)  
in all measured PS at \(\sqrt{s}  = 8\), \(13\)~\TeV\
is given in  Table~\ref{tab:average_nch_energy_13-8TeV}.
The primary charged-particle multiplicity density increases by a factor of \(2.2\) 
when \(\sqrt{s}\)  increases by a factor of about \(14\)  from  \(0.9\)~\TeV\ to \(13\)~\TeV. 

These extrapolated results 
are
from  Table~\ref{tab:average_nch_energy_13-8TeV},
are shown in Fig.~\ref{fig_energy}(a)  \cite{STDM-2010-06,STDM-2014-19}  and
compared to  predictions of  the MC models
\textsc{Pythia\,8} \textsc{A2}, \textsc{Pythia\,8} \textsc{Monash}, \textsc{EPOS} \textsc{LHC} and
\textsc{QGSJET-II}  for \(\sqrt{s}\) from \(0.9\) to \(13\)~\TeV\ \cite{STDM-2015-17}. 
The predictions  of \textsc{EPOS}   and  \textsc{Pythia\,8} \textsc{MONASH}  match the data well
at \(\sqrt{s} = 13\)~\TeV\   for events with  \(n_{\mathrm{ch}} \ge 2\), \(p_{\mathrm{T}} >100\)~\MeV. 
For \textsc{Pythia\,8} \textsc{A2},  
the agreement is not as good as that observed when measuring particles with
\(p_{\mathrm{T}} >500\)~\MeV\   \cite{STDM-2015-02}.
For events with   \(n_{\mathrm{ch}} \ge 1\), \(p_{\mathrm{T}} >500\)~\MeV\
at \(\sqrt{s} = 13\)~\TeV\  \textsc{EPOS}  and \textsc{Pythia\,8} \textsc{A2}  
describe the dependence on  \(\sqrt{s}\) very well,  while
\textsc{Pythia\,8} \textsc{Monash} and  \textsc{QGSJET-II}  
predict a steeper rise in multiplicity with  \(\sqrt{s}\).

In order to make consistent comparisons  of pseudorapidity  density at \(8\)~\TeV\  
\cite{STDM-2014-19}
with  other measurements,   these results are corrected to the earlier \(\tau  > 30\)~ps definition of 
stable particles,  using  the factor 
\(1.012 \pm 0.004\) in the \( p_{\mathrm{T}} > 100 \)~\MeV\ PS
and 
\(1.025 \pm 0.008\) in the \( p_{\mathrm{T}} > 500 \)~\MeV\  PS
derived from predictions of the \textsc{EPOS} \textsc{LHC}  tune 
with uncertainties following comparisons of the predictions of different MC models. 
Results at \(8\)~\TeV\  are shown in Fig.~\ref{fig_energy}(a) for the PS
(\( p_{\mathrm{T}} > 500 \)~\MeV, \( n_{\mathrm{ch}} \ge 1;\ 6 \))
and 
(\( p_{\mathrm{T}} > 100 \)~\MeV, \( n_{\mathrm{ch}} \ge 2 \)). 
It can be seen that the total uncertainty in the measurement at 
\(\sqrt{s}  = 8\)~\TeV\ is about 30--40\% less than for the study with 
the \(\sqrt{s}  = 7\)~\TeV\ data. 
This was achieved due to improved knowledge of the ID material distribution  \cite{Lukas:2016cge},
which reduced the dominant source of systematic uncertainty by more than 50\% with
respect to the  \(\sqrt{s}  = 0.9,\ 2.36,\ 7\)~\TeV\ measurements. 
The best description of the data is given by \textsc{EPOS}. 
The predictions of the \textsc{Pythia\,8} tunes provide a fair description of the shape of the 
multiplicity dependence with CM energy. 
As in the case of the other presented distributions,  \textsc{QGSJET-II} calculations
give the worst description.

The values for  three PS regions  are shown in  Fig.~\ref{fig_energy}(b)  with comparison of  
\textsc{Pythia\,6} \textsc{AMBT1},
\textsc{Pythia\,6} \textsc{MC09}, 
\textsc{Pythia\,6} \textsc{DW}, 
\textsc{Pythia\,8} 
and 
\textsc{PHOJET}
predictions  for \(\sqrt{s}\) from \(0.9\) to \(7\)~\TeV\ and in 
Table~\ref{tab:average_nch_energy_7TeV}
\cite{STDM-2010-06}. 

\begin{table*}[t!]
\centering 
\caption{
Central primary charged-particle multiplicity density,    
\(1/N_{\mathrm{ev}}\cdot d N_{\mathrm{ch}}/d \eta |_{\eta = 0}\),   
for events with centre-of-mass energies  at \(\sqrt{s} = 0.9,\ 2.36,\ 7\)~\TeV\
for three different phase spaces \(n_{\mathrm{ch}} \ge 2\), \(p_{\mathrm{T}} >100\)~\MeV\
and \(n_{\mathrm{ch}} \ge 1;\ 6\),  \(p_{\mathrm{T}} >500\)~\MeV\ 
\cite{STDM-2010-06}. 
The results for  primary charged-particle  average total multiplicity density 
are
denoted by the  symbol \(^{(\ast)}\)
for  the phase space \(|\eta| < 2.5\), \(n_{\mathrm{ch}} \ge 2\), \(p_{\mathrm{T}} >100\)~\MeV.
The results for  the total    multiplicity density  of 
primary charged particles
are
denoted by the  symbol \(^{(\dagger)}\)
for  the phase space \(|\eta| < 2.5\), \(n_{\mathrm{ch}} \ge 2\), \(p_{\mathrm{T}} >0\)~\MeV.
For MC  sufficient statistics were generated such that the statistical uncertainty is smaller than the
last digit quoted.
The results  were taken from Ref.\ \cite{STDM-2010-06}.
}
\label{tab:average_nch_energy_7TeV}
\medskip
 \begin{tabular}{lcll@{~~~~}cll}
\hline
\hline

 \(\sqrt{s}\) 
&
& \multicolumn{2}{l}{Phase Space}
&
& \multicolumn{2}{l}{Multiplicity Density 
}
\\
\cline{3-4}
\cline{6-7}

	[TeV]
&
& \(n_{\mathrm{ch}}\ge \)
&  \(p_{\mathrm{T}}^{\mathrm{min}}\)  [MeV]
&
&  Experimental Results 
&  \textsc{Pythia\,6} \textsc{AMBT1} 
\\

\hline
 7	&			&2	& 100	&&5.630\(\pm\)0.003 (stat)	\(\pm\)0.169 (syst) & 4.93	\\
 
0.9 	&			& 		&  		&&3.483\(\pm\)0.009 (stat)	\(\pm\)0.106 (syst) & 3.01	\\[1mm]

\hline
7   &\(^{(\ast)}\)	&2	& 100	&&5.881\(\pm\)0.002 (stat)	\(\pm\)0.276 (syst) &        \\
0.9 &\(^{(\ast)}\) 	& 	&  		&&3.614\(\pm\)0.006 (stat)	\(\pm\)0.170 (syst) & 		\\[1mm]
	
\hline
7 	&\(^{(\dagger)}\)&2	& 0		&&6.252\(\pm\)0.002 (stat)	\(\pm\)0.304 (syst) &        \\
0.9 	&\(^{(\dagger)}\)& 	&  		&&3.849\(\pm\)0.006 (stat)	\(\pm\)0.185 (syst) & 		\\[1mm]
	
\hline
7	&			&1	&500	&&2.423\(\pm\)0.001 (stat)	\(\pm\)0.050 (syst) & 2.36	\\
2.36&			&	&		&&1.740\(\pm\)0.019 (stat)	\(\pm\)0.058 (syst) & 1.70	\\	
0.9 	&			& 	& 		&&1.343\(\pm\)0.004 (stat)	\(\pm\)0.027 (syst) & 1.28	\\[1mm]
	
\hline
7	&			&6	&500	&&3.647\(\pm\)0.002 (stat)	\(\pm\)0.052 (syst) & 3.63	\\
0.9	&			&  	& 		&&2.380\(\pm\)0.009 (stat)	\(\pm\)0.027 (syst) & 2.33	\\

\hline
\hline
\end{tabular}
\end{table*}

The  PS region with the largest minimum  \( p_{\mathrm{T}}\) and 
the highest minimum multiplicity
(\( p_{\mathrm{T}} > 500 \)~\MeV, \( n_{\mathrm{ch}} \ge 6 \)), 
which is the region with the least amount of diffraction, 
is the one where the models vary the least and the energy extrapolations of most
models  
are
in the best agreement  with the data. 
For the most inclusive measurements,  none of the models agree with the data, 
and the spread at \(\sqrt{s}  = 7\)~\TeV\ 
of the expected values is almost  one-third  of the mean predicted value. 
The observed value is significantly higher at this energy than  in any of the models.

The  total multiplicity density of charged particles with  \( p_{\mathrm{T}} > 100 \)~\MeV\ 
within the  \(\mid\eta\mid < 2.5 \)  are  computed as the mean of the distributions 
shown in  Fig.~\ref{fig_7_eta}(a) and Fig.~\ref{fig_09_eta}(a).
They are found to be  
\(5.881 \pm 0.002\ \mathrm{(stat)} \pm 0.276\ \mathrm{(syst)} \)   at \(\sqrt{s}  = 7\)~\TeV\ 
and 
\(3.614 \pm 0.006\ \mathrm{(stat)} \pm  0.170\ \mathrm{(syst)} \)  at \(\sqrt{s}  = 0.9\)~\TeV\
(see Table~\ref{tab:average_nch_energy_7TeV}). 
%
These charged-particle  total  multiplicities density in the full pseudorapidity region,  \(-2.5 < \eta < 2.5\), 
are
\(29.04 \pm 0.01\ \mathrm{(stat)} \pm 1.38\ \mathrm{(syst)} \)   at \(\sqrt{s}  = 7\)~\TeV\ 
and
\(18.07 \pm 0.03\ \mathrm{(stat)} \pm  0.85\ \mathrm{(syst)} \)  at \(\sqrt{s}  = 0.9\)~\TeV\ 
and are in good agreement with the results presented in  Table~\ref{tab:average_nch}.

With extrapolation to  \( p_{\mathrm{T}} = 0 \)~\MeV,
these numbers were multiplied by the model-dependent scale factors.  
The averaged inclusive charged-particle multiplicity for events with two or more particles 
for  the kinematic region with  \( p_{\mathrm{T}} \ge 0 \)~\MeV\
is found to be 
\(6.252 \pm 0.002\ \mathrm{(stat)} \pm 0.304\ \mathrm{(syst)} \) 
at \(\sqrt{s}  = 7\)~\TeV\ 
and
\(3.849 \pm 0.006\ \mathrm{(stat)} \pm 0.185\ \mathrm{(syst)} \) 
at \(\sqrt{s}  = 0.9\)~\TeV\
(see Table~\ref{tab:average_nch_energy_7TeV}). 
%
These are \(\approx 6\)\% higher than  average multiplicities for  \( p_{\mathrm{T}} > 100 \)~\MeV.
This result is interpreted as the average total inelastic multiplicity 
for events with two or more particles  within  \(\mid\eta\mid < 2.5 \).


\begin{figure*}[t!]
\centering
\begin{minipage}[h]{0.45\textwidth} 
\center{\includegraphics[width=1.0\linewidth]{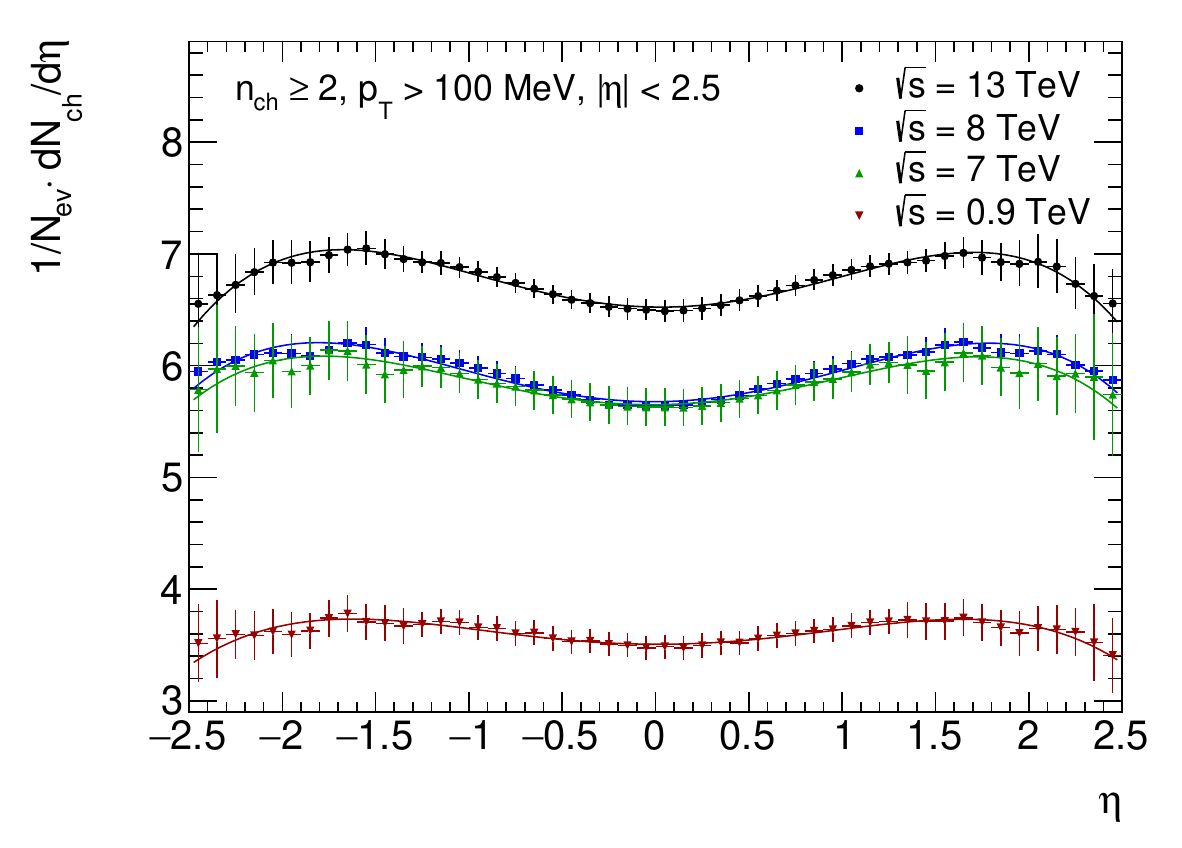}}  
(a)
\\
\end{minipage}
\hspace{2mm}
\begin{minipage}[h]{0.45\textwidth} 
\center{\includegraphics[width=1.0\linewidth]{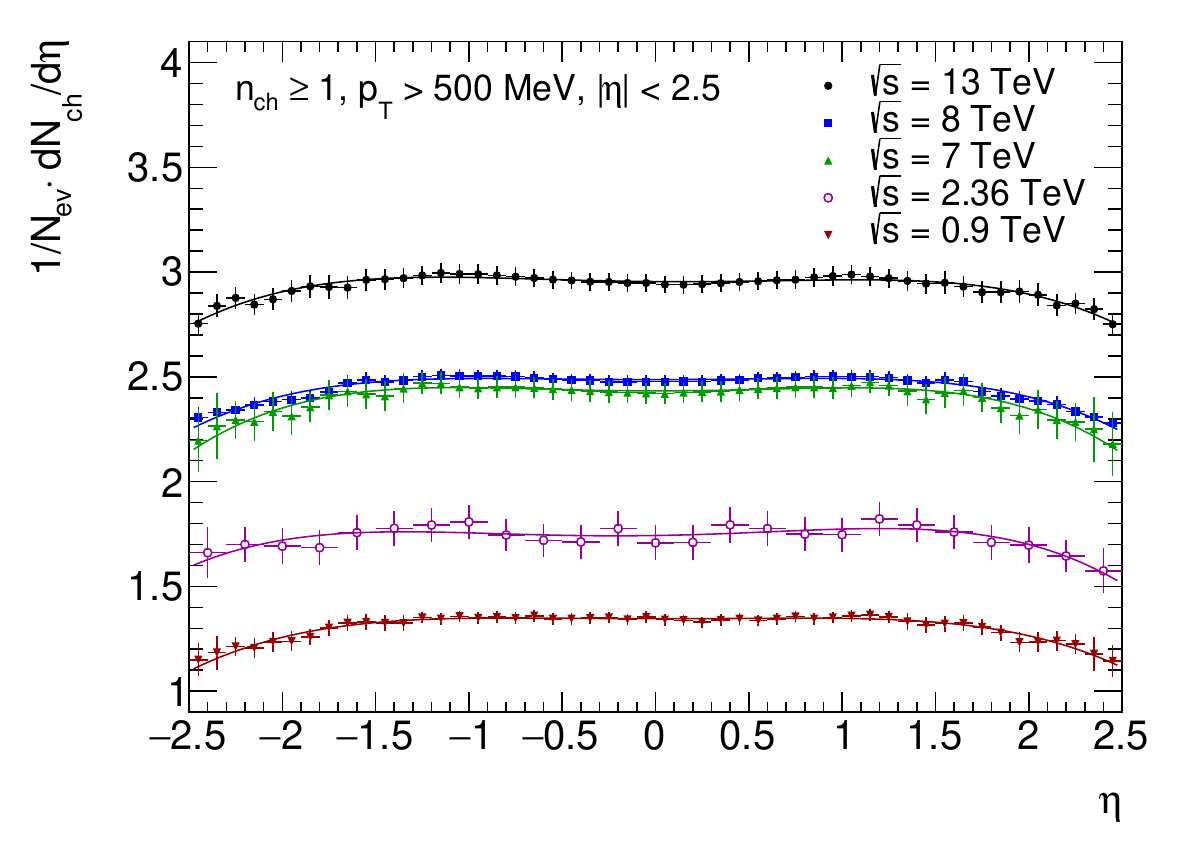}}  
(b)
\\
\end{minipage}
\caption{
The  primary charged-particle  average total multiplicity density, 
\( 1/N_{\mathrm{ev}} \cdot \sum \mathrm{d} N_{\mathrm{ch}}/ \mathrm{d} \eta \),  
dependence on pseudorapidity region \(-2.5 < \eta < 2.5\)  for the ATLAS 
results  for the charged-particle  with  
(a) \(n_{\mathrm{ch}} \ge 2\), \(p_{\mathrm{T}} > 100\)~\MeV\   and 
(b)  \(n_{\mathrm{ch}} \ge 1\), \(p_{\mathrm{T}} > 500\)~\MeV\ 
at  the centre-of-mass energies  \(\sqrt{s} = 0.9\), \(2.36\), \(7\), \(8\) and \(13\)~\TeV\
\cite{STDM-2010-06,STDM-2014-19,STDM-2015-02,STDM-2015-17}.
The coloured symbols represent the data. 
The vertical bars represent statistical and systematic uncertainties added in quadrature.
The black curves show the results of the fits with the fourth-degree polynomial function. 
Taken from Ref.~\cite{Kulchitsky:2022gkm}.
}
\label{fig_events_average_eta}
\end{figure*}

For a correct comparison of  charged-particle multiplicity  and  average transverse momentum distributions 
for different energies or  PS regions,  the scaled multiplicity is introduced as follows:
\begin{equation}
\label{eq_mch}
z =
\frac{ n_{\mathrm{ch}} (\sqrt{s}, p_{\mathrm{T}}^{\mathrm{min}}) }{
\langle 
n_{\mathrm{ch}} (\sqrt{s}, p_{\mathrm{T}}^{\mathrm{min}}) 
\rangle }. 
\end{equation}
For example, a  comparison of results for different  PS regions
with two \( p_{\mathrm{T}}^{\mathrm{min}} \) thresholds,   was presented in  
Ref.~\cite{ATLAS:2022wvk}.

\begin{table*}[t!]
\centering 
\caption{
The   average total multiplicity,  
\( \langle n_{\mathrm{ch}} (\sqrt{s}, p_{\mathrm{T}}^{\mathrm{min}})\rangle  \),
and relative uncertainty, 
\(\frac{\delta \langle n_{\mathrm{ch}} (\sqrt{s}, p_{\mathrm{T}}^{\mathrm{min}})\rangle}{ \langle n_{\mathrm{ch}} (\sqrt{s}, p_{\mathrm{T}}^{\mathrm{min}})\rangle }\),
as the  results of  the  fits with a polynomial function of the primary  charged-particle average  
multiplicity distributions  in pseudorapidity region \(-2.5 < \eta < 2.5\)  
for the events samples with  \(p_{\mathrm{T}} > 100\)~\MeV\   and \(p_{\mathrm{T}} > 500\)~\MeV\ 
at  centre-of-mass energies \(\sqrt{s} = 0.9\), \(2.36\), \(7\), \(8\) and \(13\)~\TeV\
using the ATLAS  results
\cite{STDM-2010-06,STDM-2014-19,STDM-2015-02,STDM-2015-17}.
Taken from Ref.~\cite{Kulchitsky:2022gkm}.
}
\label{tab:average_nch}
\medskip
 \begin{tabular}{lllll}
\hline
\hline

\(\sqrt{s}\) [TeV]
&\(n_{\mathrm{ch}}\ge\) 
&\(p_{\mathrm{T}}^{\mathrm{min}}\) [MeV]
& Average Total Multiplicity 
& Relative Uncertainty
\\

\hline
13		&\(2\)	& \( 100\) 	& 	33.88\(\pm\)0.11	& 0.0032\\
		&\(1\)	& \( 500\) 	& 	14.66\(\pm\)0.04	& 0.0027\\[1mm]
		
\hline
8		&\(2\)	& \( 100\) 	& 	29.81\(\pm\)0.10	& 0.0034\\
		&\(1\)	& \( 500\)	& 	12.25\(\pm\)0.03	& 0.0024\\[1mm]
		
\hline
7		&\(2\)	& \( 100\) 	& 	29.40\(\pm\)0.19	& 0.0065\\
		&\(1\)	& \( 500\) 	& 	11.98\(\pm\)0.05	& 0.0042\\[1mm]
		
\hline
2.36	&\(1\)	& \( 500\) 	& 	8.66\(\pm\)0.51	& 0.0589\\[1mm]

\hline
0.9		&\(2\)	& \( 100\) 	& 	18.06\(\pm\)0.12	& 0.0066\\
		&\(1\)	& \( 500\) 	& 	6.53\(\pm\)0.03	& 0.0046\\
\hline
\hline
\end{tabular}
\end{table*}

A fit with a fourth-degree polynomial function  of the  primary charged-particle 
multiplicity  density distributions in the  pseudorapidity region \(-2.5 < \eta < 2.5\) 
was used in  \cite{Kulchitsky:2022gkm} for  the calculation of an 
average total multiplicity, 
\( \langle n_{\mathrm{ch}} ( \sqrt{s},  p_{\mathrm{T}}^{\mathrm{min}} ) \rangle \),
for different   CM energies and \(p_{\mathrm{T}}^{\mathrm{min}} \) 
using the ATLAS results  \cite{STDM-2010-06,STDM-2014-19,STDM-2015-02,STDM-2015-17}.
The \( 1/N_{\mathrm{ev}} \cdot d N_{\mathrm{ch}}/d\eta \) distributions 
over pseudorapidity are shown  in  Fig.~\ref{fig_events_average_eta}.
The average multiplicity, 
\( \langle n_{\mathrm{ch}} ( \sqrt{s},  p_{\mathrm{T}}^{\mathrm{min}} ) \rangle \),
resulting from 
the
fit  of these distributions  with  the fourth-degree polynomial function 
is
presented in  Table~\ref{tab:average_nch}. 

The average multiplicities from Table~\ref{tab:average_nch}  were used for  
the
calculation  of 
horizontal axes using  Eq.~(\ref{eq_mch}) 
for 
the
correct comparison of  primary charged-particle multiplicity distributions
and multiplicity dependences of an average transverse momentum  in 
Sec.\ \ref{nch_energy},  and  for 
the
KNO scaling study in Sec.\ \ref{KNO_scaling}.

%
\subsubsection{Energy dependence of the multiplicity density of the  LHC experiments}
\label{nch_energy_LHC}

\begin{figure*}[t!]
\centering
\begin{minipage}[h]{0.45\textwidth} 
\center{\includegraphics[width=1.0\linewidth]{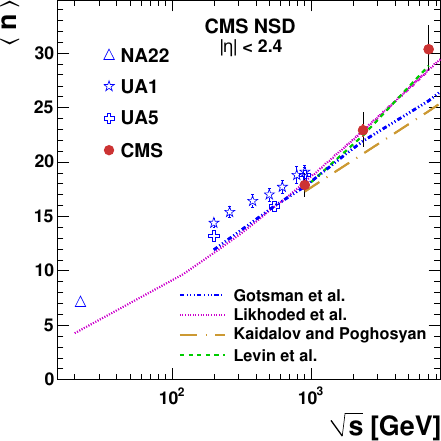}}
(a) 
\\
\end{minipage}
\hspace{2mm}
\begin{minipage}[h]{0.45\textwidth} 
\center{\includegraphics[width=1.0\linewidth,height=0.33\textheight]{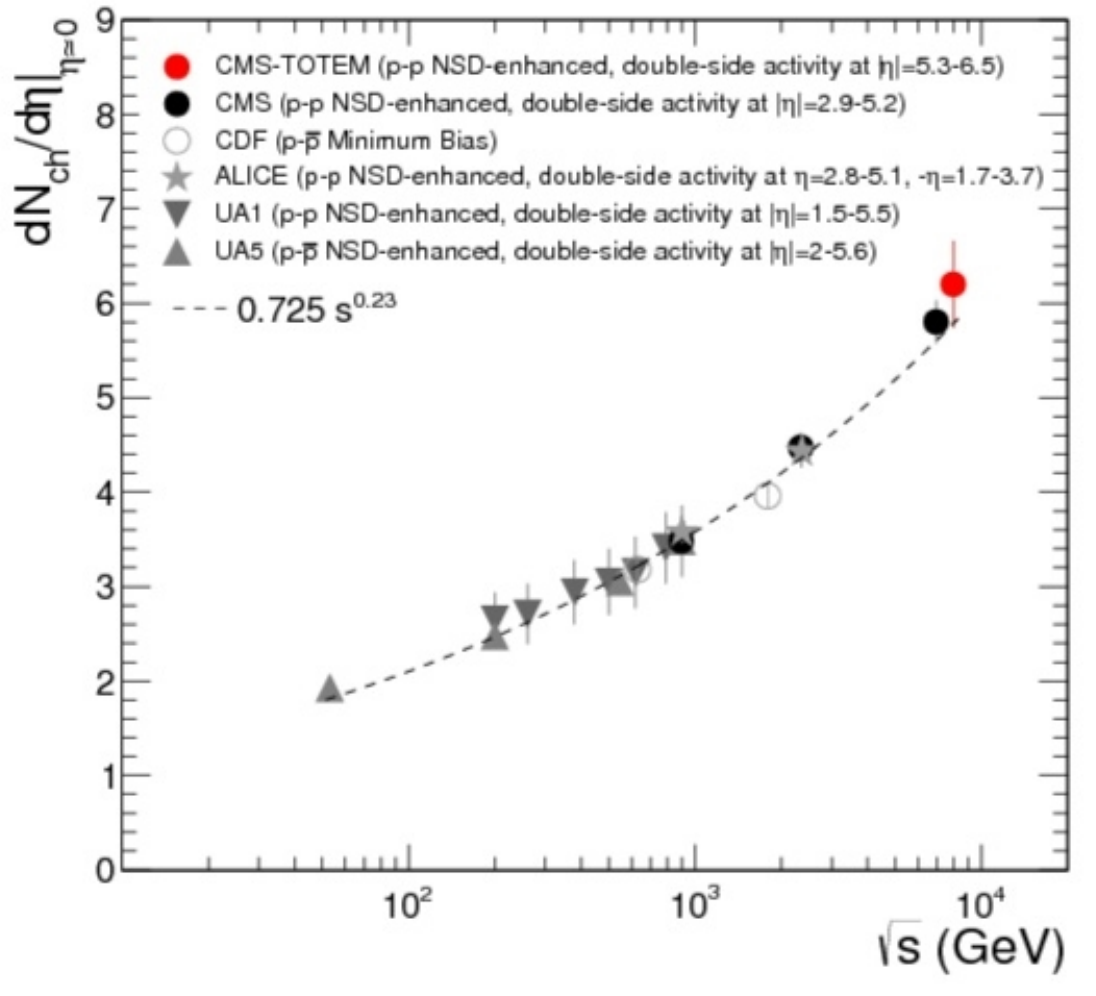}}
(b) 
\\
\end{minipage}
\caption{
(a)
The evolution of the  average total primary charged-particle multiplicity  in 
\(pp\) interactions  with a centre-of-mass energy for \( \mid \eta \mid < 2.4\),  including 
the data from lower-energy experiments 
NA22 \cite{EHSNA22:1987syd}, 
UA1  \cite{UA1:1989bou},  
and  
UA5   \cite{UA5:1985hzd,UA5:1988gup}  
for \( \mid \eta \mid < 2.5\).
The data are compared with predictions from three analytical Regge-inspired models 
\cite{Gotsman:2009xm,Likhoded:2010pc,Kaidalov:2010ybc} 
and from a saturation model  \cite{Levin:2010dw}. 
Taken from Ref.~\cite{CMS:2010qvf}.
(b)
Primary charged-particle multiplicity density 
\( \mathrm{d} N_{\mathrm{ch}} / \mathrm{d} \eta  |_{\eta =0}\) 
as a function of the  centre-of-mass energy in  \( p p \) and \( p \bar{p} \)  collisions. 
Shown are measurements performed with different  NSD  event selections from
the 
UA1  \cite{UA1:1989bou}, 
UA5  \cite{UA5:1986yef},
CDF  \cite{CDF:2009cxa,CDF:1989nkn},
ALICE  \cite{ALICE:2010cin}, 
CMS  \cite{CMS:2010tjh}, 
and 
CMS--TOTEM \cite{CMS:2014kix}.
The dashed line is a power-law fit to the data.
Taken from Ref.~\cite{CMS:2014kix}.
}
\label{fig_energy_CMS}
\end{figure*}

The  average total primary charged-particle  multiplicity,  
\(\langle n_{\mathrm{ch}} \rangle \), 
is equal to the integral of the corresponding single-particle inclusive density in the  \(\eta\)
interval considered. 
The   \(\langle n_{\mathrm{ch}} \rangle \)  is observed to rise with increasing  CM energy
in hadron-hadron collisions 
\cite{Aachen:1977izz,UA1:1982yyh,UA1:1989bou,UA5:1984zxi,UA5:1985fid,EHSNA22:1987syd,Ames:1983cqw,Rimondi:1993wq}.
The same behaviour is also observed in  \( e^{+} e^{-} \)  collisions, 
in deep-inelastic scattering  \cite{H1:1996ovs},
and in heavy ion collisions  \cite{Kittel:2005fu}.

The CMS  measured  
the 
average total  primary charged-particle multiplicity  for 
\(\mid\eta\mid < 2.4\),   
presented in  Table~\ref{tab:average_nch_energy_CMS}
and  shown  in  Fig.\ \ref{fig_energy_CMS}(a),   where  the CMS data are
compared with experimental data obtained at lower energies
and various theoretical predictions.
Recent Regge-inspired models  \cite{Gotsman:2009xm,Likhoded:2010pc,Kaidalov:2010ybc} 
predict a power-like behaviour, 
among which only  Ref.~\cite{Likhoded:2010pc} 
describes the highest energy data very well. 
Parton saturation models (such as \cite{Levin:2010dw})
predict a strong rise of the central rapidity plateau as well. 

\begin{table*}[t!]
\centering 
\caption{
The CMS average total primary charged-particle multiplicity 
in \(pp\) interactions  with the centre-of-mass energy  for  the inclusive pseudorapidity region  
\( |\eta| < 2.4\) for the data, 
\textsc{Pythia} \textsc{D6T} tune,  \textsc{Pythia\,8} and 
\textsc{PHOJET} events generators at the centre-of-mass energies  \(\sqrt{s} =0.9\),  \(2.36\),
  and  \(7\) \TeV. 
The results were extrapolated to \(p_{\mathrm{T}}^{\mathrm{min}} = 0 \)~\MeV.
For  the  data, the quoted uncertainties are first statistical, then upward and downward systematic.
Taken from Ref.~\cite{CMS:2010qvf}.
}
\label{tab:average_nch_energy_CMS}
\medskip
 \begin{tabular}{ll@{~~~~}cllll}
\hline
\hline

\(\sqrt{s}\) 
&\multicolumn{1}{l}{Phase Space}
& \multicolumn{4}{l}{Average Total Multiplicity}
\\
\cline{2-2}
\cline{4-7}

[TeV]
&	\(n_{\mathrm{ch}}\ge\)
&
& Experimental  Results 
& \textsc{Pythia} \textsc{D6T}
& \textsc{Pythia\,8}
& \textsc{PHOJET} 
\\

\hline

7		&1 
&&30.4\(\pm\)0.2 (stat) \(^{+2.2}_{-2.0}\) (syst) &21.2 &25.8 &23.2\\[2mm]

2.36	&1 
&&22.9\(\pm\)0.5 (stat) \(^{+1.6}_{-1.5}\) (syst) &16.7 &17.8 &18.7\\[2mm]

0.9 	 	&1 
&&17.9\(\pm\)0.1 (stat) \(^{+1.1}_{-1.1}\) (syst) &14.7 &14.9 &17.1\\[1mm]

\hline
\hline
\end{tabular}
\end{table*}

The \textsc{Pythia\,6}   \cite{Sjostrand:2006za} generator and its fragmentation model tuned to CDF data 
\cite{Bartalini:2008zz,Moraes:2007rq}, called  \textsc{Pythia} \textsc{D6T}, 
are
used as a baseline model to simulate inelastic \(pp\) collisions.
At \(7\) \TeV\ a dedicated  \textsc{Pythia} tune  \cite{Moraes:2007rq}
better 
describes
the high multiplicities 
used for correcting the data. 
Alternative tunings that differ mainly in the modelling of MPI have also been considered 
\cite{Bartalini:2008zz,Buckley:2009bj,Skands:2010ak}.
\textsc{PHOJET} \cite{Engel:1994vs,Engel:1995yda}
is used as an alternative event generator that differs mainly 
in the underlying dynamical model for particle production.

Table~\ref{tab:average_nch_energy_CMS} gives an overview of  the
average total primary charged-particle multiplicity
for the data and for the  \textsc{Pythia} \textsc{D6T} tune,  \textsc{Pythia\,8}
and  \textsc{PHOJET} models.
The \textsc{Pythia} \textsc{D6T} tune produces on average too few particles
per event at all energies. 
\textsc{PHOJET} is consistent with the data within uncertainties for  \(\sqrt{s} = 0.9\)~\TeV, 
but is not able to 
properly predict 
the  average total multiplicity  at higher energies. 
\textsc{Pythia\,8} describes best the \(\sqrt{s} = 7\)~\TeV\ data, 
but underestimates \(\langle n_{\mathrm{ch}} \rangle \)  systematically at  all  energies.
The CMS results at \(\sqrt{s} = 0.9\) and \(7\)~\TeV\ presented in 
Table~\ref{tab:average_nch_energy_CMS}  are in  agreement  within the error bars 
with  the ATLAS results at the same energies  with \( p_{\mathrm{T}} > 100 \)~\MeV\ 
in Table~\ref{tab:average_nch}.

The  CM energy dependence of the pseudorapidity distribution at  \(\eta = 0\)
is shown in  Fig.\ \ref{fig_energy_CMS}(b),  which includes data from  various  
experiments for NSD events in  \(pp\) and \( p \bar{p}\) collisions. 
The different experiments do not use identical event selection criteria;
they all include a large fraction of NSD events. 
Particle production at  \(\eta = 0\)  is expected to follow a power-law dependence, 
\begin{equation}
\label{eq_nch_average}
\mathrm{d} N_{\mathrm{ch}} / \mathrm{d} \eta \mid_{\eta =0}\ \
\propto s^{\Delta},
\end{equation}
where  \({\Delta}\) 
is the  Pomeron intercept   \cite{Capella:2000pe} and  the effective  Pomeron intercept  
is defined as 
\begin{equation}
\label{eq_nch_alpha}
\alpha_{\mathrm{eff}} (0) = 1 + \Delta
\end{equation}
with  \(\Delta\)  in the range  \(0.14\) -- \(0.24\)  \cite{dEnterria:2011twh}.
The result of fitting the high-energy  \(pp\) and  \( p  \bar{p}\)  central-pseudorapidity 
particle densities with this function is shown in  Fig.\ \ref{fig_energy_CMS}(b). 
The value of  \( \Delta = 0.23\pm 0.01 \)  is obtained.

\begin{figure*}[t!]
\centering
\begin{minipage}[h]{0.45\textwidth} 
\center{\includegraphics[width=1.0\linewidth,height=0.32\textheight]{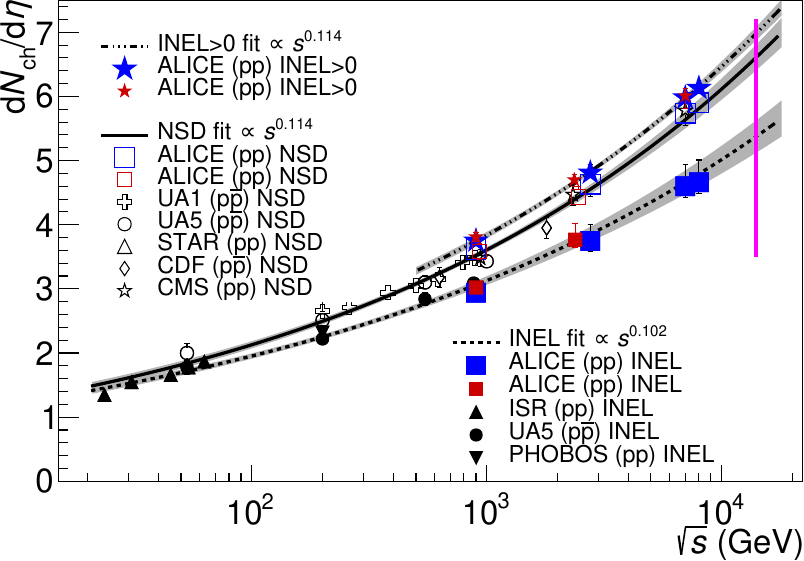}}
(a) 
\\
\end{minipage}
\hspace{2mm}
\begin{minipage}[h]{0.45\textwidth} 
\center{\includegraphics[width=1.0\linewidth,height=0.32\textheight]{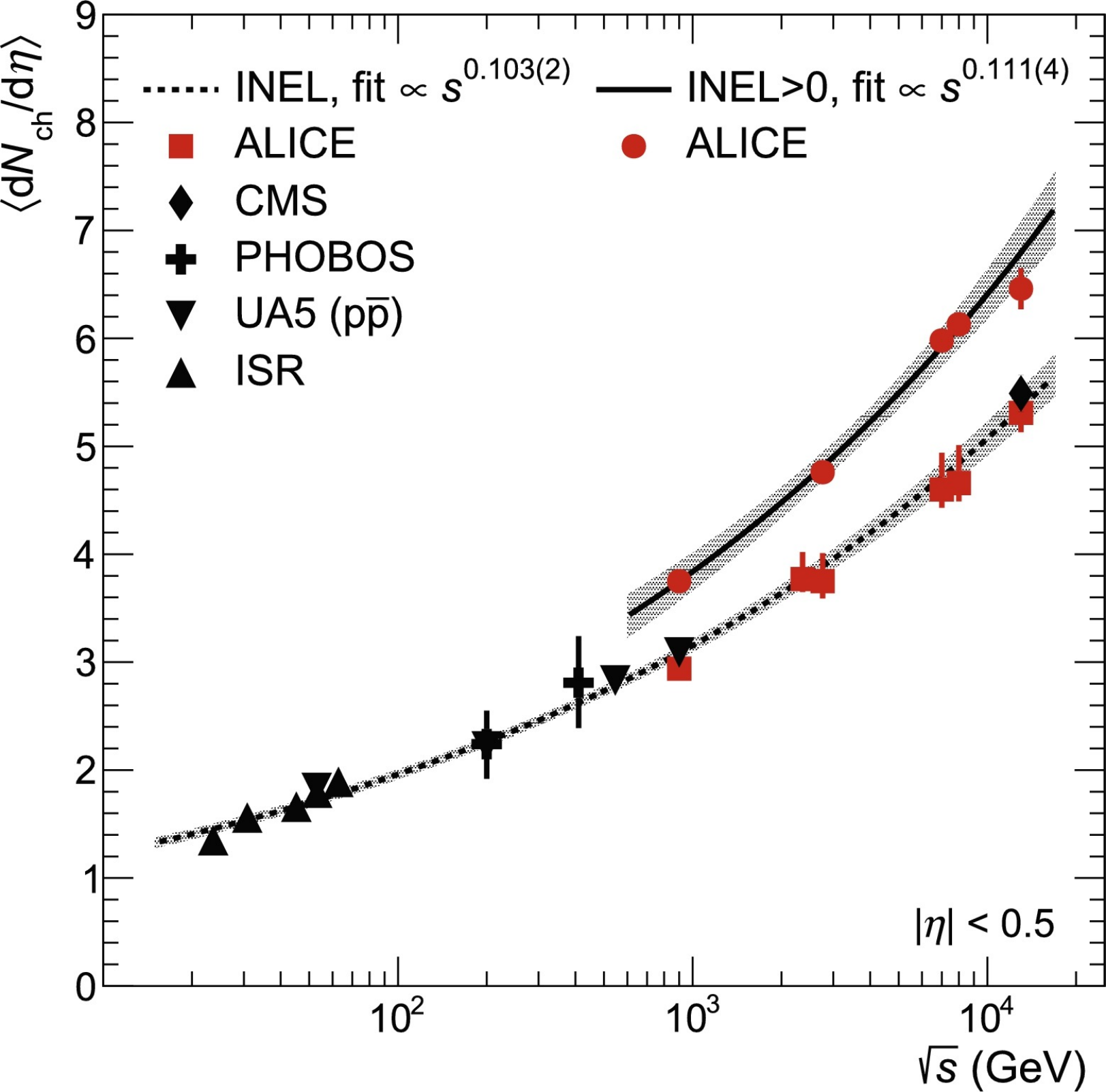}}
(b) 
\\
\end{minipage}
\caption{
(a)
Charged-particle pseudorapidity density in the pseudorapidity region  \( \mid\eta\mid< 0.5\),
\( \mathrm{d} N_{\mathrm{ch}}/  \mathrm{d} \eta \mid_{\eta =0}\). 
Results are given for three conventional event classes:  inelastic (INEL) events,  
non-single diffractive (NSD)  events, 
 and   
events with at least one charged particle in 
\( \mid\eta\mid <1 \)  (INEL$>$0) as a function of a centre-of-mass energy. 
Lines indicate fits with a power-law dependence on \( \sqrt{s} \). 
Grey bands represent  one standard deviation range. 
Data points at the same energy  are shifted horizontally for visibility. 
The nominal  centre-of-mass LHC energy   is indicated by a vertical line. 
Data  are taken
from
the 
ALICE \cite{ALICE:2010cin,ALICE:2010mty}, 
CMS \cite{CMS:2010tjh},
CDF \cite{CDF:1989nkn}, 
ISR  \cite{Aachen:1977izz,Ames:1983cqw}, 
UA1 \cite{UA1:1989bou},
UA5 \cite{UA5:1986yef,UA5:1982ygd,UA5:1987rzq},  
STAR \cite{STAR:2008med},
PHOBOS \cite{PHOBOS:2004xnp}, 
and  
CNPS \cite{CERN:1981mjm}.
Taken from Ref.~\cite{ALICE:2015olq}.
(b)
Charged-particle pseudorapidity density measured in the central pseudorapidity region 
\( \mid\eta\mid< 0.5\)   for INEL and INEL$>$0 events  measured  by  
ALICE  \cite{ALICE:2010cin,ALICE:2010mty,ALICE:2015olq,ALICE:2015qqj},
CMS \cite{CMS:2015zrm}, ACHM \cite{Aachen:1977izz}, UA5 \cite{UA5:1986yef,UA5:1982ygd,UA5:1987rzq}
and  PHOBOS \cite{PHOBOS:2010eyu}.
The uncertainties are the quadratic sum of statistical and systematic contributions. 
The lines are power-law fits of the energy dependence of the data,  and the grey 
bands represent the standard deviation of the fits.
Taken from Ref.~\cite{ALICE:2015qqj}.
}
\label{fig_energy_ALICE}
\end{figure*}

In ALICE, 
the  definition for multiplicity density  in \(pp\) collisions,  
\( 1/N_{\mathrm{ev}} \cdot \mathrm{d} N_{\mathrm{ch}}/ \mathrm{d} \eta \mid_{\eta =0}\),  
 is an integral of the data over the pseudorapidity range  \( \mid\eta\mid< 0.5\).
The results of the measurements of  multiplicity density are  shown in
Fig.\ \ref{fig_energy_ALICE} and  given  in   Table~\ref{tab:average_nch_energy_ALICE}.
Results are given for three conventional event classes:  inelastic (INEL) events,  
non-single diffractive (NSD)  events,  
and  
events with at least one charged particle in  \( \mid\eta\mid < 1\) (INEL$>$0).

\begin{table*}[t!]
\centering 
\caption{
Summary of the ALICE  measurements  and extrapolations of primary charged-particle multiplicity density,
\(  1/N_{\mathrm{ev}} \cdot  \mathrm{d} N_{\mathrm{ch}} / \mathrm{d} \eta |_{\eta =0} \). 
The experimental results were obtained  for  centre-of-mass energies 
\(\sqrt{s}= 0.9\),  \(2.76\), \(7\), 
and \(8\)~\TeV\  \cite{ALICE:2015olq}, \(\sqrt{s}= 13\)
\cite{ALICE:2015qqj} and  \(\sqrt{s} = 2.36\) \cite{ALICE:2010cin}. 	
The results were extrapolated to \(p_{\mathrm{T}}^{\mathrm{min}} = 0 \)~\MeV.
Extrapolations of primary charged-particle multiplicity density
were done  for  centre-of-mass energies  \(\sqrt{s}= 13,\ 13.5\) and \(14\)~\TeV\ 
\cite{ALICE:2015olq}. 
The  results are given for three conventional event classes: 
inelastic (INEL) events,   non-single diffractive (NSD)  events,  
and  
events with at least one charged particle in \( \mid\eta\mid <1\) (INEL$>$0).
The errors shown are systematic errors. 
Statistical errors are negligible.
Taken from Ref.~\cite{ALICE:2015olq,ALICE:2015qqj}.
}
\label{tab:average_nch_energy_ALICE}
\medskip
 \begin{tabular}{lllll}
\hline
\hline

  \(\sqrt{s}\) [TeV]
& \(n_{\mathrm{ch}} \ge \)
& INEL  
& NSD  
& INEL$>$0  
\\

\hline

13		
		&  1
		&5.15\(\pm\)0.18				&---							&6.48\(\pm\)0.19\\[2mm]

8
		&  1
		&4.66\(^{+0.35}_{-0.17}\)	&5.90\(^{+0.15}_{-0.13}\)	&6.13\(^{+0.10}_{-0.08}\)\\[2mm]

7 		 	
		&  1
		&4.60\(^{+0.34}_{-0.17}\)	&5.74\(^{+0.15}_{-0.15}\)	&5.98\(^{+0.09}_{-0.07}\)\\[2mm]

2.76 	 	
		&  1
		&3.75\(^{+0.26}_{-0.16}\)	&4.63\(^{+0.30}_{-0.19}\)	&4.76\(^{+0.08}_{-0.07}\)\\[2mm]

2.36
		&  1
		&3.77\(^{+0.25}_{-0.12}\)	&4.43\(^{+0.17}_{-0.12}\)	&--- 						\\[2mm]

0.9  	
		&  1
		&2.94\(^{+0.11}_{-0.05}\)	&3.61\(^{+0.17}_{-0.16}\)	&3.75\(^{+0.06}_{-0.05}\)\\[2mm]

\cline{3-5}
\multicolumn{2}{c}{}
&
\multicolumn{3}{l}{Extrapolations in Ref.\ \cite{ALICE:2015olq}}
\\
\cline{3-5}

14 	
		&  1
		&5.37\(\pm\)0.24				&6.62\(\pm\)0.20				&6.98\(\pm\)0.10 \\

13.5 
		&  1
		&5.33\(\pm\)0.25				&6.56\(\pm\)0.20				&6.92\(\pm\)0.10 \\

13 	
		&  1
		&5.30\(\pm\)0.24				&6.50\(\pm\)0.20				&6.86\(\pm\)0.10 \\

\hline
\hline
\end{tabular}
\end{table*}

The fit
is  
based on  Eq.~(\ref{eq_nch_average})  to 
the
combination of 
the ALICE data with other data at the LHC experiments  and
other experiments at lower energies in Fig.\ \ref{fig_energy_ALICE}  yield
\(\Delta   =  0.102\pm 0.003\) for INEL events,  
\(\Delta   =  0.114 \pm 0.003\) for NDS events
and 
\(\Delta   =  0.114 \pm 0.002\) 
for INEL$>$0  events.
These results   are compared to  \(\Delta  = 0.15\) for central Pb--Pb collisions
\cite{ALICE:2010khr}. 
This is clear evidence that the charged-particle  multiplicity density  increases  with energy 
in Pb--Pb collisions  faster  than in \( p p \) collisions. 
Fit results  are shown  in   Fig.~\ref{fig_energy_ALICE}(a).
The results of the extrapolations to  CM energies of  \(13\), \(13.5\) and \(14\)~\TeV\
are presented in Table~\ref{tab:average_nch_energy_ALICE}. 

The  multiplicity densities  \( \langle \mathrm{d} N_{\mathrm{ch}} / \mathrm{d} \eta \rangle \) 
measured in the   INEL and INEL$>$0 events in the pseudorapidity range 
\( \mid\eta\mid< 0.5\) at \(\sqrt{s}= 13\) are shown in Fig.\ \ref{fig_energy_ALICE}(b) 
\cite{ALICE:2015qqj} and  are  \(5.31\pm 0.18\)
and \(6.46\pm 0.19\), respectively. 
The  multiplicity density for the INEL$>$0 events is also measured in \( \mid\eta\mid< 1\) 
for direct comparison with  the INEL$>$0 results of  ALICE at lower energies 
and is found to be   \(6.61\pm 0.20\)   \cite{ALICE:2010mty}. 

Figure~\ref{fig_energy_ALICE}(b)    shows  
a
compilation of results on  
the
multiplicity density of 
charged particles measured in  \( \mid\eta\mid< 0.5\)   for the INEL and INEL$>$0 results at
different \(p p\) energies  by 
ALICE  \cite{ALICE:2010cin,ALICE:2010mty,ALICE:2015olq,ALICE:2015qqj},
CMS \cite{CMS:2015zrm}, 
ACHM \cite{Aachen:1977izz},
UA5 \cite{UA5:1986yef,UA5:1982ygd,UA5:1987rzq}, 
and  
PHOBOS \cite{PHOBOS:2010eyu}.
The energy dependence of 
\( \langle \mathrm{d} N_{\mathrm{ch}} / \mathrm{d} \eta \rangle \) 
is 
parametrized 
by the power law  (\ref{eq_nch_average})  fitted to 
the
data.
By combining the data at lower energies with the ALICE and CMS results at \( \sqrt{s} = 13 \) \TeV, 
it  was obtained  that  \( \Delta = 0.103 \pm 0.002\) for INEL events
and  \(\Delta = 0.111 \pm 0.004\) for INEL$>$0 events.
These fit results are  in agreement  within error bars  with  the
results obtained in  Fig.\ \ref{fig_energy_ALICE}(a).
 
The CMS obtained  value  \( \Delta = 0.23 \pm 0.01 \)  in Fig.\ \ref{fig_energy_CMS}(b)  
is higher  than ALICE result  \(\Delta   =  0.114 \pm 0.003\)  in  Fig.\ \ref{fig_energy_ALICE}(a)
by \( 0.12\pm 0.01 \)  for  
the
NSD  event class.
Note that  a more complete data sample was used for the ALICE fit 
than
for the CMS one.

The  measurement of average multiplicity density  at \(13\)~\TeV\ by the CMS  \cite{CMS:2015zrm}
for  the  pseudorapidity region \( \mid\eta\mid < 2.4\)   resulted in 
\( \mathrm{d} N_{\mathrm{ch}}/ \mathrm{d} \eta  \mid_{ \mid\eta\mid <0.5} 
= 5.49\pm0.01 \mathrm{(stat)} \pm 0.17  \mathrm{(syst)} \) 
for inelastic events,  which is consistent with the ALICE extrapolation of 
\(5.30 \pm 0.24\)  in Table~\ref{tab:average_nch_energy_ALICE}. 

Over the LHC energy range  from  \(0.9\) to \(14\) \TeV,  while the
CM energy increases by a factor  of \(15.5\),  extrapolation of  the present data for 
\(\mathrm{d} N_{\mathrm{ch}}/ \mathrm{d} \eta \mid_{ \mid\eta\mid =0}\) 
shows an increase by  a factor of 
\(1.75 \pm 0.03\) for the INEL event class, 
\(1.87 \pm 0.03\) for the NSD event class, 
and 
\(1.87 \pm 0.01\) for the INEL$>$0 event class. 
The multiplicity increase is similar for  the NSD and INEL$>$0 classes but slightly lower for the INEL class.

The ALICE results at  \( \sqrt{s} = 0.9,\ 7 \) and \(8\)~\TeV\  and 
extrapolation at  \( \sqrt{s} = 13 \)~\TeV\  for  the average multiplicity density for  
the NSD events in  Table~\ref{tab:average_nch_energy_ALICE} are in agreement 
within uncertainties  with the ATLAS results presented  in  Table~\ref{tab:average_nch_energy_13-8TeV} 
at  \( \sqrt{s} =8 \) and \(13\)~\TeV\  and in  Table~\ref{tab:average_nch_energy_7TeV}
at  \( \sqrt{s} = 0.9 \) and \(7\)~\TeV\  for  inelastic  events with 
\(p_{\mathrm{T}} > 100\)~\MeV\ and \(n_{\mathrm{ch}} \ge 2\).


The multiplicity   pseudorapidity distributions of the  
charged particles  multiplicity 
density at mid-rapidity  (\(\mid\eta\mid < 0.2\))  measured at several  \(\sqrt{s}\)  points 
were found to be well described by  the \textsc{Pythia\,8}  \textsc{Monash}  and 
\textsc{EPOS} models for three event selections.
For \( p_{\mathrm{T}} > 100\)~\MeV\ at  the highest energies,  the predictions from 
\textsc{EPOS}  and  \textsc{Pythia\,8}  \textsc{Monash} match the data well.
For  the predictions from  \textsc{Pythia\,8}  \textsc{A2},
the match is not as good as was observed when measuring particles with  \( p_{\mathrm{T}} > 500\)~\MeV.
For \( p_{\mathrm{T}} > 500\)~\MeV\ at  the  highest energies,  the predictions from  \textsc{EPOS} 
and  \textsc{Pythia\,8}  \textsc{A2} match the data well.
The energy dependence of the particle density 
\( 1/N_{\mathrm{ev}} \cdot \mathrm{d} N_{\mathrm{ch}} / \mathrm{d} \eta \mid_{\eta =0}\)
is shown 
in  Fig.~\ref{fig_energy} for ATLAS,  
in Fig.~\ref{fig_energy_CMS}(b) for the CMS--TOTEM  
and  
in Fig.~\ref{fig_energy_ALICE} for ALICE.

\clearpage
\subsection{Transverse momentum dependence of charged-particle multiplicity }
\label{Nev_pT}
\subsubsection{ATLAS  distributions of multiplicity over \(p_{\mathrm{T}}\)}
\label{Nev_pT_data}
\begin{figure*}[t!]
\centering
\begin{minipage}[h]{0.45\textwidth} 
\center{\includegraphics[width=1.0\linewidth]{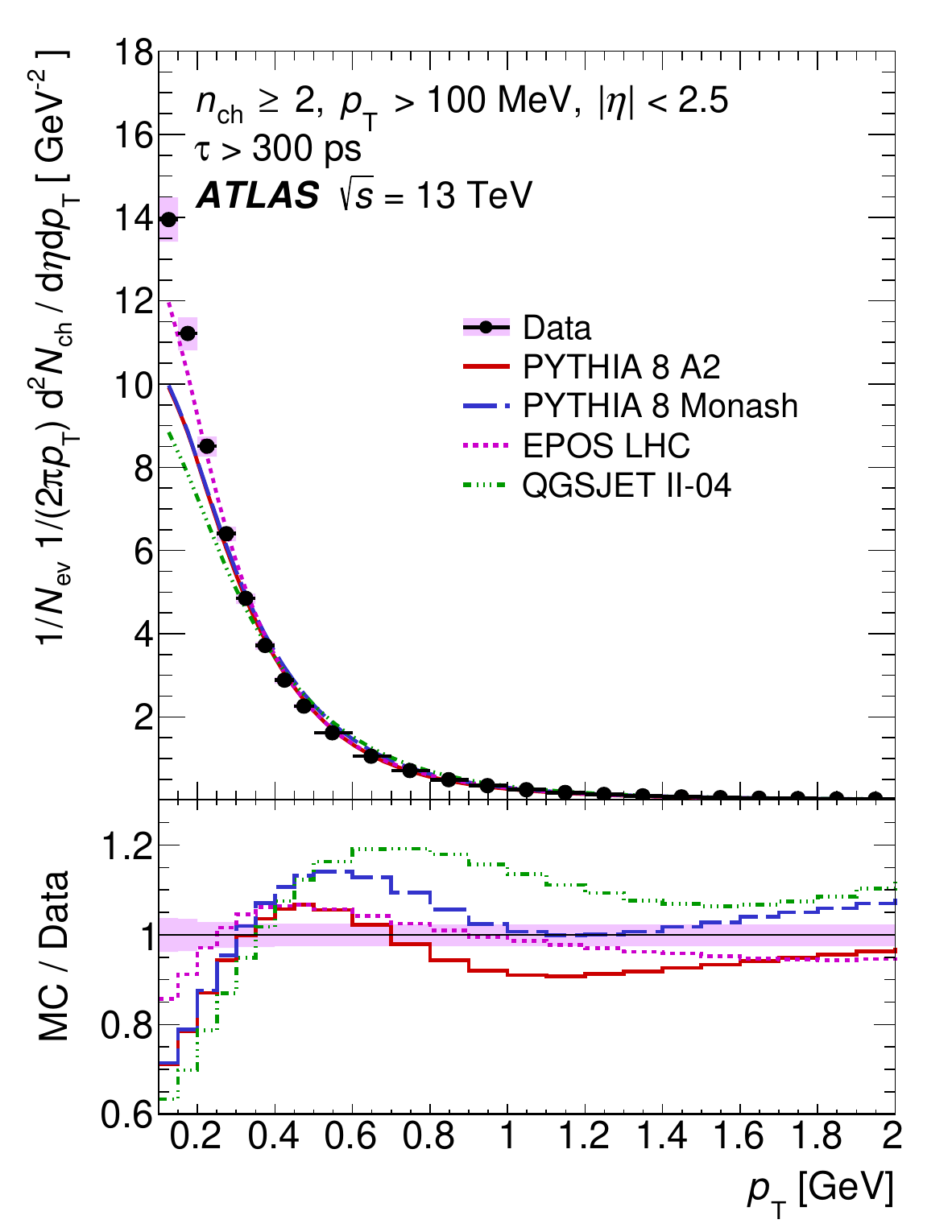}}
(a) 
\\
\end{minipage}
\hspace{2mm}
\begin{minipage}[h]{0.45\textwidth}
\center{\includegraphics[width=1.0\linewidth]{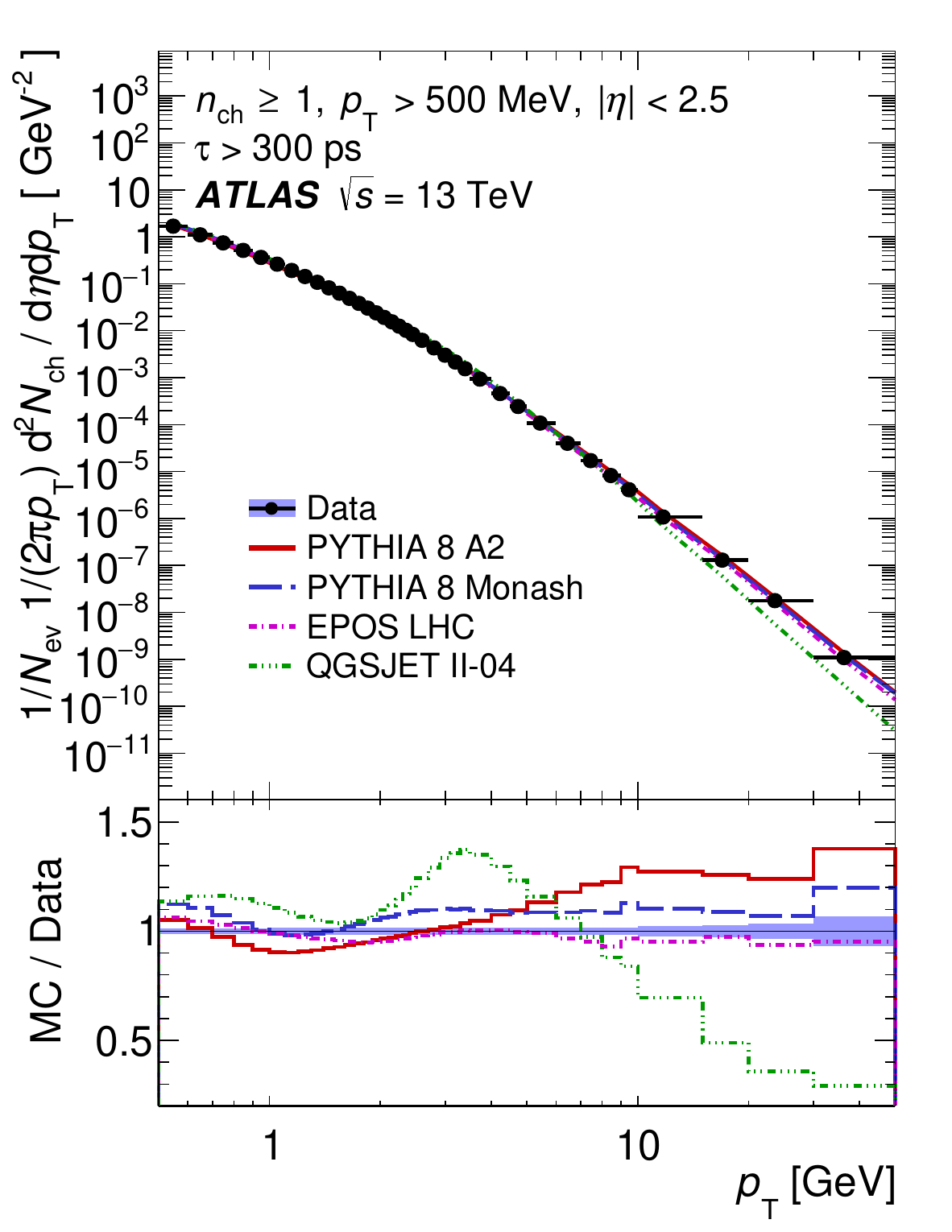}}
(b) 
\\
\end{minipage}
\caption{
Top panel: 
Primary charged-particle multiplicities as  a function of  the transverse momentum
measured by ATLAS at the centre-of-mass energy  \(\sqrt{s}= 13\)~\TeV\ 
with 
(a) \(n_{\mathrm{ch}} \ge 2\) and \(p_{\mathrm{T}} >100\)~\MeV\ \cite{STDM-2015-17} 
and
(b) \(n_{\mathrm{ch}} \ge 1\)  and  \(p_{\mathrm{T}} >500\)~\MeV\ \cite{STDM-2015-02}. 
The data represented by dots  
is
compared to various particle-level MC predictions,  
which are shown by curves. 
The shaded areas around the data points represent the total statistical and systematic uncertainties added in quadrature.
Bottom panel: 
The ratios of the MC predictions to the experimental results are shown. 
Bands represent the uncertainties of the experimental results.
Taken from Refs.~\cite{STDM-2015-02,STDM-2015-17}.
}
\label{fig_13_pT}
\end{figure*}

\begin{figure*}[t!]
\centering
\begin{minipage}[h]{0.32\textwidth} 
\center{\includegraphics[width=1.0\linewidth]{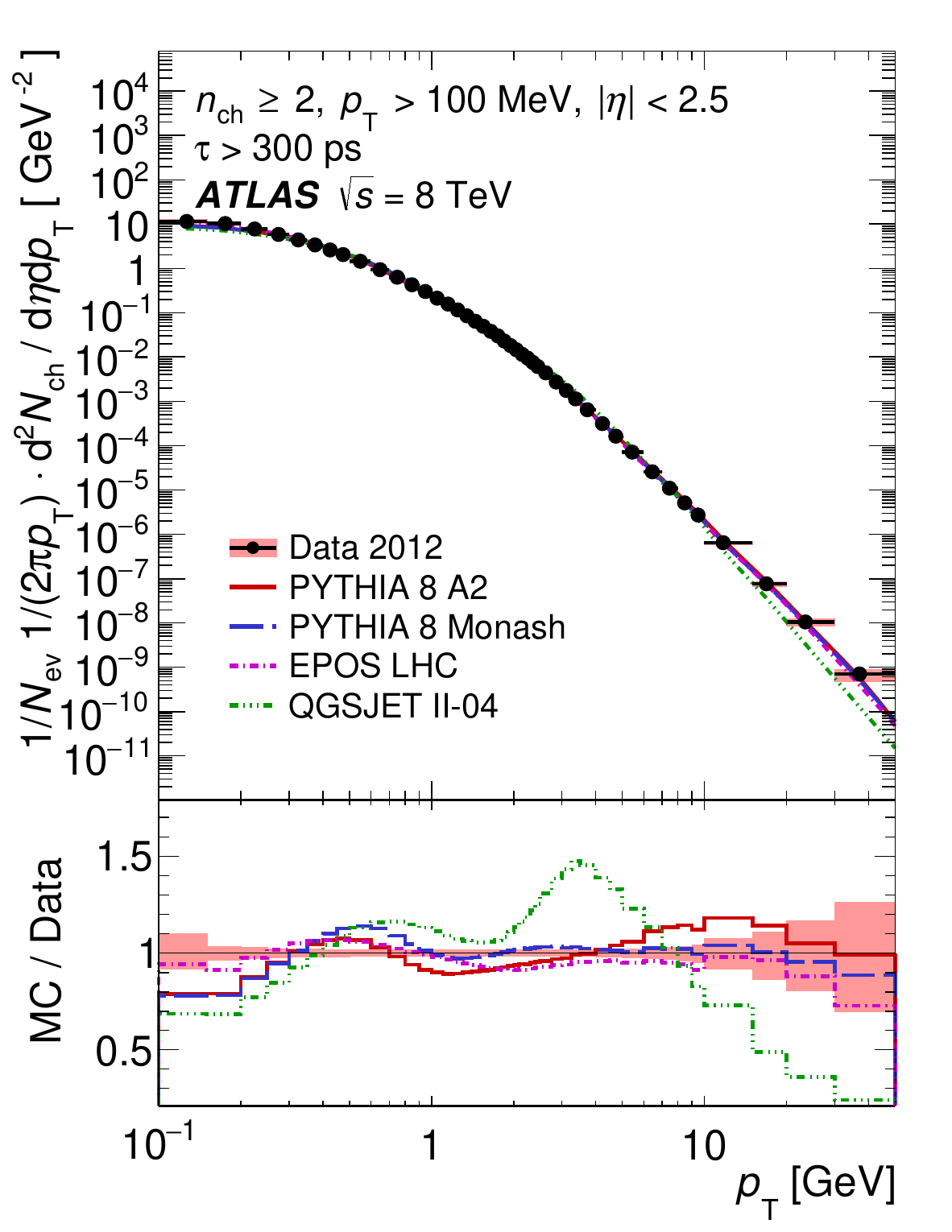}}
(a) 
\\
\end{minipage}
\hfill
\begin{minipage}[h]{0.32\textwidth}
\center{\includegraphics[width=1.0\linewidth]{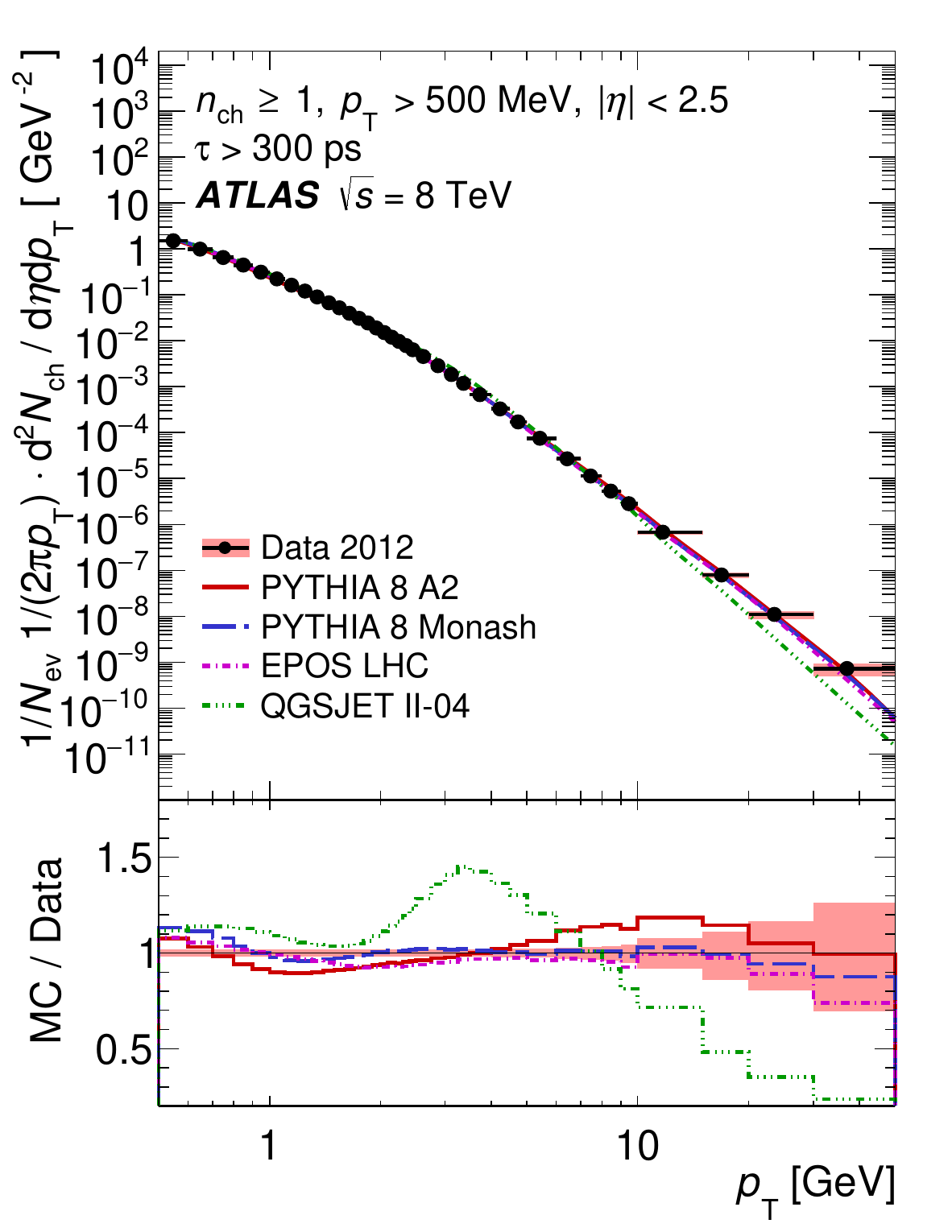}}
(b) 
\\
\end{minipage}
\hfill
\begin{minipage}[h]{0.32\textwidth} 
\center{\includegraphics[width=1.0\linewidth]{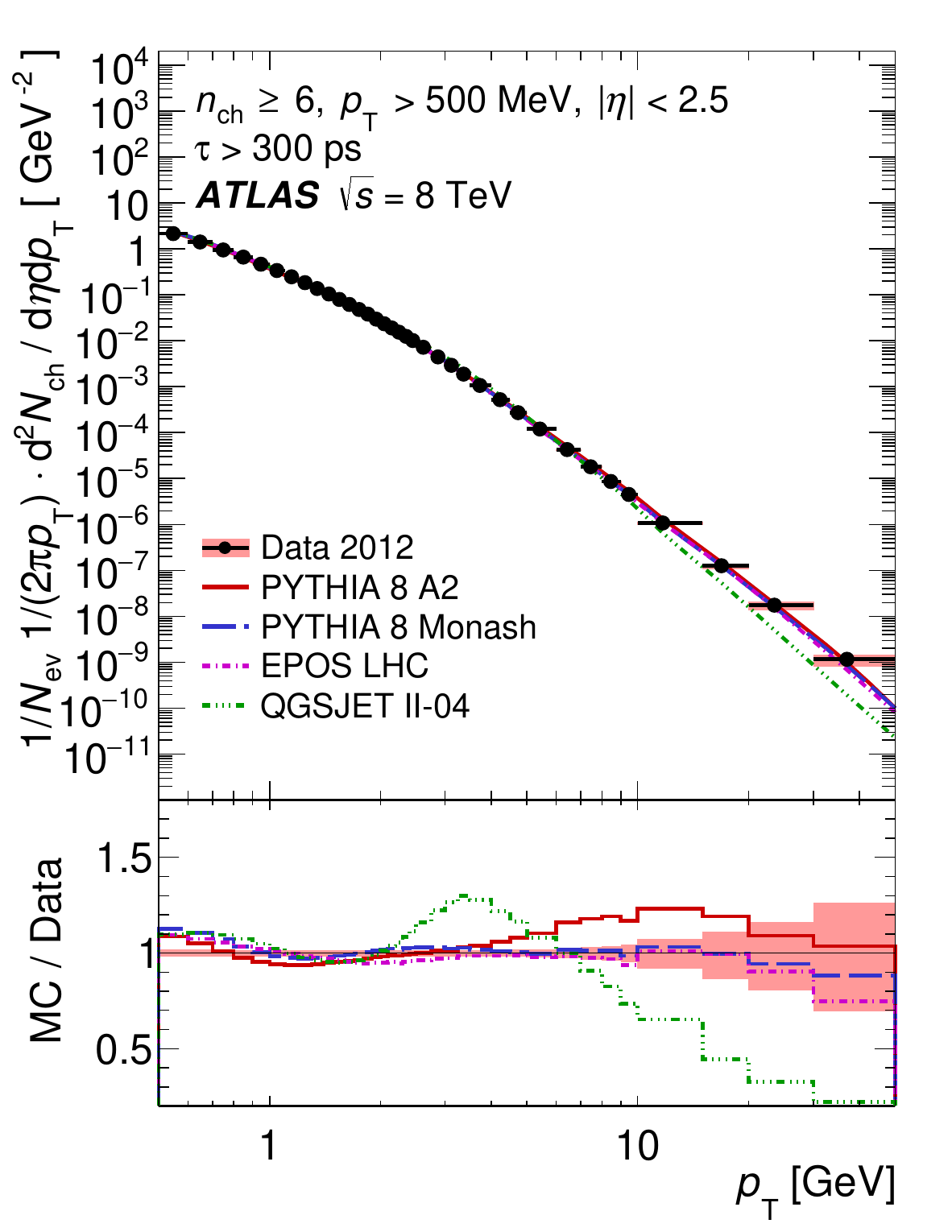}}
(c) 
\\
\end{minipage}
\vfill
\begin{minipage}[h]{0.32\textwidth}
\center{\includegraphics[width=1.0\linewidth]{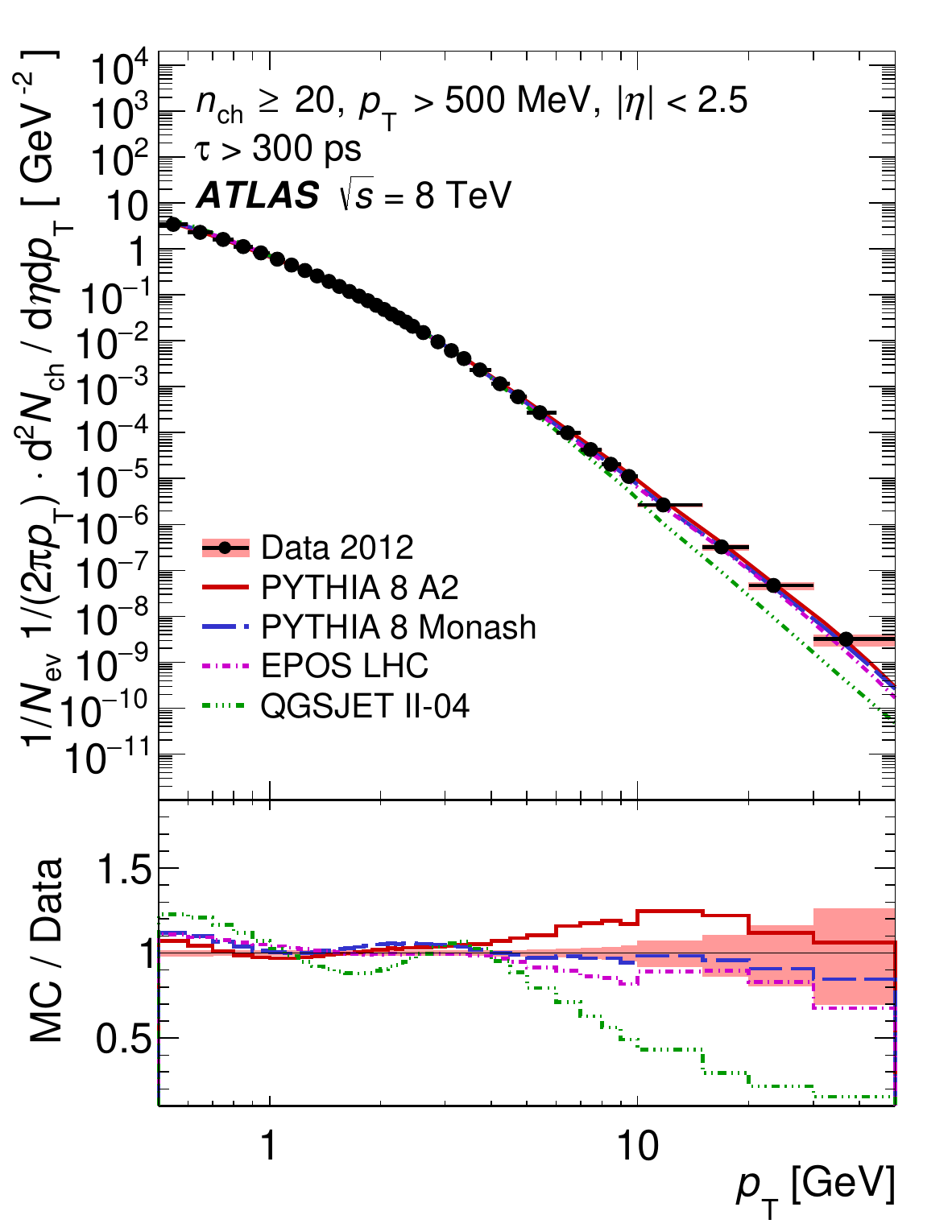}}
(d) 
\\
\end{minipage}
\hspace{2mm}
\begin{minipage}[h]{0.32\textwidth} 
\center{\includegraphics[width=1.0\linewidth]{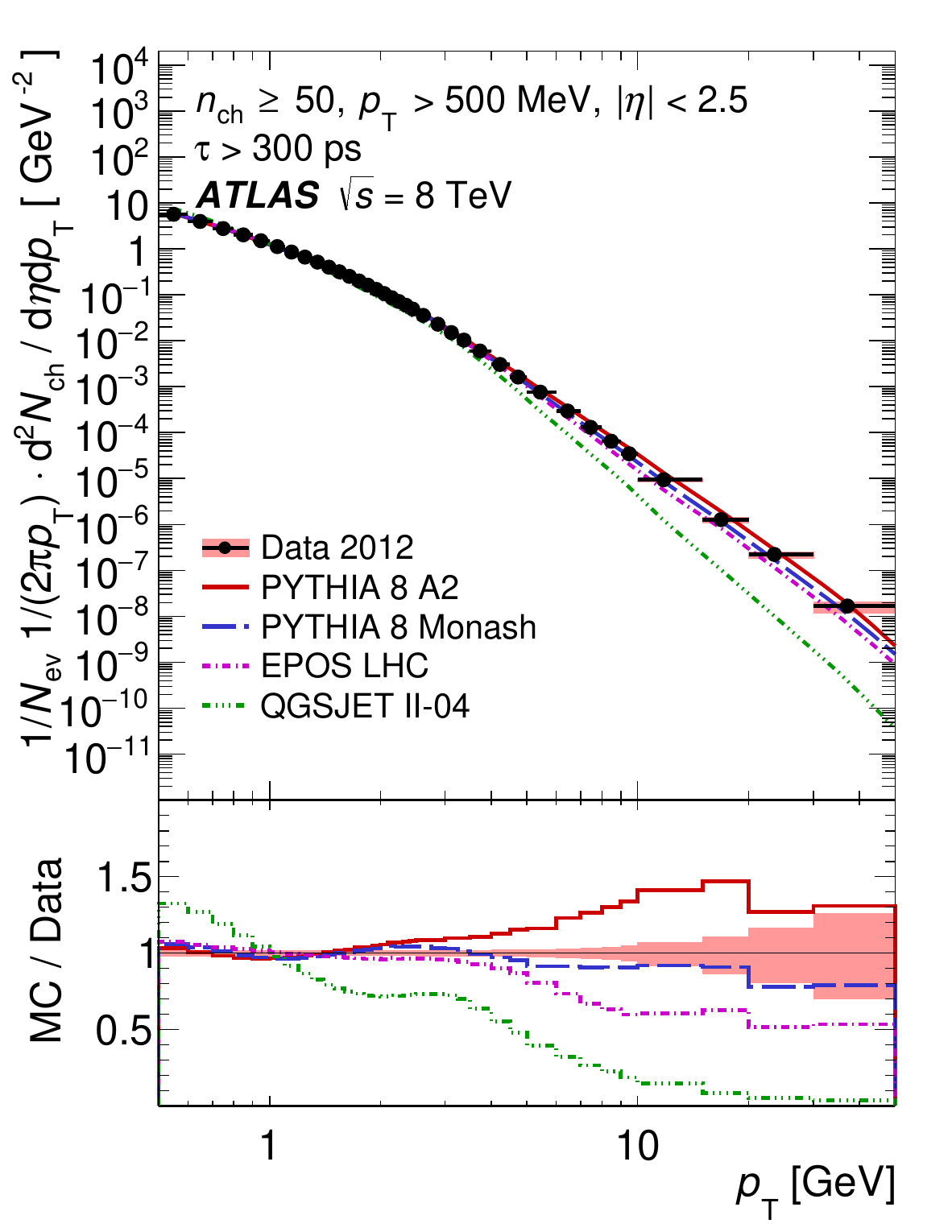}}
(e) 
\\
\end{minipage}
\caption{
Top panel: 
Primary charged-particle multiplicities as  a function of the  transverse momentum measured by ATLAS
at the centre-of-mass energy  \(\sqrt{s}=  8\)~\TeV\ \cite{STDM-2014-19}   with 
(a) \(n_{\mathrm{ch}} \ge 2\) and \(p_{\mathrm{T}} >100\)~\MeV\ 
and 
for \(p_{\mathrm{T}} >500\)~\MeV\ with 
(b) \(n_{\mathrm{ch}} \ge 1\),
(c)  \(n_{\mathrm{ch}} \ge 6\), 
(d)  \(n_{\mathrm{ch}} \ge 20\)
and 
(e)  \(n_{\mathrm{ch}} \ge 50\).
The data represented by dots 
is
compared to various particle-level MC predictions,  which are shown by curves. 
The shaded areas around the data points represent the total statistical and systematic uncertainties added in quadrature.
Bottom panel: 
The ratios of the MC predictions to  the experimental results are shown. 
Bands represent the uncertainties of the experimental results.
Taken from Ref.~\cite{STDM-2014-19}.
}
\label{fig_8_pT}
\end{figure*}

\begin{figure*}[t!]
\centering
\begin{minipage}[h]{0.32\textwidth} 
\center{\includegraphics[width=1.0\linewidth]{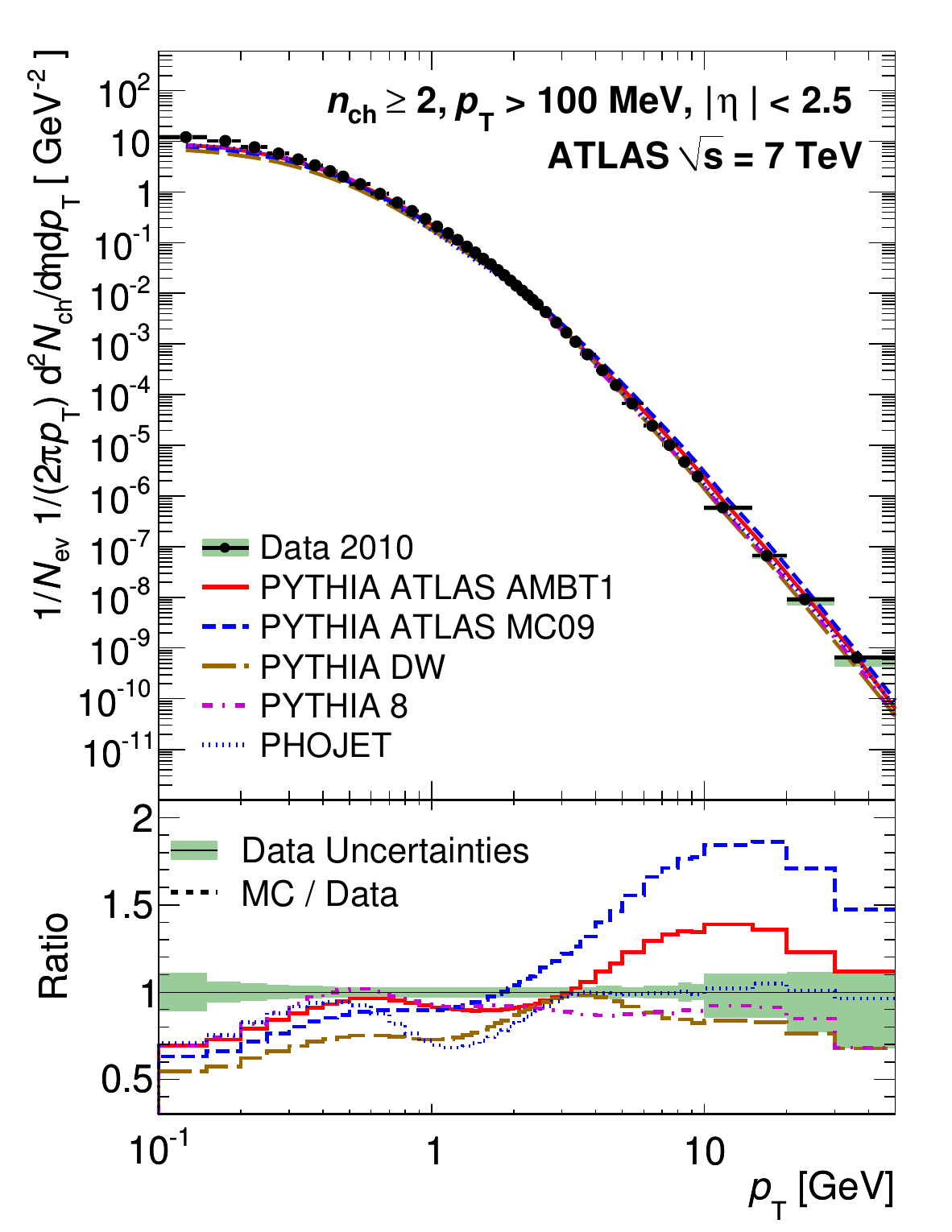}}
(a) 
\\
\end{minipage}
\hfill
\begin{minipage}[h]{0.32\textwidth}
\center{\includegraphics[width=1.0\linewidth]{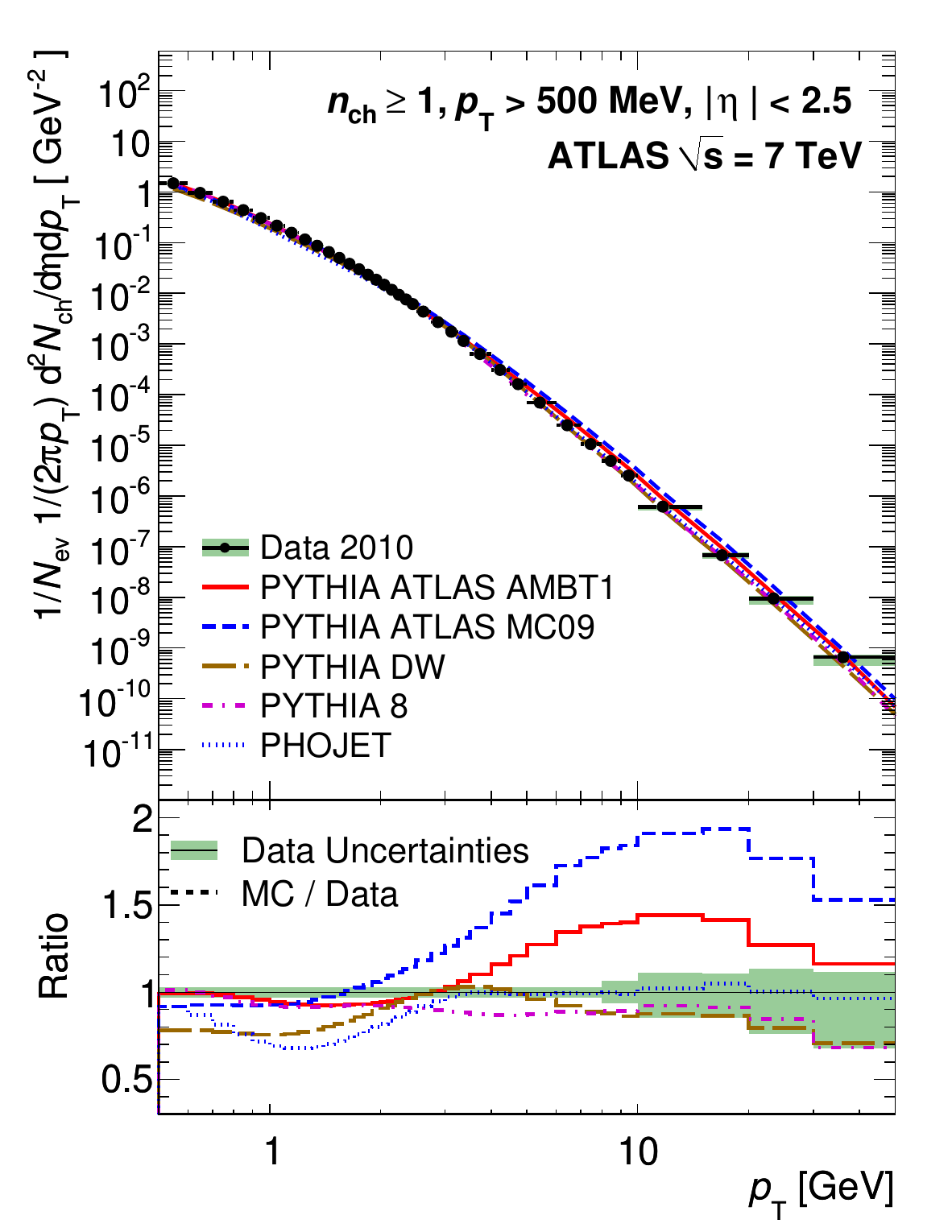}}
(b) 
\\
\end{minipage}
\hfill
\begin{minipage}[h]{0.32\textwidth} 
\center{\includegraphics[width=1.0\linewidth]{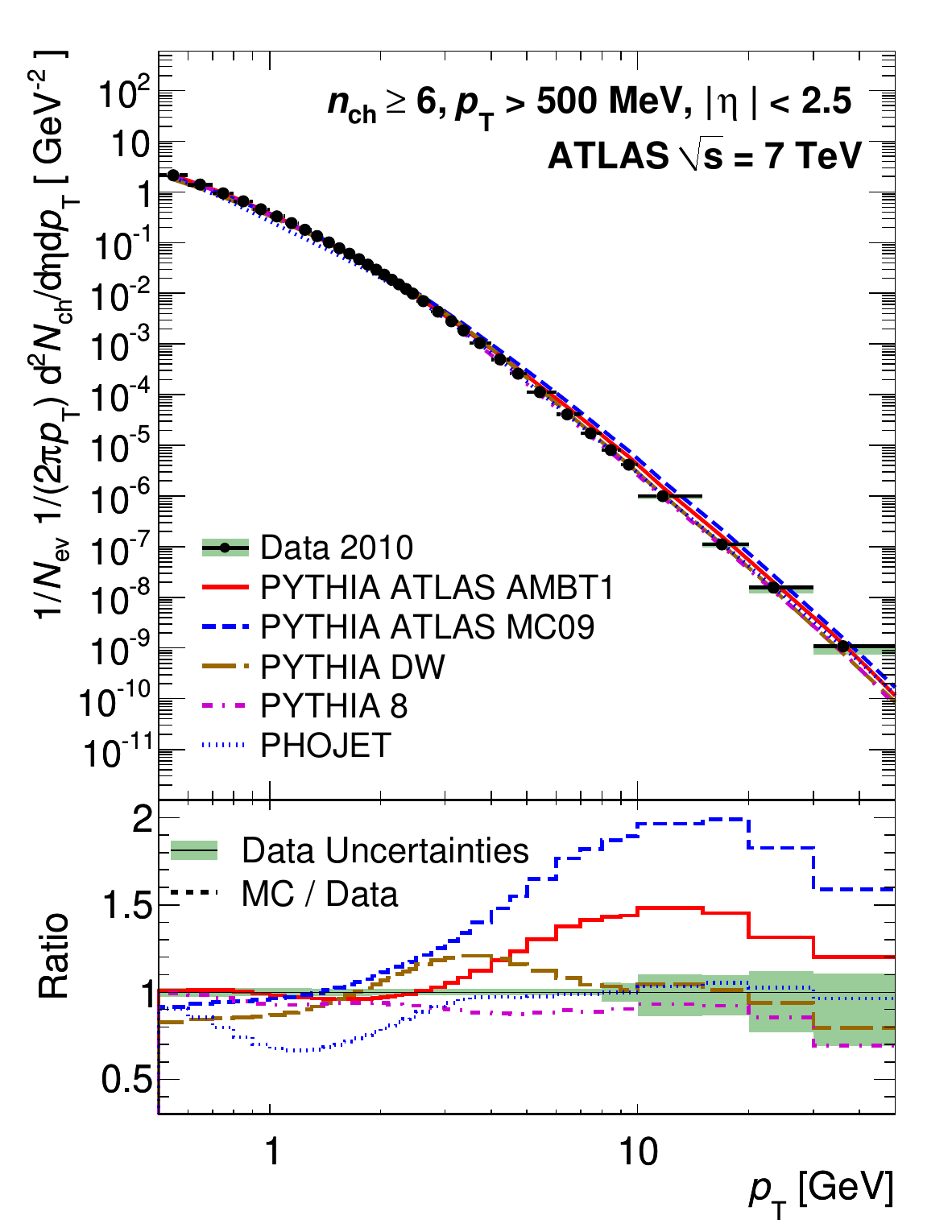}}
(c) 
\\
\end{minipage}
\caption{
Top panel: 
Primary charged-particle multiplicities as  a function of the  transverse momentum measured by ATLAS
at the centre-of-mass energy  \(\sqrt{s}= 7\)~\TeV\ \cite{STDM-2010-06}	   with 
(a) \(n_{\mathrm{ch}} \ge 2\) and \(p_{\mathrm{T}} >100\)~\MeV\  and 
for \(p_{\mathrm{T}} >500\)~\MeV\ with 
(b) \(n_{\mathrm{ch}} \ge 1\) and
(c)  \(n_{\mathrm{ch}} \ge 6\).
The data represented by dots 
is
compared to various particle-level MC predictions,  which are shown by curves. 
The shaded areas around the data points represent the total statistical and systematic uncertainties added in quadrature.
Bottom panel: 
The ratios of the MC predictions to  the experimental results are shown. 
Bands represent the uncertainties of the experimental results.
Taken from Ref.~\cite{STDM-2010-06}.
}
\label{fig_7_pT}
\end{figure*}

\begin{figure*}[t!]
\centering
\begin{minipage}[h]{0.45\textwidth} 
\center{\includegraphics[width=1.0\linewidth]{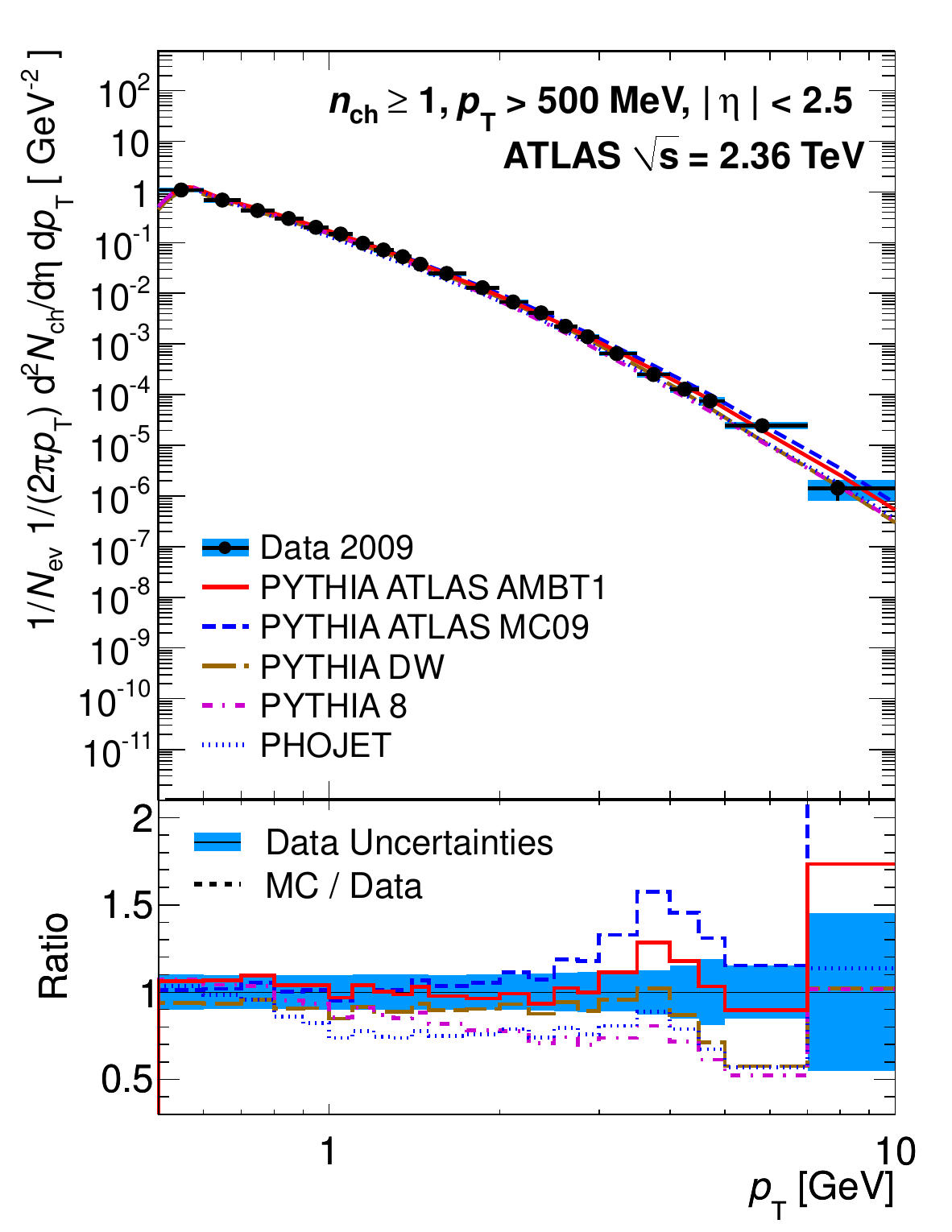}}
(a)
\\
\end{minipage}
\hspace{2mm}
\begin{minipage}[h]{0.42\textwidth} 
\center{\includegraphics[width=1.0\linewidth]{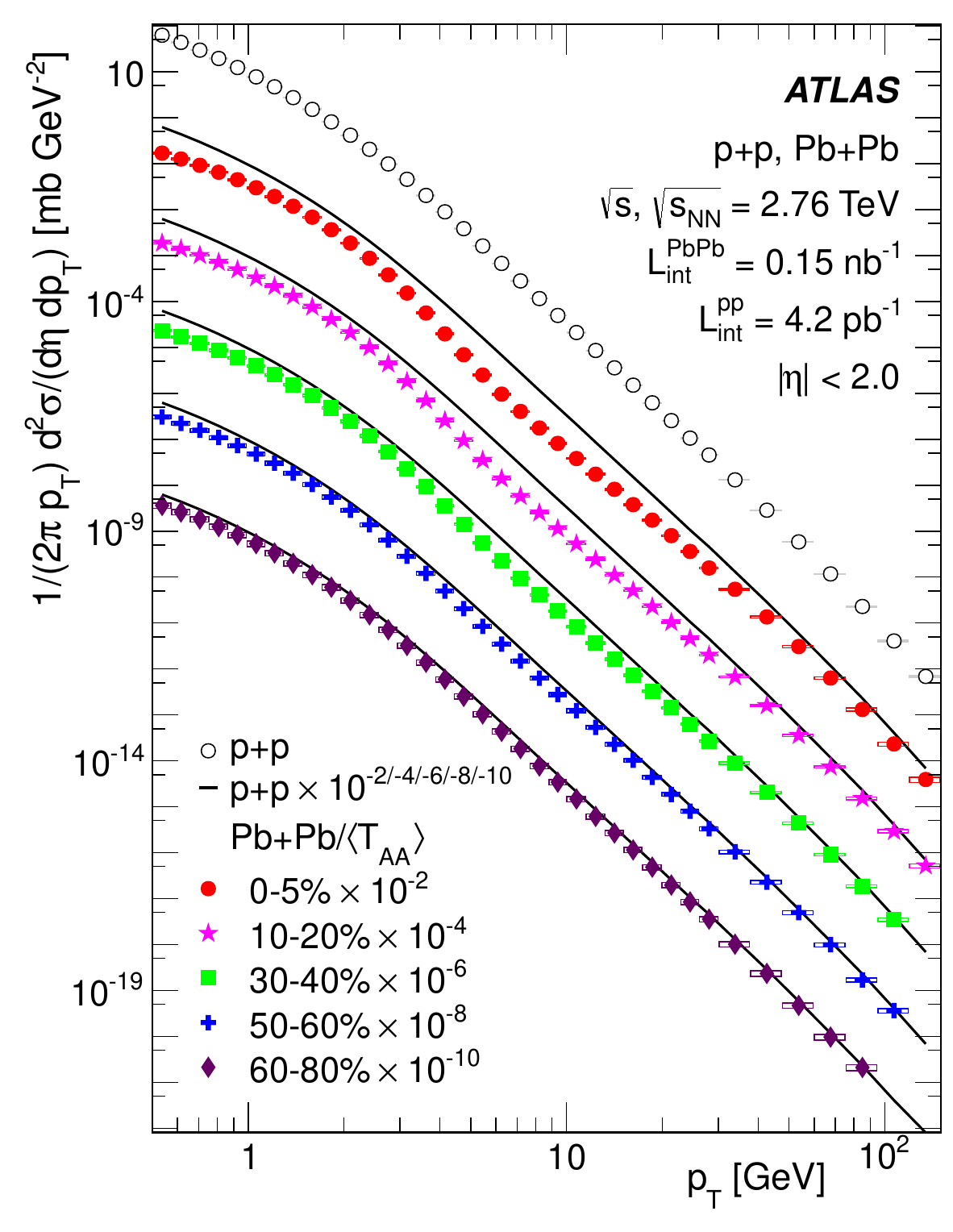}}
(b)
\\
\end{minipage}
\caption{
(a)
Top panel: 
Primary charged-particle multiplicities as  a function of the  transverse momentum measured by ATLAS
at the centre-of-mass energy  \(\sqrt{s}=  2.36\)~\TeV\ \cite{STDM-2010-06}	   with 
\(n_{\mathrm{ch}} \ge 1\) and \(p_{\mathrm{T}} >500\)~\MeV.
The data represented by dots 
is
compared to various particle-level MC predictions,  which are shown by curves. 
The shaded areas around the data points represent the total statistical and systematic uncertainties added in quadrature.
Bottom panel: 
The ratios of the MC predictions to  the experimental results are shown. 
Bands represent the uncertainties of the experimental results.
Taken from Ref.~\cite{STDM-2010-06}.
(b)
Primary charged-particle multiplicities as  a function of the  transverse momentum for 
\(Pb + Pb\) interactions at \(2.76\) \TeV\ for the pseudorapidity range 
\(\mid\eta\mid <2\)  shown with filled  symbols in five centrality intervals: 
0--5\%, 10--20\%, 30--40\%, 50--60\%, 
and 
60--80\% 
as well as the  primary charged-particle multiplicities as  a function of the  transverse momentum
for
fully corrected charged-particle transverse momentum  for \(pp\)  interactions 
shown by open circles.
Statistical uncertainties are smaller than the  size of the symbols.
%
Systematic uncertainties are shown by open boxes. 
The different centrality intervals are scaled down by powers of ten for  clarity.
Each centrality interval is divided by the corresponding  \(\langle T_{\mathrm{AA}}  \rangle\) 
(see text) and plotted together with the  \(pp\)
cross section scaled by the same factor shown with solid lines. 
The total systematic uncertainty on the \(Pb + Pb\) spectra includes the uncertainty of 
\(\langle T_{\mathrm{AA}}  \rangle\).
Taken from Ref.~\cite{ATLAS:2015qmb}.
}
\label{fig_236_pT}
\end{figure*}

\begin{figure*}[t!]
\centering
\begin{minipage}[h]{0.32\textwidth} 
\center{\includegraphics[width=1.0\linewidth]{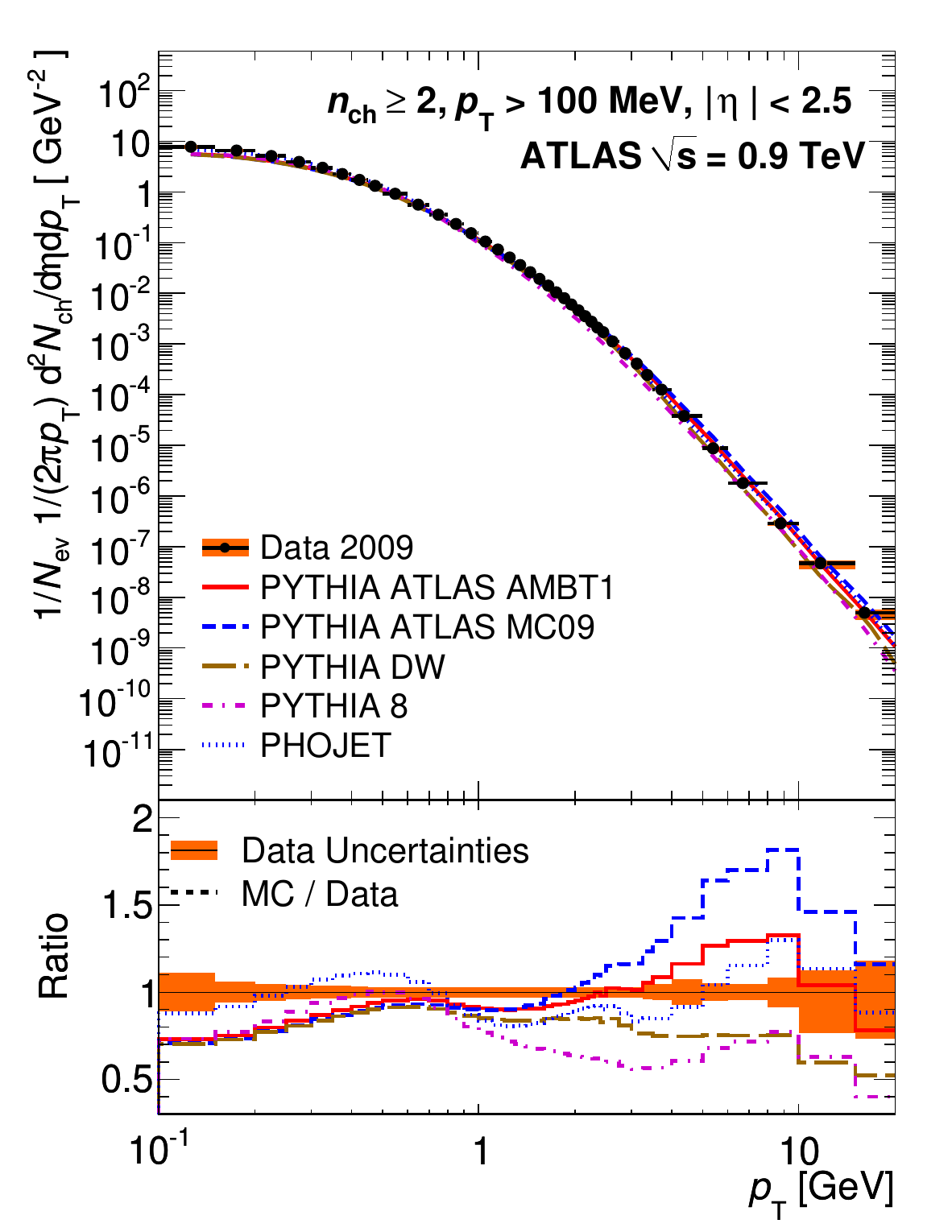}}
(a) 
\\
\end{minipage}
\hfill
\begin{minipage}[h]{0.32\textwidth}
\center{\includegraphics[width=1.0\linewidth]{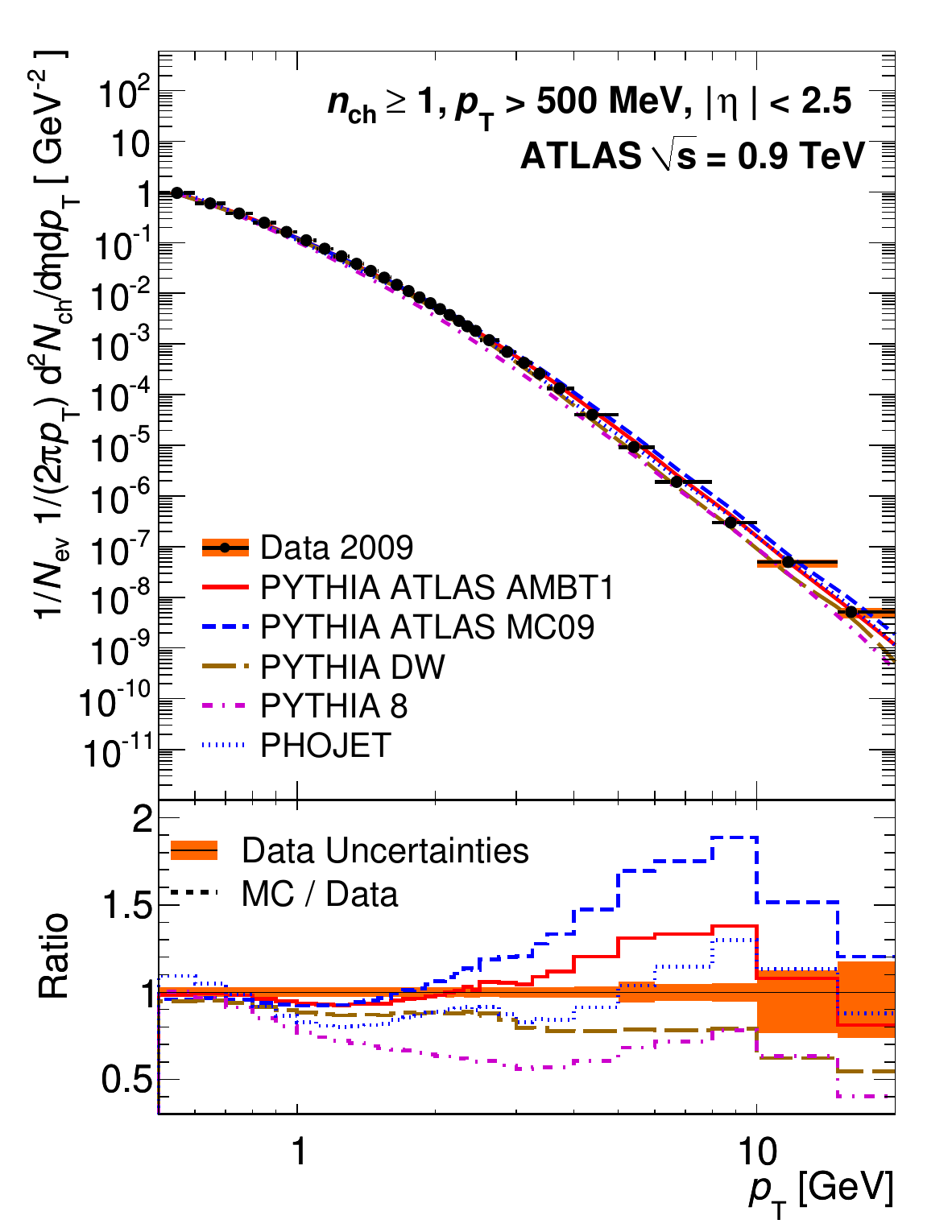}}
(b) 
\\
\end{minipage}
\hfill
\begin{minipage}[h]{0.32\textwidth} 
\center{\includegraphics[width=1.0\linewidth]{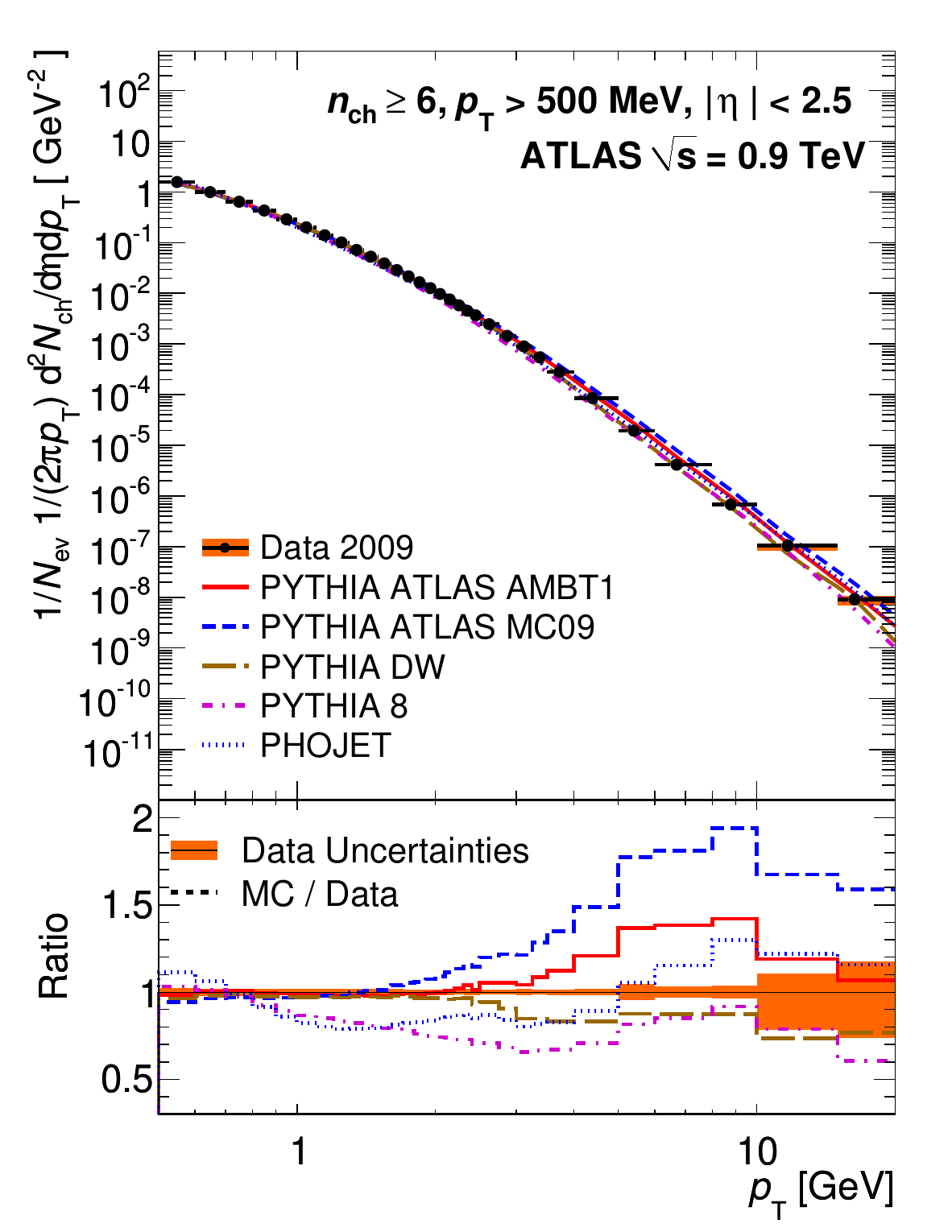}}
(c) 
\\
\end{minipage}
\caption{
Top panel: 
Primary charged-particle multiplicities as  a function of the  transverse momentum
measured by ATLAS
at the centre-of-mass energy  \(\sqrt{s}=  0.9\)~\TeV\ \cite{STDM-2010-06}	  
with 
(a) \(n_{\mathrm{ch}} \ge 2\) and \(p_{\mathrm{T}} >100\)~\MeV\ 
and 
for \(p_{\mathrm{T}} >500\)~\MeV\ with 
(b) \(n_{\mathrm{ch}} \ge 1\)
and
(c)  \(n_{\mathrm{ch}} \ge 6\).
The data represented by dots 
is
compared to various particle-level MC predictions,  
which are shown by curves.
The shaded areas around the data points represent the total statistical and systematic uncertainties added in quadrature.
Bottom panel: 
The ratios of the MC predictions to  the experimental results are shown. 
Bands represent the uncertainties of the experimental results.
Taken from Ref.~\cite{STDM-2010-06}.
}
\label{fig_09_pT}
\end{figure*}

The  transverse momentum distributions  of  
charged particles   
measured by ATLAS
are shown in Figs.~\ref{fig_13_pT} -- \ref{fig_09_pT} at  
the CM energies  \(\sqrt{s}= 0.9,\ 2.36,\ 7,\ 8,\) and  \(13\)~\TeV. 

Figure~\ref{fig_13_pT}(a)  shows the charged-particle transverse momentum distribution 
at \(\sqrt{s}= 13\)~\TeV\ for \(p_{\mathrm{T}} >100\)~\MeV\ \cite{STDM-2015-17}.
The \textsc{EPOS} describes the data well for \(p_{\mathrm{T}} >300\)~\MeV. 
For  lower \(p_{\mathrm{T}}\) the data are underestimated by up to 15\%. 
The other generators show similar mis-modelling at low  momenta but with larger  discrepancies up to 35\%
for  \textsc{QGSJET-II}. 
MC models  mostly overestimate the charged-particle multiplicity for 
 \(p_{\mathrm{T}} >400\)~\MeV;  \textsc{Pythia\,8} \textsc{A2}  yields overestimated 
results only in the intermediate  \(p_{\mathrm{T}}\) region  and  slightly underestimates  the data 
for  \(p_{\mathrm{T}} >800\)~\MeV.
Figure~\ref{fig_13_pT}(b)  shows the charged-particle transverse momentum distribution 
at \(\sqrt{s}= 13\)~\TeV\ for \(p_{\mathrm{T}} >500\)~\MeV\ \cite{STDM-2015-02}.
\textsc{EPOS} describes the data well over the entire \(p_{\mathrm{T}}\) spectrum. 
The \textsc{Pythia\,8} tunes describe the data reasonably well,  but  they are slightly above
the data in the high-\(p_{\mathrm{T}}\) region. 
\textsc{QGSJET-II}  gives a poor prediction over the entire spectrum, overshooting the
data in the low-\(p_{\mathrm{T}}\)  region and undershooting it in the  high-\(p_{\mathrm{T}}\)  region.

Figures~\ref{fig_8_pT}(a) -- \ref{fig_8_pT}(e)  
show   charged-particle multiplicities as a function of the transverse momentum; 
see  Eq.~(\ref{eq_eta_pT}),  for various  PS at  the CM energy 
 \(\sqrt{s}= 8\)~\TeV\  \cite{STDM-2014-19}. 
No model is fully consistent with the distributions.
Above \(1\)~\GeV\  \textsc{Pythia\,8} \textsc{Monash} predictions agree well with the data. 
This model  is  the only one
that  gives a fair description of the data corresponding to the highest multiplicity
threshold with \(n_{\mathrm{ch}} \ge 50\) and \(p_{\mathrm{T}} >500\)~\MeV, 
where all other models show large deviations as \(p_{\mathrm{T}}\) increases.
The \textsc{EPOS}  predictions give the best description of the data corresponding to the  PS
\(n_{\mathrm{ch}} \ge 2\) and \(p_{\mathrm{T}} >100\)~\MeV, 
particularly at transverse 
momenta
below \(1\)~\GeV, 
while the other models underestimate the data at the lowest \(p_{\mathrm{T}}\) values. 
The \textsc{EPOS}   provides fair predictions for the  PS
\(n_{\mathrm{ch}} \ge 1; 6 \) and \(p_{\mathrm{T}} >500\)~\MeV,
but for the higher multiplicity thresholds,   \(n_{\mathrm{ch}} \ge 20; 50 \), 
deviations from the data are seen at high transverse momenta. 
\textsc{Pythia\,8} \textsc{A2} gives fair descriptions of the data below \(6\)~\GeV,
yet shows deviations of up to 30\% around  \(p_{\mathrm{T}} \sim 10\)~\GeV. 
In all measured  PS the  \textsc{QGSJET-II}  approach shows large disagreements with the data as 
\(p_{\mathrm{T}}\) increases.

Figures~\ref{fig_7_pT},  
\ref{fig_236_pT}(a), 
and \ref{fig_09_pT} show 
the charged-particle multiplicities  as a function of the transverse momentum,  Eq.~(\ref{eq_eta_pT}). 
Figures~\ref{fig_7_pT}(b),  \ref{fig_236_pT}(a),  and \ref{fig_09_pT}(b)  
show  three 
CM energies considered in the  PS region 
\(n_{\mathrm{ch}} \ge 1\), \(p_{\mathrm{T}} >500\)~\MeV\ and \(\mid\eta\mid< 2.5\). 
The observed \(p_{\mathrm{T}}\) spectrum is not described by  any of the models over the whole range. 
The region  that  is most difficult for the models  to  describe  is the region above \(1\)~\GeV.
Figures~\ref{fig_7_pT}(a) and \ref{fig_09_pT}(a)  
show the charged-particle multiplicities in the most inclusive 
PS region \(n_{\mathrm{ch}} \ge 2\), \(p_{\mathrm{T}} >100\)~\MeV\ and \(\mid\eta\mid< 2.5\). 
At \(\sqrt{s} = 0.9\)~\TeV\ \textsc{PHOJET} 
describes the data 
is
best over the whole range, 
even though the agreement is still not excellent. 
The other models tend to under-predict  the number of low-\(p_{\mathrm{T}}\) particles, 
while at higher \(p_{\mathrm{T}}\) the models vary widely. 
At \(\sqrt{s} = 7\)~\TeV\ the effect at  low \(p_{\mathrm{T}}\) is more pronounced, whereas at 
high \(p_{\mathrm{T}}\)  the agreement of \textsc{Pythia\,8} and \textsc{PHOJET} 
with the data is quite good. 
The \textsc{AMBT1} and \textsc{MC09} tunes of \textsc{Pythia\,6}  
predict too many particles at higher \(p_{\mathrm{T}}\).
Figures ~\ref{fig_7_pT}(c) and  \ref{fig_09_pT}(c)
show the charged-particle multiplicities with the smallest contribution from diffractive events. 
This distribution carried the most weight in the \textsc{Pythia\,6} \textsc{AMBT1} tune.
Considerable improvement in 
agreement with  the data is seen between the older 
\textsc{Pythia\,6} \textsc{MC09}  and \textsc{AMBT1}    
but the parameters varied in this tune 
and
were not sufficient to describe the full spectrum.

The  charged-particle multiplicities as a function of the transverse momentum 
measured in \(pp\) collisions at \(\sqrt{s} = 2.76\) \TeV\ and
in \(Pb+Pb\) collisions at \(\sqrt{s_{\mathrm{NN}}} = 2.76\) \TeV\ are shown in 
Fig.\ \ref{fig_236_pT}(b)  for the pseudorapidity range \(\mid\eta\mid <2\) 
and for five centrality intervals in \(Pb+Pb\) collisions: 
0--5\%, 10--20\%, 30--40\%, 50--60\%, and 60--80\%  
in the \(0.5 < p_{\mathrm{T}} < 150\)  \GeV. 
This figure shows  the  \(Pb + Pb\)  spectra divided by the  \(\langle T_{\mathrm{AA}}  \rangle\)
(which is estimated as the number of nucleon--nucleon collisions over their cross section)
of the corresponding centrality interval compared with the charged-particle production cross sections measured 
in  \(pp\) collisions at \(\sqrt{s} = 2.76\) \TeV.
The charged-particle multiplicities as a function of the transverse momentum  combine 
the measurement of the  soft regime at   low \( p_{\mathrm{T}}\)  with the hard regime at  
high \( p_{\mathrm{T}}\)  which can be calculated in pQCD. 
While early measurements could focus only on the regime up to a few \GeV,  distributions 
were later  measured up to \(\approx 200\) \GeV\ as presented  in 
Fig.\ \ref{fig_236_pT}(b)  \cite{ATLAS:2015qmb}
and in  \(pp\) collisions at \(\sqrt{s} = 5.02\) \TeV\ \cite{ATLAS:2016kvp}.
%
A similar result of the CMS  is presented  in   Ref.~\cite{CMS:2011mry}.


For \( p_{\mathrm{T}} > 100\)~\MeV\ at  the highest energies \textsc{EPOS} 
describes the data well for  \( p_{\mathrm{T}} > 300\)~\MeV,  while for 
\( p_{\mathrm{T}} < 300\)~\MeV,  the data are underestimated by up to \(15\)\%.
MCs show similar mis-modelling at low momentum but with larger discrepancies, 
up to \(35\)\% for  
\textsc{QGSJET-II}.
MCs mostly overestimate the charged-particle multiplicity for \( p_{\mathrm{T}} > 400\)~\MeV.
\textsc{Pythia\,8}  \textsc{A2}  overestimates  the data only in the intermediate  \( p_{\mathrm{T}}\) 
region and  slightly  underestimates  them for  \( p_{\mathrm{T}} > 800\)~\MeV.
For \( p_{\mathrm{T}} > 500\)~\MeV\ at  the highest energies,  
the  measurement spans 10 orders of magnitude; \textsc{EPOS} and 
\textsc{Pythia\,8}  \textsc{Monash} give remarkably good predictions.

\begin{figure*}[t!]
\centering
\begin{minipage}[h]{0.32\textwidth} 
\center{\includegraphics[width=1.0\linewidth]{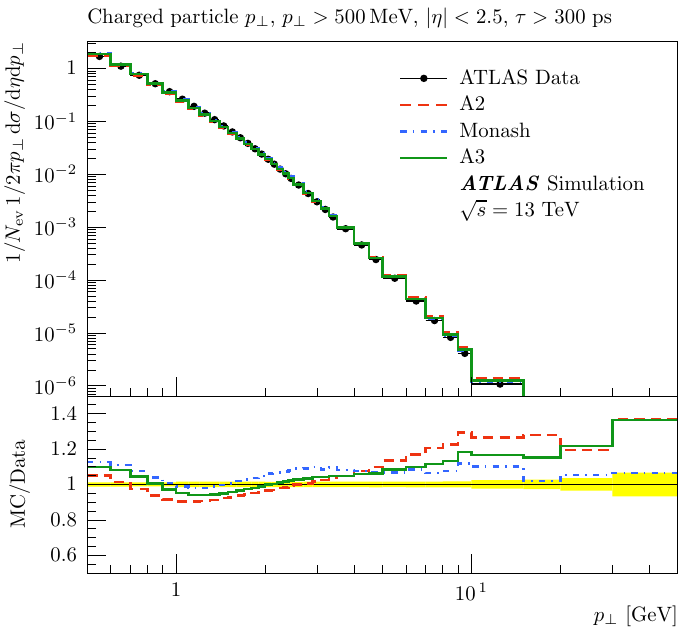}}
(a) 
\\
\end{minipage}
\hfill
\begin{minipage}[h]{0.32\textwidth}
\center{\includegraphics[width=1.0\linewidth]{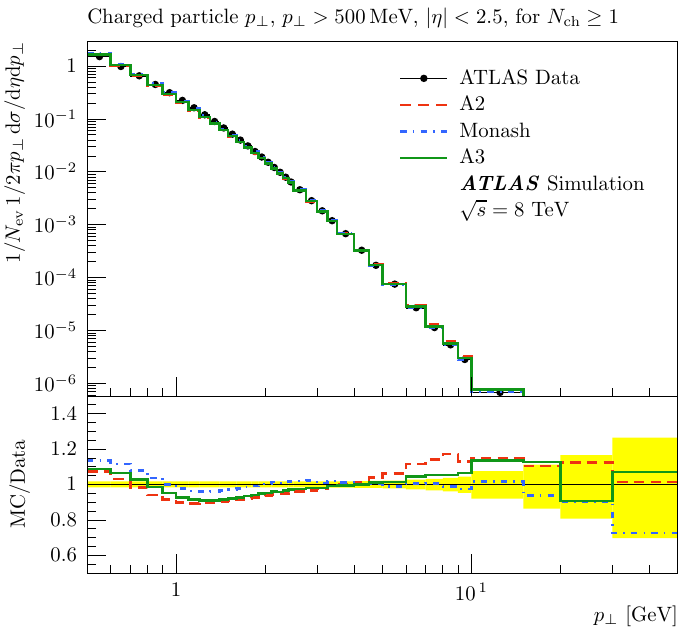}}
(b) 
\\  
\end{minipage}
\hfill
\begin{minipage}[h]{0.32\textwidth} 
\center{\includegraphics[width=1.0\linewidth]{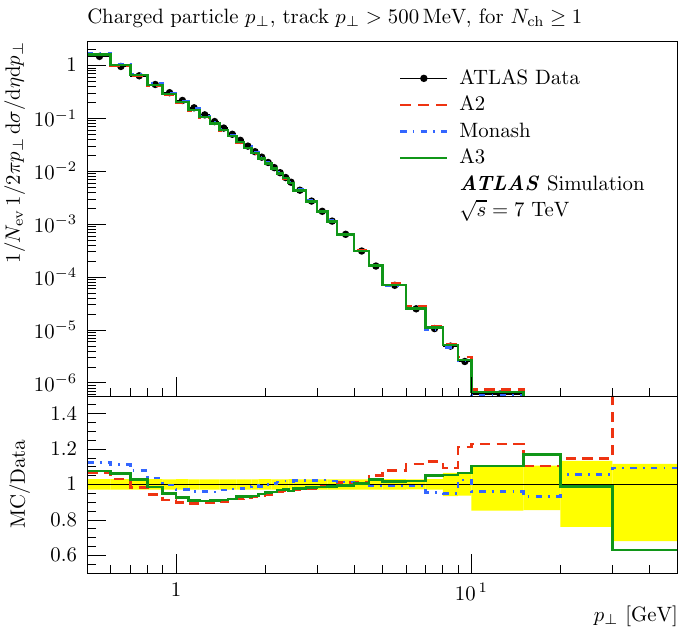}}
(c) 
\\
\end{minipage}
\vfill
\begin{minipage}[h]{0.32\textwidth} 
\center{\includegraphics[width=1.0\linewidth]{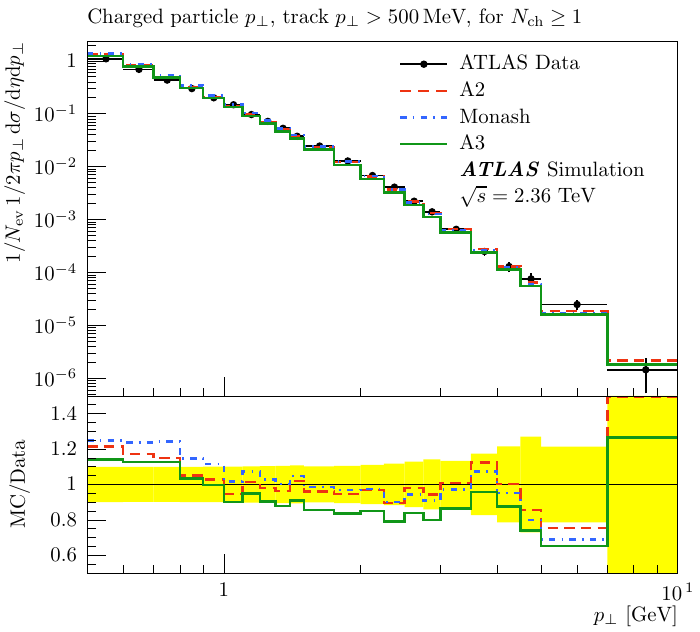}}
(d) 
\\
\end{minipage}
\hspace{2mm}
\begin{minipage}[h]{0.32\textwidth}
\center{\includegraphics[width=1.0\linewidth]{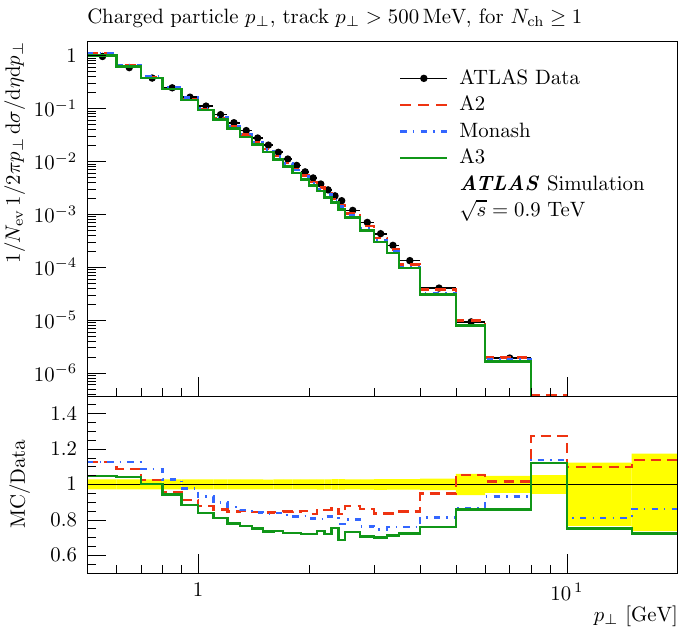}}
(e) 
\\
\end{minipage}
\caption{
Top panel:
The \textsc{Pythia\,8 A3}, \textsc{A2} and \textsc{Monash}  tune predictions 
\cite{ATLAS:2016puo}
compared with 
the
ATLAS 
primary charged-particle multiplicities as  a function of 
the 
transverse momentum
distributions for events with 
\(n_{\mathrm{ch}} \ge 1\) with \(p_{\mathrm{T}} >500\)~\MeV\ 
at 
centre-of-mass energies (a) \(13\)~\TeV, (b) \(8\)~\TeV, (c) \(7\)~\TeV,
(d) \(2.36\)~\TeV\ and (e) \(0.9\)~\TeV. 
The yellow-shaded areas represent the measurement uncertainty.
Bottom panel: 
The ratios of the MC predictions to  
the experimental results  are shown. 
Bands represent the uncertainties of the experimental results.
Taken from Ref.\ \cite{ATLAS:2016puo}.
}
\label{fig_A3_pT}
\end{figure*}

Compared to \textsc{Pythia\,8} \textsc{A2}, \textsc{Pythia\,8}  \textsc{A3} 
provides a slightly worse description of the charged particle multiplicity distribution, 
which coincides with 
the
improved charged-particle  
\(p_{\mathrm{T}}\) distribution that performs similarly to 
\textsc{Pythia\,8} \textsc{Monash},  as shown by Fig.~\ref{fig_A3_pT}.
In all cases, \(\sqrt{s} = 8\)~\TeV\  results are very similar to those at \(\sqrt{s} = 7\)~\TeV.

\begin{figure*}[t!]
\centering  
\begin{minipage}[h]{0.45\textwidth} 
\center{\includegraphics[width=1.0\linewidth]{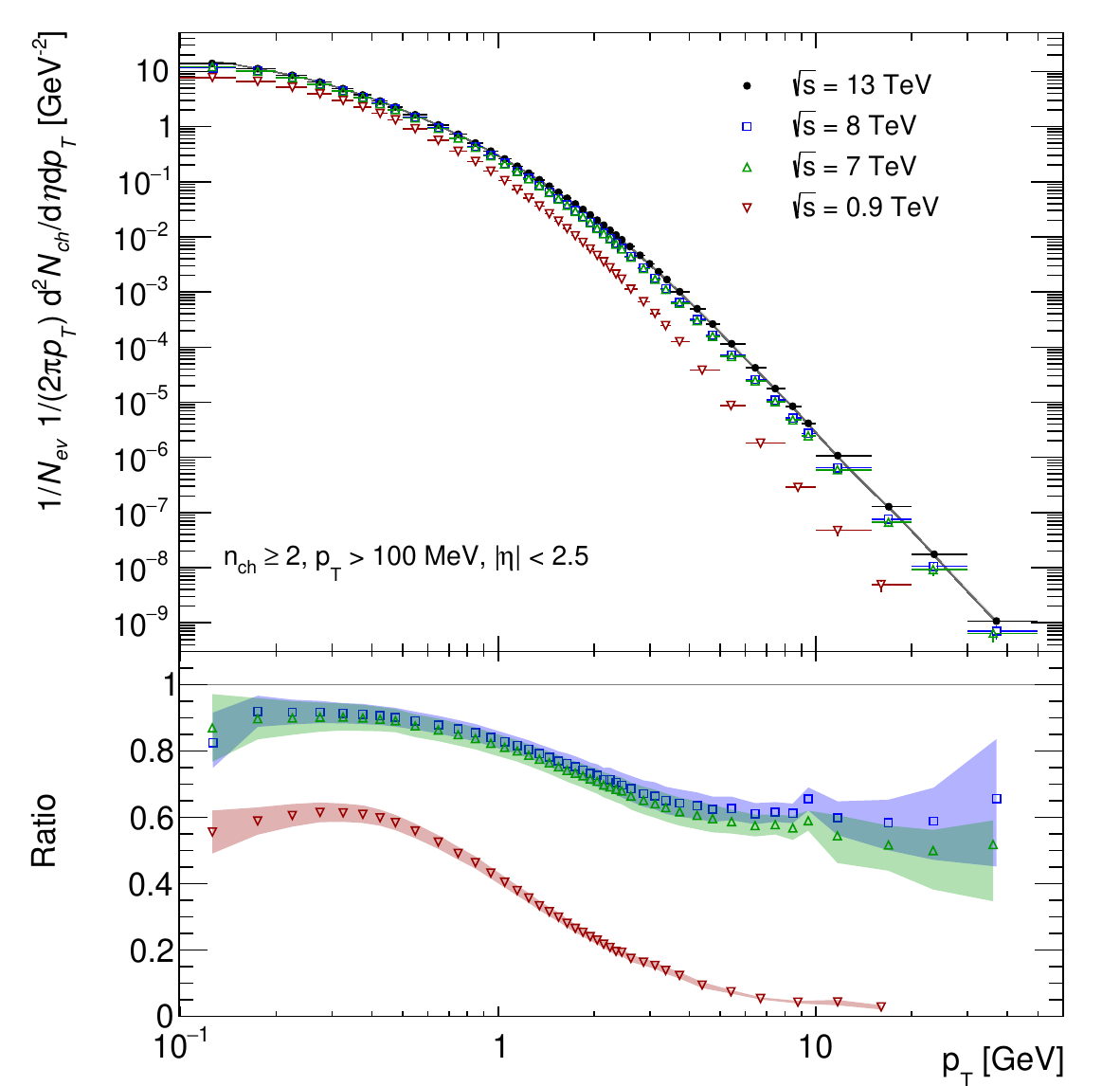}}
(a) 
\\
\end{minipage}
\hspace{2mm}
\begin{minipage}[h]{0.45\textwidth}
\center{\includegraphics[width=1.0\linewidth]{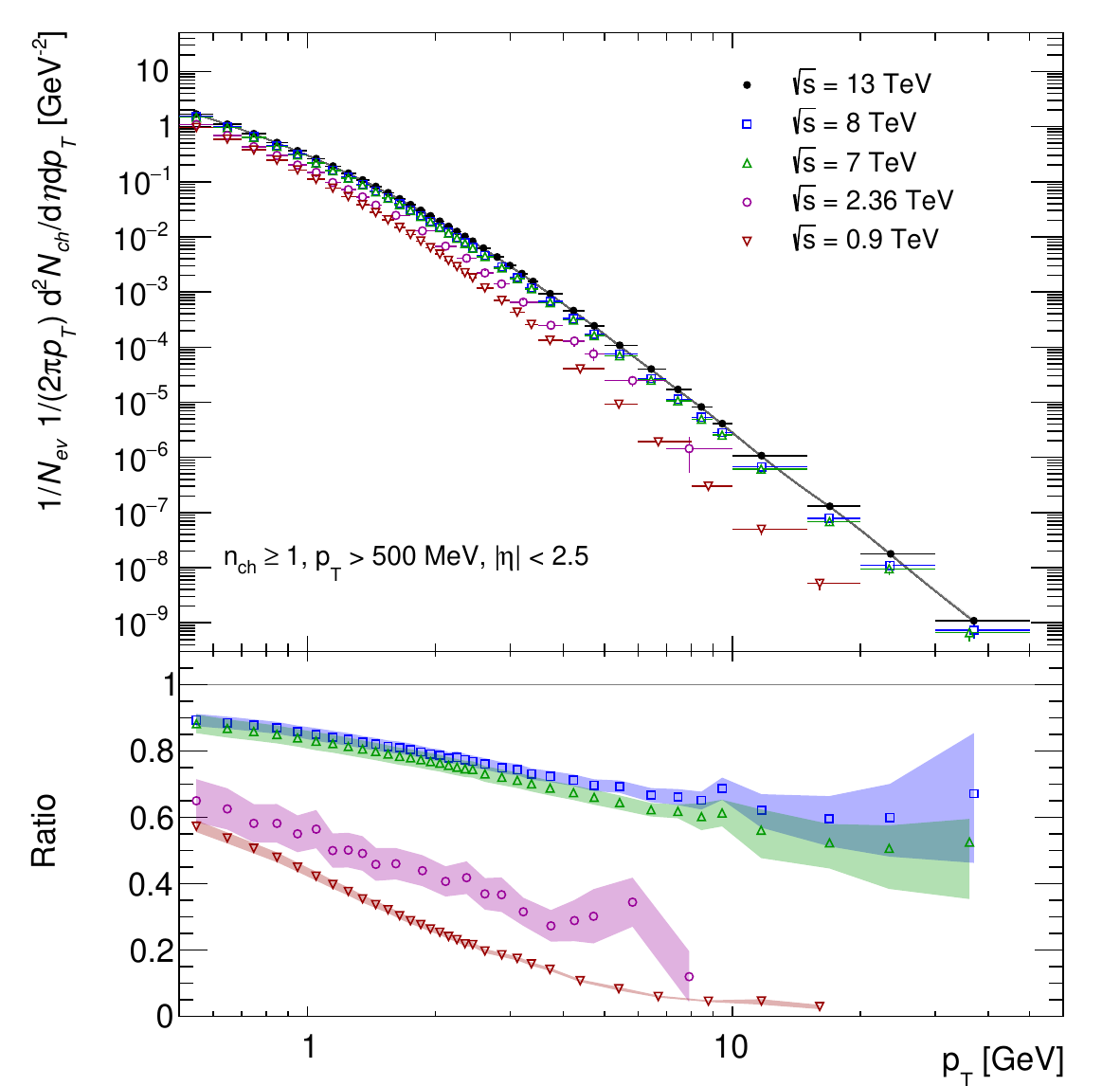}}
(b) 
\\
\end{minipage}
\caption{
Top panel: 
Primary charged-particle multiplicities  as  a function of the   transverse momentum
for pseudorapidity region  \(\mid\eta\mid < 2.5 \) at the 
centre-of-mass energies from \( \sqrt{s} = 0.9\) to  \(\sqrt{s}= 13\)~\TeV\ 
\cite{STDM-2010-06,STDM-2014-19,STDM-2015-02,STDM-2015-17}
for events with 
(a) \(n_{\mathrm{ch}} \ge 2\), \(p_{\mathrm{T}} >100\)~\MeV\ 
and
(b) \(n_{\mathrm{ch}} \ge 1\), \(p_{\mathrm{T}} >500\)~\MeV.
The shaded areas around the data points represent the total statistical and systematic uncertainties added in quadrature.
The grey curve and  the band  of 
uncertainties are the result of the interpolation 
of the charged-particle multiplicity distribution at  \(13\)~\TeV.
Bottom panel: 
The ratios of the  lower energy distribution at  \(\sqrt{s}= 0.9,\ 2.36,\ 7,\  8\)~\TeV\ 
to the distribution  at \(\sqrt{s}= 13\)~\TeV\  are shown. 
Bands represent the uncertainties for the ratios 
as 
the
results of statistical and systematic uncertainties added in quadrature for both distributions.
Taken from Ref.~\cite{Kulchitsky:2022gkm}.
}
\label{fig_energy_pT_100}
\end{figure*}

The comparison  of the primary charged-particle multiplicities   as a function of the transverse momentum 
for  \(\mid\eta\mid  < 2.5\)  measured at the   CM energies from \(0.9\)  to  \(13\)~\TeV\  by
ATLAS  \cite{STDM-2010-06,STDM-2014-19,STDM-2015-02,STDM-2015-17}
are presented  for events with  PS
\(n_{\mathrm{ch}} \ge 2\), \(p_{\mathrm{T}} >100\)~\MeV\ in Fig.\ \ref{fig_energy_pT_100}(a)
and with 
\(n_{\mathrm{ch}} \ge 1\), \(p_{\mathrm{T}} >500\)~\MeV\ in Fig.\ \ref{fig_energy_pT_100}(b). 

Figures~\ref{fig_energy_pT_100}(a) and (b) show  an increase 
in
the primary charged-particle 
multiplicity distributions  with the transverse momentum.
As expected, 
the    distributions acquire higher values  at higher collision energies,
and an increase  by \( \approx 40 \)\%  and  \(\approx 10\)\%  
is observed in the region of \( p_{\mathrm{T}} < 1\)~\GeV\    as the
energy increases  from \(0.9\) to \(13\)~\TeV\  
for  \(p_{\mathrm{T}} >100\)~\MeV\ and \(p_{\mathrm{T}} >500\)~\MeV,  
respectively. 
The results at \(7\) and \(8\)~\TeV\ are in agreement within error bars. 
The particle multiplicity in the transverse momentum region of   \(p_{\mathrm{T}} > 5\)~\GeV\
increases by  \( \approx 40 \)\% for particle  \(p_{\mathrm{T}}\)  
thresholds of 
\(100\)~\MeV\ 
and for that of \(500\)~\MeV\ when energy rises from  \(7\) to \(13\)~\TeV. 

%
\subsubsection{Distributions of multiplicity over \(p_{\mathrm{T}}\) of the   LHC experiments}
\label{Nev_pT_LHC}

\begin{figure*}[t!]
\centering
\begin{minipage}[h]{0.45\textwidth}
\center{\includegraphics[width=1.0\linewidth]{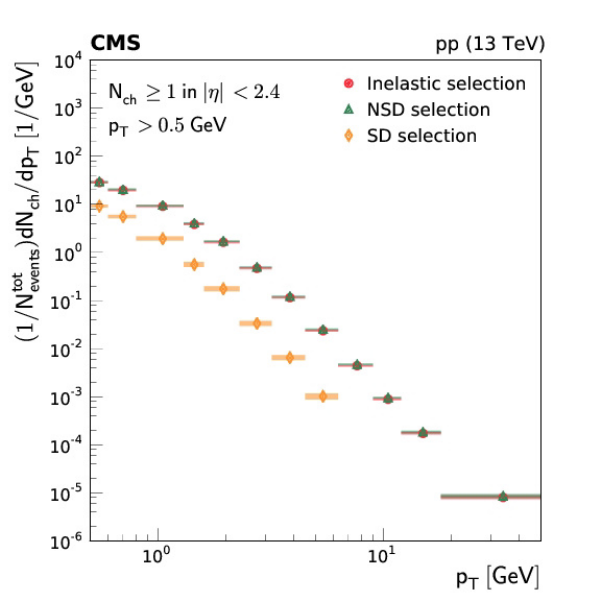}}
(a) 
\\
\end{minipage}
\hspace{2mm}
\begin{minipage}[h]{0.45\textwidth}
\center{\includegraphics[width=1.0\linewidth]{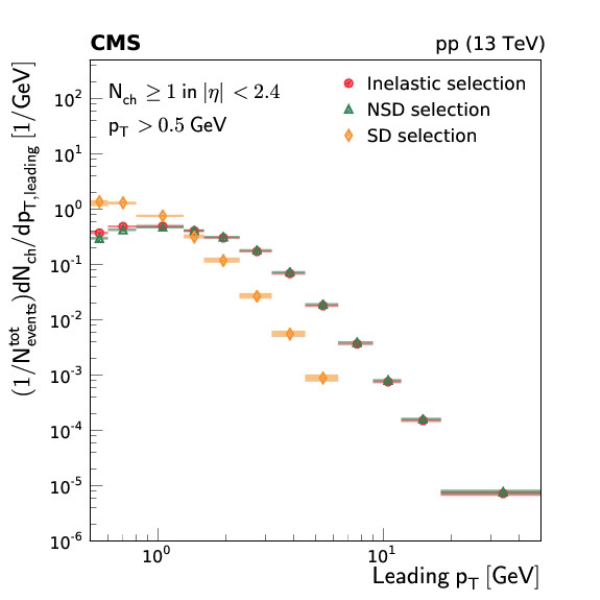}}
(b) 
\\
\end{minipage}
\caption{
Primary charged-particle multiplicities as a function of 
(a) 
the  transverse   momentum and
(b)
the leading transverse  momentum,   \(p_{\mathrm{T},\, \mathrm{leading}}\),   
from the most inclusive (inelastic)  sample,  the
sample dominated by non-single diffractive dissociation events  (NSD-enhanced sample), 
and   the sample enriched by single diffractive dissociation  events (SD-enhanced event samples) 
for events   at a centre-of-mass energies  \(\sqrt{s}= 8\)~\TeV\ 
with   \(n_{\mathrm{ch}} \ge 1\) and  \(p_{\mathrm{T}} > 500\)~\MeV.
The error bars represent the statistical plus uncorrelated systematics uncertainties between 
neighbouring bins,  and the bands show the combined systematic   and statistical uncertainties. 
Taken from Ref.~\cite{CMS:2018nhd}.
}
\label{fig_13_pT_CMS}
\end{figure*}

The CMS  results for  primary charged-particle multiplicities as a function of 
the  transverse   momentum,  \(p_{\mathrm{T}}\),    and
a leading transverse   momentum,  
\(p_{\mathrm{T},\, \mathrm{leading}}\),    for events for  \(\mid\eta\mid < 2.4\) at 
the CM energy   \(\sqrt{s}= 13\)~\TeV\  with  \(n_{\mathrm{ch}} \ge 1\) and 
\(p_{\mathrm{T}} >500\)~\MeV\ \cite{CMS:2018nhd}
are shown in Fig.\ \ref{fig_13_pT_CMS}. 
The measured distributions are presented for three different event data sets:  an  inelastic (INEL) sample, 
an  NSD-enhanced sample,  and  an  SD-enhanced sample. 
The \(p_{\mathrm{T}}\)  distributions  (i.\ e.,
\(p_{\mathrm{T}}\)  and  \(p_{\mathrm{T},\, \mathrm{leading}}\))
of the SD-enhanced event sample fall very steeply for large \(p_{\mathrm{T}}\)  values. 

\begin{figure*}[t!]
\centering
\begin{minipage}[h]{0.45\textwidth} 
\center{\includegraphics[width=1.0\linewidth]{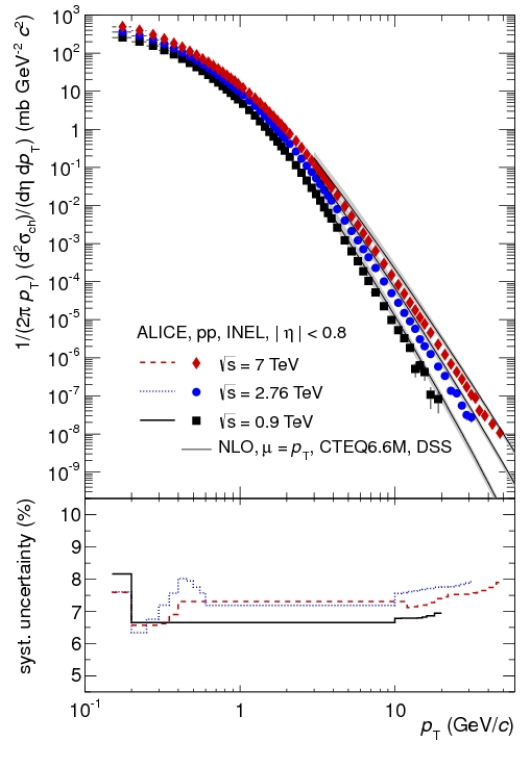}}
(a) 
\\
\end{minipage}
\hspace{2mm}
\begin{minipage}[h]{0.45\textwidth}
\center{\includegraphics[width=1.0\linewidth]{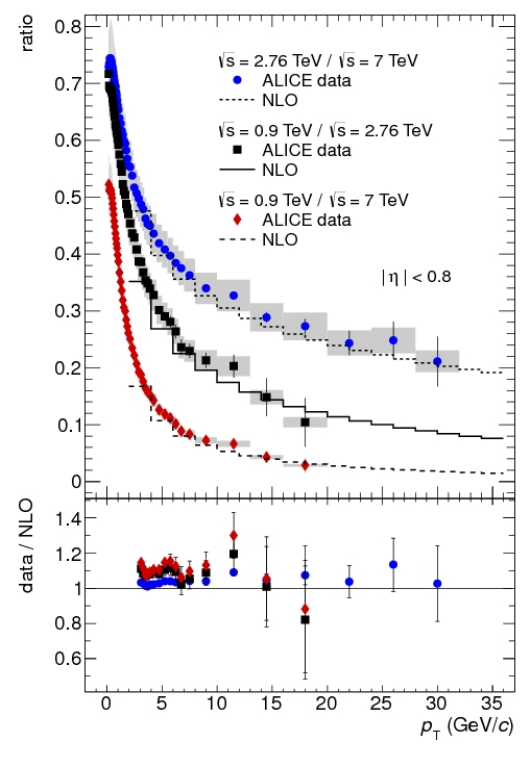}}
(b) 
\\
\end{minipage}
\caption{
(a)
Top panel: 
Differential cross section of charged particles in INEL \(pp\) collisions 
for particles in the pseudorapidity range  \(\mid\eta\mid < 0.8\) with \(p_{\mathrm{T}} > 150 \)~\MeV\
at  \( \sqrt{s} = 0.9,\ 2.76\) and \(7\)~\TeV\ as a function of  \(p_{\mathrm{T}}\)
compared to  the next-to-Leading-Order pQCD (NLO-pQCD)  calculation 
\cite{Sassot:2010bh} at the same energy. 
Only statistical uncertainties are shown. 
Bottom panel: 
Systematic uncertainties as a function of  \(p_{\mathrm{T}}\)  for all three energies. 
The uncertainty on the normalisation  of the spectra is not included (see colour figure online).
(b)
Top panel: 
Ratio of differential cross sections of charged particles in INEL \(pp\)  collisions  for 
\(\mid\eta\mid < 0.8\) at different collision energies as a function of  \(p_{\mathrm{T}}\).
%
Grey
boxes denote  \(p_{\mathrm{T}}\)-dependent  systematic uncertainties. 
Normalisation uncertainties are not shown.
The histograms show the same ratio determined from NLO calculations. 
Bottom panel: 
Ratio of data and NLO calculations derived from  the top panel. 
A variation of the 
renormalization 
and factorization scale of the NLO calculation
gives a systematic uncertainty on the double ratio of  \(0.5\)--\(23.6\)\% 
for
\(0.9\)~\TeV/\(2.76\)~\TeV, 
\(1.0\)--\(37.8\)\% 
for 
\(0.9\)~\TeV/\(7\)~\TeV, 
and 
\(2.4\)--\(12.3\)\% 
for
\(2.76\)~\TeV/\(7\)~\TeV. 
Taken from Ref.~\cite{ALICE:2013txf}.
}
\label{fig_energy_pT_ALICE}
\end{figure*}

The ALICE  measurement  of primary charged particle transverse momentum spectra in
\(pp\) collisions at \(\sqrt{s} = 0.9,\ 2.76\), \(7\)~\TeV\ 
was
presented in 
Ref.~\cite{ALICE:2013txf}. 
The measurement is performed in the pseudorapidity range  \(\mid\eta\mid < 0.8\)
for particles with \(p_{\mathrm{T}} > 150 \)~\MeV. 
The differential cross section  for the INEL \(pp\) collisions as a function of  \(p_{\mathrm{T}}\) 
measured by ALICE  is shown in Fig.\ \ref{fig_energy_pT_ALICE}(a)
for  three measured collision energies  \cite{ALICE:2013txf}. 
At high \(p_{\mathrm{T}}\)  a clear evolution of the slope from
\(\sqrt{s} = 0.9\)  to \(7\)~\TeV\  can be observed. 
The next-to-Leading-Order pQCD (NLO-pQCD) calculation  \cite{Sassot:2010bh}
for \(p_{\mathrm{T}} > 3\)~\GeV\ is compared to the spectra.
The calculation shows a similar evolution of the high-\(p_{\mathrm{T}}\) 
dependence with \(\sqrt{s}\) but over-predicts the data by a factor of  two
\cite{ALICE:2012wos,CMS:2011mry}.
The low systematic uncertainties demonstrate the accuracy  of the measurements for all energies over the full
\(p_{\mathrm{T}}\)  range.

Though the \(p_{\mathrm{T}}\)   dependence of the cross section for a single \(\sqrt{s}\)
is not well described by NLO-pQCD, the relative dependence on  
\(p_{\mathrm{T}}\)   of cross sections of two collision energies is described better. 
Figure \ref{fig_energy_pT_ALICE}(b)  shows the ratio between the differential cross section in
 INEL \(pp\)  collisions at
\(\sqrt{s} = 2.76\)  to \(7\)~\TeV, 
\(\sqrt{s} = 0.9\)  to \(2.76\)~\TeV\ 
and 
\(\sqrt{s} = 0.9\)  to \(7\)~\TeV\
as a function of  \(p_{\mathrm{T}}\)   in comparison to the same ratio calculated with NLO-pQCD. 
The total  \(p_{\mathrm{T}}\)-dependent  systematic uncertainties on the ratios are 
evaluated  with allowance for correlated contributions, and amount to 
\(8.1\)--\(9.8\)\% 
for
\(0.9\)~\TeV/\(2.76\)~\TeV, 
\(7.8\)--\(9.9\)\% 
for 
\(0.9\)~\TeV/\(7\)~\TeV, 
and 
\(7.9\)--\(9.9\)\% 
for
\(2.76\)~\TeV/\(7\)~\TeV. 
The corresponding normalisation uncertainties amount  to 
\(+5.4\)\%/\(-4.4\)\%, 
\(+6.2\)\%/\(-5.4\)\%, 
and
\(\pm 4.1\)\%, 
and are calculated assuming that the normalisation uncertainties on the 
\(p_{\mathrm{T}}\)   spectra  are uncorrelated.
In all  ratios, 
good agreement between the data and  the NLO-pQCD calculations is found, 
which can be seen in the double ratio of data and NLO-pQCD for the three energy ratios in
the lower panel of  Fig.\ \ref{fig_energy_pT_ALICE}(b).

\subsection{Charged-particle multiplicity dependence}
\label{Nev_nch}
\subsubsection{ATLAS multiplicity  distributions}
\label{Nev_nch_data}
\begin{figure*}[t!]
\centering
\begin{minipage}[h]{0.45\textwidth} 
\center{\includegraphics[width=1.0\linewidth]{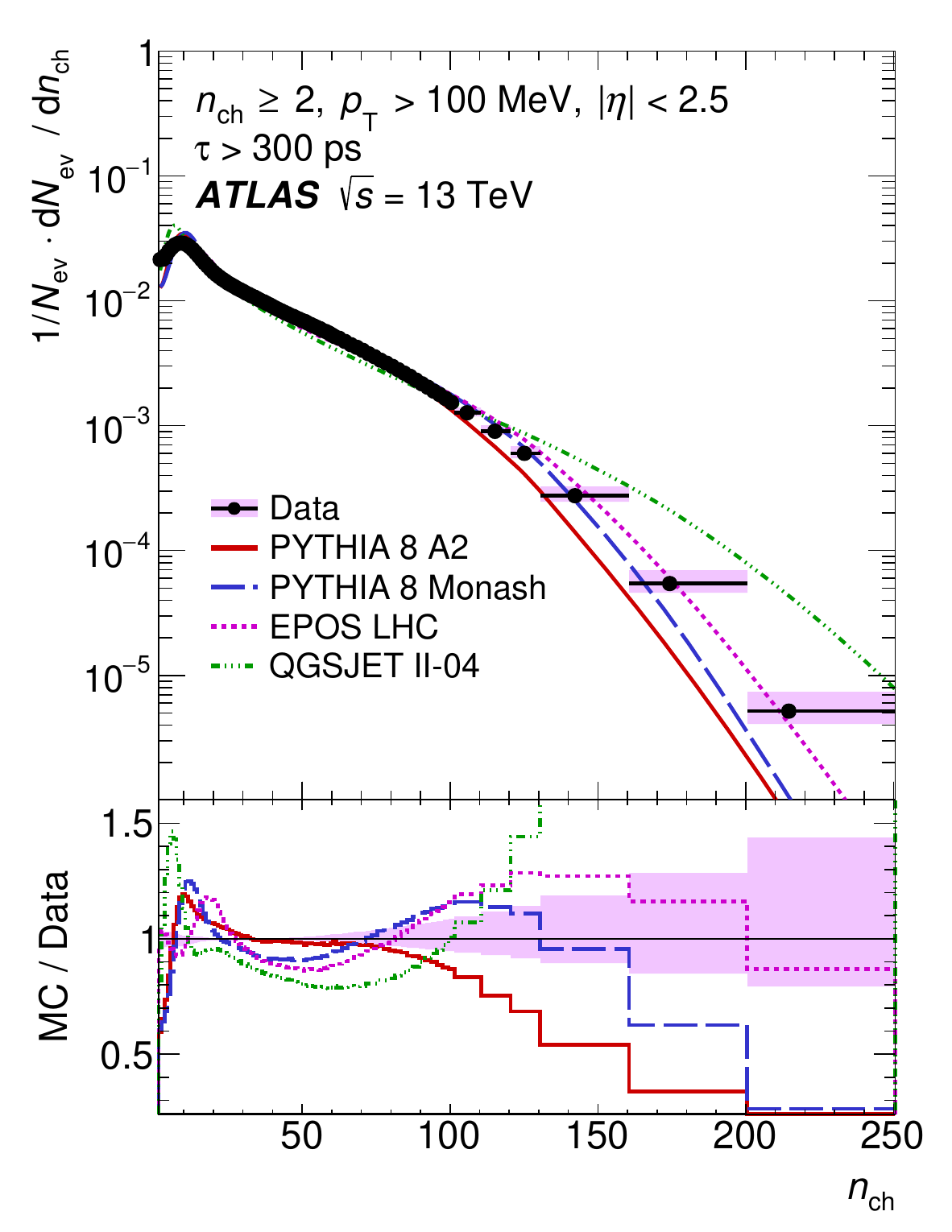}}
(a) 
\\
\end{minipage}
\hspace{2mm}
\begin{minipage}[h]{0.45\textwidth}
\center{\includegraphics[width=1.0\linewidth]{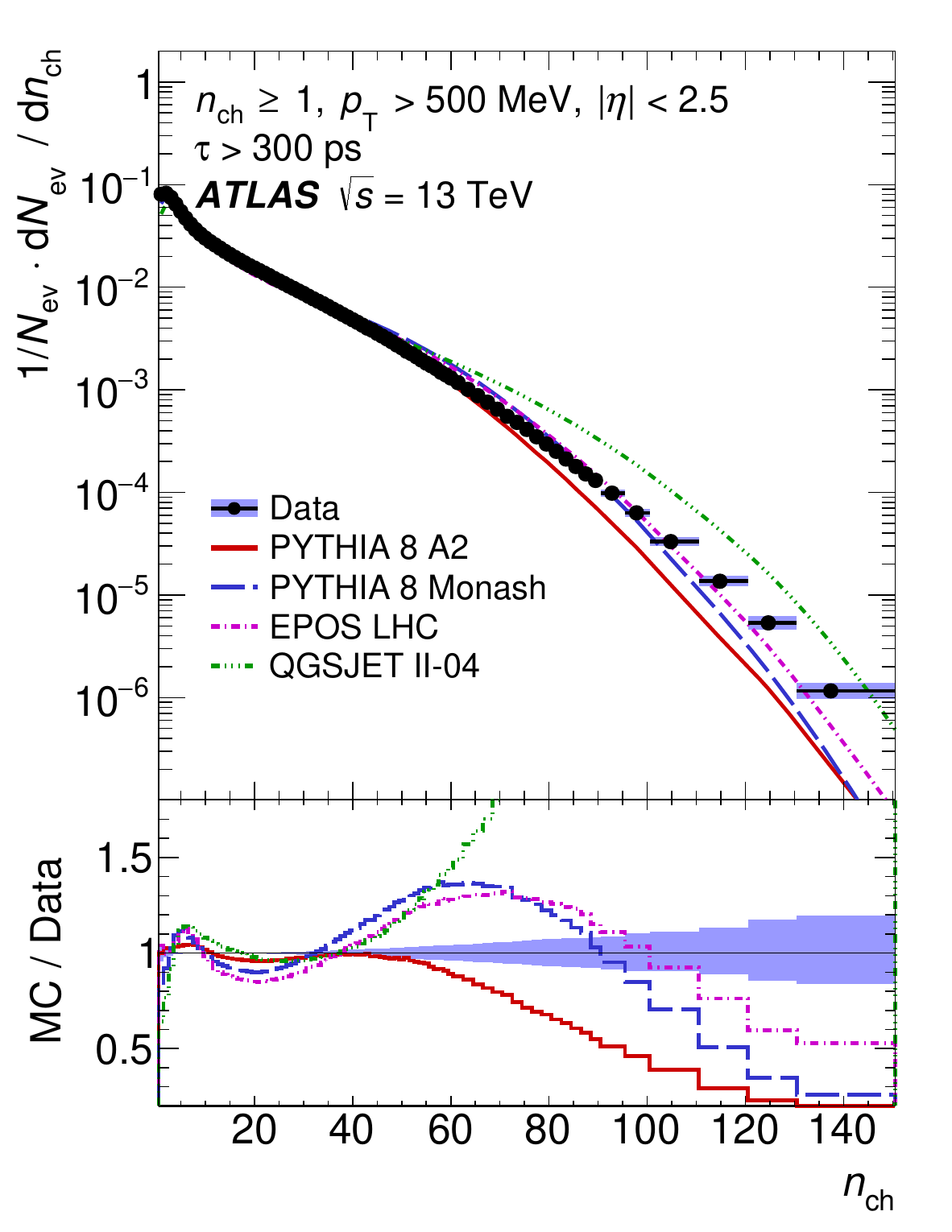}}
(b) 
\\
\end{minipage}
\caption{
Top panel: 
Primary charged-particle multiplicities as  a function of  the  multiplicity
at the  centre-of-mass energy \(\sqrt{s}= 13\)~\TeV\ 
\cite{STDM-2015-02,STDM-2015-17}  with 
(a) \(n_{\mathrm{ch}} \ge 2\), \(p_{\mathrm{T}} >100\)~\MeV\ 
and
(b) \(n_{\mathrm{ch}} \ge 1\),  \(p_{\mathrm{T}} >500\)~\MeV. 
The data represented by dots
is
compared 
to various particle-level MC predictions,  which are shown by curves.
The shaded areas around the data points represent the total statistical and systematic uncertainties added in quadrature.
Bottom panel: 
The ratios of the MC predictions to  the experimental results are shown. 
Bands represent the uncertainties of the experimental results.
Taken from Refs.~\cite{STDM-2015-02,STDM-2015-17}.
}
\label{fig_13_nch}
\end{figure*}

\begin{figure*}[t!]
\centering
\begin{minipage}[h]{0.45\textwidth} 
\center{\includegraphics[width=1.0\linewidth]{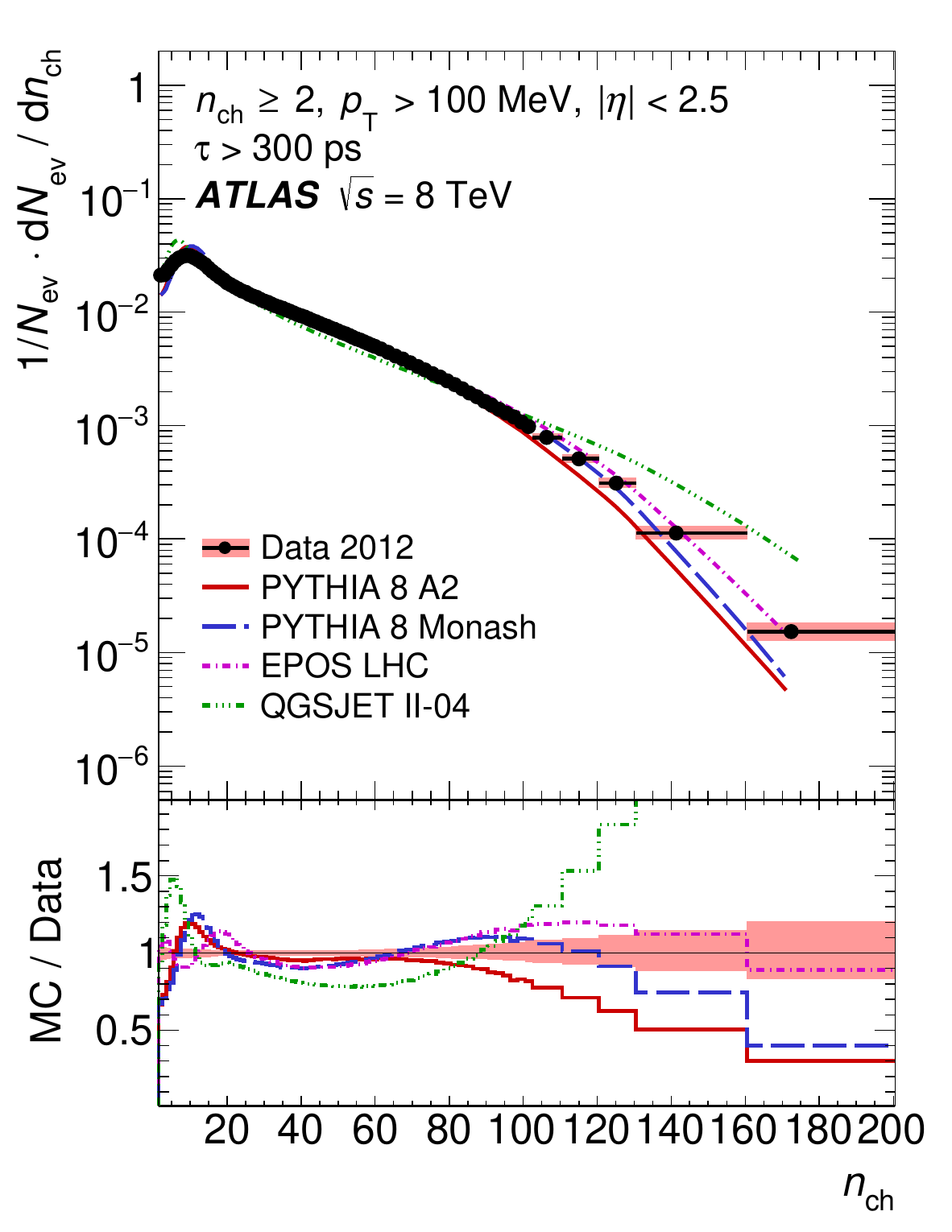}}
(a) 
\\
\end{minipage}
\hspace{2mm}
\begin{minipage}[h]{0.45\textwidth}
\center{\includegraphics[width=1.0\linewidth]{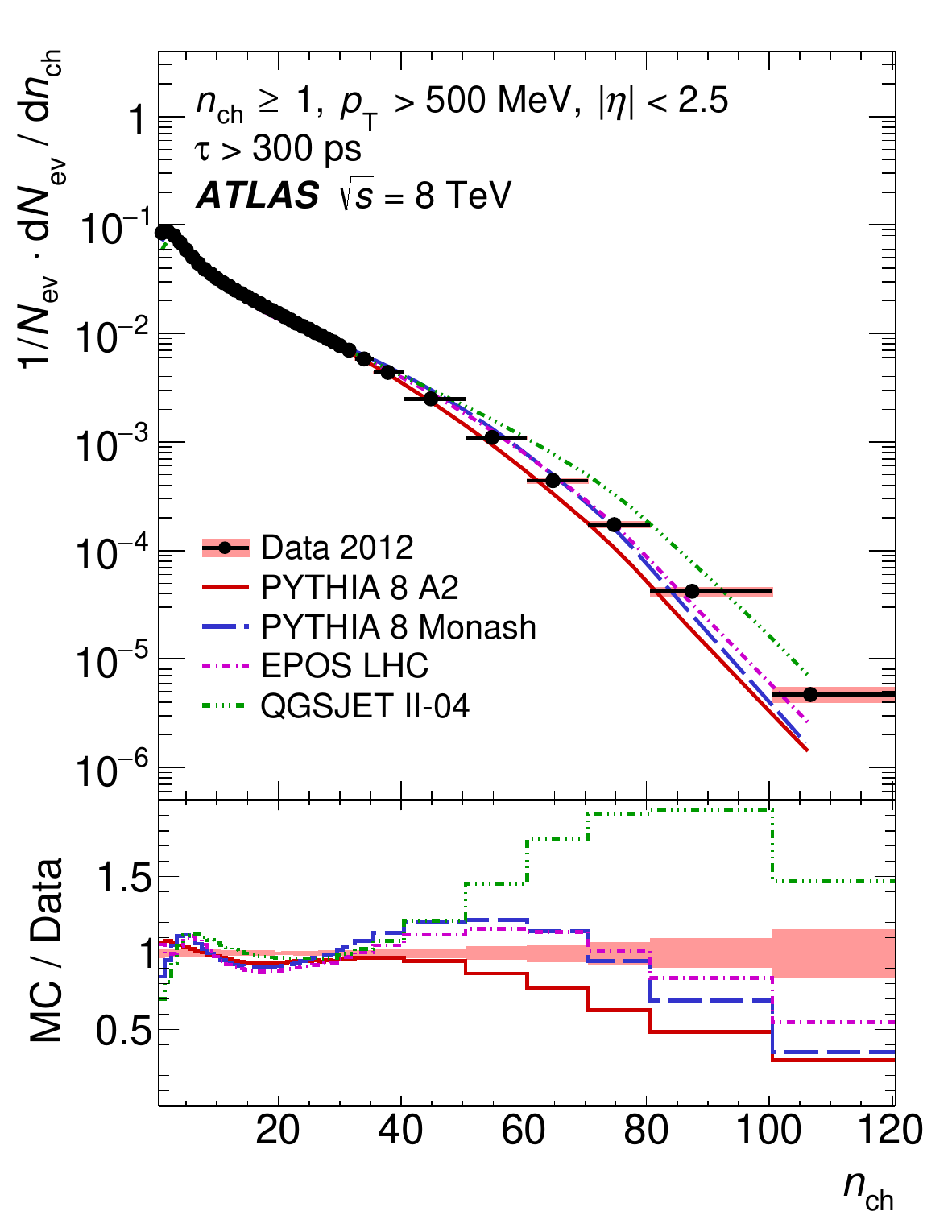}}
(b) 
\\
\end{minipage}
\caption{
Top panel: 
Primary charged-particle multiplicities as  a function of the  multiplicity
at the centre-of-mass energy  \(\sqrt{s}=  8\)~\TeV\ \cite{STDM-2014-19}  
with 
(a) \(n_{\mathrm{ch}} \ge 2\) and \(p_{\mathrm{T}} >100\)~\MeV\ 
and 
(b) \(n_{\mathrm{ch}} \ge 1\) and \(p_{\mathrm{T}} >500\)~\MeV.
The data represented by dots 
is
compared 
to various particle-level MC predictions,  which are shown by curves. 
The shaded areas around the data points represent the total statistical and systematic uncertainties added in quadrature.
Bottom panel: 
The ratios of the MC predictions to  the experimental results are shown. 
Bands represent the uncertainties of the experimental results.
Taken from Ref.~\cite{STDM-2014-19}.
}
\label{fig_8_nch}
\end{figure*}

\begin{figure*}[t!]
\centering
\begin{minipage}[h]{0.32\textwidth} 
\center{\includegraphics[width=1.0\linewidth]{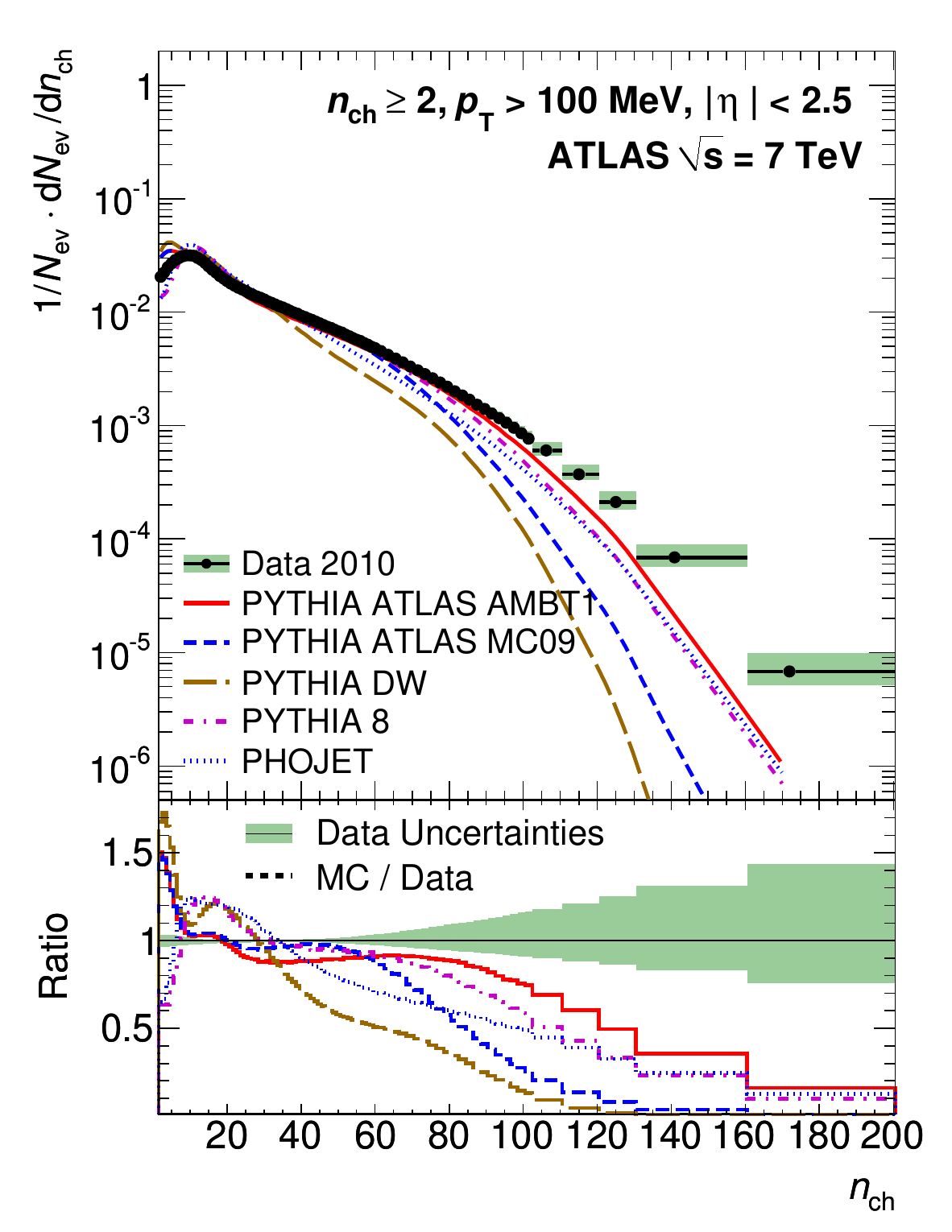}}
(a) 
\\
\end{minipage}
\hfill
\begin{minipage}[h]{0.32\textwidth}
\center{\includegraphics[width=1.0\linewidth]{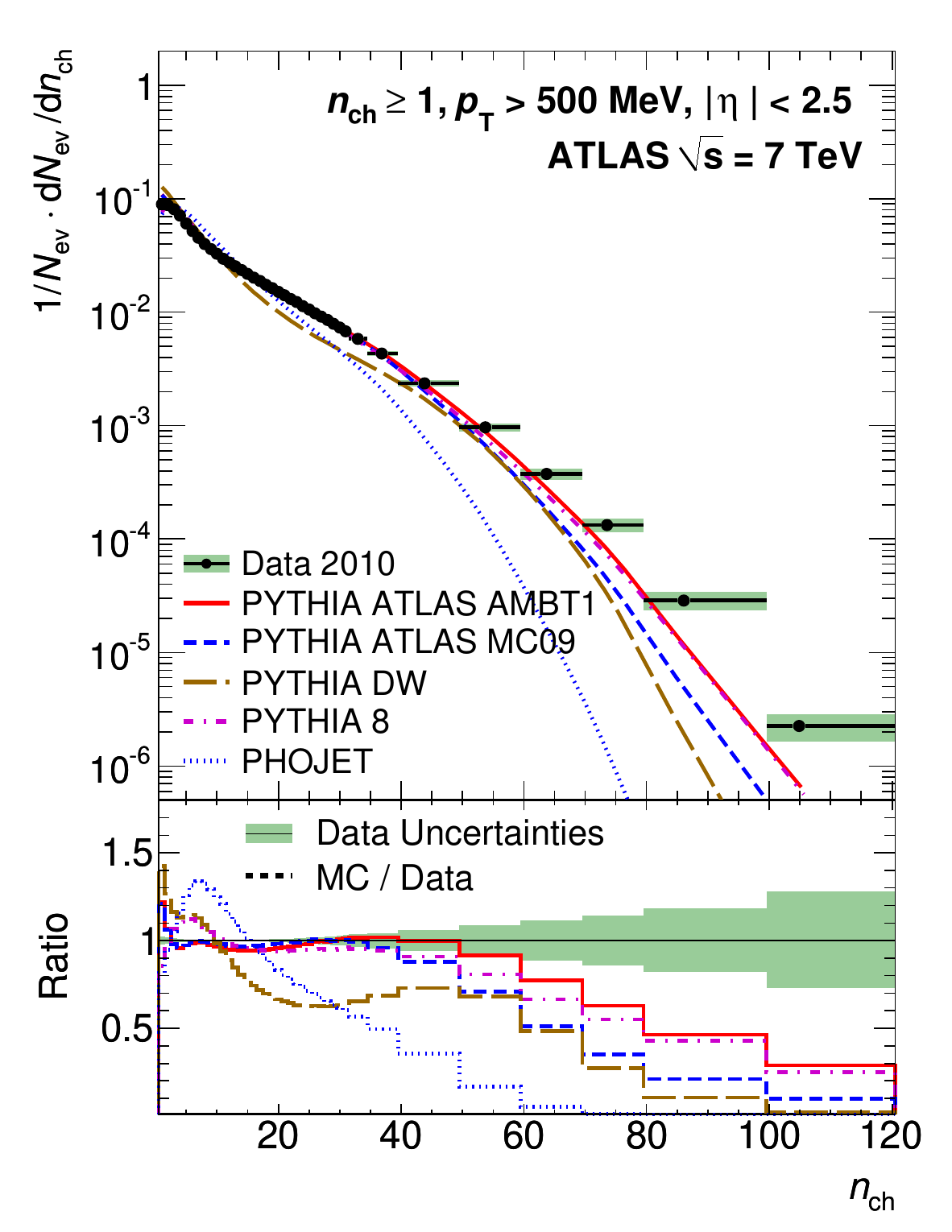}}
(b) 
\\
\end{minipage}
\hfill
\begin{minipage}[h]{0.32\textwidth} 
\center{\includegraphics[width=1.0\linewidth]{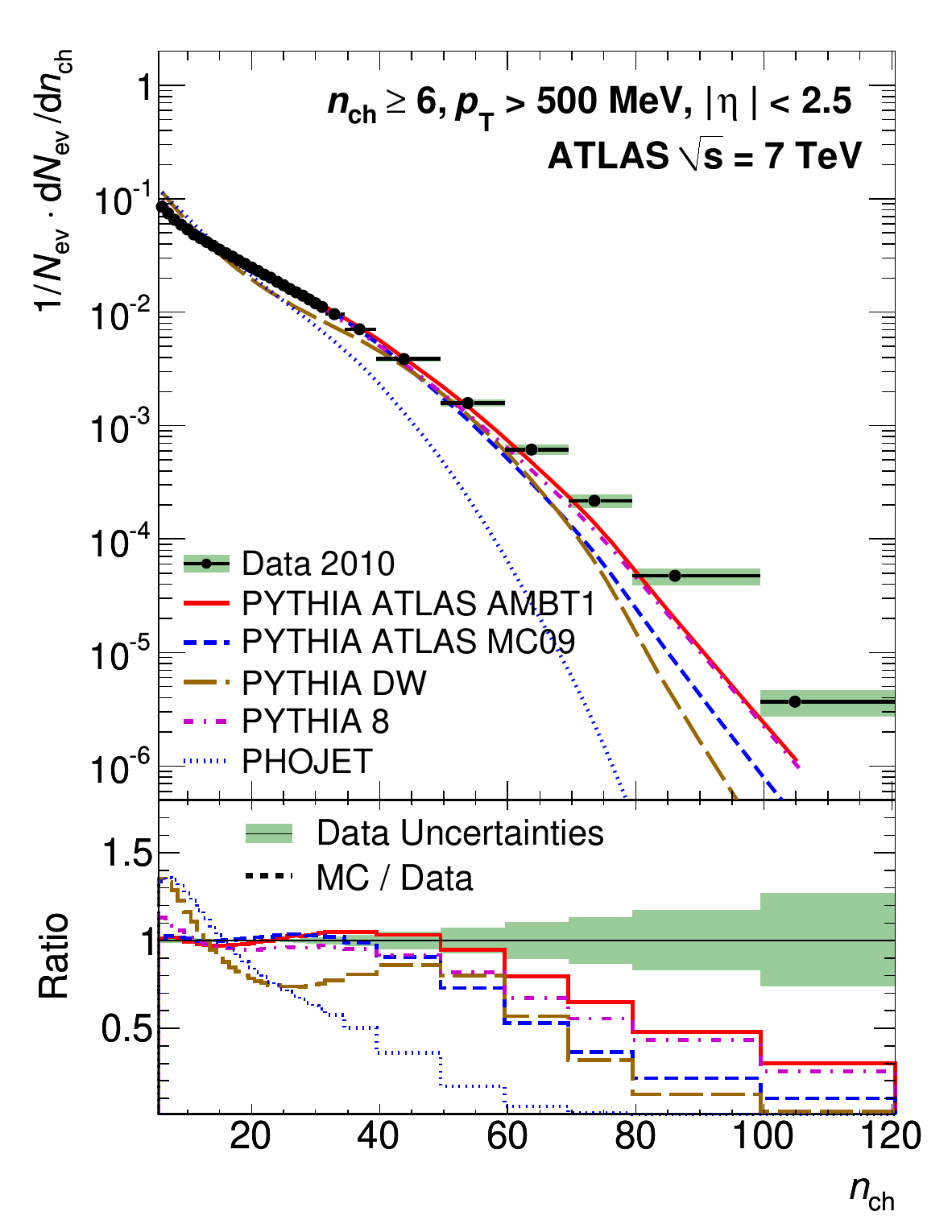}}
(c) 
\\
\end{minipage}
\caption{
Top panel: 
Primary charged-particle multiplicities as  a function of  the  multiplicity
at the centre-of-mass energy  \(\sqrt{s}= 7\)~\TeV\ \cite{STDM-2010-06}	   with 
(a) \(n_{\mathrm{ch}} \ge 2\) and \(p_{\mathrm{T}} >100\)~\MeV\ 
and 
for \(p_{\mathrm{T}} >500\)~\MeV\ with 
(b) \(n_{\mathrm{ch}} \ge 1\)
and
(c)  \(n_{\mathrm{ch}} \ge 6\).
The data represented by dots 
is
compared to various particle-level MC predictions,  which are shown by curves. 
The shaded areas around the data points represent the total statistical and systematic uncertainties added in quadrature.
Bottom panel: 
The ratios of the MC predictions to  the experimental results are shown. 
Bands represent the uncertainties of the experimental results.
Taken from Ref.~\cite{STDM-2010-06}.
}
\label{fig_7_nch}
\end{figure*}

\begin{figure*}[t!]
\centering
\begin{minipage}[h]{0.45\textwidth} 
\center{\includegraphics[width=1.0\linewidth]{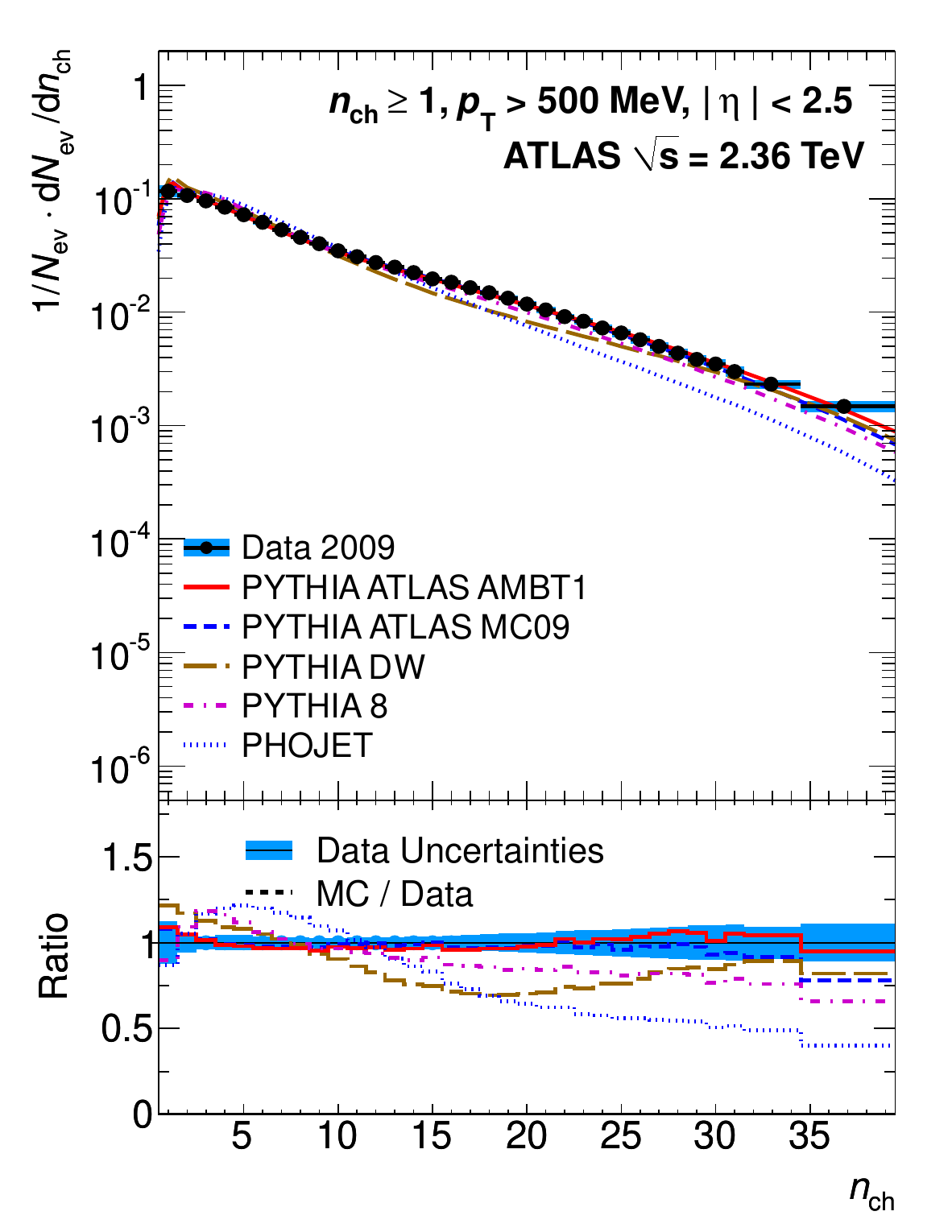}}
\\
\end{minipage}
\caption{
Top panel: 
Primary charged-particle multiplicities as  a function of the  multiplicity
at the centre-of-mass energy  \(\sqrt{s}=  2.36\)~\TeV\ \cite{STDM-2010-06}	   with 
\(n_{\mathrm{ch}} \ge 1\) and \(p_{\mathrm{T}} >500\)~\MeV.
The data represented by dots 
is
compared 
to various particle-level MC predictions,  which are shown by curves. 
The shaded areas around the data points represent the total statistical and systematic uncertainties added in quadrature.
Bottom panel: 
The ratios of the MC predictions to  the experimental results are shown. 
Bands represent the uncertainties of the experimental results.
Taken from Ref.~\cite{STDM-2010-06}.
}
\label{fig_236_nch}
\end{figure*}

\begin{figure*}[t!]
\centering
\begin{minipage}[h]{0.32\textwidth} 
\center{\includegraphics[width=1.0\linewidth]{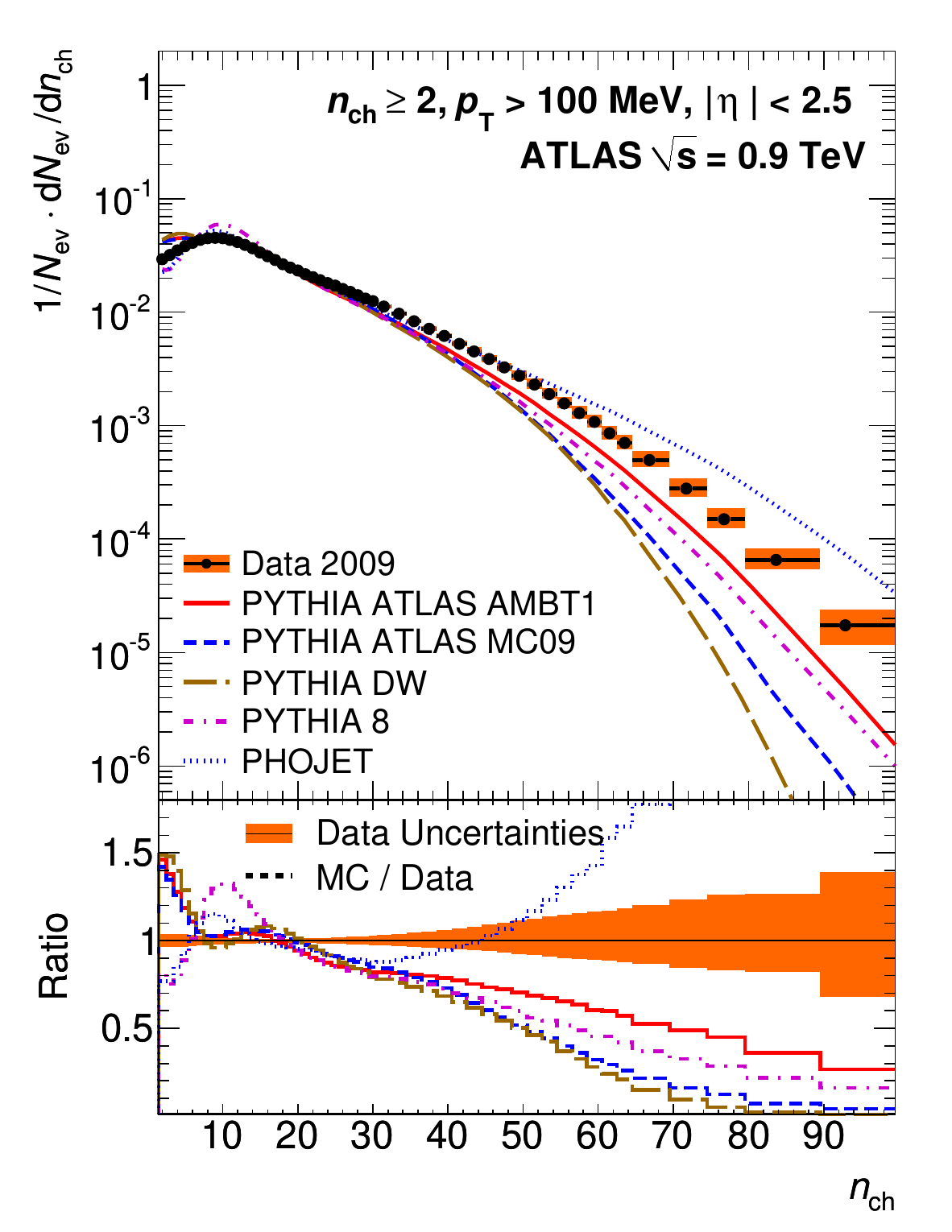}}
(a) 
\\
\end{minipage}
\hfill
\begin{minipage}[h]{0.32\textwidth}
\center{\includegraphics[width=1.0\linewidth]{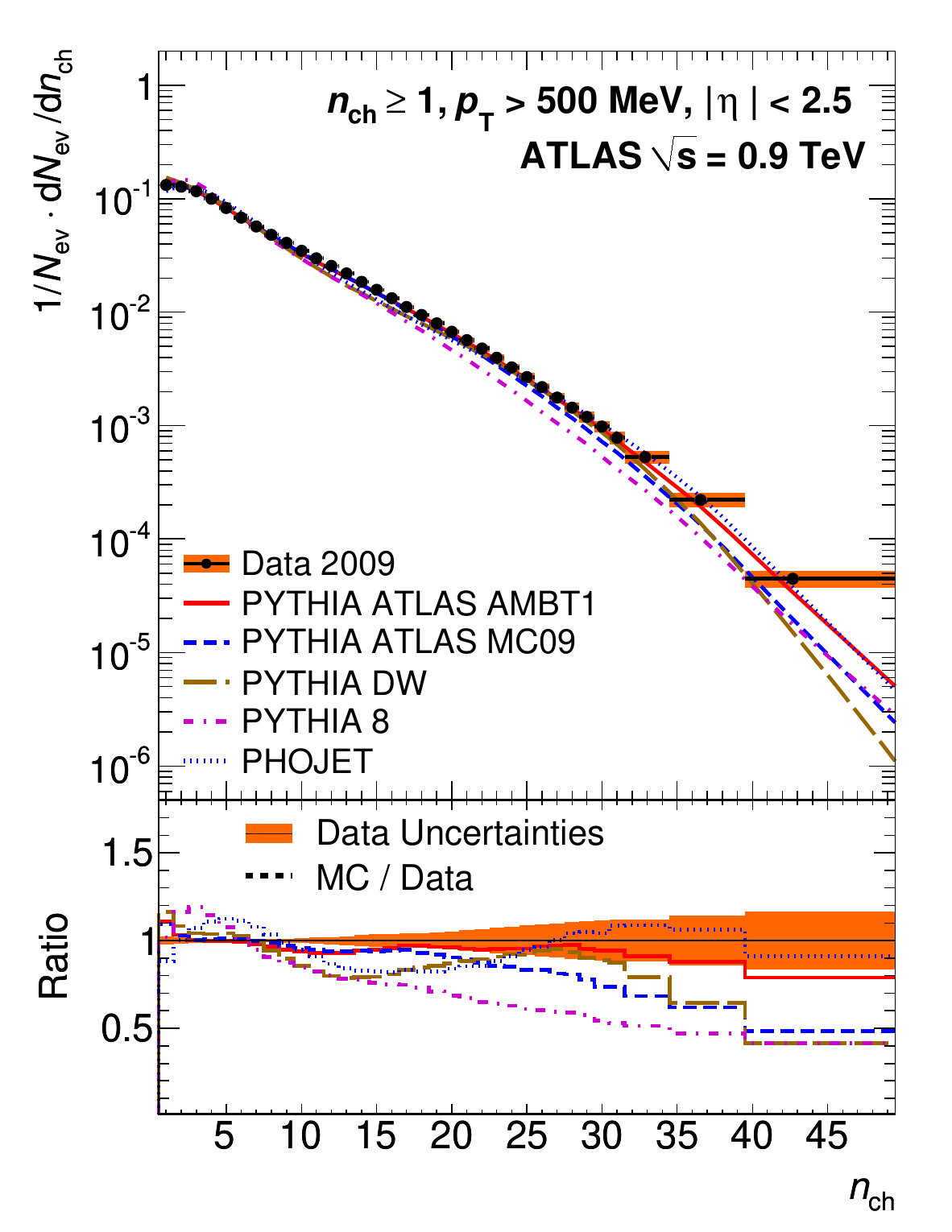}}
(b) 
\\
\end{minipage}
\hfill
\begin{minipage}[h]{0.32\textwidth} 
\center{\includegraphics[width=1.0\linewidth]{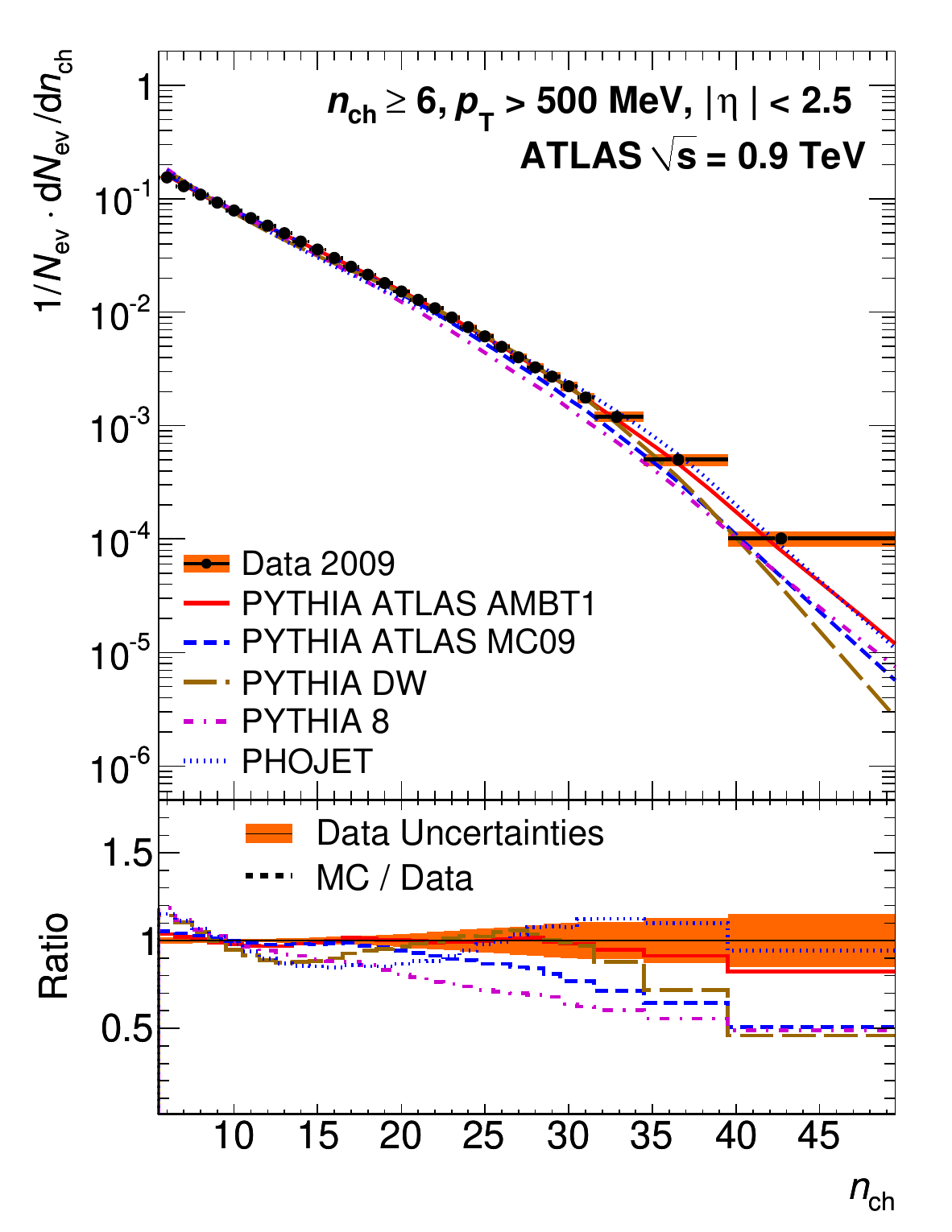}}
(c) 
\\
\end{minipage}
\caption{
Top panel: 
Primary charged-particle multiplicities as  a function of  the multiplicity
at the centre-of-mass energy  \(\sqrt{s}=  0.9\)~\TeV\ \cite{STDM-2010-06}	   with 
(a) \(n_{\mathrm{ch}} \ge 2\) and \(p_{\mathrm{T}} >100\)~\MeV\ 
and 
for \(p_{\mathrm{T}} >500\)~\MeV\ with 
(b) \(n_{\mathrm{ch}} \ge 1\)
and
(c)  \(n_{\mathrm{ch}} \ge 6\).
The data represented by dots  
is
compared 
to various particle-level MC predictions,  which are shown by curves. 
The shaded areas around the data points represent the total statistical and systematic uncertainties added in quadrature.
Bottom panel: 
The ratios of the MC predictions to  the experimental results are shown. 
Bands represent the uncertainties of the experimental results.
Taken from Ref.~\cite{STDM-2010-06}.
}
\label{fig_09_nch}
\end{figure*}

The charged-particle multiplicity distributions are  shown in  Figs.~\ref{fig_13_nch} -- \ref{fig_09_nch} at 
the CM energies  \(\sqrt{s}= 0.9,\ 2.36,\ 7,\ 8,\)
and  \(13\)~\TeV. 

Figures~\ref{fig_13_nch}(a) and (b)   show the  charged-particle multiplicity distributions at 
the CM energy  \(\sqrt{s}= 13\)~\TeV\  for events with 
\(n_{\mathrm{ch}} \ge 2\), \(p_{\mathrm{T}} >100\)~\MeV\ 
\cite{STDM-2015-17} and
\(n_{\mathrm{ch}} \ge 1\),  \(p_{\mathrm{T}} >500\)~\MeV\ 
\cite{STDM-2015-02},  respectively. 

In Fig.~\ref{fig_13_nch}(a) for events with 
\(n_{\mathrm{ch}} \ge 2\), \(p_{\mathrm{T}} >100\)~\MeV\ 
at \(\sqrt{s}= 13\)~\TeV\ 
the form of the measured distribution is reproduced reasonably by all models. 
\textsc{Pythia\,8} \textsc{A2} describes the data well for \(30 < n_{\mathrm{ch}} < 80\)
but underestimates  them for higher  \(n_{\mathrm{ch}}\). 
For this multiplicity region,  \textsc{Pythia\,8} \textsc{Monash}, \textsc{EPOS}  and 
\textsc{QGSJET-II}  underestimate the data by up to \(20\)\%. 
\textsc{Pythia\,8} \textsc{Monash} and  \textsc{EPOS} 
overestimate the data for  the multiplicity region \(n_{\mathrm{ch}} > 80\)  and drop below
the measurement in the high-\(n_{\mathrm{ch}}\) region, 
starting from \(n_{\mathrm{ch}} > 130\) and  \(n_{\mathrm{ch}} > 200\),  respectively. 
\textsc{QGSJET-II}  significantly  overestimates the data  for  the multiplicity region 
\(n_{\mathrm{ch}} > 100\).
%
Figure~\ref{fig_13_nch} (b) shows the charged-particle multiplicity distribution for events with 
\(n_{\mathrm{ch}} \ge 1\), \(p_{\mathrm{T}} >500\)~\MeV\  at \(\sqrt{s}= 13\)~\TeV. 
The high-\(n_{\mathrm{ch}}\) region has significant contributions from events 
with numerous MPI. 
\textsc{Pythia\,8} \textsc{A2}  describes  
the 
data well 
in the multiplicity  region 
\(n_{\mathrm{ch}} < 50\) but predicts too few events at larger \(n_{\mathrm{ch}}\).
\textsc{Pythia\,8} \textsc{Monash},  \textsc{EPOS}  and 
\textsc{QGSJET-II}   describe the data reasonably well in the multiplicity region  
\(n_{\mathrm{ch}} < 30\) but predict too many events in
mid-\(n_{\mathrm{ch}}\)  region, 
with \textsc{Pythia\,8} \textsc{Monash} and \textsc{EPOS} 
predicting too few events in 
region  \(n_{\mathrm{ch}} > 100\) while 
\textsc{QGSJET-II}    continues to be above the data.

In Figs.~\ref{fig_8_nch}(a) and (b)  
show  
the distributions of 
primary charged-particle multiplicity
for the minimum transverse momentum thresholds of 
\(100\)~\MeV\ and \(500\)~\MeV\ at \(\sqrt{s}=  8\)~\TeV\  \cite{STDM-2014-19},  respectively. 
For the lower threshold, the distribution rises until  \(n_{\mathrm{ch}} \sim 9\) before falling steeply. 
For the higher threshold, the distribution peaks at  \(n_{\mathrm{ch}} \sim 2\). 
The models are consistent with the data, 
although the  \textsc{EPOS} model provides a fair description. 
The two \textsc{Pythia\,8} calculations predict distribution peaks 
that
are at higher \(n_{\mathrm{ch}}\) 
than those observed and underestimate  the event yield at low and high  multiplicities. 
The 
\textsc{QGSJET-II}  tune overestimates the data at low and high  \(n_{\mathrm{ch}}\) 
values and underestimates the data for intermediate  \(n_{\mathrm{ch}}\) values.

In Figs.~\ref{fig_7_nch}(a) and  \ref{fig_09_nch}(a)  
shown
the distributions of primary charged-particle multiplicity
for  the most inclusive PS region
\(n_{\mathrm{ch}} \ge 2\),  \(p_{\mathrm{T}} >100\)~\MeV\ and \(\mid\eta\mid < 2.5\) 
at the CM energies \(\sqrt{s} = 7\)~\TeV\ and \(\sqrt{s} = 0.9\)~\TeV, respectively.
Here, 
the variations between models at both  low \(n_{\mathrm{ch}}\)  and 
high \(n_{\mathrm{ch}}\)   are increased, 
and no model predicts the observed spectra.
Due to the normalisation,  \( 1 / N_{\mathrm{ev}}\), 
the  deviation observed in one region needs to be compensated for  by 
the one in the other direction somewhere else. 
%
Figures~\ref{fig_7_nch}(b),  \ref{fig_236_nch},  and  \ref{fig_09_nch}(b)
show  the primary charged-particle multiplicity distributions for 
\(n_{\mathrm{ch}} \ge 1\),  \(p_{\mathrm{T}} >500\)~\MeV\ and \(\mid\eta\mid < 2.5\)
at  the CM energies  \(\sqrt{s} = 7\)~\TeV,   \( 2.36\)~\TeV, and  \(0.9\)~\TeV, respectively. 
At low  \(n_{\mathrm{ch}}\), all models predict more events than observed in 
the data,  which is compensated for by an under-prediction in the tails of the distributions. 
The predictions of \textsc{PHOJET} at  \(\sqrt{s} = 0.9\)~\TeV\  
model  the data reasonably well,  but at \(\sqrt{s} = 2.36\)~\TeV\  and 
\(\sqrt{s} = 7\)~\TeV\   they do not model the observed spectrum so well. 
The \textsc{Pythia\,6} \textsc{AMBT1} tune seems to provide the best agreement with 
the data.
%
Figures ~\ref{fig_7_nch}(c)  and   \ref{fig_09_nch}(c)   show the distribution for the diffraction-reduced 
PS region for events with  
\(n_{\mathrm{ch}} > 6\),   \(p_{\mathrm{T}} >500\)~\MeV.
The distributions are very similar to those in  Figs.~\ref{fig_7_nch}(c)
and \ref{fig_09_nch}(c) with a cut at \(n_{\mathrm{ch}} > 6\);  only the normalisation is different.

\begin{figure*}[t!]
\centering
\begin{minipage}[h]{0.32\textwidth}     
\center{\includegraphics[width=1.0\linewidth]{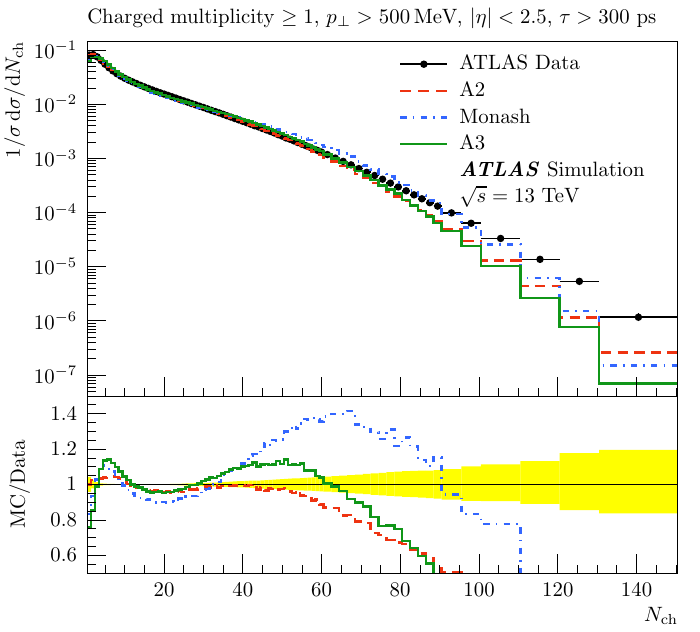}}
(a) 
\\
\end{minipage}
\hfill
\begin{minipage}[h]{0.32\textwidth}
\center{\includegraphics[width=1.0\linewidth]{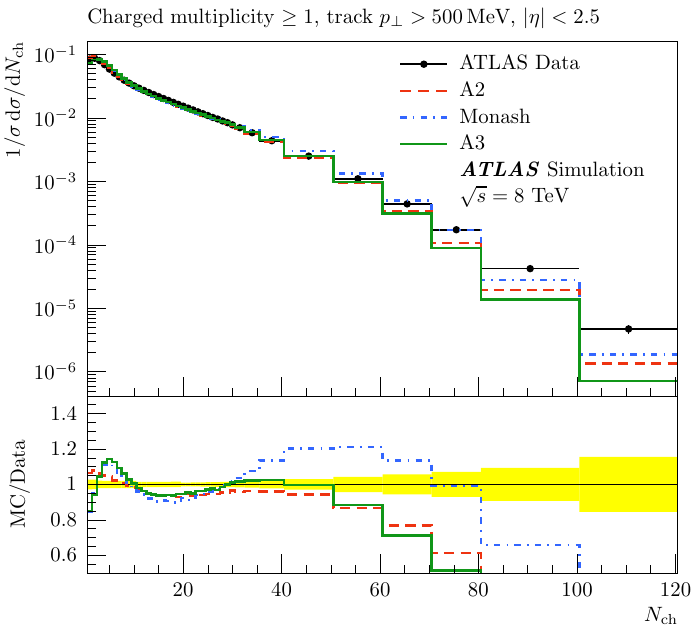}}
(b) 
\\
\end{minipage}
\hfill
\begin{minipage}[h]{0.32\textwidth} 
\center{\includegraphics[width=1.0\linewidth]{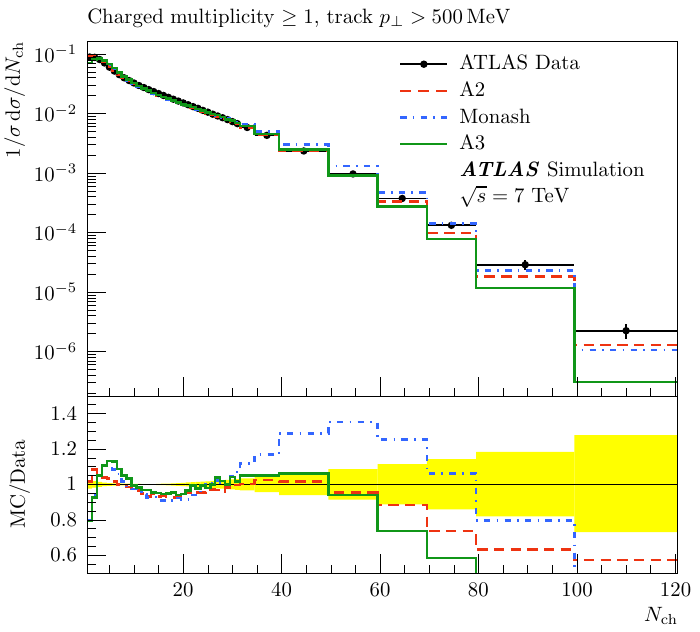}}
(c) 
\\
\end{minipage}
\vfill
\begin{minipage}[h]{0.32\textwidth} 
\center{\includegraphics[width=1.0\linewidth]{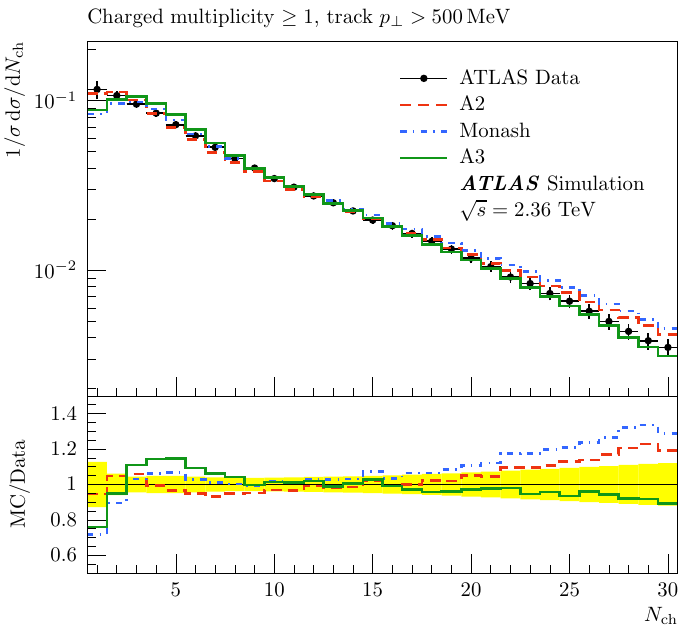}}
(d) 
\\
\end{minipage}
\hspace{2mm}
\begin{minipage}[h]{0.32\textwidth}
\center{\includegraphics[width=1.0\linewidth]{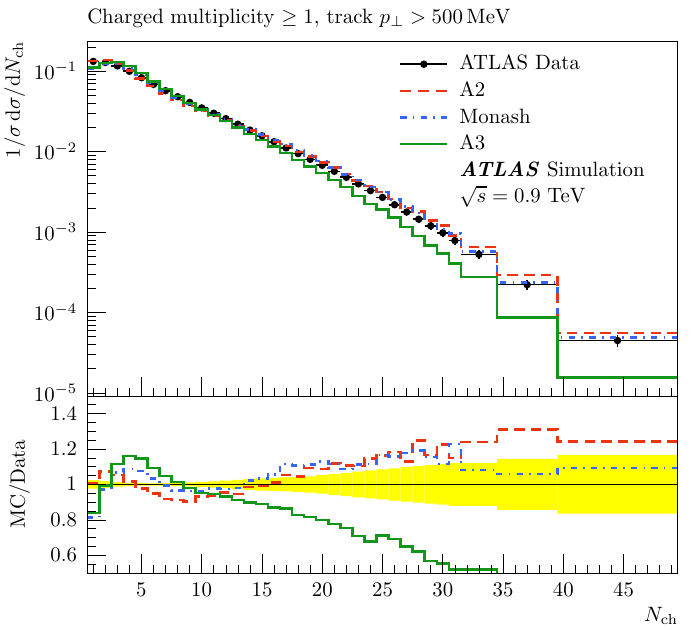}}
(e) 
\\
\end{minipage}
\caption{
Top panel:
The \textsc{Pythia\,8} \textsc{A3}, \textsc{A2} and \textsc{Monash}  tune predictions 
\cite{ATLAS:2016puo} compared with  the ATLAS 
primary charged-particle multiplicities as  a function of  the  multiplicity
distributions for events with  \(n_{\mathrm{ch}} \ge 1\) with \(p_{\mathrm{T}} >500\)~\MeV\ 
at centre-of-mass energies 
(a) \(\sqrt{s} = 13\)~\TeV, 
(b) \(\sqrt{s} = 8\)~\TeV, 
(c) \(\sqrt{s} = 7\)~\TeV,
(d) \(\sqrt{s} = 2.36\)~\TeV\ 
and 
(e) \(\sqrt{s} = 0.9\)~\TeV. 
The yellow-shaded areas represent the measurement uncertainty.
Bottom panel: 
The ratios of the MC predictions to  the experimental results  are shown. 
Bands represent the uncertainties of the experimental results.
Taken from Ref.~\cite{ATLAS:2016puo}.
}
\label{fig_A3_nch}
\end{figure*}

In Fig.~\ref{fig_A3_nch},   for   the  charged-particle multiplicity, 
the ATLAS \textsc{Pythia\,8} \textsc{A3}  is comparable to other tunes.
At \(\sqrt{s} = 13\)~\TeV\  \textsc{Pythia\,8} \textsc{A2}  
describes the low multiplicity part better than  \textsc{Pythia\,8} \textsc{A3} 
in the range of  \(40\)--\(60\) charged particles.
The shape of the distribution predicted by the  \textsc{Pythia\,8} \textsc{A3} tune 
is consistent across the 
CM energies. 

\begin{figure*}[t!]
\centering
\begin{minipage}[h]{0.45\textwidth} 
\center{\includegraphics[width=1.0\linewidth]{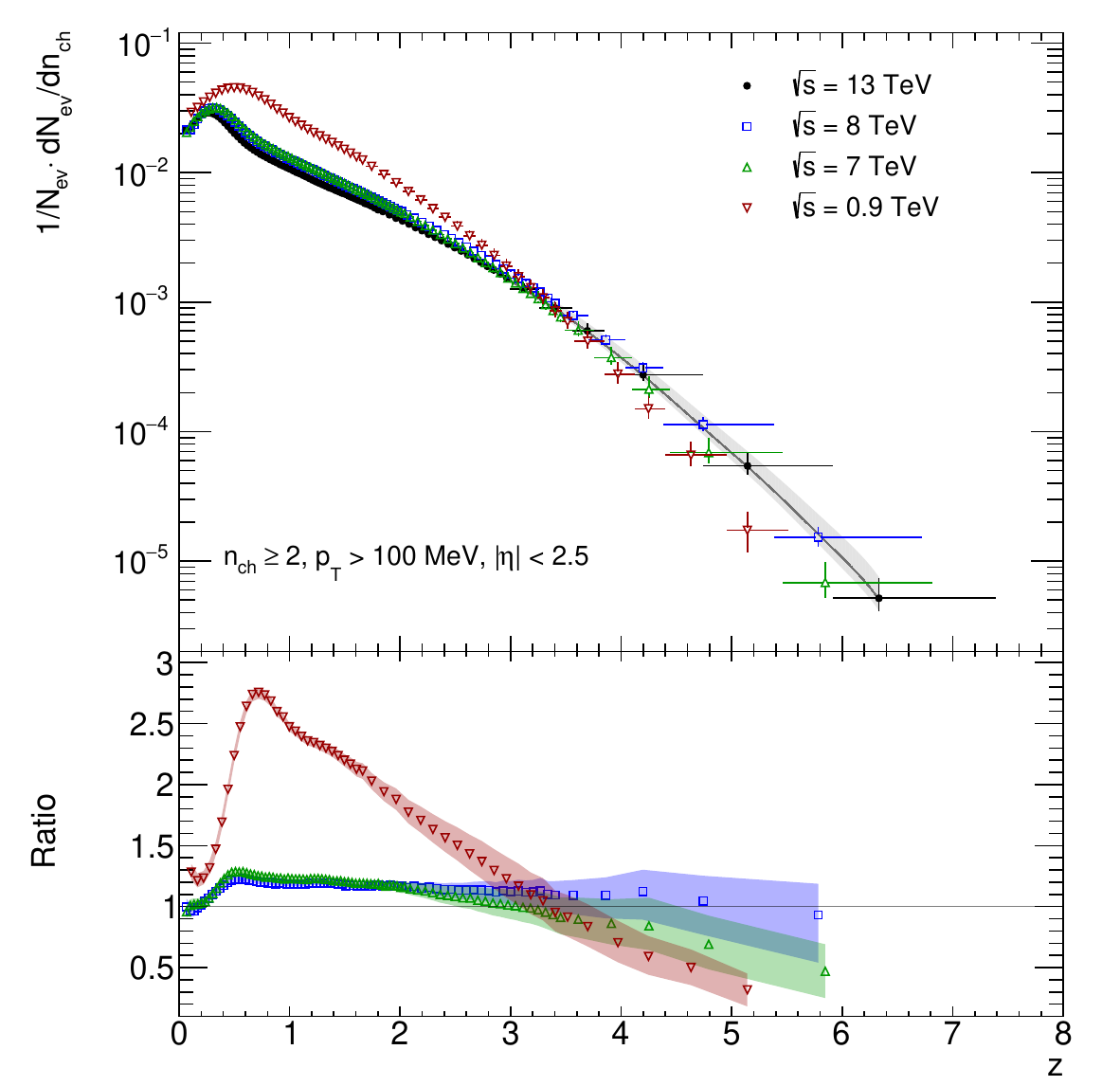}} 
(a) 
\\
\end{minipage}
\hspace{2mm}
\begin{minipage}[h]{0.45\textwidth} 
\center{\includegraphics[width=1.0\linewidth]{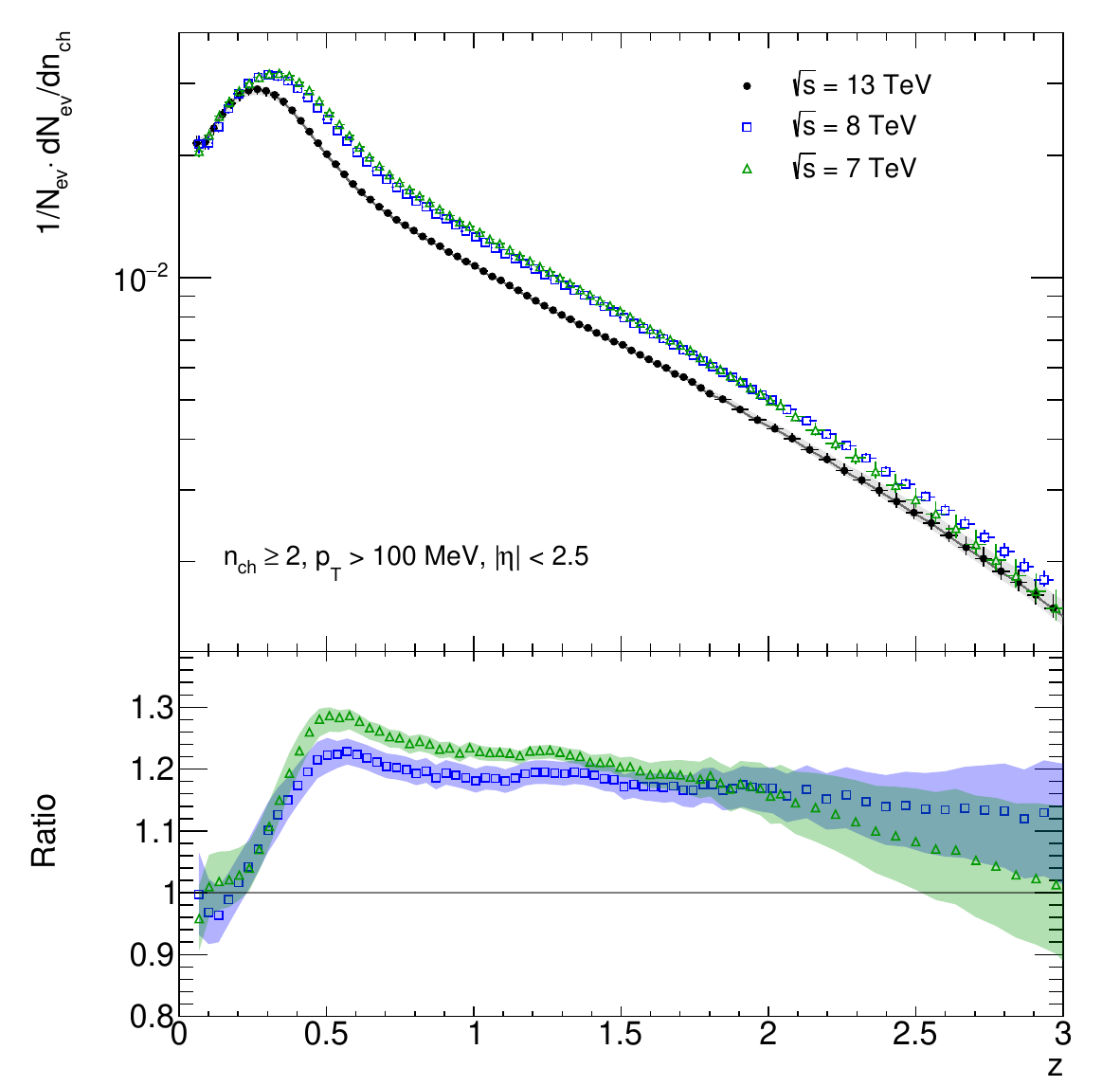}} 
(b)
\\
\end{minipage}
\caption{
Top panel: 
Primary charged-particle  multiplicity distributions     as a function of the scaled multiplicity 
\(z\), defined in Eq.~(\ref{eq_mch}), for events with   
\(n_{\mathrm{ch}} \ge 2\), \(p_{\mathrm{T}} >100\)~\MeV\  and \(\mid\eta\mid  < 2.5\) 
measured  at the centre-of-mass energies  \(0.9\),  \(7\),  \(8\) and \(13\)~\TeV\  by ATLAS 
\cite{STDM-2010-06,STDM-2014-19,STDM-2015-02,STDM-2015-17}
in 
(a) the complete multiplicity region and 
(b) the zoom multiplicity region with \(z \le 3\) 
at  \( \sqrt{s} = 7\),  \(8\) and \(13\)~\TeV.
The grey curve and  the band of 
uncertainties are the result of the interpolation 
of the charged-particle multiplicity distribution at  \(13\)~\TeV.
The error bars and boxes represent the statistical and systematic contributions, respectively.
Bottom panel: 
The ratios of the charged-particle  multiplicity distributions to the interpolated distribution 
at \( \sqrt{s}  = 13\)~\TeV\ are shown. 
Bands represent the uncertainties for the ratios 
as the 
results of statistical and systematic uncertainties added in quadrature for both distributions.
Taken from Ref.~\cite{Kulchitsky:2022gkm}.
}
\label{fig_events_pT100_mch}
\end{figure*}

\begin{figure*}[t!]
\centering
\begin{minipage}[h]{0.45\textwidth} 
\center{\includegraphics[width=1.0\linewidth]{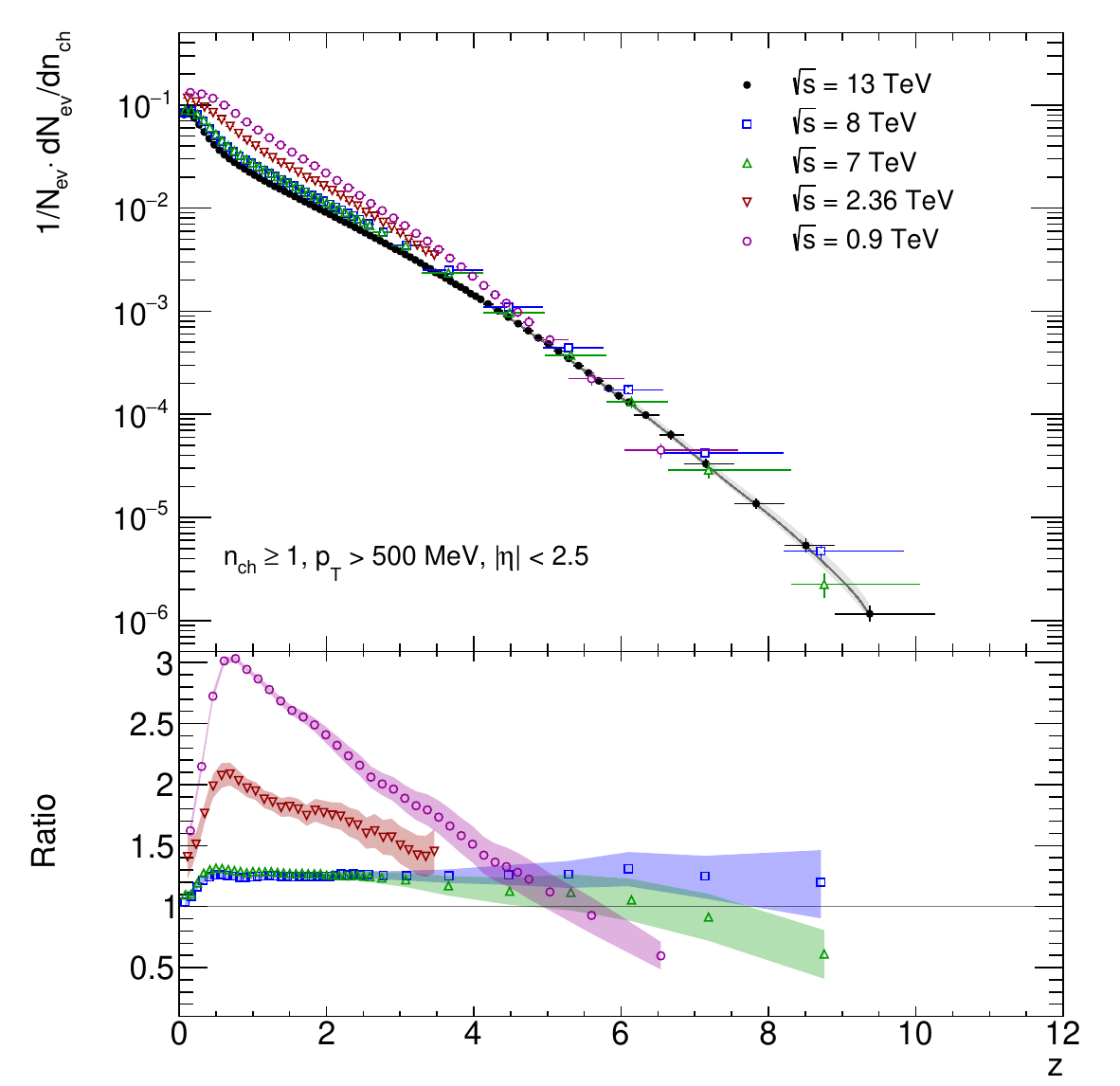}} 
(a) 
\\
\end{minipage}
\hspace{2mm}
\begin{minipage}[h]{0.45\textwidth} 
\center{\includegraphics[width=1.0\linewidth]{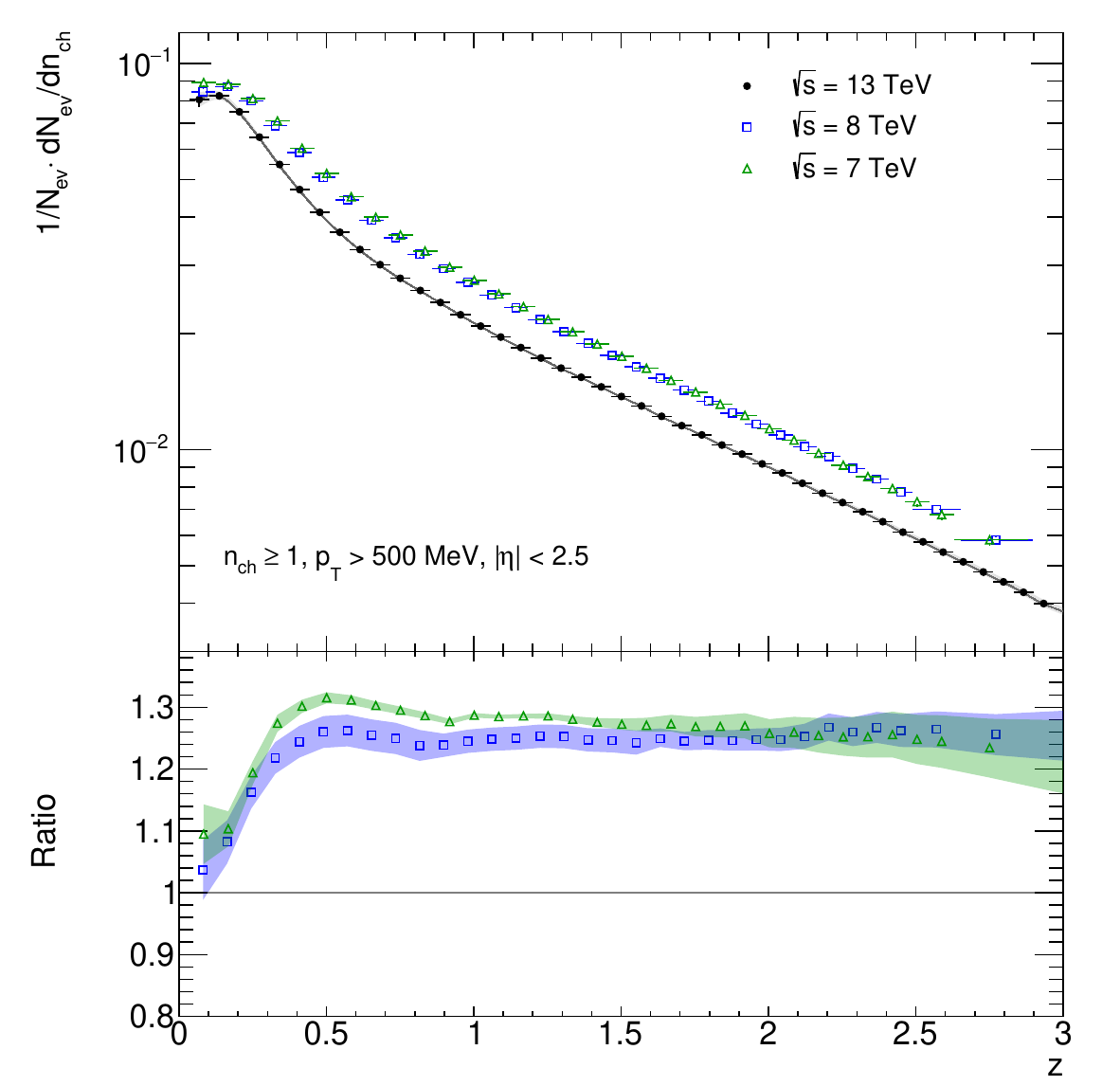}} 
(b)
\\
\end{minipage}
\caption{
Top panel: 
Primary charged-particle  multiplicity distributions  as a function of the scaled multiplicity 
\(z\), defined in Eq.~(\ref{eq_mch}),
for events with  \(n_{\mathrm{ch}} \ge 1\), \(p_{\mathrm{T}} >500\)~MeV and \(\mid\eta\mid  < 2.5\) 
measured   at the centre-of-mass energies  \(0.9\),  \(2.36\), \(7\),  \(8\) and \(13\)~\TeV\  by  
ATLAS 
\cite{STDM-2010-06,STDM-2014-19,STDM-2015-02,STDM-2015-17} in
(a) the complete multiplicity region and 
(b) the zoom multiplicity region with \(z \le 3\) 
at    \( \sqrt{s} = 7\),  \(8\) and \(13\)~\TeV.
The grey curve and  the band of 
uncertainties are the result of the interpolation 
of the charged-particle multiplicity distribution at  \(13\)~\TeV.
The error bars and boxes represent the statistical and systematic contributions, respectively.
Bottom panel: 
The ratios of the charged-particle  multiplicity distributions to the interpolated distribution 
at \( \sqrt{s}  = 13\)~\TeV\ are shown. 
Bands represent 
uncertainties for the ratios 
as 
the
results of statistical and systematic uncertainties added in quadrature for both distributions.
Taken from Ref.~\cite{Kulchitsky:2022gkm}.
}
\label{fig_events_pT500_mch}
\end{figure*}

For correct comparison of  the  charged-particle multiplicity and average transverse momentum distributions 
for different energies or kinematic regions,  the scaled multiplicity  \(z\), 
usually called  the KNO variable, see Eq.~(\ref{eq_mch}) is introduced. 
For example,   a comparison of  the results for different kinematic regions
with two \( p_{\mathrm{T}}^{\mathrm{min}} \) thresholds   was presented in 
Ref.~\cite{ATLAS:2022wvk}.
The comparison of the primary charged-particle multiplicities as a function of  the scaled multiplicity
\(z\) or  the KNO scale for events with  
\(n_{\mathrm{ch}} \ge 2\) and \(p_{\mathrm{T}} >100\)~\MeV;
\(n_{\mathrm{ch}} \ge 1\) and \(p_{\mathrm{T}} >500\)~\MeV\ 
for  \(\mid\eta\mid  < 2.5\) 
measured  by  
ATLAS  at  
\( \sqrt{s} \) from \(0.9\)  to  \(13\)~\TeV\ 
\cite{STDM-2010-06,STDM-2014-19,STDM-2015-02,STDM-2015-17}
are presented  in Fig.~\ref{fig_events_pT100_mch}  and   Fig.~\ref{fig_events_pT500_mch} 
\cite{Kulchitsky:2022gkm},  respectively. 
For these figures, 
the multiplicity axis was compressed  by the factor 
\( \langle n_{\mathrm{ch}} ( \sqrt{s},  p_{\mathrm{T}}^{\mathrm{min}} ) \rangle \).
The KNO scale is the same, 
and therefore it is the correct scale for comparing
distributions at different  \( \sqrt{s} \) or distributions in different PS regions.
The scaled multiplicity regions are up to   \(7.5\)  of the  average total multiplicity  
for  \(p_{\mathrm{T}} >100\)~\MeV\   and up to  \(10.5\)  of the average total multiplicity  
for  \(p_{\mathrm{T}} >500\)~\MeV\ as shown  in
Figs.~\ref{fig_events_pT100_mch}(a) and \ref{fig_events_pT500_mch}(a), respectively. 

In Table~\ref{tab:average_nch}, 
the relative uncertainty,  \( \delta\langle n_{\mathrm{ch}} \rangle / \langle n_{\mathrm{ch}}\rangle \), 
is presented for  average total multiplicities.
Relative uncertainties are small and equal to 
\(0.32\)--\(0.66\)\% for  \(p_{\mathrm{T}} >100\)~\MeV\  and
\(0.24\)--\(0.46\)\% for  \(p_{\mathrm{T}} >500\)~\MeV, 
except of  the result at  \(\sqrt{s}=2.36\)~\GeV\  which was measured with 
a lower accuracy.

In the bottom panels in Figs.~\ref{fig_events_pT100_mch} and \ref{fig_events_pT500_mch}
ratios of  the charged-particle distributions at \(0.9\) -- \(8\)~\TeV\ 
to the distribution at \( 13\)~\TeV\ are shown. 
These ratios  and their uncertainties    are obtained by interpolation.
For the interpolation procedure,  the  \textsc{Interpolator}  method  
of the \textsc{Root} statistical analysis framework \cite{Antcheva:2009zz} was used. 
In Figs.~\ref{fig_events_pT100_mch} -- \ref{fig_averagepT_pT100_mch},  the grey curve and  the band of 
uncertainties   are the result of the interpolation of the   distribution  at  \(13\)~\TeV.

Figures~\ref{fig_events_pT100_mch} and \ref{fig_events_pT500_mch}
show that primary charged-particle multiplicity distributions decrease 
as  the collision energy  increases from \(0.9\)  to  \(13\)~\TeV\  by 
a factor of  \(\approx 3\)  for maximum  of the functions  at \( z \approx 0.7 \).
The results for  \(\sqrt{s} = 7\), \(8\) and \(13\)~TeV and \(z \le 3\) 
are presented in  Fig.~\ref{fig_events_pT100_mch}(b)  for \(p_{\mathrm{T}} >100\)~\MeV\  
and in Fig.~\ref{fig_events_pT500_mch}(b)  for \(p_{\mathrm{T}} >500\)~\MeV. 
The distributions at \( \sqrt{s} = 7\) and \(8\)~\TeV\  are in agreement within error bars except for  
region  \( 0.5 < z < 1.5 \). 
The multiplicity distribution at  \(8\)~\TeV\ is  \(\approx  20\)\% larger  than at \(13\)~\TeV\ for   
region \( z < 3\) in both cases.


For \( p_{\mathrm{T}} > 100\)~\MeV\ and  \( p_{\mathrm{T}} > 500\)~\MeV\  at  
the highest energies, 
the form of the measured distribution is reproduced reasonably by all models.
\textsc{Pythia\,8}  \textsc{A2}  describe
the data well for middle \( n_{\mathrm{ch}}\)
but underestimates it for higher \( n_{\mathrm{ch}}\).
For middle \( n_{\mathrm{ch}}\) \textsc{Pythia\,8}  \textsc{Monash}, \textsc{EPOS}, 
\textsc{QGSJET-II}  underestimate the data by up to \(10\)--\(20\)\%.
\textsc{Pythia\,8}  \textsc{Monash}, \textsc{EPOS} overestimate the data for higher \( n_{\mathrm{ch}}\) and
drop below the measurement in the very high-\( n_{\mathrm{ch}}\) region.
\textsc{QGSJET-II}   overestimates the data significantly.
The high-\( n_{\mathrm{ch}}\) region has significant contributions from events with numerous MPI.

%
\subsubsection{Multiplicity  distributions of the LHC experiments}
\begin{figure*}[t!]
\centering
\begin{minipage}[h]{0.45\textwidth}
\center{\includegraphics[width=1.0\linewidth]{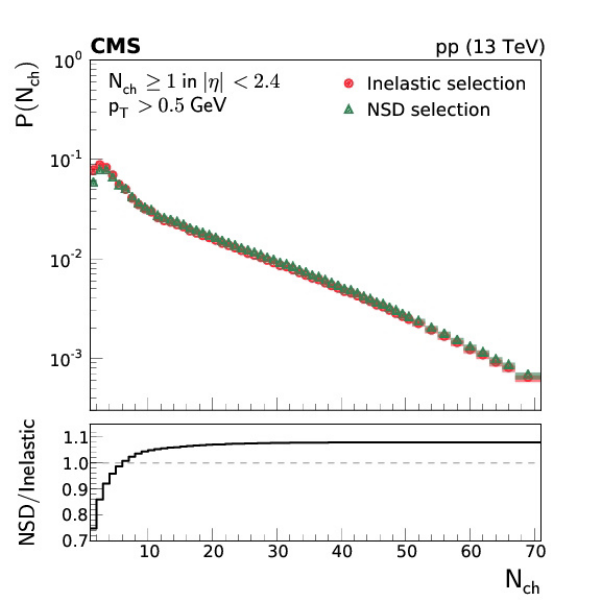}} 
\\
\end{minipage}
\caption{
Top panel:
The CMS probability density of charged-particle multiplicity for the most inclusive (inelastic)  sample and
the sample dominated by non-single diffractive dissociation events  (NSD-enhanced sample)
for events   at a centre-of-mass energies  \(\sqrt{s}= 8\)~\TeV\  with  
\(n_{\mathrm{ch}} \ge 1\) and  \(p_{\mathrm{T}} > 500\)~\MeV.
The error bars represent the statistical plus uncorrelated systematics between 
neighbouring bins,  and the bands show the combined systematic and statistical uncertainties. 
Bottom panel:
The ratio  of the  NSD-enhanced sample to the inelastic sample results  is presented. 
Taken from Ref.~\cite{CMS:2018nhd}.
}
\label{fig_13_Nev_nch_CMS}
\end{figure*}

The CMS  results for  primary charged-particle multiplicities as a function of  
the multiplicity  for events with  \(\mid\eta\mid < 2.4\) at  the  CM energy  
\(\sqrt{s}= 13\)~\TeV\  with  \(n_{\mathrm{ch}} \ge 1\) and \(p_{\mathrm{T}} >500\)~\MeV\
\cite{CMS:2018nhd} are shown in Fig.\ \ref{fig_13_Nev_nch_CMS}. 
The measured distributions are presented for  two different event data sets:  an INEL sample
and  an  NSD-enhanced sample.
The charged particle multiplicity distribution of the NSD-enhanced event sample shows 
a depletion of low-\(n_{\mathrm{ch}}\)  events and an increase of high-\(n_{\mathrm{ch}}\) 
multiplicity  events  compared to that of the inelastic sample.

\begin{figure*}[t!]
\centering
\begin{minipage}[h]{0.32\textwidth} 
\center{\includegraphics[width=1.0\linewidth]{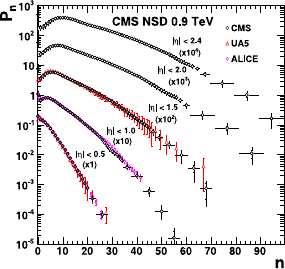}} 
(a) 
\\
\end{minipage}
\hfill
\begin{minipage}[h]{0.32\textwidth} 
\center{\includegraphics[width=1.0\linewidth]{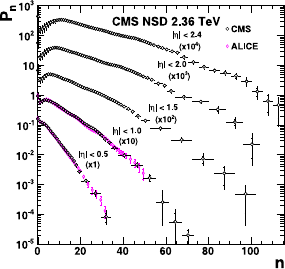}} 
(b)
\\
\end{minipage}
\hfill
\begin{minipage}[h]{0.32\textwidth} 
\center{\includegraphics[width=1.0\linewidth]{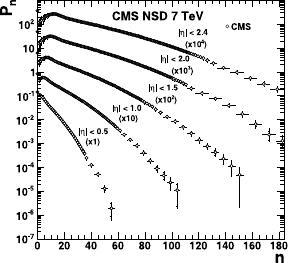}} 
(c)
\\
\end{minipage}
\caption{
The  charged-particle multiplicity distributions
for 
\( \mid\eta\mid < 0.5\),  
\( \mid\eta\mid < 1.5\), 
\( \mid\eta\mid < 1.5\), 
\( \mid\eta\mid < 2.0\) 
and   
\( \mid\eta\mid < 2.4\)
at
(a) \(\sqrt{s} = 0.9\)~\TeV, 
(b) \(\sqrt{s} = 2.36\)~\TeV, 
and
(c) \(\sqrt{s} = 7\)~\TeV.
The CMS results were compared with  the measurements  of 
ALICE  \cite{ALICE:2010cin,ALICE:2010mty} 
and
the UA5 \cite{UA5:1984zxi,UA5:1985fid}
in the same \( \eta \) interval  and  at the same centre-of-mass energy. 
For clarity, results in different pseudorapidity intervals are scaled by powers of \(10\) 
as given in the plots. 
The error bars are the statistical and systematic uncertainties added in quadrature.
Taken from Ref.~\cite{CMS:2010qvf}.
}
\label{fig_Nev_nch_CMS}
\end{figure*}

\begin{figure*}[t!]
\centering
\begin{minipage}[h]{0.45\textwidth} 
\center{\includegraphics[width=1.0\linewidth]{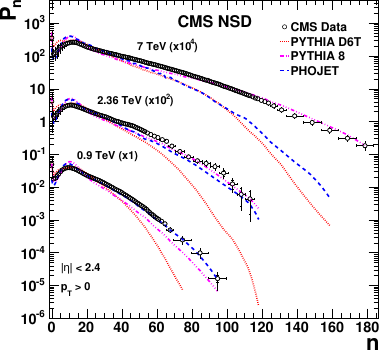}} 
(a) 
\\
\end{minipage}
\hspace{2mm}
\begin{minipage}[h]{0.45\textwidth} 
\center{\includegraphics[width=1.0\linewidth]{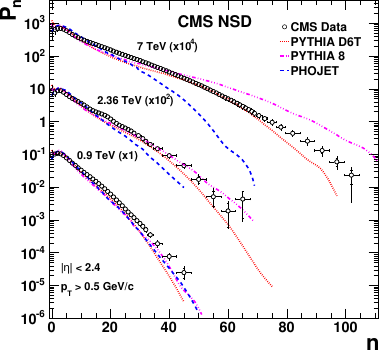}} 
(b)
\\
\end{minipage}
\caption{
The  charged-particle multiplicity distributions
for 
(a) \( \mid\eta\mid < 2.4\), \(p_{\mathrm{T}} > 0\)~\MeV\
and 
(b) \( \mid\eta\mid < 2.4\), \(p_{\mathrm{T}} > 500\)~\MeV\
at
\(\sqrt{s} = 0.9\)~\TeV, 
\(\sqrt{s} = 2.36\)~\TeV, 
and
\(\sqrt{s} = 7\)~\TeV. 
The CMS results were compared with  the  predictions  in the same \( \eta \) interval 
and  at the same centre-of-mass energy of the MC models 
\textsc{Pythia} \textsc{D6T} tune,
\textsc{Pythia\,8}
and
\textsc{PHOJET}. 
For clarity,  the results in different pseudorapidity intervals are scaled by powers of \(10\) 
as given in the plots. 
The error bars are the statistical and systematic uncertainties added in quadrature.
Taken from Ref.~\cite{CMS:2010qvf}.
}
\label{fig_Nev_nch_CMS_2}
\end{figure*}

The  NSD charged hadron multiplicity distributions are measured 
in increasing ranges of pseudorapidity from  \( \mid\eta\mid < 0.5 \) to  \( \mid\eta\mid < 2.4\). 
The fully corrected results at \(\sqrt{s} = 0.9,\ 2.36\) and \(7\)~\TeV\  
are compared in Fig.~\ref{fig_Nev_nch_CMS}  with  the measurements in the same pseudorapidity ranges
performed by the UA5 \cite{UA5:1984zxi,UA5:1985fid} and ALICE \cite{ALICE:2010cin,ALICE:2010mty}.
The CMS measurements were also compared with  the results obtained  from the 
CMS cross-check analysis  of the data at  \(\sqrt{s} = 0.9\) and \(7\)~\TeV\
using a tracklet-based tracking algorithm as in Ref.~\cite{CMS:2010wcx}.
With a reconstruction efficiency exceeding \(90\)\% for   \(p_{\mathrm{T}} > 50\)~\MeV, 
the latter provided a cross-check of the extrapolation for tracks below 
\(p_{\mathrm{T}} < 100\)~\MeV,  including the use of  the data without  the
magnetic field at  \(\sqrt{s} = 7\)~\TeV. 
All measurements agree well within their total uncertainties.
In the largest pseudorapidity interval  \( \mid\eta\mid < 2.4 \), 
there is a change of slope in \(P_{\mathrm{n}}\)  for  \(n_{\mathrm{ch}} > 20\),
indicating a multicomponent structure,  as was discussed in 
Refs.~\cite{Alexopoulos:1998bi,Giovannini:1997ce} in terms of multiple-soft-Pomeron exchanges. 
This feature becomes more pronounced with increasing CM energies, notably at  \(\sqrt{s} = 7\)~\TeV.

An extensive range of tunes  \cite{Bartalini:2008zz,Moraes:2007rq,Buckley:2009bj,Skands:2010ak}
based on the \textsc{Pythia\,6}  fragmentation model have been developed. 
They differ mainly in their  parametrization  of the multiple-parton interaction model. 
Some reproduce the charged hadron multiplicities better than others,
but none is able to give a good description simultaneously at all  \( \sqrt{s} \) and in all pseudorapidity ranges. 
For clarity, only the baseline tune  \textsc{Pythia} \textsc{D6T}  \cite{Bartalini:2008zz,Moraes:2007rq}
is shown in comparison with other models having a  different physical description 
of soft-particle production,  such as  \textsc{PHOJET}   \cite{Engel:1994vs,Engel:1995yda}
and the  fragmentation model of  \textsc{Pythia\,8} \cite{Sjostrand:2007gs}.
A comparison of the CMS  measurements with  three classes of models is shown in
Fig.\ \ref{fig_Nev_nch_CMS_2} for all charged hadrons  and  for those with \(p_{\mathrm{T}} > 500\)~\MeV. 
\textsc{Pythia} \textsc{D6T} drastically  underestimates  the multiplicity at all measured energies 
but improves when \(p_{\mathrm{T}} > 500\)~\MeV\    is required. 
\textsc{Pythia\,8}  is the only model that gives a reasonable description 
of the multiplicity distribution at all energies  but tends to overestimate the multiplicity at 
\(\sqrt{s} = 7\)~\TeV\  when \(p_{\mathrm{T}} > 500\)~\MeV\ is required. 
\textsc{PHOJET}  produces too few charged hadrons overall but gives a good description of the 
average transverse momentum  \(\langle p_{\mathrm{T}} \rangle\) at  the fixed multiplicity 
\(n_{\mathrm{ch}}\),  as illustrated in  Fig.\ \ref{fig_Nev_nch_CMS_2}.

\begin{figure*}[t!]
\centering
\begin{minipage}[h]{0.45\textwidth} 
\center{\includegraphics[width=1.0\linewidth]{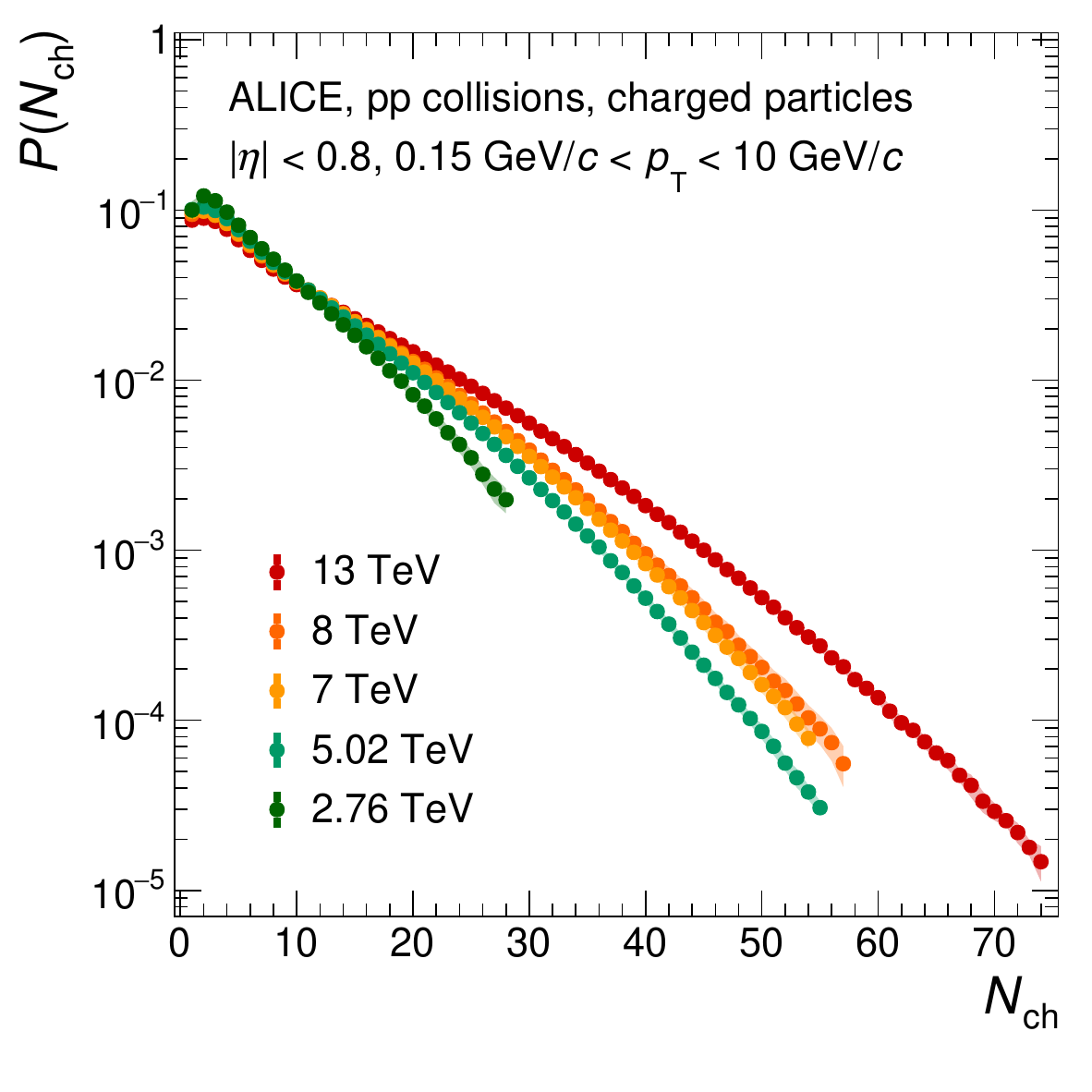}} 
(a) 
\\
\end{minipage}
\begin{minipage}[h]{0.45\textwidth} 
\center{\includegraphics[width=1.0\linewidth]{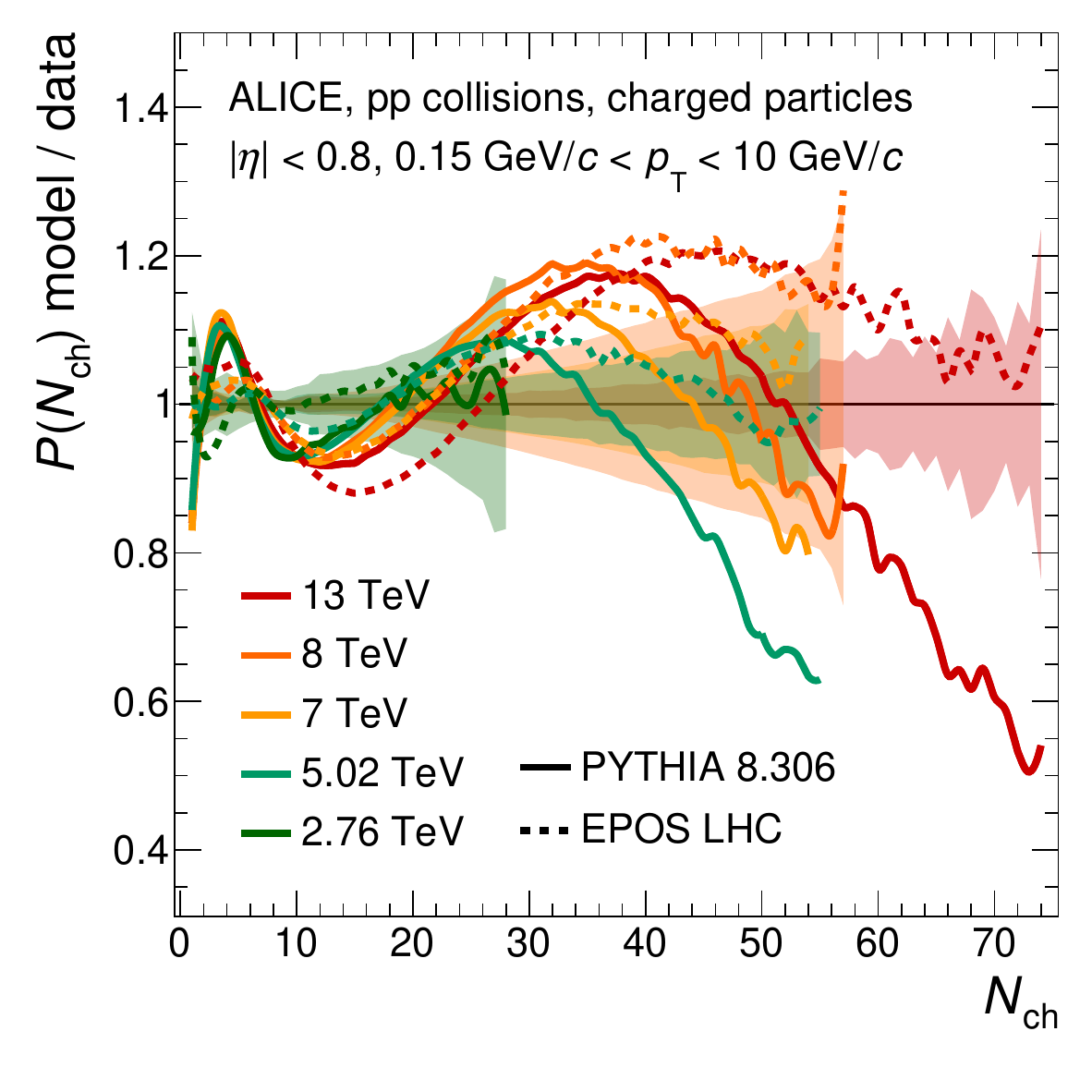}} 
(b)
\\
\end{minipage}
\caption{
(a)
The  ALICE   probability density of charged-particle multiplicity for \( pp\)  collisions at the different 
centre-of-mass energies \(\sqrt{s} = 2.36,\ 5.02,\ 7,\ 8\) and  \(13\)~\TeV\
for events in the kinematic range \( N_{\mathrm{ch}} > 0\),  \( \mid\eta\mid < 0.8 \) 
and  \( 0.15 < p_{\mathrm{T}} < 10 \)~\GeV.
Statistical and systematic uncertainties are shown as bars and semi-transparent bands, respectively.
(b)
The ratio of  \textsc{Pythia\,8} \cite{Sjostrand:2014zea} and \textsc{EPOS\ LHC}  \cite{Pierog:2013ria}
model predictions to data for \(pp\) collisions at various energies for 
the primary charged-particle  multiplicity distributions.
The semi-transparent bands indicate the relative systematic uncertainties of the data.
Taken from Ref.~\cite{ALICE:2022xip}.
}
\label{fig_Nev_nch_ALICE_1}
\end{figure*}

The ALICE  results of the study   on multiplicity  (\(N_{\mathrm{ch}}\))  distributions, 
transverse momentum  spectra, 
and KNO scaling  of inclusive primary charged particles  in the kinematic range  of 
\( \mid\eta\mid < 0.8 \)  and  \( 0.15 < p_{\mathrm{T}} < 10 \)~\GeV\ for
\(pp\), \(p\)--\(Pb\), \(Xe\)--\(Xe\) and \(Pb\)--\(Pb\)  collisions 
at CM  energies per nucleon pair ranging from
\( \sqrt{ s_{\mathrm{NN}} } = 2.76\) \TeV\ up to \(13\) \TeV\  were published in 
Ref.~\cite{ALICE:2022xip}.
The \(N_{\mathrm{ch}}\) distributions  for \( pp\)  collisions at the different centre-of-mass energies
\(\sqrt{s} = 2.36,\ 5.02,\ 7,\ 8\) and  \(13\)~\TeV\   for the kinematic region 
\( \mid\eta\mid < 0.8 \)  and  \( 0.15 < p_{\mathrm{T}} < 10 \)~\GeV\
are shown in Fig.~\ref{fig_Nev_nch_ALICE_1}(a).
These distributions reach a maximum around \( N_{\mathrm{ch}} \approx  2\)
and then fall steeply off over several orders of magnitude. 
The slope of the decay with \( N_{\mathrm{ch}}\)  decreases with increasing collision energy. 
This can be attributed to the larger  \( p_{\mathrm{T}}\) in the initial hard scattering, 
which results in larger multiplicities.
Figure~\ref{fig_Nev_nch_ALICE_1}(b)  compares 
measured results  for \(pp\) collisions  for the respective multiplicity distributions with predictions from 
\textsc{Pythia\,8}  \cite{Sjostrand:2014zea} (solid lines)  and 
\textsc{EPOS\ LHC}   \cite{Pierog:2013ria} (dashed lines).
The  \textsc{Pythia\,8.306}  event generator is used with the
\textsc{Monash-2013} tune  \cite{Skands:2014pea}  for \(pp\) collisions.
The overall shapes of the multiplicity distribution  shown in Fig.~\ref{fig_Nev_nch_ALICE_1}(b) 
are better described by  \textsc{EPOS\ LHC}, while 
\textsc{Pythia\,8}  falls sharply off above \( N_{\mathrm{ch}}/ \langle N_{\mathrm{ch}}\rangle \approx 4\). 
Both models agree with the experimental distributions within \(25\)\% 
with larger deviations at the highest multiplicities.

\subsection{Average transverse momentum multiplicity dependence}
\label{pT_nch}
\subsubsection{ATLAS average transverse momentum distributions}
\label{pT_nch_data}
\begin{figure*}[t!]
\centering
\begin{minipage}[h]{0.45\textwidth} 
\center{\includegraphics[width=1.0\linewidth]{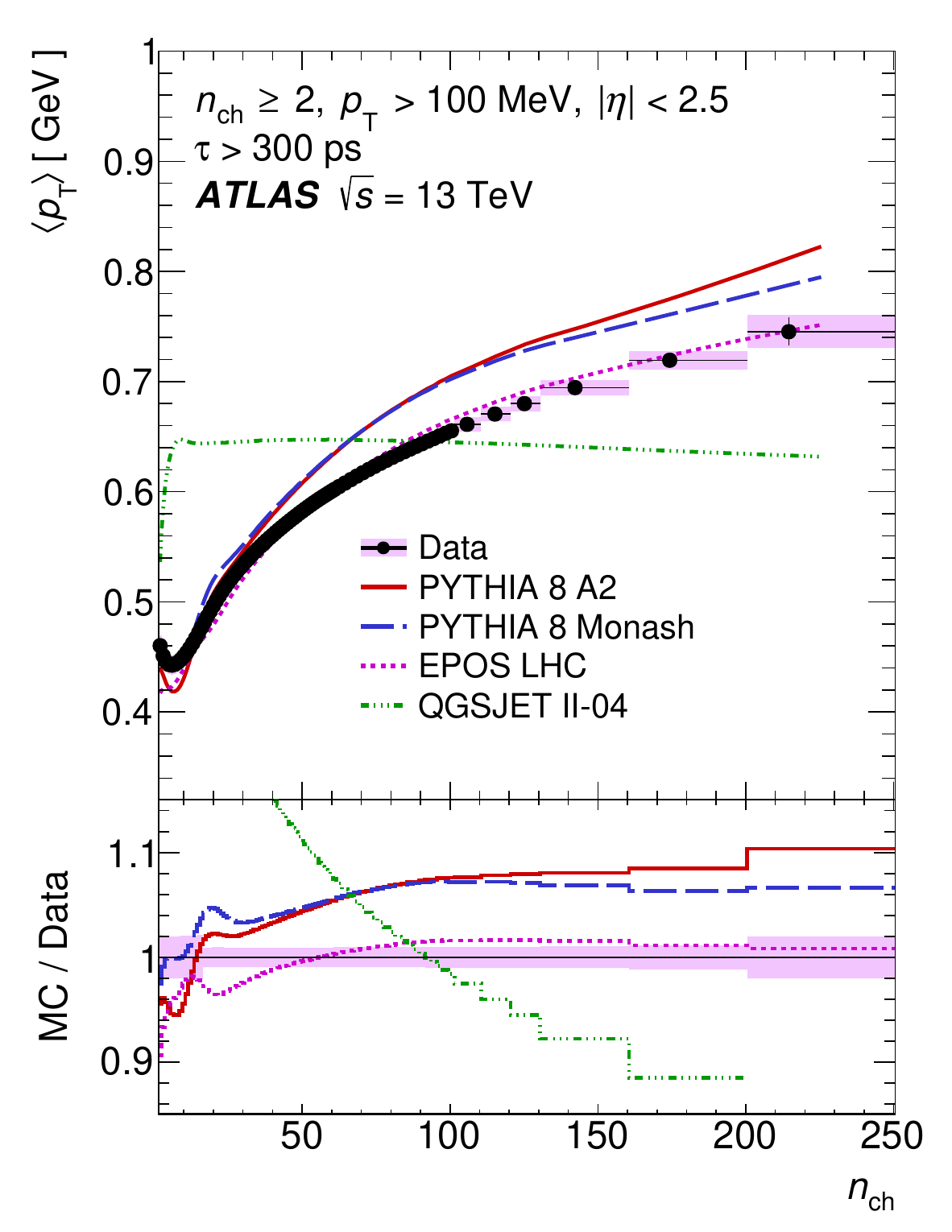}}
(a) 
\\
\end{minipage}
\hspace{2mm}
\begin{minipage}[h]{0.45\textwidth}
\center{\includegraphics[width=1.0\linewidth]{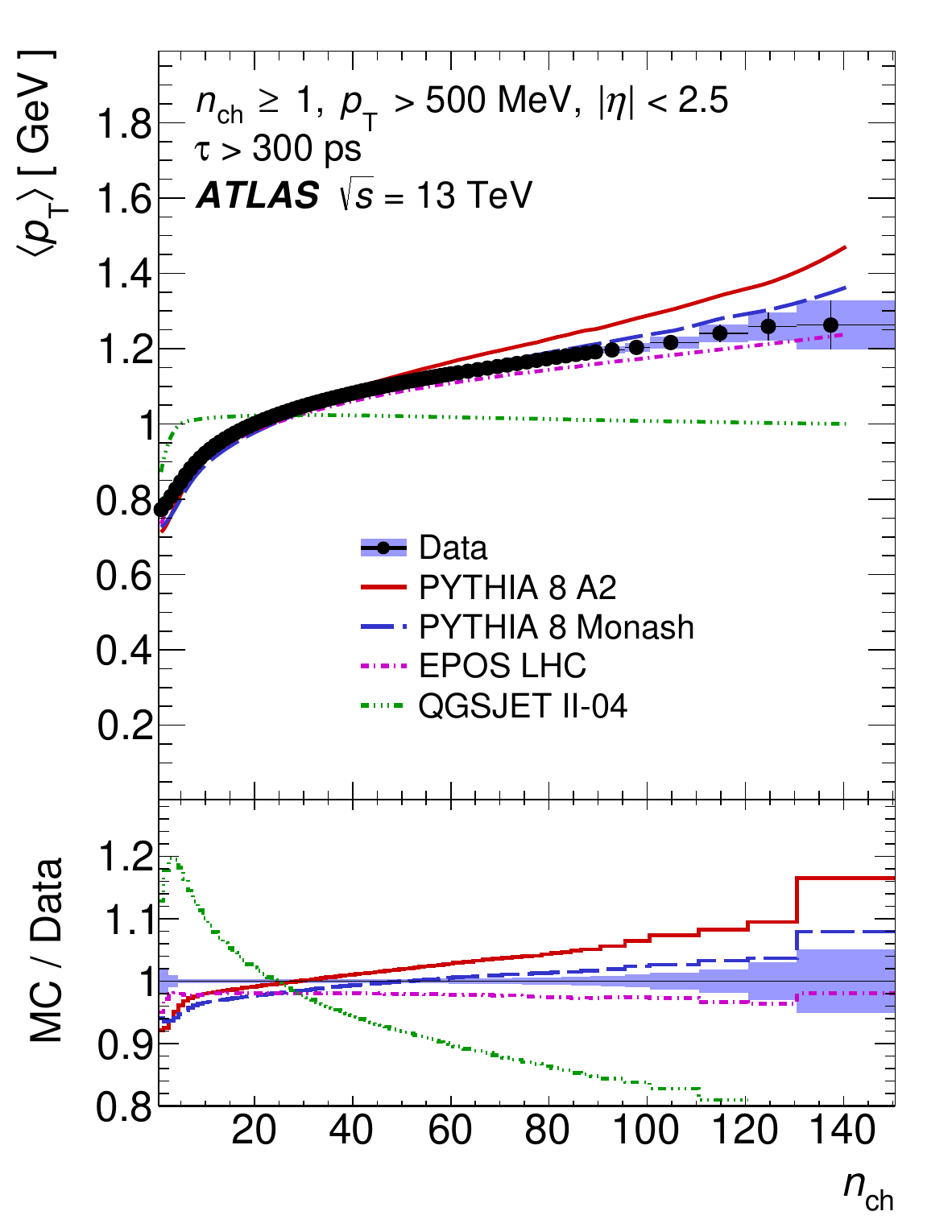}}
(b) 
\\
\end{minipage}
\caption{
Top panel: 
The average transverse momentum \(\langle p_{\mathrm{T}}\rangle \) as  a function of the multiplicity
at the centre-of-mass energy \(\sqrt{s}= 13\)~\TeV\ 
\cite{STDM-2015-02,STDM-2015-17}  with 
(a) \(n_{\mathrm{ch}} \ge 2\), \(p_{\mathrm{T}} >100\)~\MeV\ 
and
(b) \(n_{\mathrm{ch}} \ge 1\),  \(p_{\mathrm{T}} >500\)~\MeV. 
The data represented by dots 
is
compared to various particle-level MC predictions,  
which are shown by curves. 
The shaded areas around the data points represent the total statistical and systematic uncertainties added in quadrature.
Bottom panel: 
The ratios of the MC predictions to  the experimental results are shown. 
Bands represent the uncertainties of the experimental results.
Taken from Refs.\ \cite{STDM-2015-02,STDM-2015-17}.
}
\label{fig_13_pT_nch}
\end{figure*}

\begin{figure*}[t!]
\centering
\begin{minipage}[h]{0.45\textwidth} 
\center{\includegraphics[width=1.0\linewidth]{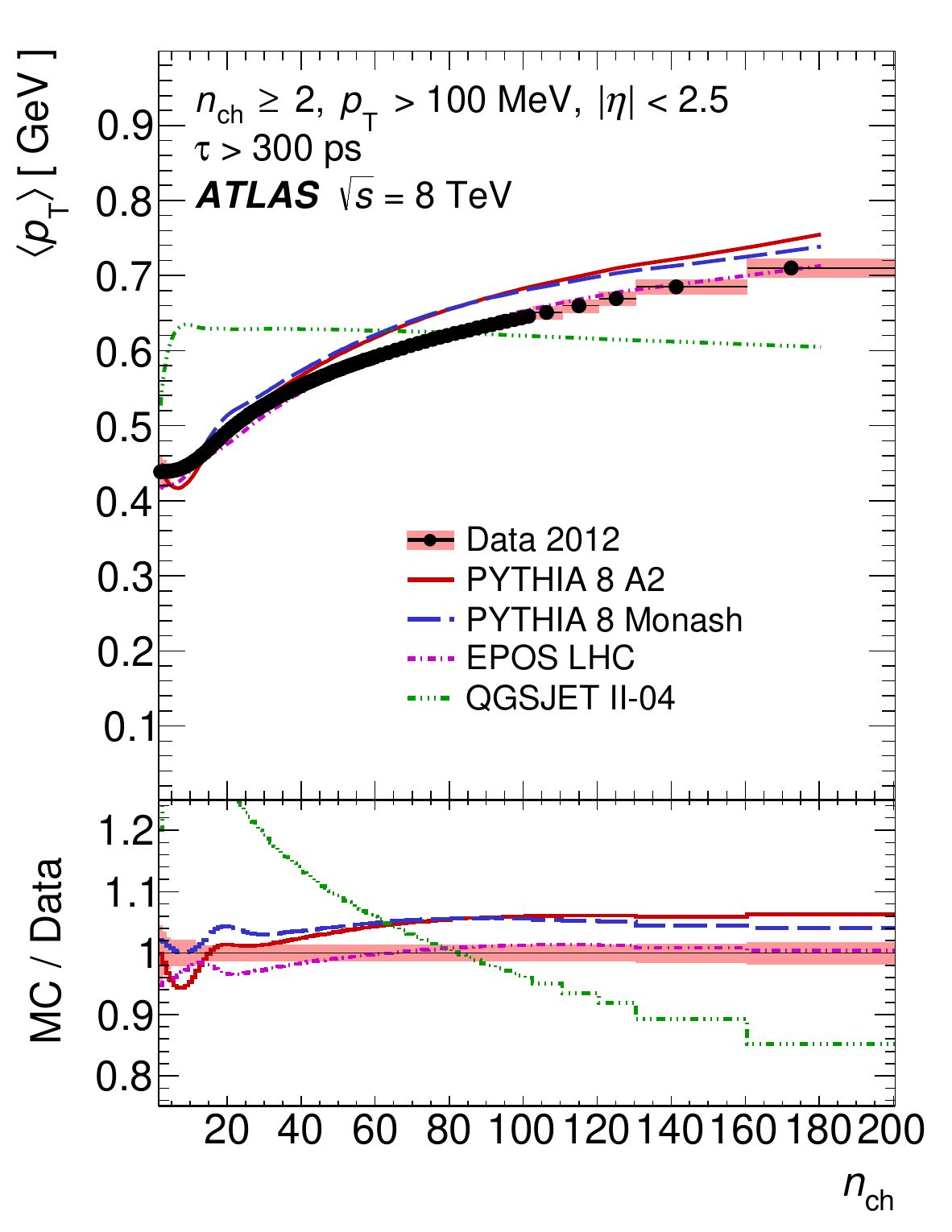}}
(a) 
\\
\end{minipage}
\hspace{2mm}
\begin{minipage}[h]{0.45\textwidth}
\center{\includegraphics[width=1.0\linewidth]{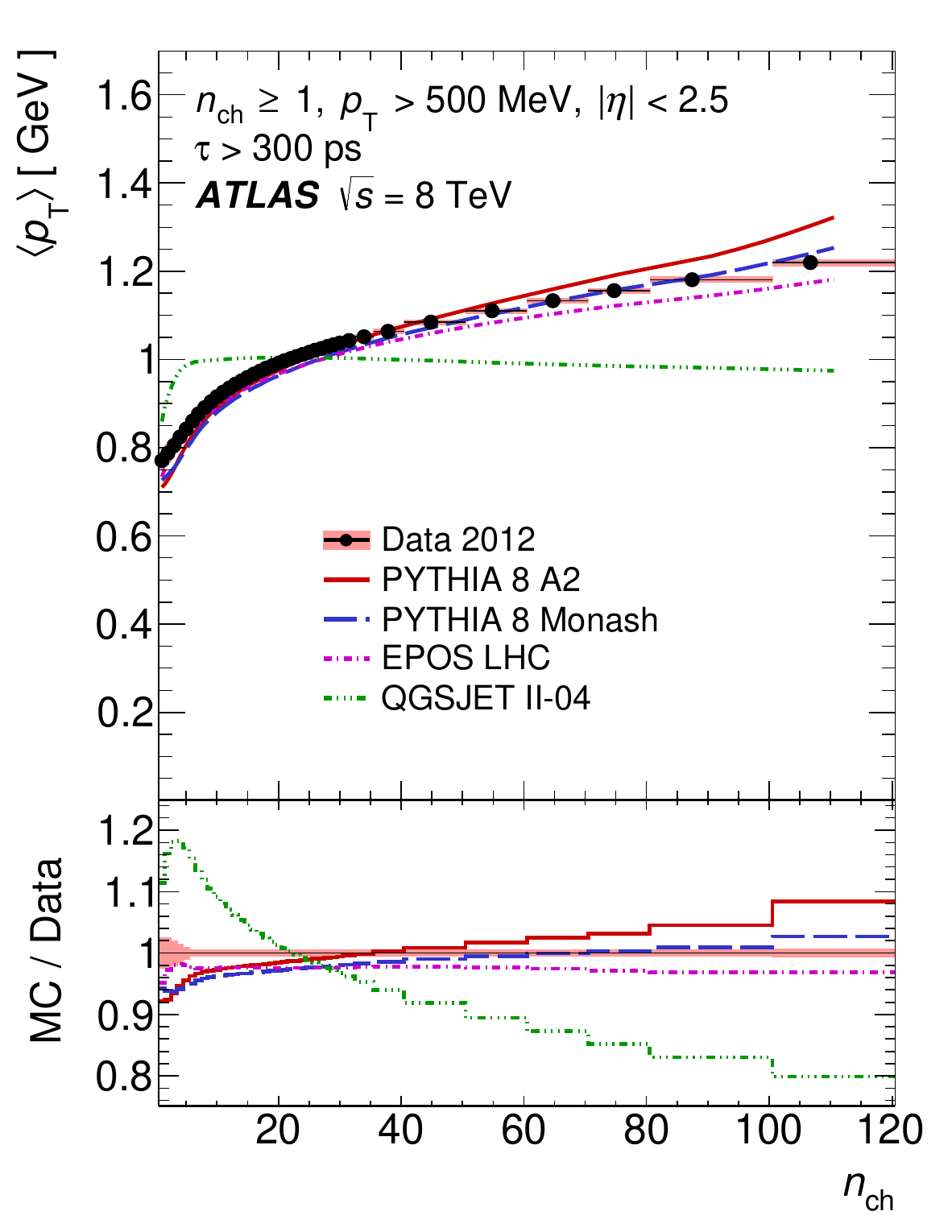}}
(b) 
\\
\end{minipage}
\caption{
Top panel: 
The average transverse momentum \(\langle p_{\mathrm{T}}\rangle \) 
as  a function of the  multiplicity at the centre-of-mass energy  \(\sqrt{s}=  8\)~\TeV\ 
\cite{STDM-2014-19}   with 
(a) \(n_{\mathrm{ch}} \ge 2\), \(p_{\mathrm{T}} >100\)~\MeV\ 
and 
(b) \(n_{\mathrm{ch}} \ge 1\), \(p_{\mathrm{T}} >500\)~\MeV.
The data represented by dots 
is
compared to various particle-level MC predictions,  
which are shown by curves. 
The shaded areas around the data points represent the total statistical and systematic uncertainties added in quadrature.
Bottom panel: 
The ratios of the MC predictions to  the experimental results are shown. 
Bands represent the uncertainties of the experimental results.
Taken from Ref.~\cite{STDM-2014-19}.
}
\label{fig_8_pT_nch}
\end{figure*}

\begin{figure*}[t!]
\centering
\begin{minipage}[h]{0.45\textwidth} 
\center{\includegraphics[width=1.0\linewidth]{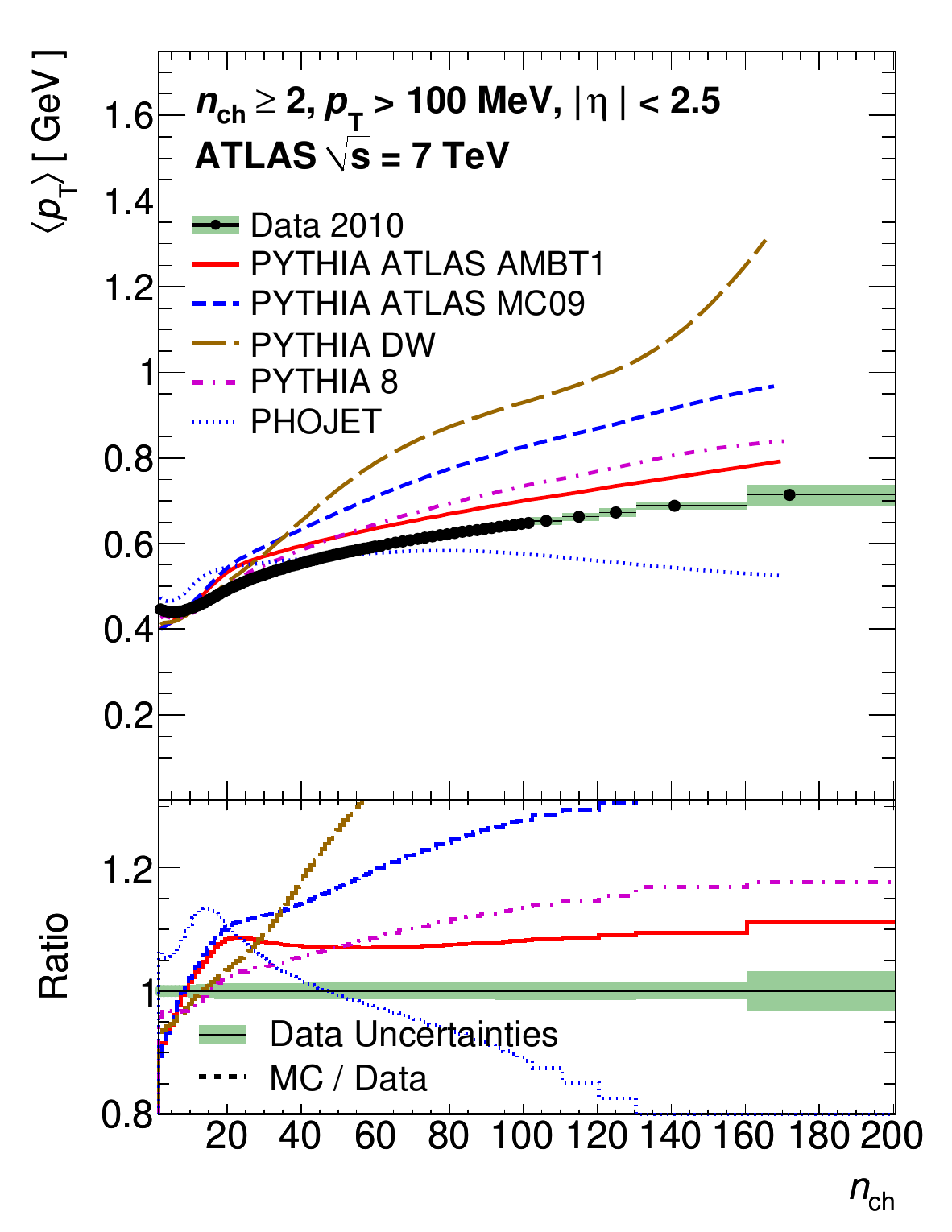}}
(a) 
\\
\end{minipage}
\hspace{2mm}
\begin{minipage}[h]{0.45\textwidth}
\center{\includegraphics[width=1.0\linewidth]{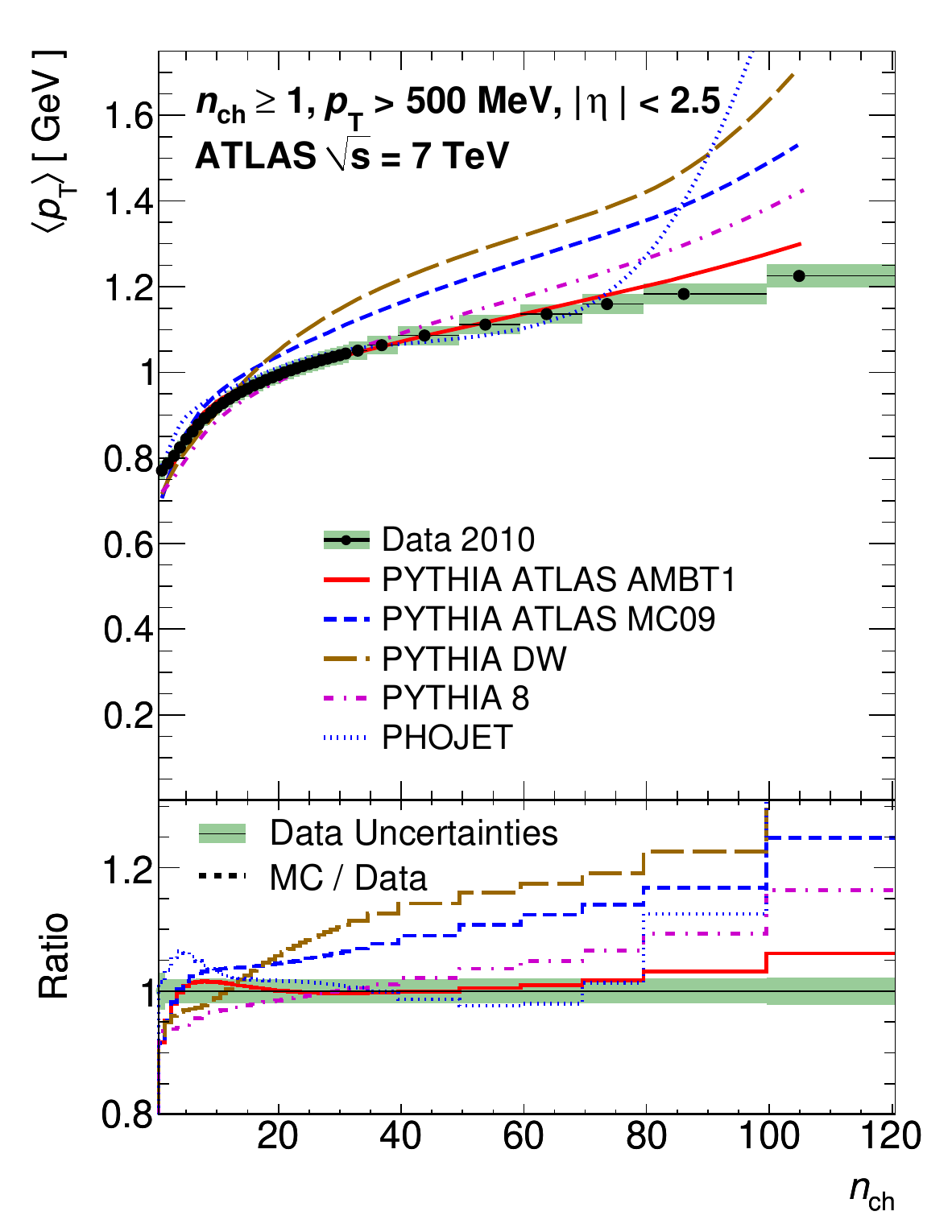}}
(b) 
\\
\end{minipage}
\caption{
Top panel: 
The average transverse momentum \(\langle p_{\mathrm{T}}\rangle \) as  
a function of the  multiplicity, at the centre-of-mass energy  \(\sqrt{s}= 7\)~\TeV\ 
\cite{STDM-2010-06}	   with 
(a) \(n_{\mathrm{ch}} \ge 2\), \(p_{\mathrm{T}} >100\)~\MeV\ 
and 
(b) \(n_{\mathrm{ch}} \ge 1\), \(p_{\mathrm{T}} >500\)~\MeV.
The data represented by dots 
is
compared to various particle-level MC predictions,  
which are shown by curves. 
The shaded areas around the data points represent the total statistical and systematic uncertainties added in quadrature.
Bottom panel: 
The ratios of the MC predictions to  the experimental results are shown. 
Bands represent the uncertainties of the experimental results.
Taken from Ref.~\cite{STDM-2010-06}.
}
\label{fig_7_pT_nch}
\end{figure*}

\begin{figure*}[t!]
\centering
\begin{minipage}[h]{0.45\textwidth} 
\center{\includegraphics[width=1.0\linewidth]{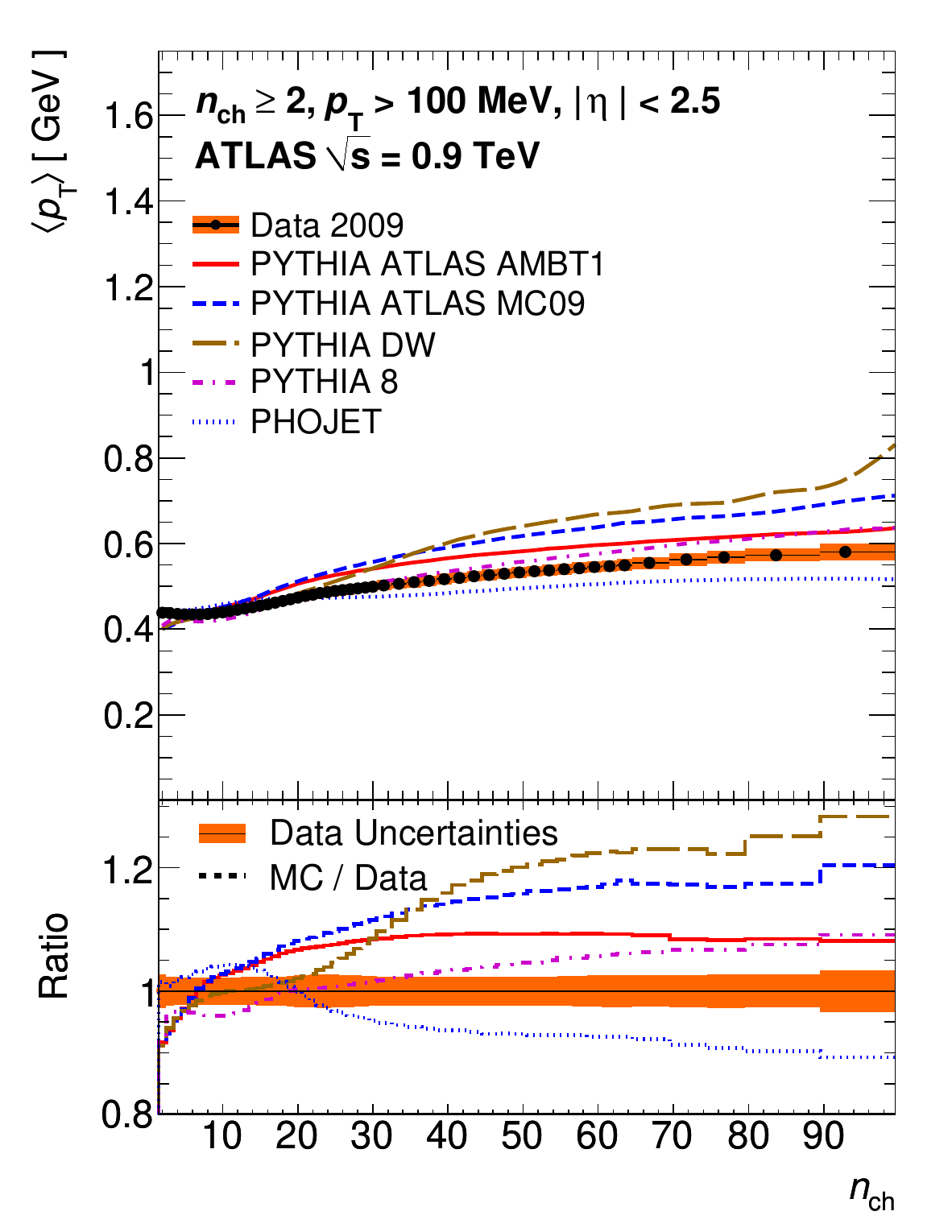}}
(a) 
\\
\end{minipage}
\hspace{2mm}
\begin{minipage}[h]{0.45\textwidth}
\center{\includegraphics[width=1.0\linewidth]{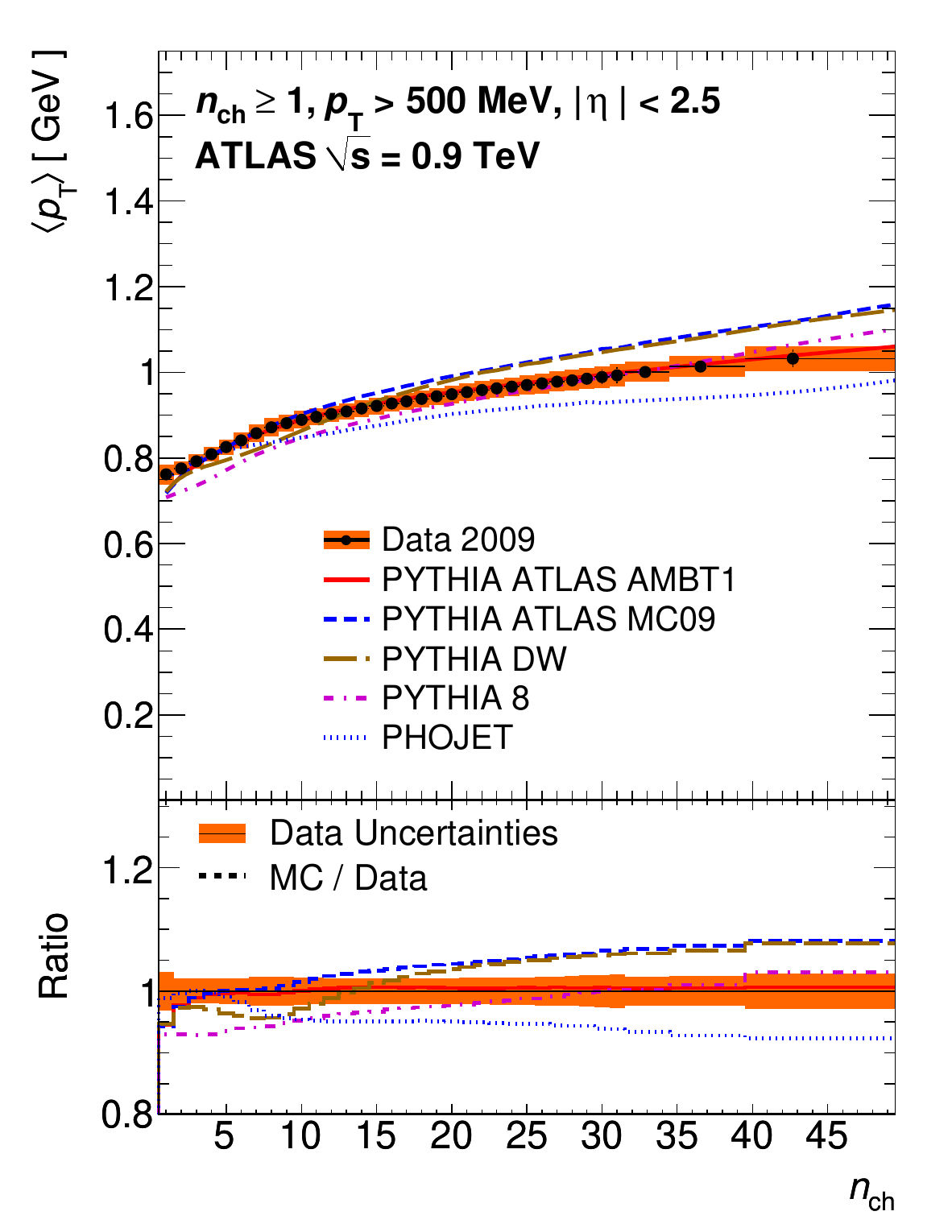}}
(b) 
\\
\end{minipage}
\caption{
Top panel: 
The average transverse momentum \(\langle p_{\mathrm{T}}\rangle \) as  
a function of the  multiplicity at the centre-of-mass energy  \(\sqrt{s}= 0.9\)~\TeV\ 
\cite{STDM-2010-06}	   with 
(a) \(n_{\mathrm{ch}} \ge 2\), \(p_{\mathrm{T}} >100\)~\MeV\ 
and 
(b) \(n_{\mathrm{ch}} \ge 1\), \(p_{\mathrm{T}} >500\)~\MeV.
The data represented by dots 
is
compared to various particle-level MC predictions,  
which are shown by curves. 
The shaded areas around the data points represent the total statistical and systematic uncertainties added in quadrature.
Bottom panel: 
The ratios of the MC predictions to  the experimental results are shown. 
Bands represent the uncertainties of the experimental results.
Taken from Ref.~\cite{STDM-2010-06}.
}
\label{fig_09_pT_nch}
\end{figure*}

The charged-particle average  transverse momentum distributions
are shown in Figs.~\ref{fig_13_pT_nch} -- \ref{fig_09_pT_nch}
at  the CM energies  \(\sqrt{s}= 0.9,\ 2.36,\ 7,\ 8,\) and  \(13\)~\TeV. 

The average  transverse momentum versus the primary charged-particle multiplicity is shown in 
Fig.~\ref{fig_13_pT_nch} at  \(\sqrt{s}= 13\)~\TeV\  for 
\(n_{\mathrm{ch}} \ge 2\), \(p_{\mathrm{T}} >100\)~\MeV\ 
\cite{STDM-2015-17} and \(n_{\mathrm{ch}} \ge 1\),   \(p_{\mathrm{T}} >500\)~\MeV\ 
\cite{STDM-2015-02},  respectively.
%
For \(p_{\mathrm{T}} >100\)~\MeV\ in Fig.~\ref{fig_13_pT_nch}(a)
it increases towards higher \(n_{\mathrm{ch}}\), 
as modelled by a colour reconnection mechanism in  \textsc{Pythia\,8} 
and by the hydrodynamical evolution model in  \textsc{EPOS}. 
The 
\textsc{QGSJET-II}  generator, which has no model for colour coherence effects,  describes the data poorly. 
For low \(n_{\mathrm{ch}}\), 
\textsc{Pythia\,8} \textsc{A2} and \textsc{EPOS} underestimate the data,
where \textsc{Pythia\,8} \textsc{Monash} agrees within the uncertainties. 
For higher \(n_{\mathrm{ch}}\) all generators overestimate the data, but for \(n_{\mathrm{ch}} > 40\), 
there is a constant offset for both \textsc{Pythia\,8} tunes, 
which describes the data to within  \(10\)\%. 
\textsc{EPOS} describes the data reasonably well and to within \(2\)\%.
%
Figure~\ref{fig_13_pT_nch}(b) for 
\(n_{\mathrm{ch}} \ge 1\),  \(p_{\mathrm{T}} >500\)~\MeV\ 
shows the mean transverse momentum versus the charged-particle multiplicity. 
The  \(\langle p_{\mathrm{T}}\rangle\) rises with \(n_{\mathrm{ch}}\),  from \(0.8\)  to \(1.2\)~\GeV. 
This increase is expected due to colour coherence effects  being important in dense parton environments and 
is modelled by the  colour reconnection mechanism in  \textsc{Pythia\,8}  or by
the hydrodynamical evolution model used in  \textsc{EPOS}. 
If the high-\(n_{\mathrm{ch}}\) region is assumed  to be dominated by events with numerous MPI, 
without colour coherence effects, 
the   \(\langle p_{\mathrm{T}}\rangle\) 
is approximately independent of  \(n_{\mathrm{ch}}\). 
Inclusion of  
colour coherence effects leads to fewer  additional charged particles produced with every
additional MPI, with an equally large \(p_{\mathrm{T}}\)  to be shared among the produced hadrons 
\cite{Sjostrand:2013cya}. 
\textsc{EPOS} predicts a slightly lower  \(\langle p_{\mathrm{T}}\rangle\)
but describes the dependence on \(n_{\mathrm{ch}}\) very well. 
The \textsc{Pythia\,8}  tunes predict a steeper rise of 
\(\langle p_{\mathrm{T}}\rangle\) 
with \(n_{\mathrm{ch}}\) than the data, predicting lower values in the low-\(n_{\mathrm{ch}}\) 
region and higher values in the high-\(n_{\mathrm{ch}}\) region. 
\textsc{QGSJET-II} 
predicts a \(\langle p_{\mathrm{T}}\rangle\) of  \( \sim  1\)~\GeV, 
with very little dependence on \(n_{\mathrm{ch}}\); this is expected
as it contains no model for colour coherence effects.

Similar plots as for \(13\)~\TeV\ are also shown for \(8\)~\TeV\ in Fig.~\ref{fig_8_pT_nch} 
for transverse momentum thresholds of \(100\)~\MeV\ and \(500\)~\MeV, respectively. 
The average \(p_{\mathrm{T}}\) rises with multiplicity, 
although the
rise becomes progressively less steep as the multiplicity increases. 
This is expected due to colour coherence effects in dense parton environments, which are modelled 
by a colour reconnection mechanism in \textsc{Pythia\,8} or by the hydrodynamical evolution model 
used in \textsc{EPOS}. 
It is assumed that numerous MPI
dominate the high-multiplicity events
and that colour coherence effects thereby lead to fewer additional
charged particles produced with every additional MPI, which share a higher average \(p_{\mathrm{T}}\). 
The \textsc{EPOS} and \textsc{Pythia\,8} models provide a fair description of the data. 
The \textsc{QGSJET-II}  
model fails to predict the mean transverse momentum over the entire multiplicity range, 
as it does not simulate colour coherence effects and therefore shows very little dependence on the multiplicity.


Figures~\ref{fig_7_pT_nch} and  \ref{fig_09_pT_nch} show  the results for events  at the 
CM energies  \(\sqrt{s}= 7\)~\TeV\ and  \(\sqrt{s}= 0.9\)~\TeV\
for 
\(n_{\mathrm{ch}} \ge 2\), \(p_{\mathrm{T}} >100\)~\MeV\ 
and
\(n_{\mathrm{ch}} \ge 1\),  \(p_{\mathrm{T}} >500\)~\MeV, 
respectively. 
Globally, 
one can say that 
at \(\sqrt{s}= 0.9\)~\TeV\ the slope versus \(n_{\mathrm{ch}}\) 
for high values of \(n_{\mathrm{ch}}\) seems to be well described by most
models,  but the absolute value is best modelled by 
\textsc{Pythia\,6} \textsc{DW}. 
At the highest  CM energy (\(8\) and \(13\)~\TeV) above  
the
multiplicity of \(20\)  
the models vary widely both in slope and in absolute value; 
at low values of \(n_{\mathrm{ch}}\)  none of the models describe the data very well. 
In the more inclusive  PS region, 
Figs.\ref{fig_7_pT_nch}(a) and  \ref{fig_09_pT_nch}(a), 
the models vary widely, especially at \(\sqrt{s}= 7\)~\TeV.
The measurement of \(\langle p_{\mathrm{T}}\rangle \) as a function of 
the charged multiplicity at  \(\sqrt{s}=  2.36\)~\TeV\ 
is not shown because different track reconstruction methods are used for determining 
\(p_{\mathrm{T}}\) and multiplicity distributions.

\begin{figure*}[t!]
\centering
\begin{minipage}[h]{0.32\textwidth} 
\center{\includegraphics[width=1.0\linewidth]{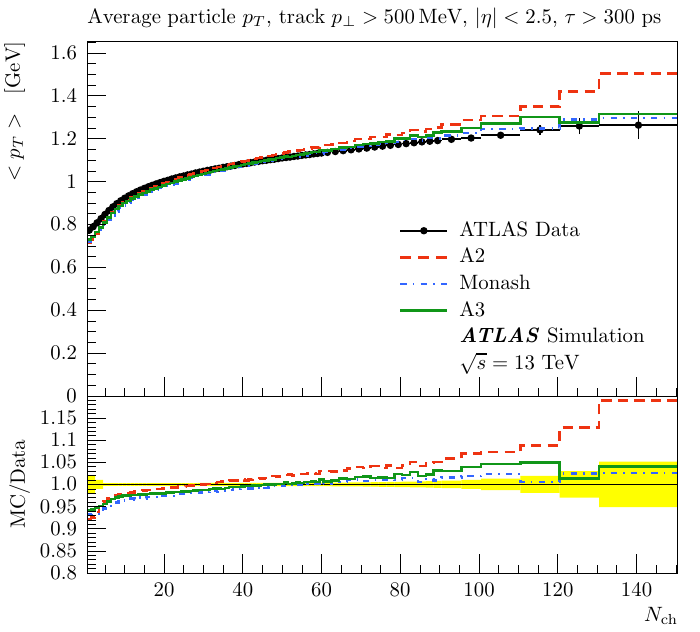}}
(a) 
\\
\end{minipage}
\hfill
\begin{minipage}[h]{0.32\textwidth}  
\center{\includegraphics[width=1.0\linewidth]{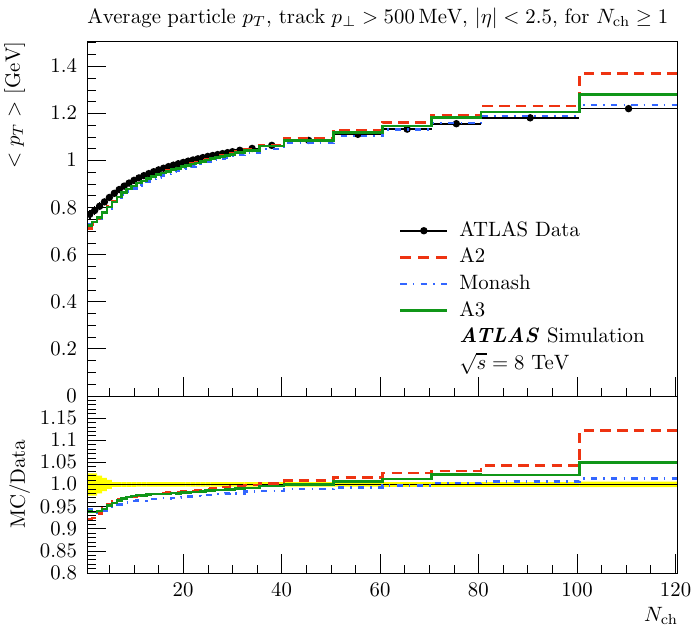}}
(b) 
\\
\end{minipage}
\hfill
\begin{minipage}[h]{0.32\textwidth} 
\center{\includegraphics[width=1.0\linewidth]{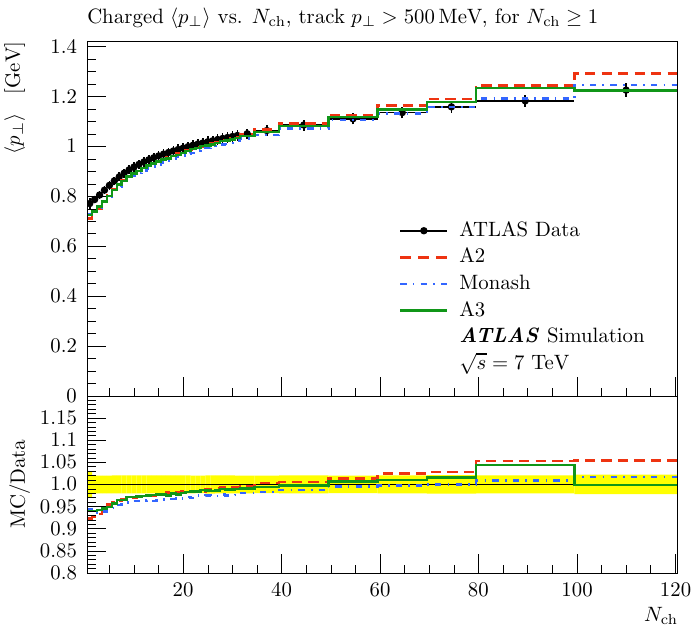}}
(c) 
\\
\end{minipage}
\vfill
\begin{minipage}[h]{0.32\textwidth}
\center{\includegraphics[width=1.0\linewidth]{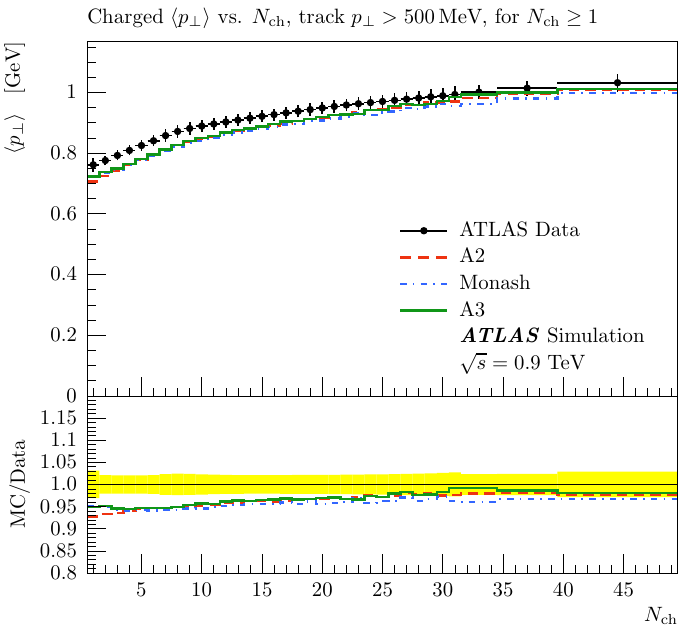}}
(d) 
\\
\end{minipage}
\caption{
Top panel:
The \textsc{Pythia\,8 A3}, \textsc{A2} and \textsc{Monash}  tune predictions  compared with  the
ATLAS charged-particle  average transverse momentum,  \(\langle p_{\mathrm{T}}\rangle \),
distributions for events with 
\(n_{\mathrm{ch}} \ge 1\) with \(p_{\mathrm{T}} >500\)~\MeV\ 
at  the centre-of-mass energies (a) \(13\)~\TeV, (b) \(8\)~\TeV, (c) \(7\)~\TeV,
and (d) \(0.9\)~\TeV. 
The yellow-shaded areas represent the measurement uncertainty.
Bottom panel: 
The ratios of the MC predictions to  the experimental results  are shown. 
Bands represent the uncertainties of the experimental results.
Taken from Ref.~\cite{ATLAS:2016puo}.
}
\label{fig_A3_pT_nch}
\end{figure*}

In Fig.~\ref{fig_A3_pT_nch},  which shows the mean transverse momentum,  
\(\langle p_{\mathrm{T}}\rangle \), against  the  charged particle multiplicity correlation 
\cite{ATLAS:2016puo}, 
the choice of lower colour reconnection strength led to slight improvement over 
\textsc{Pythia\,8} \textsc{A2}.
Although  \(\sqrt{s} = 2.36\)~\TeV\  \cite{Buckley:2014ana}
and  \(\sqrt{s} = 8\)~\TeV\    charged particle distributions 
were not used in tuning, comparisons are  made with those distributions for completeness.


In  Figs.\ \ref{fig_A3_eta},  \ref{fig_A3_pT},  \ref{fig_A3_nch}  and  \ref{fig_A3_pT_nch} 
distributions at   \(\sqrt{s} = 7\)~\TeV\ and \(\sqrt{s} = 13\)~\TeV\ 
predicted by  \textsc{Pythia\,8} \textsc{A3},  in compared to 
\textsc{Pythia\,8} \textsc{A2},  show a broadly comparable,  or better, level of agreement. 
\textsc{Pythia\,8} \textsc{A2}   demonstrates  that an acceptable description of data 
can be achieved by using the 
DL model for diffraction  and can be viewed 
as a possible starting point  for further systematic studies of soft-QCD tunes. 
The results of  \textsc{Pythia\,8} \textsc{A3}  provide good reasons to believe that an improved and more
reliable simulation of pile-up overlay can be obtained.

\begin{figure*}[t!]
\centering
\begin{minipage}[h]{0.45\textwidth}
\center{\includegraphics[width=1.0\linewidth]{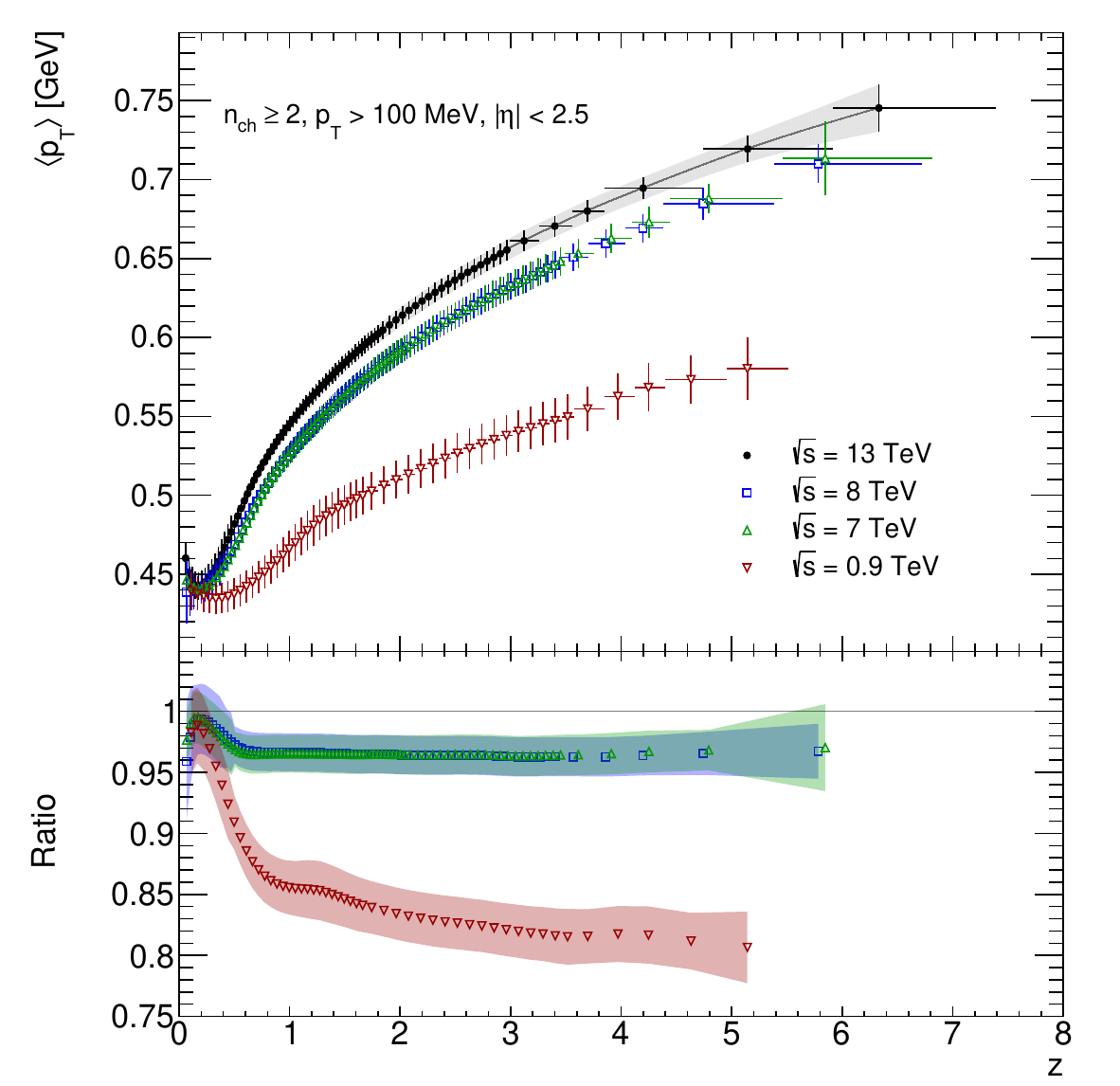}}
(a)
\\
\end{minipage}
\hspace{2mm}
\begin{minipage}[h]{0.45\textwidth}
\center{\includegraphics[width=1.0\linewidth]{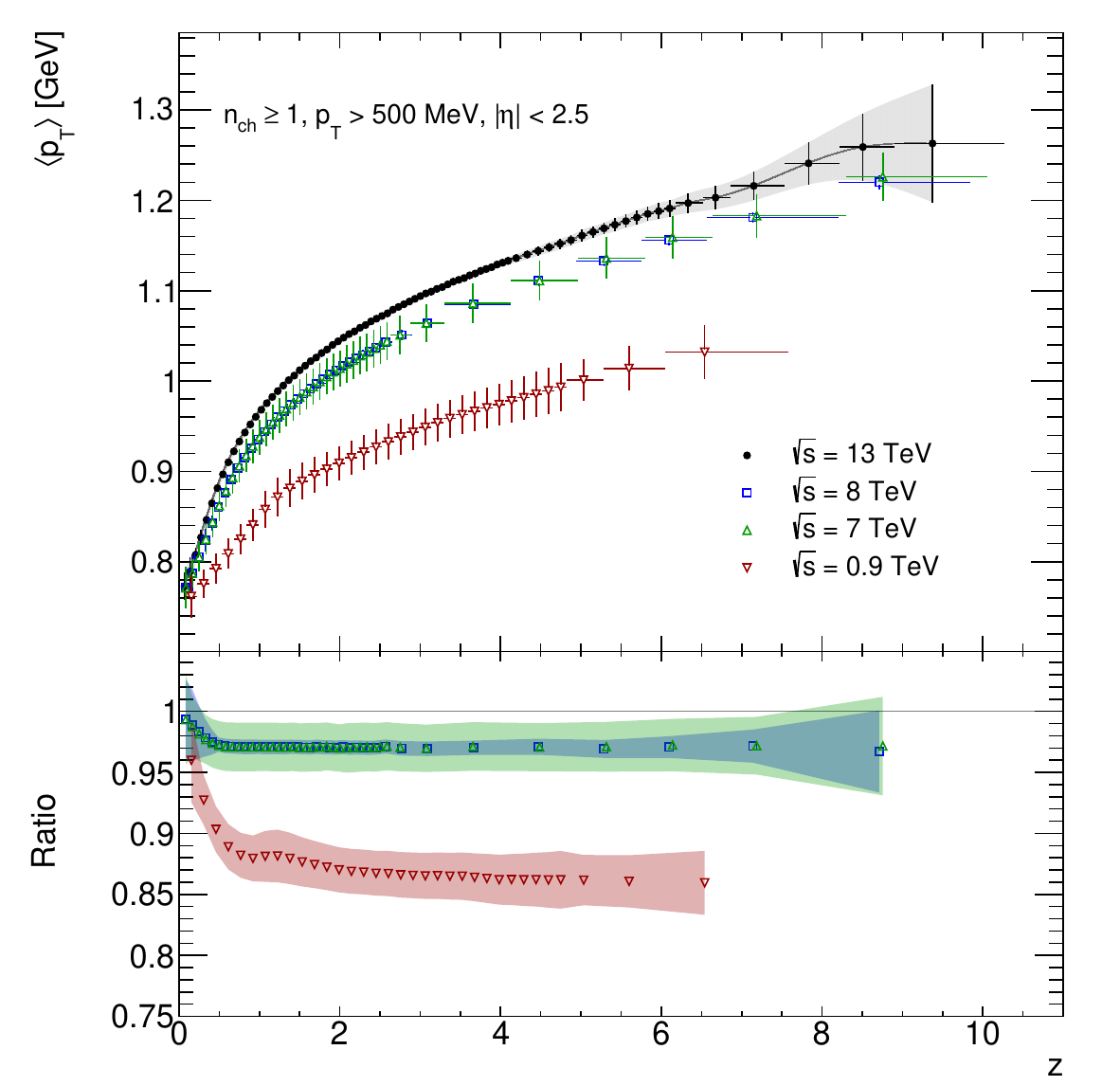}}
(b)
\\
\end{minipage}
\caption{
Top panel: 
The average transverse momentum,  \(\langle p_{\mathrm{T}}\rangle \),
as a function of the scaled multiplicity  \(z\)  defined  by  Eq.~(\ref{eq_mch}) for events with 
(a)  \(n_{\mathrm{ch}} \ge 2\),   \(p_{\mathrm{T}} >100\)~MeV  and 
(b)  \(n_{\mathrm{ch}} \ge 1\),   \(p_{\mathrm{T}} >500\)~MeV 
for  \(\mid\eta\mid  < 2.5\)  measurement   at the  centre-of-mass energies 
\(0.9\), \(7\), \(8\) and  \(13\)~\TeV\  by 
ATLAS 
\cite{STDM-2010-06,STDM-2014-19,STDM-2015-02,STDM-2015-17}. 
(a) The grey curve and  the  band of the uncertainties are the result of the interpolation 
of the charged-particle multiplicity distribution at  \(13\)~\TeV.
The grey curve and band of the uncertainties are the result of the interpolation 
of the charged-particle multiplicity distribution at  \(13\)~\TeV.
The error bars and boxes represent the statistical and systematic contributions, respectively.
Bottom panel: 
The ratios of the average transverse momentum distributions to the interpolated distribution 
at \( \sqrt{s}  = 13\)~\TeV\ are shown. 
Bands represent the uncertainties 
for the ratios 
as the results 
of statistical and systematic uncertainties added in quadrature for both distributions.
Taken from Ref.\ \cite{Kulchitsky:2022gkm}.
}
\label{fig_averagepT_pT100_mch}
\end{figure*}

The correct comparison  of the primary charged-particle average transverse momentum, 
\(\langle p_{\mathrm{T}}\rangle \),  as a function of the  scaled multiplicity 
\( z\) for events with  
\(n_{\mathrm{ch}} \ge 2\) and \(p_{\mathrm{T}} >100\)~\MeV;
\(n_{\mathrm{ch}} \ge 1\) and \(p_{\mathrm{T}} >500\)~\MeV\ 
measure for  \(\mid\eta\mid  < 2.5\)  at the   CM energies 
from \(0.9\)  to  \(13\)~\TeV\  by 
ATLAS 
\cite{STDM-2010-06,STDM-2014-19,STDM-2015-02,STDM-2015-17}
are presented in  Fig.~\ref{fig_averagepT_pT100_mch} \cite{Kulchitsky:2022gkm}. 
The \(\langle p_{\mathrm{T}}\rangle \) distribution  as a function of \( z \) acquires 
a higher value at higher collision energies.
The values of \(\langle p_{\mathrm{T}}\rangle \)  distributions increase by \( 18\)\%  and  \(13\)\%
  for \(z > 1\)   with energy increasing 
  from \(0.9\) to \(13\)~\TeV\   for 
\(p_{\mathrm{T}} >100\)~\MeV\ and \(p_{\mathrm{T}} >500\)~\MeV,  respectively. 
The results at \(7\) and \(8\)~\TeV\ are in agreement within error bars. 
The values of \(\langle p_{\mathrm{T}}\rangle \) 
distributions  increase by 
\(\approx 3\)\% for \(p_{\mathrm{T}} >100\)~\MeV\ 
and  by \(\approx 2.5\)\% for \(p_{\mathrm{T}} >500\)~\MeV\
with increase in energy from \(8\) to \(13\)~\TeV\  for  \(z > 0.5\).
The ratio of  \(\langle p_{\mathrm{T}}\rangle \) distributions
for  \(8\) to \(13\)~\TeV\  are  \(\approx 6\) times  smaller than the ratio for \(0.9\)  to  \(13\)~\TeV.

For \( p_{\mathrm{T}} > 100\)~\MeV\ and  \( p_{\mathrm{T}} > 500\)~\MeV\  at 
the highest energy
distributions  increase towards higher \( n_{\mathrm{ch}}\),  as modelled by 
the
CR mechanism in  \textsc{Pythia\,8} and by the hydrodynamical evolution model in
\textsc{EPOS}.
The \textsc{QGSJET-II}  generator describes the data poorly.
For low \( n_{\mathrm{ch}}\),  \textsc{Pythia\,8}  \textsc{A2}, \textsc{EPOS}
underestimate the data, 
and for higher \( n_{\mathrm{ch}}\) all generators overestimate the data.
\textsc{EPOS} describes the data reasonably well and to within \(2\)\%.
 
As discussed in  Ref.~\cite{ParticleDataGroup:2020ssz}, 
the  \(\langle p_{\mathrm{T}} (n) \rangle\)  of distributions of primary charged particles  
produced via jet fragmentation
slowly  increases with collision energy,  as shown in 
Fig.\ \ref{fig_averagepT_pT100_mch}.  
This is caused by the stronger absorption  (at larger \(\sqrt{s}\))  of the gluons with a smaller 
\( k_{\mathrm{T}} \)  (\(\sigma^{abs} \propto  1 / k_{\mathrm{T}}^2\)). 
The growth of  \( \langle p_{\mathrm{T}}  \rangle\)  with multiplicity  can be explained by the 
fact that events with larger  \( n_{\mathrm{ch}} \)  correspond to a smaller impact parameter, \(b\), 
where the absorption of  the low  \( k_{\mathrm{T}} \)   component is stronger,  and
larger multiplicity can be originated  by 
events with 
jets or minijets with higher  \( p_{\mathrm{T}} \).
Since  \( \langle p_{\mathrm{T}}  \rangle\)  of primary charged particles grows with \(\sqrt{s}\), 
the increase with \(\sqrt{s}\) of transverse energy flow   is a bit faster than that of  the particle density.

%
\subsubsection{Average transverse momentum distributions of the LHC experiments}
\label{Comparison_pT_nch_LHC}

\begin{figure*}[t!]
\centering
\begin{minipage}[h]{0.45\textwidth} 
\center{\includegraphics[width=1.0\linewidth]{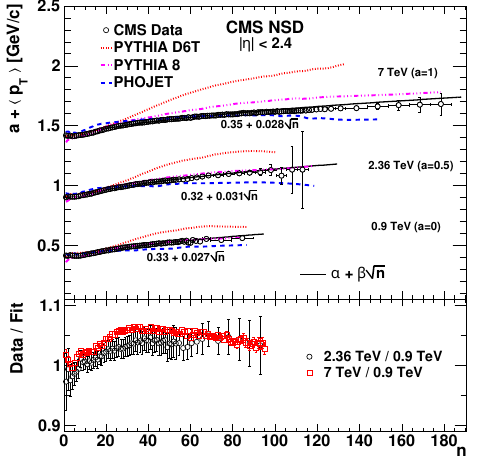}}
\\
\end{minipage}  
\caption{
Top panel: 
A comparison of CMS  results  \cite{CMS:2010qvf} for  the average transverse momentum,  
\(\langle p_{\mathrm{T}} \rangle\),   as a function of of  the charge-particle multiplicity, \(n_{\mathrm{ch}}\),  
for inclusive pseudorapidity  region  \( \mid\eta\mid < 2.4\) with  the prediction of 
the \textsc{Pythia}  \textsc{D6T} tune, 
the \textsc{Pythia\,8} 
and \textsc{PHOJET} models  at 
\(\sqrt{s} = 0.9,\ 2.36\) and \(7\)~\TeV. 
For clarity,  the results for different energies are shifted by the  values  shown in the plots. 
Fits to the high-multiplicity part  (\(n_{\mathrm{ch}} > 15 \))  with a linear form in 
\(\sqrt{n_{\mathrm{ch}}}\)  are superimposed. 
Bottom panel: 
The ratios of the higher-energy data to the fit at \(\sqrt{s} = 0.9\)~\TeV\  
indicate the approximate energy independence of 
\(\langle p_{\mathrm{T}} \rangle\)  at fixed  \(n_{\mathrm{ch}}\).
Taken from Ref.~\cite{CMS:2010qvf}.
}
\label{fig_pT_nch_CMS_2}
\end{figure*}

Figure~\ref{fig_pT_nch_CMS_2} (top)  
shows  a CMS comparison of   the
average transverse momentum,  \(\langle p_{\mathrm{T}} \rangle\),  
as a function  of  the charge-particle multiplicity, \(n_{\mathrm{ch}}\),  
for  the inclusive pseudorapidity  region  \( \mid\eta\mid < 2.4\)
with prediction of  the \textsc{Pythia}  \textsc{D6T} tune, 
the \textsc{Pythia\,8}  and \textsc{PHOJET} models  at  \(\sqrt{s} = 0.9,\ 2.36\) and \(7\)~\TeV\ 
\cite{CMS:2010qvf}. 
In Fig.\ \ref{fig_Nev_nch_CMS_2} (bottom)  the ratios of the higher-energy data to the fit at 
\(\sqrt{s} = 0.9\)~\TeV\  indicate the approximate energy independence of   
\(\langle p_{\mathrm{T}} \rangle\)  at fixed  \(n_{\mathrm{ch}}\).
%
%
These results are in disagreement with  the ATLAS results presented in
Fig.\ \ref{fig_averagepT_pT100_mch},  where a ratio  depends on the multiplicity.

The ATLAS  ratio of  \( \langle p_{\mathrm{T}} \rangle \) distributions for 
\( 7 \)~\TeV\  to  \( 0.9 \)~\TeV\ is  \(\approx 1.18\)
for \(z \gtrsim 2\) as shown in Fig.\ \ref{fig_averagepT_pT100_mch}(a).
According to the  CMS,  the same ratio 
is
shown in Fig.\ \ref{fig_Nev_nch_CMS_2}
is  \(\approx 1.05\)  for  \( n_{\mathrm{ch}} \gtrsim 30\)  or  \(z \gtrsim 1\), 
because   \(\langle n_{\mathrm{ch}} \rangle = 30.4 \) at \(7\)~\TeV\
in Table~\ref{tab:average_nch_energy_CMS}. 
That is  \(\approx 3.5\) times smaller than for ATLAS.

Among the three classes of models,  \textsc{Pythia\,8} 
gives the best overall description of the multiplicity  distribution and the dependence of the
average transverse momentum on  \(n_{\mathrm{ch}}\). 
%
%
Inspired by  \cite{McLerran:2010ex} the fit  
of
 the first-degree polynom in  \(\sqrt{ n_{\mathrm{ch}} }\)
to the multiplicity dependence of  \(\langle p_{\mathrm{T}} (n_{\mathrm{ch}} ) \rangle\) for 
\(n_{\mathrm{ch}} > 1.5\) at each energy
yielding a good description 
that
is valid at all three energies. 
The ratios of the data obtained at  \(\sqrt{s} = 7\)~\TeV\   and  \(\sqrt{s} = 2.36\)~\TeV\ 
with respect to the data at  \(\sqrt{s} = 0.9\)~\TeV\   show that the rise of the average
transverse momentum with the multiplicity  weakly depends on energy.

\begin{figure*}[t!]
\centering
\begin{minipage}[h]{0.32\textwidth} 
\center{\includegraphics[width=1.0\linewidth]{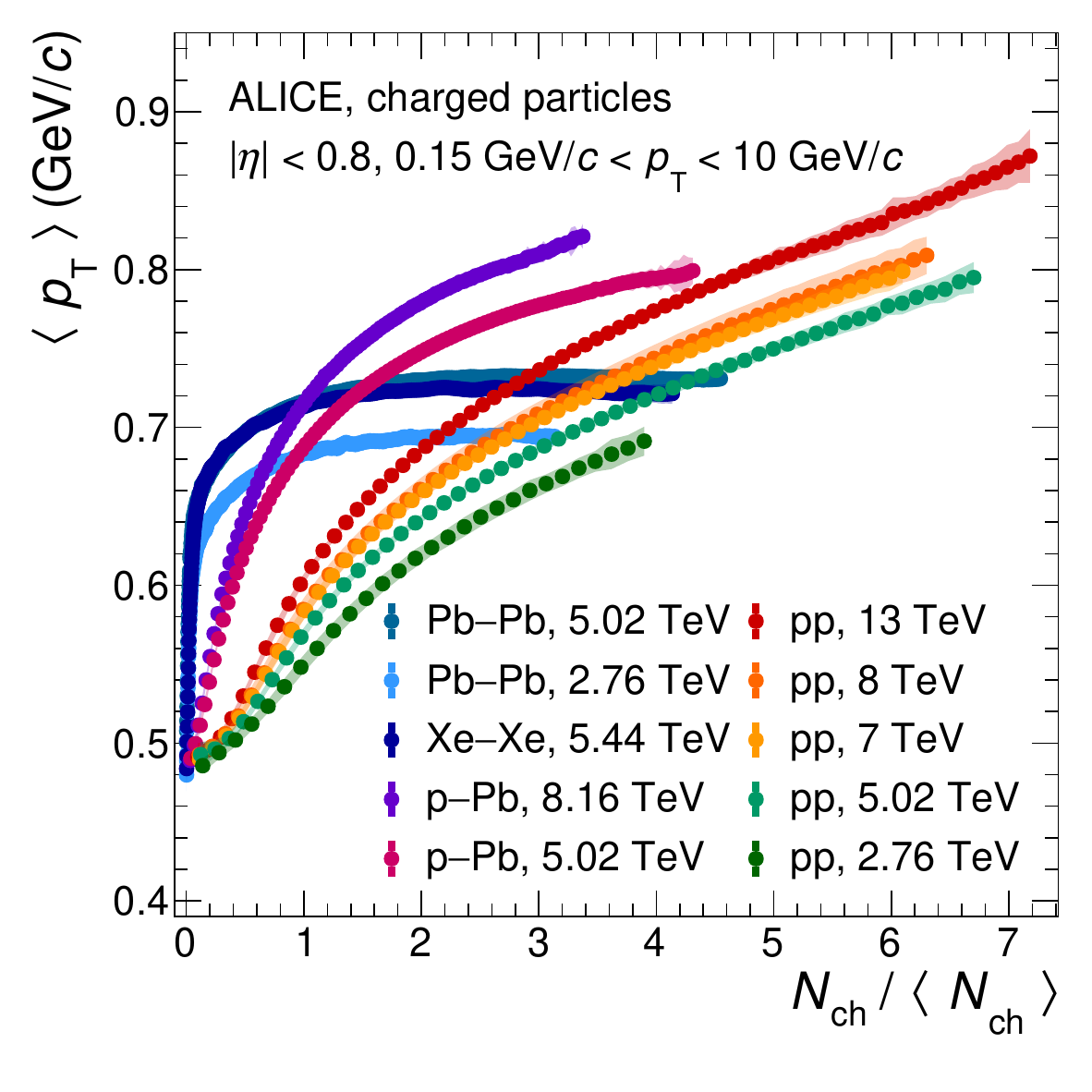}}
(a) 
\\
\end{minipage}
\begin{minipage}[h]{0.32\textwidth} 
\center{\includegraphics[width=1.0\linewidth]{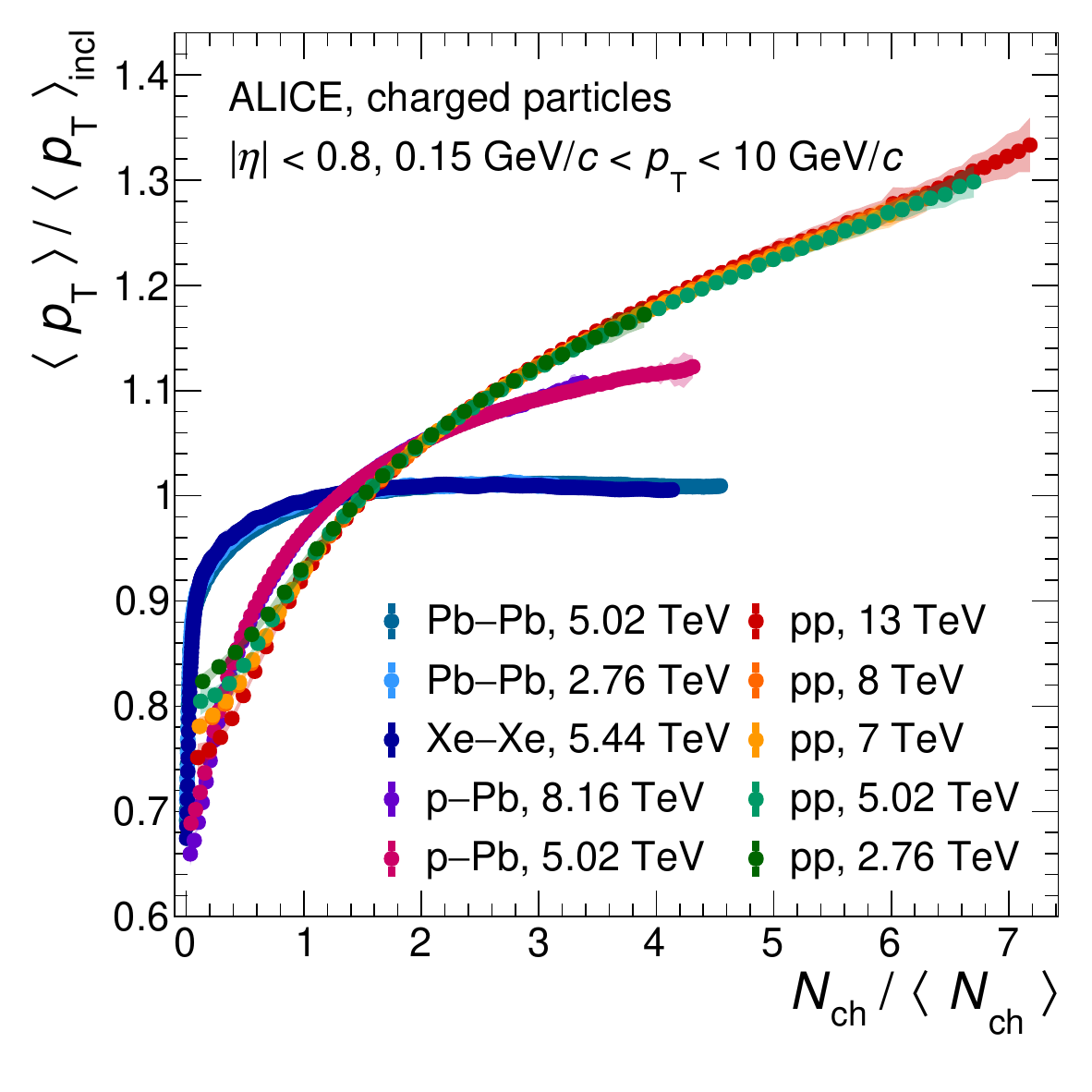}}
(b)
\\
\end{minipage}
\begin{minipage}[h]{0.32\textwidth} 
\center{\includegraphics[width=1.0\linewidth]{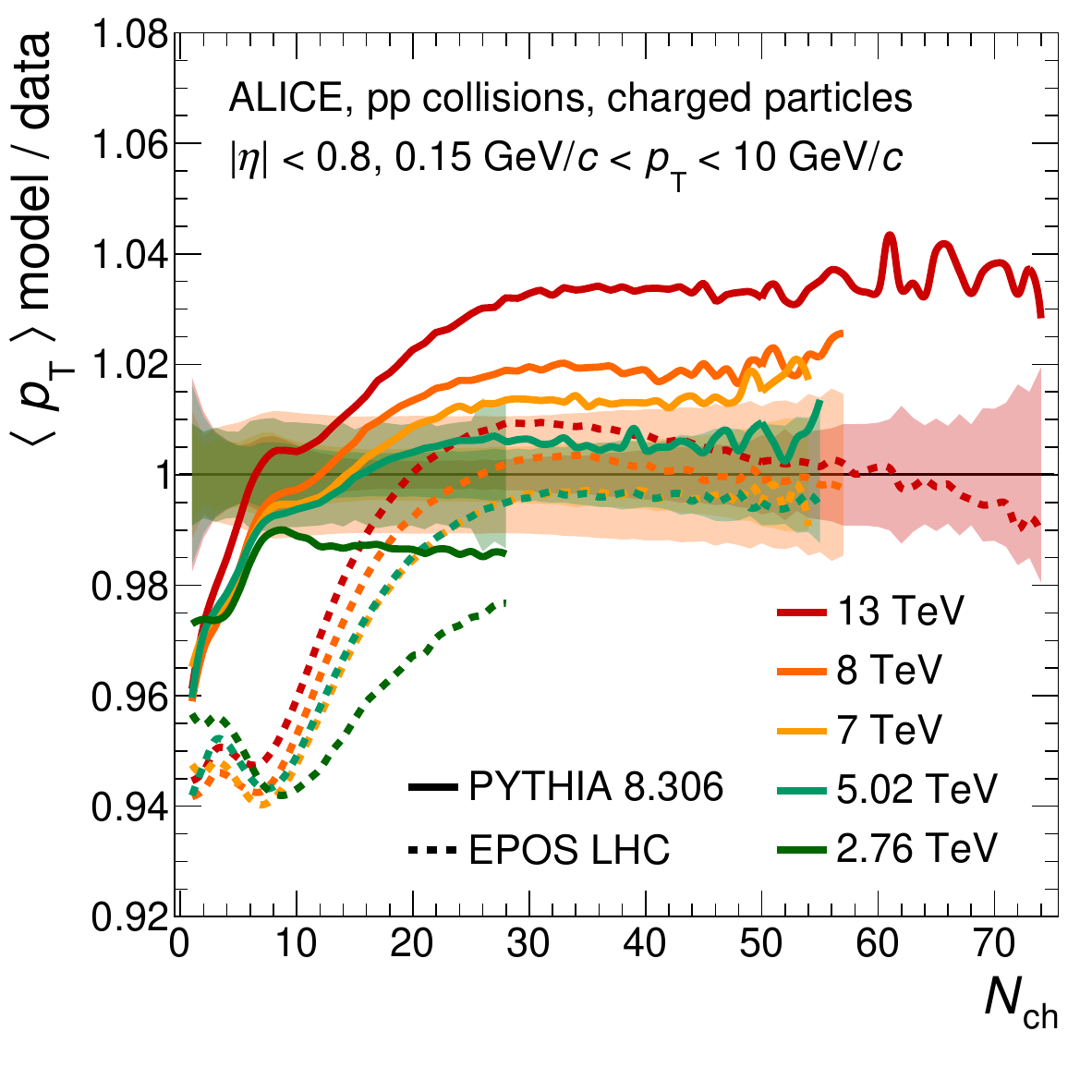}}
(c)
\\
\end{minipage}
\caption{
The 
(a) 
average charged-particle transverse momentum,  \( \langle p_{\mathrm{T}} \rangle \),  and 
(b) 
normalised on  \( \langle p_{\mathrm{T}} \rangle_{\mathrm{incl}} \), 
\( \langle p_{\mathrm{T}} \rangle / \langle p_{\mathrm{T}}\rangle_{\mathrm{incl}} \),
distributions as a function of the scaled multiplicity  \(z\) 
or \( N_{\mathrm{ch}} /\langle N_{\mathrm{ch}} \rangle \) for 
\(pp\), \(p\)-\(Pb\),  \(Xe\)-\(Xe\) and \(Pb\)-\(Pb\) collisions 
at the different centre-of-mass energies
\(\sqrt{s} = 2.36,\ 5.02,\ 7,\ 8\) and  \(13\)~\TeV\ for \(pp\),
\(\sqrt{s} = 5.02\) and  \(8.16\)~\TeV\ for \(p\)-\(Pb\),
\(\sqrt{s} = 5.44\)~\TeV\ for \(Xe\)-\(Xe\)
and \(\sqrt{s} = 2.76\) and  \(5.02\)~\TeV\ for \(Pp\)-\(Pb\)
for events in the kinematic range \( N_{\mathrm{ch}} > 0\),  \( \mid\eta\mid < 0.8 \) 
and \( 0.15 < p_{\mathrm{T}} < 10 \)~\GeV.
Statistical and systematic uncertainties are shown as bars and semi-transparent bands, respectively.
The ratio of  \textsc{Pythia\,8} \cite{Sjostrand:2014zea} and \textsc{EPOS\ LHC}  \cite{Pierog:2013ria}
model predictions to data for \(pp\) collisions at various energies are 
is
shown  for
(c) 
\( \langle p_{\mathrm{T}} \rangle \)  distributions. 
The semi-transparent bands indicate the relative systematic uncertainties of the data.
Taken from Ref.~\cite{ALICE:2022xip}.
}
\label{fig_pT_nch_ALICE_1}
\end{figure*}

The  average  charged-particle transverse momenta
for \( pp\)  collisions at the different centre-of-mass energies
\(\sqrt{s} = 2.36,\ 5.02,\ 7,\ 8\) and  \(13\)~\TeV\  
for the kinematic region 
\( \mid\eta\mid < 0.8 \)  and  \( 0.15 < p_{\mathrm{T}} < 10 \)~\GeV\
were obtained by the ALICE experiment 
\cite{ALICE:2022xip}
and are presented
in Fig.~\ref{fig_pT_nch_ALICE_1}.
In 
Fig.~\ref{fig_pT_nch_ALICE_1}(a) 
the average charged-particle transverse momentum 
\( \langle p_{\mathrm{T}} \rangle \)  spectra 
and
in Fig.~\ref{fig_pT_nch_ALICE_1}(b) 
the 
\( \langle p_{\mathrm{T}} \rangle \) spectra 
divided by their respective multiplicity-integrated values,
\( \langle p_{\mathrm{T}} \rangle_{\mathrm{incl}} \),
as a function of relative multiplicity
\( N_{\mathrm{ch}} /\langle N_{\mathrm{ch}} \rangle \),
  same as the scale variable \(z\),  
are  shown. 
The value of 
\( \langle p_{\mathrm{T}} \rangle_{\mathrm{incl}} \) 
for \(pp\)  collisions
increase from 
\( 6.05 \pm 0.17 \) at \(\sqrt{s} = 2.76\)~\TeV\ 
to 
\( 9.48 \pm 0.07 \) at \(\sqrt{s} =13\)~\TeV\
(see in  Table 2 \cite{ALICE:2022xip}).
The values for each collision system align almost perfectly for the 
\( \langle p_{\mathrm{T}} \rangle / \langle p_{\mathrm{T}} \rangle_{\mathrm{incl}} \).
In \(pp\) collisions, the overall shapes of the  \( \langle p_{\mathrm{T}} \rangle \)  
distributions are shown in Fig.~\ref{fig_pT_nch_ALICE_1}(c) 
in comparison with predictions from  \textsc{Pythia\,8}  \cite{Sjostrand:2014zea} (solid lines) 
and  \textsc{EPOS\ LHC}   \cite{Pierog:2013ria} (dashed lines).
\textsc{Pythia\,8}  underpredicts the experimental data on 
\( \langle p_{\mathrm{T}} \rangle \)  
at the lowest values of 
\(N_{\mathrm{ch}} \)   
by up to \(4\)\%. 
The \(N_{\mathrm{ch}} \)  dependent 
\( \langle p_{\mathrm{T}} \rangle \)  
values produced by 
\textsc{Pythia\,8}
increase faster than the measurements with an almost linear dependence up to 
\(N_{\mathrm{ch}} \approx 20 \), 
after which the ratio shows a flat multiplicity dependence with an offset from unity varying from 
\(0.5\)\% at\(\sqrt{s} = 5.02\)~\TeV\   up to \(4\)\% at the highest CM energy. 
\textsc{EPOS\ LHC} is further off at low multiplicities by up to \(5\)\% 
and increases slower than the measurements, underestimating themby up to \(6\)\% around 
\(N_{\mathrm{ch}} \approx 9 \). 
At higher multiplicities, the increase is faster with a linearly rising ratio up to 
\(N_{\mathrm{ch}} \approx 20 - 30 \),
reaching a plateau 
that
describes the measurements within \(\pm 2\)\%.

\section{KNO scaling }
\label{KNO_scaling}
\subsection{KNO scaling on the ATLAS results}
\label{KNO_scaling_ATLAS}

In  \(pp\) collisions at \(\sqrt{s}\) from  \(0.0304\) to \(0.0622\)~\TeV\ in the full  PS for inelastic events 
at the ISR (CERN) the deviation from  the KNO scaling was already observed
\cite{Ames:1983cqw,Aachen:1977izz}.
%
%
For NSD collisions KNO scaling was still found to be present  \cite{Ames:1983cqw}, 
suggesting that diffractive processes might also play a role in KNO scaling violations. 
Clear scaling violations become manifested  above  \( \sqrt{s} \approx  0.2\)~\TeV\ 
both for the multiplicity distributions in full  PS and  in central pseudorapidity ranges 
\cite{Alexopoulos:1998bi,EHSNA22:1987syd,NA22:1986nvb,CDF:2001hmt}.
%
In NSD  \(pp\) collisions at the LHC, at  \(\sqrt{s} = 2.36\)~\TeV, \(7\)~\TeV\ in  pseudorapidity region \(\mid\eta\mid < 0.5\),  
ALICE \cite{ALICE:2010cin,ALICE:2015olq} and  the CMS  \cite{CMS:2010qvf} observed no significant deviation from  the KNO scaling.
%
%
In \(p\bar{p}\) collisions at the SPS (CERN) at  \(\sqrt{s} = 0.2\), \(0.546\) and \(0.9\)~\TeV, 
the KNO scaling was found to be violated for NSD collisions in full  PS
\cite{UA5:1988gup,Chou:1982dn,UA5:1981ccg}.
Nevertheless, for NSD collisions, in limited central pseudorapidity intervals,  
the KNO scaling was still found to hold up to \(0.9\)~\TeV,   
and at  \(\sqrt{s} = 0.546\)~\TeV,  
the KNO scaling was found to hold in the pseudorapidity interval  \(\mid\eta\mid < 3.5\) 
\cite{UA1:1982yyh,Kam:1984wnl}.
%
%
In \(e^+ e^-\) collisions at  \(\sqrt{s}\) from  \(0.005\) to \(0.034\)~\TeV\ 
the KNO scaling was found to  hold within \(\pm 20\)\%  \cite{TASSO:1983cre}.

\begin{figure*}[t!]
\centering
\begin{minipage}[h]{0.45\textwidth}
\center{\includegraphics[width=1.0\linewidth]{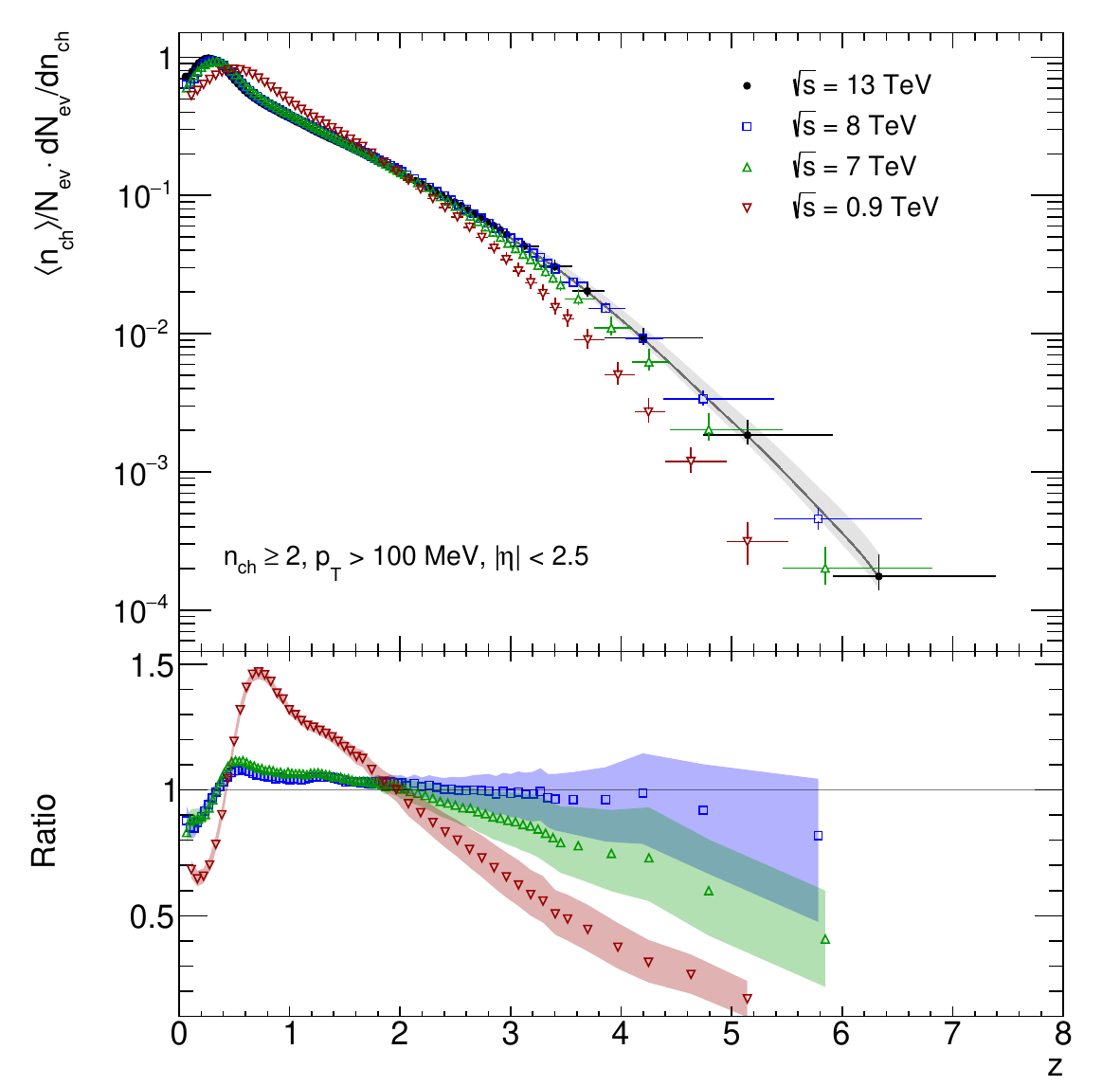}} 
(a) 
\\
\end{minipage}
\hspace{2mm}
\begin{minipage}[h]{0.45\textwidth}
\center{\includegraphics[width=1.0\linewidth]{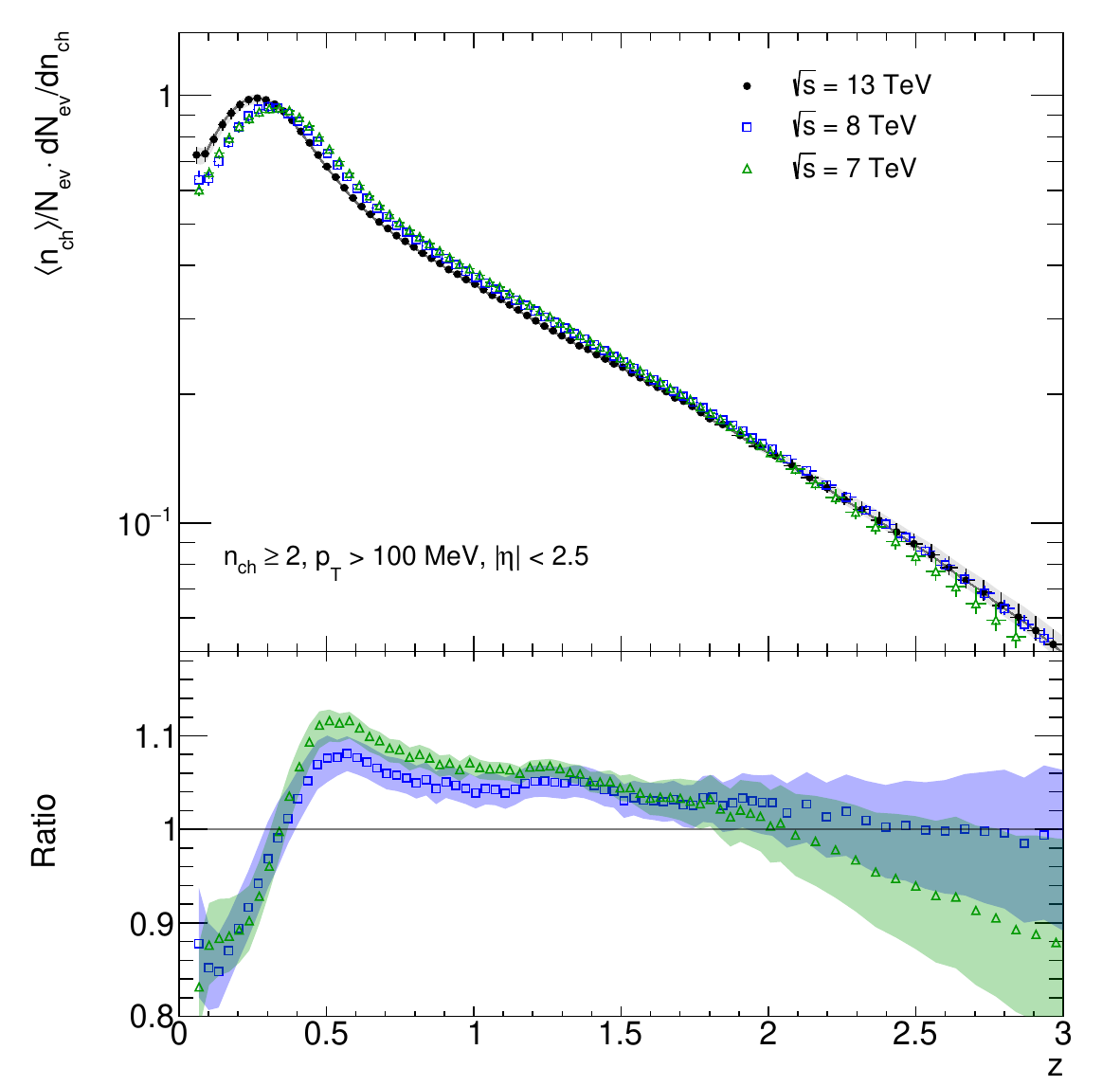}} 
(b)
\\
\end{minipage}
\caption{
Top panel: 
KNO-scaled primary charged-particle  multiplicity distributions    as a function of the scaled multiplicity  \(z\) 
for events with  \(n_{\mathrm{ch}} \ge 2\), \(p_{\mathrm{T}} >100\)~\MeV\ 
and \(\mid\eta\mid  < 2.5\)  measured   at the centre-of-mass energies 
\(0.9\),  \(7\),  \(8\) and \(13\)~\TeV\  by 
ATLAS 
\cite{STDM-2010-01,STDM-2010-06,STDM-2014-19,STDM-2015-02,STDM-2015-17}
in 
(a) the complete multiplicity region and 
(b) the zoom multiplicity region with \(z \le 3\) 
at  the  \(\sqrt{s} = 7\),  \(8\) and \(13\)~\TeV.
The grey curve and  the band of  uncertainties are the result of the interpolation 
of the charged-particle multiplicity distribution at  \(13\)~\TeV.
The uncertainties represent the sum in quadrature of the statistical and systematic contributions.
Bottom panel: 
The ratios of the KNO-scaled primary charged-particle distributions 
to the interpolated distribution at \( \sqrt{s}  = 13\)~\TeV\ are shown. 
Bands represent  uncertainties for the ratios  as  the
results of statistical and systematic uncertainties added in quadrature for both distributions.
Taken from Ref.~\cite{Kulchitsky:2022gkm}.
}
\label{fig_events_pT100_mch_KNO}
\end{figure*}

\begin{figure*}[t!]
\centering
\begin{minipage}[h]{0.45\textwidth}
\center{\includegraphics[width=1.0\linewidth]{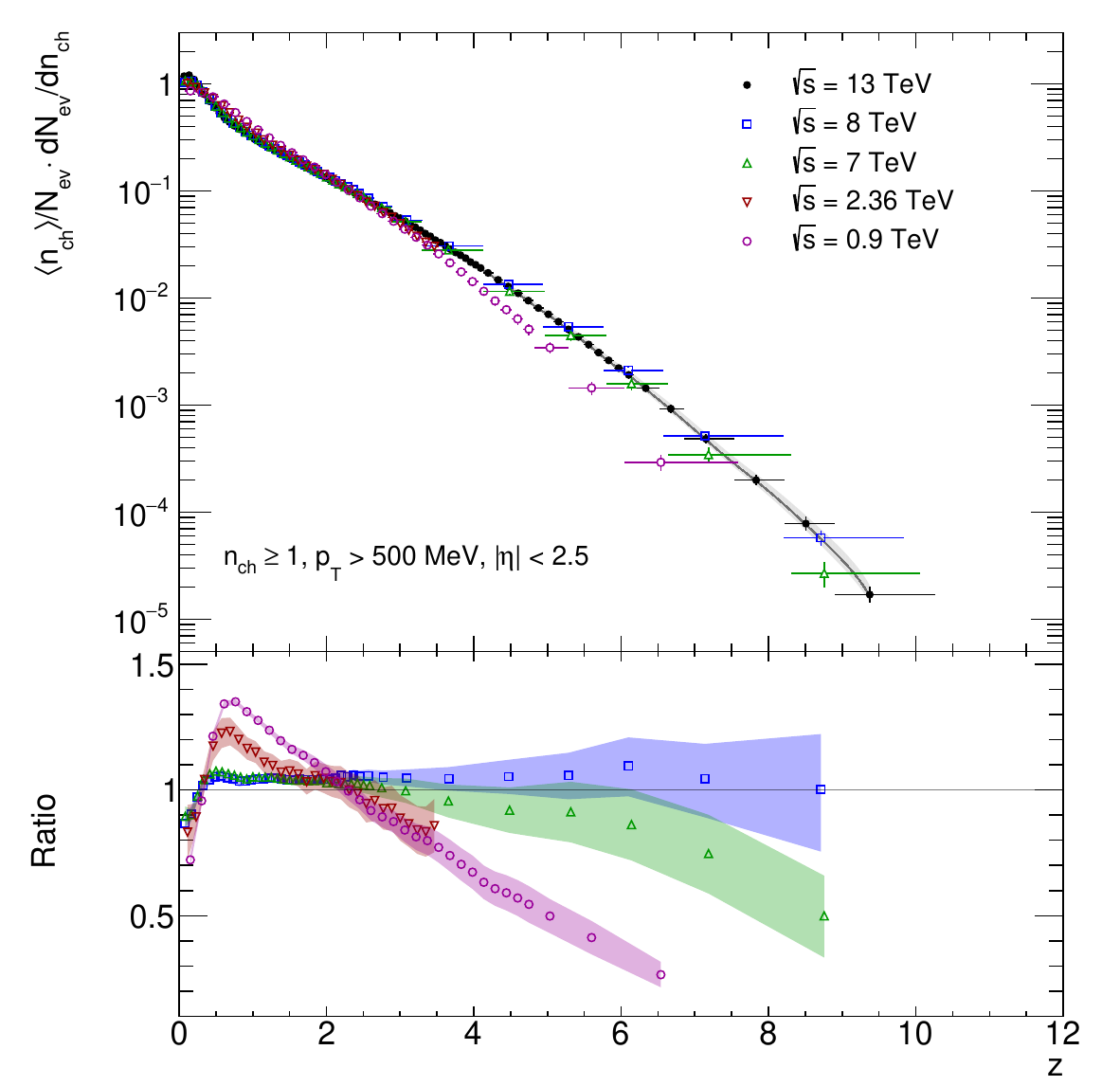}} 
(a) 
\\
\end{minipage}
\hspace{2mm}
\begin{minipage}[h]{0.45\textwidth}
\center{\includegraphics[width=1.0\linewidth]{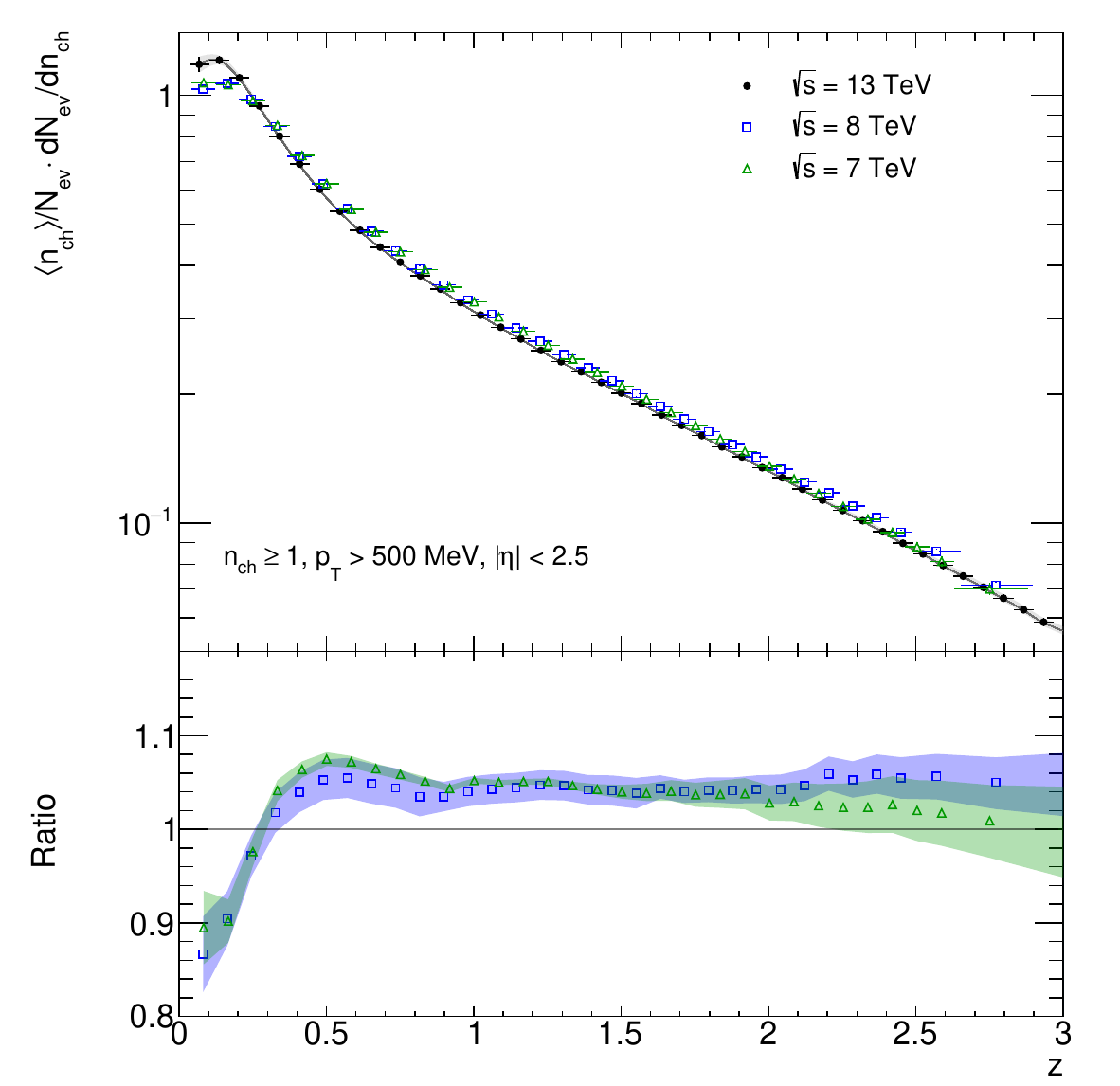}} 
(b)
\\
\end{minipage}
\caption{
Top panel: 
KNO-scaled primary charged-particle multiplicity distributions  as a function of the scaled multiplicity 
\(z\) for events with  \(n_{\mathrm{ch}} \ge 1\), \(p_{\mathrm{T}} >500\)~\MeV\ 
and \(\mid\eta\mid  < 2.5\)  measurement  at the centre-of-mass energies 
\(0.9\),  \(2.36\), \(7\),  \(8\) and \(13\)~\TeV\  by 
ATLAS 
\cite{STDM-2010-01,STDM-2010-06,STDM-2014-19,STDM-2015-02,STDM-2015-17}
in 
(a) the complete multiplicity region and 
(b) the zoom multiplicity region with \(z \le 3\)  at  the  \(\sqrt{s} = 7\),  \(8\) and \(13\)~\TeV.
The grey curve and band of the uncertainties are the result of the interpolation 
of the charged-particle multiplicity distribution at  \(13\)~\TeV.
The uncertainties represent the sum in quadrature of the statistical and systematic contributions.
Bottom panel: 
The ratios of the KNO-scaled primary charged-particle 
distributions to the interpolated distribution at \( \sqrt{s}  = 13\)~\TeV\ are shown. 
Bands represent the uncertainties for the ratios 
as 
the
results of statistical and systematic uncertainties added in quadrature for both distributions.
Taken from Ref.~\cite{Kulchitsky:2022gkm}.
}
\label{fig_events_pT500_mch_KNO}
\end{figure*}

%
For the verification of  the KNO scaling hypothesis  the following  equation with  dependence  on the CM energy and a kinematic region, 
\( p_{\mathrm{T}}^{\mathrm{min}} \),   was used in Ref.~\cite{Kulchitsky:2022gkm}:
\begin{equation}
\label{eq_Psi_nch}
\Psi ( z , \sqrt{s})  
= { \langle n_{\mathrm{ch}} (\sqrt{s}, p_{\mathrm{T}}^{\mathrm{min}}) \rangle }  
\cdot  
P (n_{\mathrm{ch}}, \sqrt{s}, p_{\mathrm{T}}^{\mathrm{min}})  
= \frac{ \langle n_{\mathrm{ch}} (\sqrt{s}, p_{\mathrm{T}}^{\mathrm{min}}) \rangle }{ N_{\mathrm{ev}} (\sqrt{s},  p_{\mathrm{T}}^{\mathrm{min}} ) } \cdot \frac{d N_{\mathrm{ev}} (\sqrt{s}, p_{\mathrm{T}}^{\mathrm{min}} )}{ d n_{\mathrm{ch}}},
\end{equation}
where 
\( z \) is  defined in  Eq.~(\ref{eq_mch}),
\(n_{\mathrm{ch}}\)  is the number of primary charged particles within the kinematic acceptance in an event,
\( P (n_{\mathrm{ch}}, \sqrt{s}) \)  is   the  probability distributions  of producing 
\( n_{\mathrm{ch}} \)  particles,  \(N_{\mathrm{ev}} \)  is the number of events with  primary 
charged particles in  the kinematic acceptance,  \( \langle n (\sqrt{s}) \rangle \)   is the average multiplicity of primary particles at 
the  CM energy,  and  \( \Psi ( z ) \) is the  particle distribution as a function of the scaled multiplicity.

The KNO scale variable  \(z \)
provides  a way to study  the evolution of shapes of the KNO charged-particle
multiplicity distributions (see Eq.~(\ref{eq_Psi_nch})) with varying  CM energy and kinematic region,  
for example,  the    \(p_{\mathrm{T}}^{\mathrm{min}}\) threshold.
The  KNO distributions and their ratios, studied using ATLAS results,  
are presented in Fig.~\ref{fig_events_pT100_mch_KNO}  for charged particles with \(p_{\mathrm{T}} >100\)~\MeV\ and 
in Fig.~\ref{fig_events_pT500_mch_KNO} for those with \(p_{\mathrm{T}} >500\)~\MeV.
These figures are similar to  Fig.~\ref{fig_events_pT100_mch} and Fig.~\ref{fig_events_pT500_mch} 
but  the vertical axis is stretched by the factor  \( \langle n_{\mathrm{ch}} (\sqrt{s}, p_{\mathrm{T}}^{\mathrm{min}})\rangle \).
The quantities of interest are derived from the original set of   KNO distributions and the  ratios  of  these distributions 
to the one at \(13\)~\TeV. 
The high-multiplicity tail of  the  distributions is pushed up,  and the maximum of the distribution 
is shifted towards small values of \(z\) with  increasing collision energy.

Ratios of the KNO distributions between the smallest  CM energy  \(0.9\)~\TeV\ to  \(13\)~\TeV\ 
reach the maximum  value at  \( z \approx 0.8 \) and the 
minimum value for the highest multiplicity  at \( z \approx 5.5 \) 
for \(p_{\mathrm{T}} >100\)~\MeV, as can be seen in    Fig.~\ref{fig_events_pT100_mch_KNO}(a), and 
\( z \approx 6.5 \) for \(p_{\mathrm{T}} >500\)~\MeV, in  Fig.~\ref{fig_events_pT500_mch_KNO}(a).
There is an intersection point for all distributions at  \( z \approx 2\).

A test of the KNO scaling distributions between  \(\sqrt{s} = 0.9\) and \(13\)~\TeV\ 
confirms that KNO scaling violations increase with decreasing collision energy.
The ratios of the KNO distributions  between the highest energies  \(8\)  and \(13\)~\TeV\ 
exceed   the maximum value of  \(+8\)\% at  \( z \approx 0.5 \)  and   
the minimum value of   \(-15\)\% at \( z \approx 0.1 \)    for \(p_{\mathrm{T}} >100\)~\MeV,  
as can be seen in  Fig.~\ref{fig_events_pT100_mch_KNO}(b), and  
the maximum value  of  \(+5\)\% at  \( z \approx 0.5 \)  and  \( -13 \)\% at \( z \approx 0.1 \) 
for \(p_{\mathrm{T}} >500\)~\MeV,  in  Fig.~\ref{fig_events_pT500_mch_KNO}(b).
For the high multiplicity tail, these ratios are in agreement within error bars with the KNO distribution at 
\(13\)~\TeV.
Single-diffractive and double-diffractive processes  make an important  contribution only for the 
low-multiplicity  region,  \(z \lesssim 0.3\).
The typologies of diffractive and non-diffractive events are different,  and their  KNO  behaviour  may also be  different. 
The negative spread, \( \lesssim -8 \)\%, for the low multiplicity may  be the  result 
of  the contribution from  diffractive processes.
The KNO scaling   tends  to be  valid in the energy region from \(\sqrt{s} = 7\) to \(\sqrt{s} =13\)~\TeV\ within   
\( \approx \mbox{}^{\mbox{~}+8}_{-15} \)\% for \( z \lesssim 2 \) 
and  within error bars for \( z \gtrsim 2  \)   for events with  the charged-particle transverse momentum  
\(p_{\mathrm{T}} >100\)~\MeV\   (Fig.~\ref{fig_events_pT100_mch_KNO}(b)), 
and  within   \( \mbox{}^{\mbox{~}+5}_{-13} \)\% for \( z \lesssim 3 \)  and  within error bars for \( z \gtrsim 3  \)   
for events with  the charged-particle transverse momentum  \(p_{\mathrm{T}} >500\)~\MeV\ 
(Fig.~\ref{fig_events_pT500_mch_KNO}(b)). 
The tendency of the KNO scaling to  hold  for the highest collision energies is observed.  

The MC QGSM predictions are made for the KNO  non-diffractive charged-particle multiplicity distributions 
for  \(pp\) collisions including at  the highest LHC  CM energy  \(\sqrt{s}= 14\)~\TeV\ 
for \(\mid\eta\mid <2.4\)  in Fig.~12 in Ref.~\cite{Bleibel:2010ar}.
These distributions have the same qualitative behaviour as those presented in  Fig.~\ref{fig_events_pT100_mch_KNO}(a).
The MC QGSM described the KNO  distributions as the contribution of the cylinder diagram and diagrams with multi-Pomeron scattering. 
The pronounced peak in the low \(z\)  arises solely due to a single Pomeron exchange,
and the maxima of  the distributions for multi-Pomeron processes are moved in the direction of 
high \(z\)  thus  pushing  up the tail \cite{Bleibel:2010ar}.

\begin{figure*}[t!]
\centering
\begin{minipage}[h]{0.32\textwidth} 
\center{\includegraphics[width=1.0\linewidth]{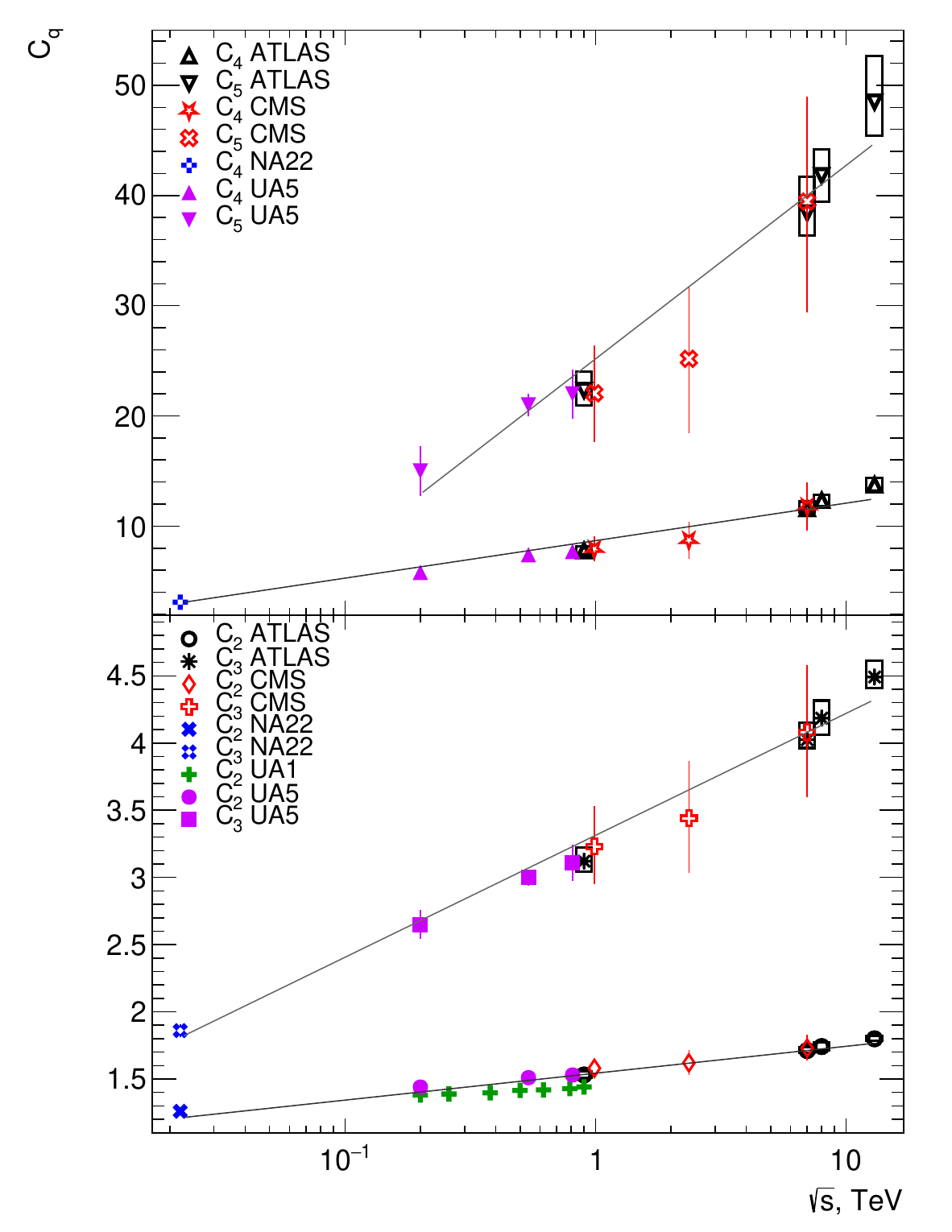}} 
(a)
\\
\end{minipage}
\hfill
\begin{minipage}[h]{0.32\textwidth} 
\center{\includegraphics[width=1.0\linewidth]{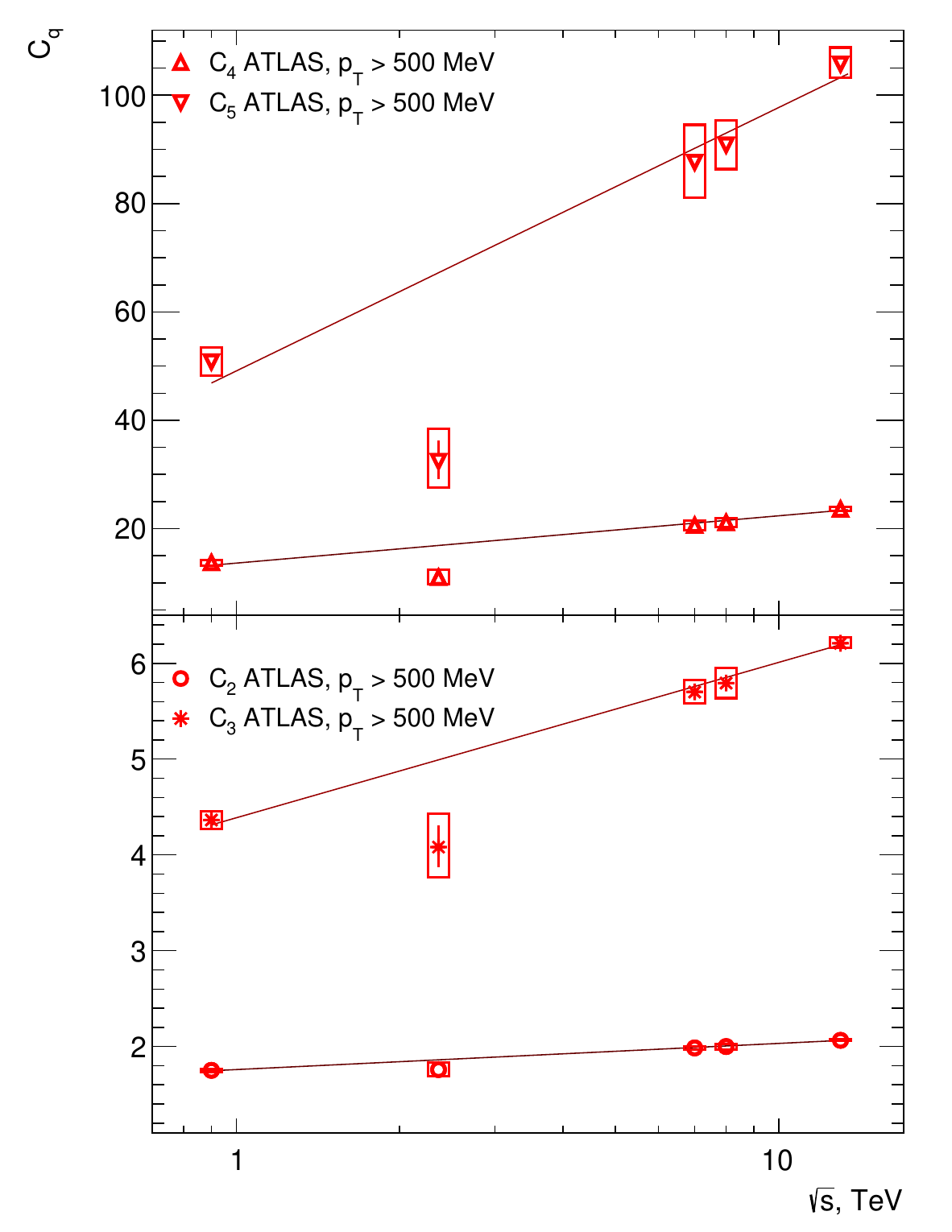}} 
(b)
\\
\end{minipage}
\hfill
\begin{minipage}[h]{0.32\textwidth} 
\center{\includegraphics[width=1.0\linewidth]{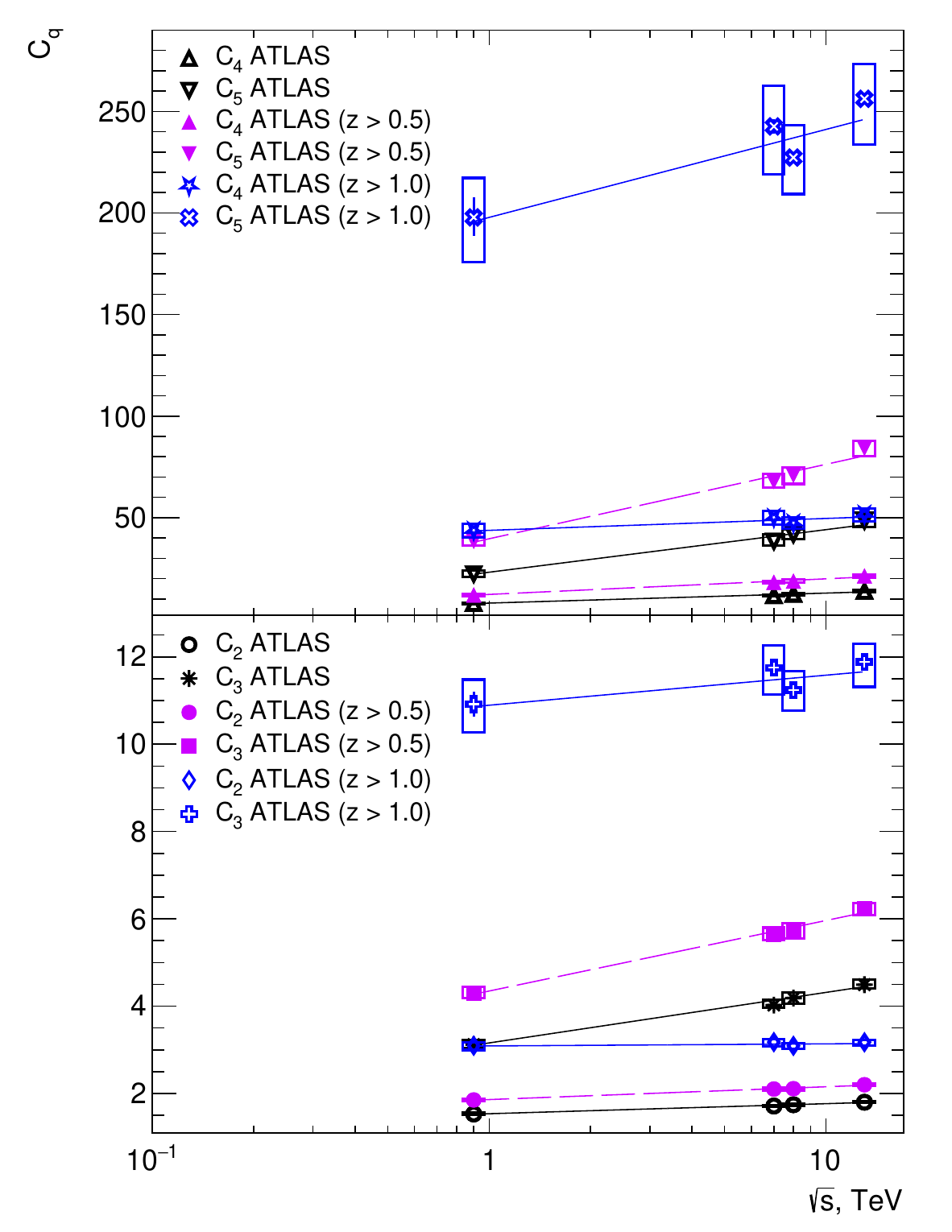}} 
(c)
\\
\end{minipage}
\caption{
The  normalised order-\(q\) moments   \(C_{\mathrm{q}}(\sqrt{s})\)
of the primary charged-particle multiplicity distributions   measured by  the ATLAS  experiment
for events  collected at \( \sqrt{s} = 0.9,\ 2.36,\ 7, 8\) and \(13\)~\TeV\ for 
(a) the  pseudorapidity region  \( \mid\eta\mid < 2.5\).  
The results of  the CMS   \cite{CMS:2010qvf} and lower-energy experiments 
NA22   \cite{EHSNA22:1987syd},
UA1   \cite{UA1:1989bou},   and 
UA5  \cite{UA5:1985hzd,UA5:1988gup}  are included. 
(b) 
The ATLAS results for
\( \mid\eta\mid < 2.5\), \(n_{\mathrm{ch}} \ge 1\),  \(p_{\mathrm{T}} >500\)~\MeV.
(c) 
The ATLAS results for
\( \mid\eta\mid < 2.5\), \(n_{\mathrm{ch}} \ge 2\), \(p_{\mathrm{T}} >100\)~MeV
with additional scaled multiplicity thresholds:  \( z  > 0.5 \) and  \( z > 1.0 \).
The  
\(C_{\mathrm{2}}\)
and
\(C_{\mathrm{3}}\)   
results are shown in the  bottom panel,
and  
\(C_{\mathrm{4}}\)
and
\(C_{\mathrm{5}}\)  
results are shown in the  top panels. 
The vertical bars  are the statistical uncertainties,  and  the  squares  are the systematic uncertainties.
The coloured symbols  are the data. 
Fits of the  \( \log{\sqrt{s}}\)   dependence of the  \( C_{\mathrm{q}} (\sqrt{s}) \) 
of the multiplicity distribution   (assuming linear dependence)   are shown.
In (a) for \(\sqrt{s} = 0.9\)~\TeV,   
data from experiments other than ATLAS were drawn shifted to lower  \(\sqrt{s}\) for clarity.
The  lines show the results of the fits for  \(C_{\mathrm{q}}  (\sqrt{s}) \)  with statistical and systematic uncertainties added in quadrature.
Taken from Ref.~\cite{Kulchitsky:2023fqd}.
}
\label{fig_Cq_ATLAS_1}
\end{figure*}

The energy independence of the moments of  the probability distributions  defined as 
\( P (n_{\mathrm{ch}}, \sqrt{s}) \)  
\begin{equation}
\label{eq_Cq}
C_{\mathrm{q}} (\sqrt{s}) = 
\frac{ \sum_{n=1}^{n_{max}} n_{\mathrm{ch}}^\mathrm{q} (\sqrt{s}) P (n_{\mathrm{ch}}, \sqrt{s}) 
}{
\left( \sum_{n=1}^{n_{max}} n_{\mathrm{ch}}  (\sqrt{s}) P (n_{\mathrm{ch}}, \sqrt{s}) \right)^\mathrm{q}
}
\end{equation}
in the energy asymptotic was the precise finding of the KNO scaling   \cite{Koba:1972ng}.
The analysis results for the validity of KNO scaling  is shown quantitatively in   
Fig.~\ref{fig_Cq_ATLAS_1}  by the  \( C_{\mathrm{q}} (\sqrt{s}) \)  
of the multiplicity distributions measured by 
ATLAS and complemented with 
the CMS  measurements at  \(\sqrt{s} =0.9,\ 2.36\) and \(7\)~\TeV\ 
\cite{CMS:2010qvf} and  the  results of the  lower-energy experiments by
NA22  \cite{EHSNA22:1987syd}, UA1  \cite{UA1:1989bou},  
and  UA5  \cite{UA5:1985hzd,UA5:1988gup}. 
%
The \( C_{\mathrm{q}} (\sqrt{s}) \)  calculations based on  the ATLAS results  for  the kinematic region  
\( \mid\eta\mid < 2.5\), \(n_{\mathrm{ch}} \ge 2\) and  \(p_{\mathrm{T}} >100\)~\MeV\
are shown in  Fig.~\ref{fig_Cq_ATLAS_1}(a). 
The  ATLAS and CMS  results agree within the error range.
The values of  \( C_{\mathrm{q}} (\sqrt{s}) \)    for all  experiments   linearly increase with 
\( \log{\sqrt{s}}\) as illustrated by the fits in   Fig.~\ref{fig_Cq_ATLAS_1}(a).
Since, as mentioned above,  the KNO scaling requires  that \( C_{\mathrm{q}} (\sqrt{s}) \) 
be independent of  energy,  one can state that the KNO scaling is violated at least for 
the full region of scaled multiplicity.
%
Figure~\ref{fig_Cq_ATLAS_1}(b)  shows for the first time the values of  \( C_{\mathrm{q}} (\sqrt{s}) \) 
calculated using multiplicity distributions  measured by ATLAS for the kinematic region
\( \mid\eta\mid < 2.5\), \(n_{\mathrm{ch}} \ge 1\) and  \(p_{\mathrm{T}} >500\)~\MeV.
Similarly as in Fig.~\ref{fig_Cq_ATLAS_1}(a)  the values of  \( C_{\mathrm{q}} (\sqrt{s}) \)  
linearly  increase  with \( \log{\sqrt{s}}\).

The \( C_{\mathrm{q}} \) values at \(\sqrt{s} = 2.36\)~TeV  in Fig.~\ref{fig_Cq_ATLAS_1}(b) 
are much smaller  than those for other energies.
This is because the  region of primary charged-particle multiplicity distributions at  \(2.36\)~\TeV\ 
is  smaller (up to \(z \approx 3.5\))  than that for higher  CM  energies (up to \(z \approx 9\)) 
\cite{Kulchitsky:2022gkm}.
Therefore,   the \( C_{\mathrm{q}} \) values at \(\sqrt{s} = 2.36\)~\TeV\   were noted  in the fits. 

The  \( C_{\mathrm{q}} (\sqrt{s}) \)    for  \(p_{\mathrm{T}} >500\)~\MeV\ 
have  higher   bias  
and  slope 
of the fits than those for   
minimum \(p_{\mathrm{T}}\) threshold,  the bias   increasing  from \(1.1\) at  \(q=2\) up to  \(2.1\) at  \(q=5\), 
and  the slope   increasing from \(1.4\) at  \(q=2\) up to  \(2.6\) at  \(q=5\).
This is the result of  stronger interactions with  a higher \(p_{\mathrm{T}}\) threshold. 
%
Figure~\ref{fig_Cq_ATLAS_1}(c)  shows moments  \(C_{\mathrm{q}}\)   for events with 
\(n_{\mathrm{ch}} \ge 2\), \(p_{\mathrm{T}} >100\)~\MeV\  and for  \( z > 0.5 \)
without  the fraction of single and double diffraction events, which was accepted  by 
the ATLAS minimum-bias trigger
\cite{STDM-2010-01,STDM-2010-06,STDM-2014-19,STDM-2015-02,STDM-2015-17}.
In this case,  the values of  \( C_{\mathrm{q}} (\sqrt{s}) \)  are systematically higher than those for 
full distributions  with  \( z > 0\) and   show a similar linear increase with \( \log{\sqrt{s}}\)  
as is  illustrated in   Fig.~\ref{fig_Cq_ATLAS_1}(c).
For multiplicity distributions for \( z > 1.0 \)  the  values of  \( C_{\mathrm{q}} (\sqrt{s}) \) 
at the highest energies  \( \sqrt{s} =7,\ 8\) and \(13\)~\TeV\ are  in agreement within error uncertainties,
as can be seen  in Fig.~\ref{fig_Cq_ATLAS_1}(c).
Therefore,  the energy independence of  the moments  of various orders can be considered 
as a confirmation of  the KNO scaling.

%
\subsection{KNO scaling at the LHC experiments}

\begin{figure*}[t!]
\centering
\begin{minipage}[h]{0.32\textwidth} 
\center{\includegraphics[width=1.0\linewidth]{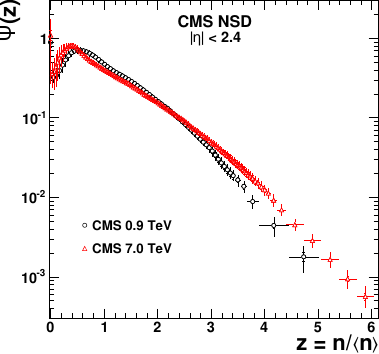}}
(a) 
\\
\end{minipage}
\hfill
\begin{minipage}[h]{0.32\textwidth} 
\center{\includegraphics[width=1.0\linewidth]{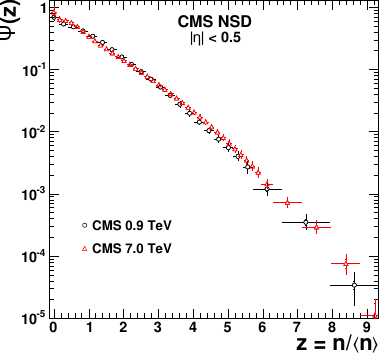}}
(b)
\\
\end{minipage}
\hfill
\begin{minipage}[h]{0.32\textwidth} 
\center{\includegraphics[width=1.0\linewidth,height=0.22\textheight]{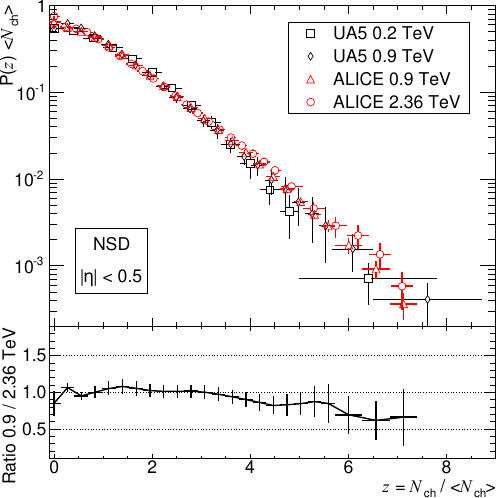}}
(c)
\vspace{1mm}
\\
\end{minipage}
\caption{
The CMS KNO-scaled primary charged-particle  multiplicity distributions    \cite{CMS:2010qvf}
as a function of the scaled multiplicity  \(z\)  at  the centre-of-mass energies \(\sqrt{s} = 0.9\) and \(7\)~\TeV\  
in two pseudorapidity intervals: 
(a) \( \mid\eta\mid < 2.4\)
and 
(b) \( \mid\eta\mid < 0.5\).
Taken from Ref.~\cite{CMS:2010qvf}.
(c)
Top panel:
Comparison of multiplicity distributions in KNO variables measured by  UA5 \cite{UA5:1985hzd,UA5:1988gup} 
in \(  p \bar{p}\)  collisions at   \( \sqrt{s} = 0.2 \)  and  \(0.9\)~\TeV\    and 
by ALICE   \cite{ALICE:2010cin} at   \(\sqrt{s} = 0.9 \)  and  \(2.36\)~\TeV,    for NSD events in  \( \mid\eta\mid < 0.5 \). 
Bottom panel: 
The ratio between  the ALICE measurements at \( \sqrt{s} = 0.9 \)  and \( 2.36 \)~\TeV\  is shown. 
The error bars represent the combined statistical and systematic uncertainties.
Taken from Ref.~\cite{ALICE:2010cin}.
}
\label{fig_KNO_CMS}
\end{figure*}

The KNO scaling violations   were studied   for  different pseudorapidity ranges  
in  LHC experiments by the CMS 
\cite{CMS:2010qvf}  and  ALICE  \cite{ALICE:2010cin,ALICE:2015olq}  at  
the CM energies  from \( \sqrt{s} = 0.9 \) to \(8\)~\TeV.
%
The multiplicity distributions obtained by the  CMS detector   are shown in  the KNO form 
\cite{CMS:2010qvf}  for  the pseudorapidity interval of   \(\mid\eta\mid < 2.4\)
in  Fig.~\ref{fig_KNO_CMS}(a),  which is close to  the similar ATLAS results with \(\mid\eta\mid < 2.5\), 
and for a more  central pseudorapidity interval  \(\mid\eta\mid < 0.5\)   in  Fig.~\ref{fig_KNO_CMS}(b). 
The variation of the ratio for  the central region  of \(0.9\) to \(7\)~\TeV\ with \(\mid\eta\mid < 0.5\) is 
about \(\pm 15\)\%  and agree with  \(1\)  within error bars;    therefore,  the KNO scaling  holds.
The variation of the ratio for  the full region with \(\mid\eta\mid < 2.4\) is  twice wider  \(\approx \pm 30\)\% 
and does not agree with \(1\) in error bars,   therefore  the KNO scaling is violated, similar to  the ATLAS  data
in Fig.~\ref{fig_events_pT100_mch_KNO}(a). 
Scaling is a characteristic property of the multiplicity distribution in  cascade processes of a single jet 
with self-similar branching  and   a fixed coupling constant 
\cite{Polyakov:1970lyy,Polyakov:1971gx,Orfanidis:1974bd,Cohen-Tannoudji:1987uvy,Dokshitzer:1993dc,Carius:1990cz,Szwed:1989mm,Gazdzicki:1990bp}.

A similar conclusion  about the shape evolution of  the multiplicity distributions  like  from Fig.~\ref{fig_KNO_CMS}(b) 
can be  extracted from  Fig.~\ref{fig_KNO_CMS}(c), where are compared  the ALICE measurements   plotted in terms of KNO variables
at the two energies and UA5 \(p \bar{p} \)  data at  \(\sqrt{s} = 0.2\) and \(0.9\)~\TeV, 
for NSD collisions and pseudorapidity interval  \(\mid\eta\mid < 0.5\). 
While the KNO scaling gives a reasonable description of the data from  \(\sqrt{s} = 0.2\) and \(2.36\)~\TeV,  
the ratio between the  \(\sqrt{s} = 0.9\) and \(2.36\)~\TeV\ 
data shows a slight departure from unity above  \(z = 4\), 
but  it is in agreement with unit within error bars.

\begin{figure*}[t!]
\centering
\begin{minipage}[h]{0.90\textwidth} 
\center{\includegraphics[width=1.0\linewidth]{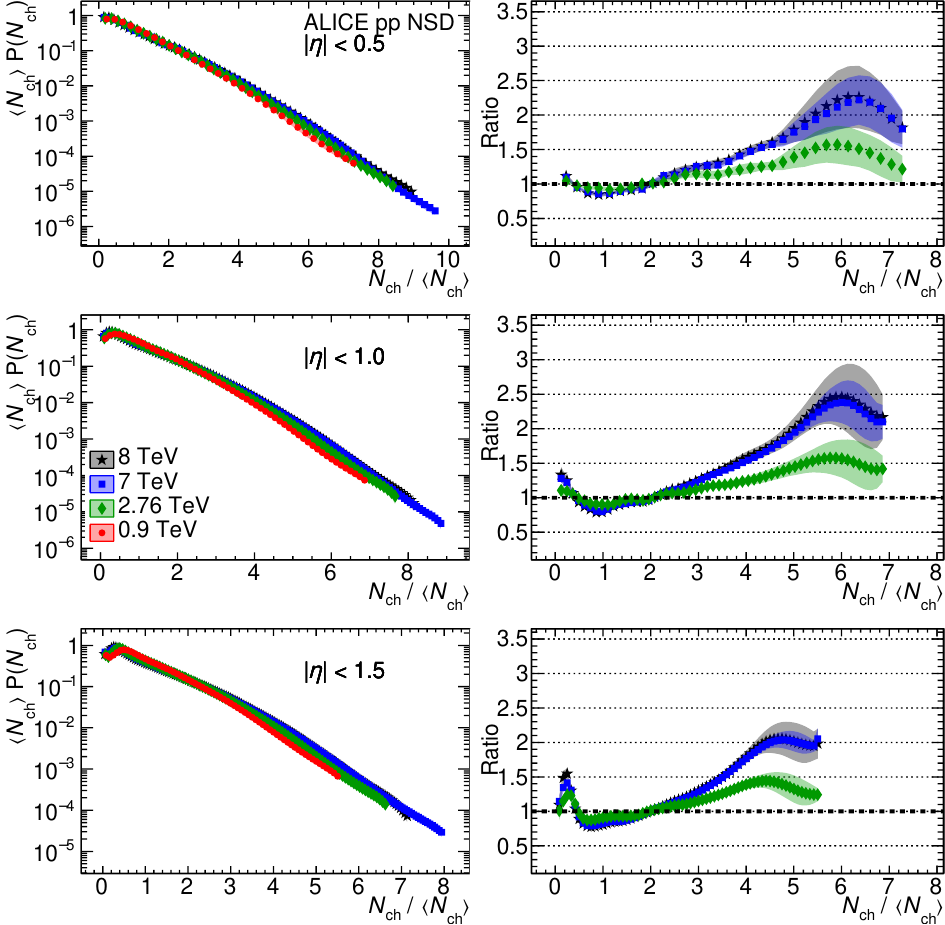}}
\\
\end{minipage}
\caption{
Left side panel: 
The ALICE  KNO scaled primary charged-particle  multiplicity distributions   
as a function of the scaled multiplicity  \(z\) at  the centre-of-mass energies 
\(\sqrt{s} = 0.9,\ 2.36,\ 7\) and \(8\)~\TeV\   in pseudorapidity intervals: 
(top) 		\( \mid\eta\mid < 0.5\),
(middle) 	\( \mid\eta\mid < 1.0\)
and 
(bottom) 	\( \mid\eta\mid < 1.5\)
\cite{ALICE:2015olq}.
The uncertainties represent the sum in quadrature of the statistical and systematic contributions.
Right side panel: 
The ratios of the KNO-scaled primary charged-particle distributions to the interpolated distribution 
at  \( \sqrt{s}  = 0.9 \)~\TeV\  are shown. 
Bands represent the uncertainties for the ratios 
as 
the
results of statistical and systematic uncertainties added in quadrature for both distributions.
Taken from Ref.~\cite{ALICE:2015olq}.
}
\label{fig_KNO_ALICE}
\end{figure*}

The KNO test on the ALICE  results in the range   of \(0.9\) to \(8\)~\TeV\ 
\cite{ALICE:2015olq} is presented in Fig.~\ref{fig_KNO_ALICE}.
The KNO-scaled distributions and their ratios were obtained for each of the available combinations of 
corrections with the same procedure used for multiplicity distribution measurements.
Bin-to-bin correlations were ignored when comparing KNO distributions and \( C_{\mathrm{q}} \)-moments at various CM energies.
Consequently, the relative errors obtained  from the ratios are somewhat overestimated. 

In Fig.~\ref{fig_KNO_ALICE}(right)  \cite{ALICE:2015olq} the ratios of the KNO distributions between the 
two highest  energies and \( \sqrt{s} = 0.9\)~\TeV\  exceed the maximum  value multiplicity at   
\( z \approx 0.2 \)  
and   the value of \(\approx 2\) for the   multiplicity  at  \( z \approx 4.8 \),  \( z \approx 6.0 \) and  
\( z \approx 6.4 \)
for  the pseudorapidity intervals 
\( \mid\eta\mid < 1.5 \),  \( \mid\eta\mid < 1.0 \)  and  \( \mid\eta\mid < 0.5\),  respectively.
There is an intersection point for all distributions at \( z \approx 2\).
The shapes at \( \sqrt{s} =7 \) and \( 8 \) \TeV\  are similar and agree within error bars. 
Therefore, the  ALICE results show  a tendency for  the KNO scaling   to be independent of 
energy for the highest energies. 
%
%
This confirms that KNO scaling violations  increase with the size of   the increasing pseudorapidity interval. 
The shape of the KNO scaling violations  reflects the fact that the high-multiplicity tail  of 
the distribution increases 
with energy and with  the size of  the pseudorapidity interval  faster  than that for  the low-multiplicity tail  
(\( n_{\mathrm{ch}} \le 20\)). 
A test of  the KNO scaling between  \(\sqrt{s} = 0.9\) to \(8\)~\TeV\ confirms that KNO scaling violations  increase
with increasing \(\sqrt{s}\) and,  at a given  CM energy,  with increasing width of   pseudorapidity intervals.
%
%
This is similar to  the ATLAS result in Fig.~\ref{fig_events_pT100_mch_KNO}(a).

\begin{figure*}[t!]
\centering
\begin{minipage}[h]{0.32\textwidth} 
\center{\includegraphics[width=1.0\linewidth]{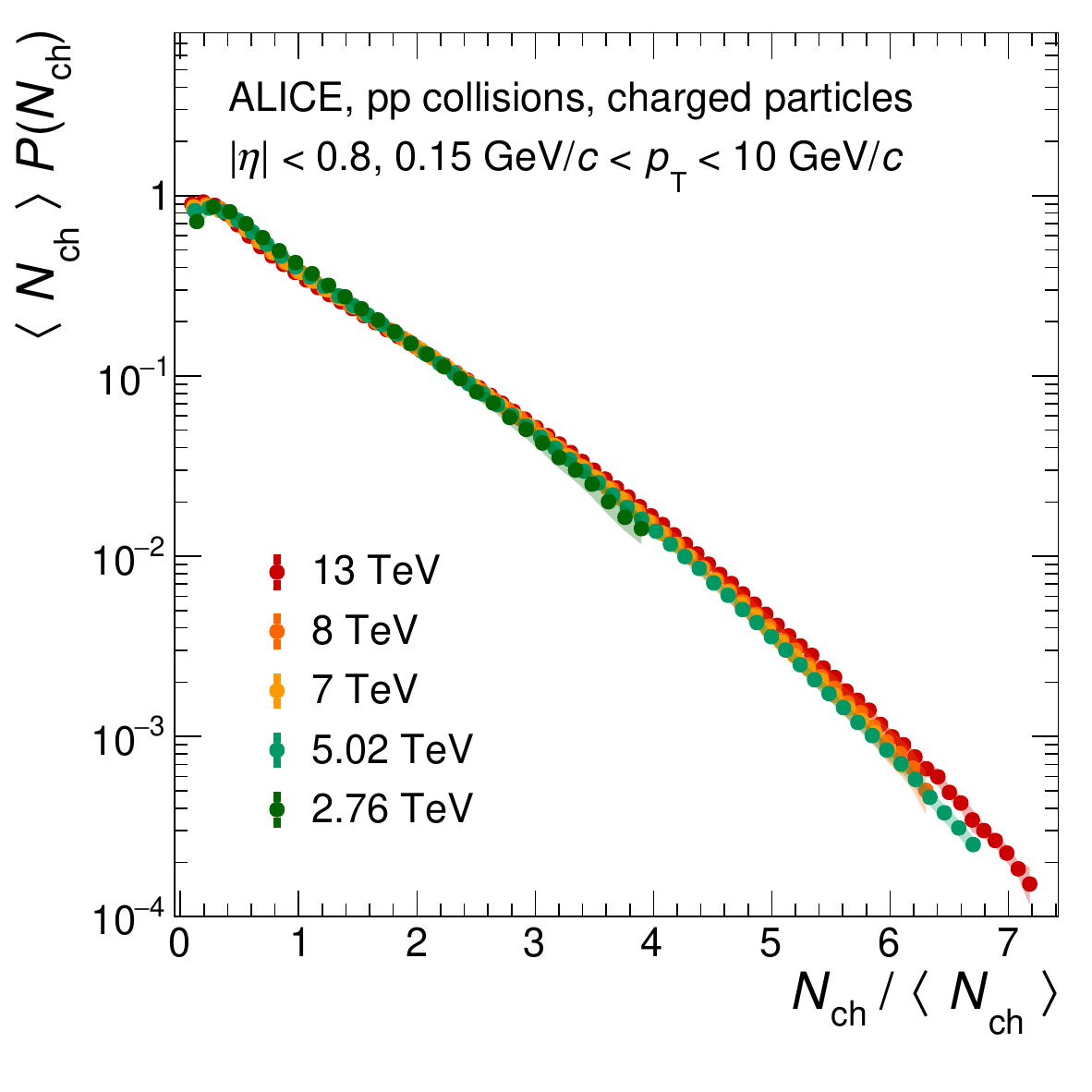}}
(a) 
\\
\end{minipage}
\hfill
\begin{minipage}[h]{0.32\textwidth} 
\center{\includegraphics[width=1.0\linewidth]{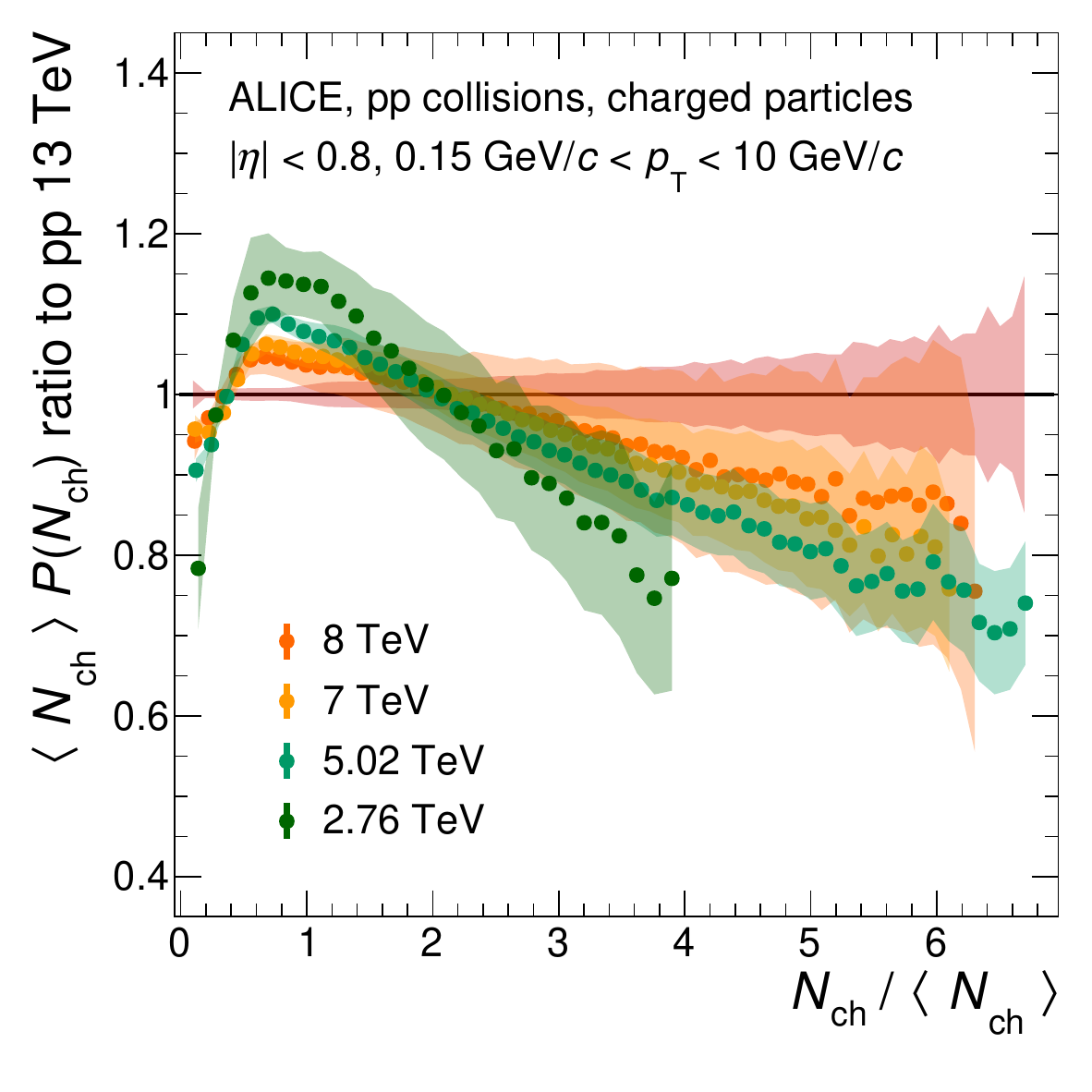}}
(b)
\\
\end{minipage}
\hfill
\begin{minipage}[h]{0.32\textwidth} 
\center{\includegraphics[width=1.0\linewidth]{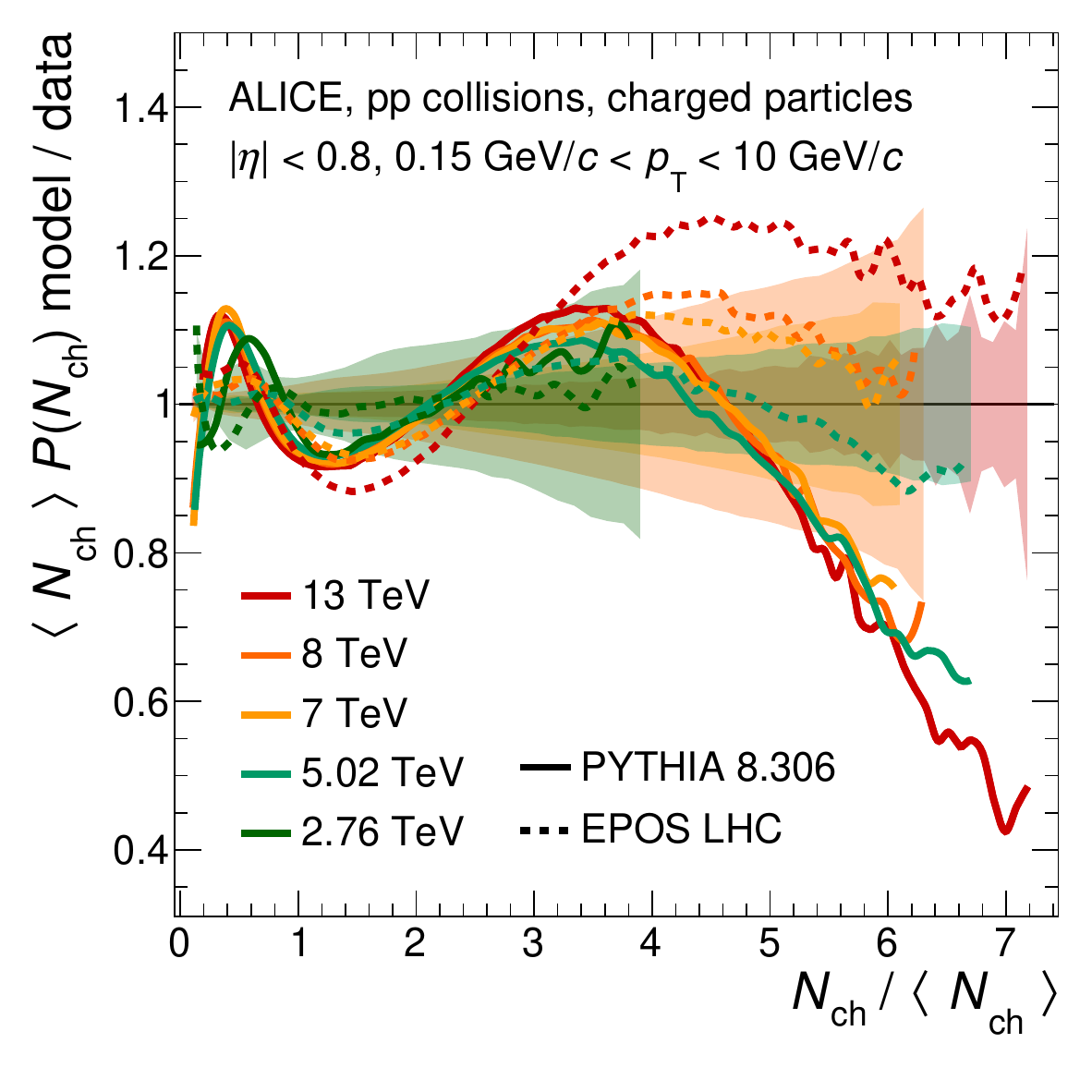}}
(c)
\\
\end{minipage}
\caption{
(a)
The ALICE KNO scaled primary charged-particle  multiplicity distributions   
as a function of the scaled multiplicity  \(z\) for \( pp\)  collisions at the different centre-of-mass energies
\(\sqrt{s} = 2.36,\ 5.02,\ 7,\ 8\) and  \(13\)~\TeV\ for events in the kinematic range
\( N_{\mathrm{ch}} > 0\),  \( \mid\eta\mid < 0.8 \)  and
 \( 0.15 < p_{\mathrm{T}} < 10 \)~\GeV.
Statistical and systematic uncertainties are shown as bars and semi-transparent bands, respectively.
(b)
The ratios of the KNO-scaled primary charged-particle distributions to the interpolated distribution 
at   \( \sqrt{s}  = 13 \)~\TeV\   are shown. 
Statistical and systematic uncertainties are shown as bars and semi-transparent bands, respectively.
(c)
The ratio of  \textsc{Pythia\,8} \cite{Sjostrand:2014zea} and \textsc{EPOS\ LHC}  \cite{Pierog:2013ria}
model predictions to data for \(pp\) collisions  at various energies for 
the KNO scaling of primary charged-particle  multiplicity distributions.
The semi-transparent bands indicate the relative systematic uncertainties of the data.
Taken from Ref.~\cite{ALICE:2022xip}.
}
\label{fig_KNO_ALICE_3}
\end{figure*}

The KNO test on the ALICE results  for \( pp\)  collisions at the different CM energies
\(\sqrt{s} = 2.36,\ 5.02,\ 7,\ 8\) and  \(13\)~\TeV\  for \( 0.15 < p_{\mathrm{T}} < 10 \)~\GeV\
and for  the pseudorapidity  region \( \mid\eta\mid < 0.8 \) 
is presented in Fig.~\ref{fig_KNO_ALICE_3}(a).
Figure~\ref{fig_KNO_ALICE_3}(b)   shows the corresponding ratios of the KNO scaled multiplicity distributions 
at various CM energies  relative to \(\sqrt{s} = 13\)~\TeV.  
The KNO scaling apparently holds  within \(\approx 30\)\% for 
CM energies from \(2.36\) to \(8\) \TeV\ in relative to  \(\sqrt{s} = 13\)~\TeV.
Figure~\ref{fig_KNO_ALICE_3}(c)  compares measured results 
for the respective KNO scaled multiplicity distributions with predictions from 
\textsc{Pythia\,8}  \cite{Sjostrand:2014zea} (solid lines) 
and  \textsc{EPOS\ LHC}   \cite{Pierog:2013ria} (dashed lines).
Like the multiplicity distributions in Fig.~\ref{fig_Nev_nch_ALICE_1}(b), 
the overall shapes of the  KNO-scaled distribution shown in Fig.~\ref{fig_KNO_ALICE_3}(c) 
are better described by  \textsc{EPOS\ LHC},  while  \textsc{Pythia\,8}  falls sharply off above
\( N_{\mathrm{ch}}/ \langle N_{\mathrm{ch}}\rangle \approx 4\)
and these models within \(25\)\% agree with the experimental distributions  
with larger deviations at highest multiplicities. 

\begin{figure*}[t!]
\centering
\begin{minipage}[h]{0.32\textwidth} 
\center{\includegraphics[width=1.0\linewidth]{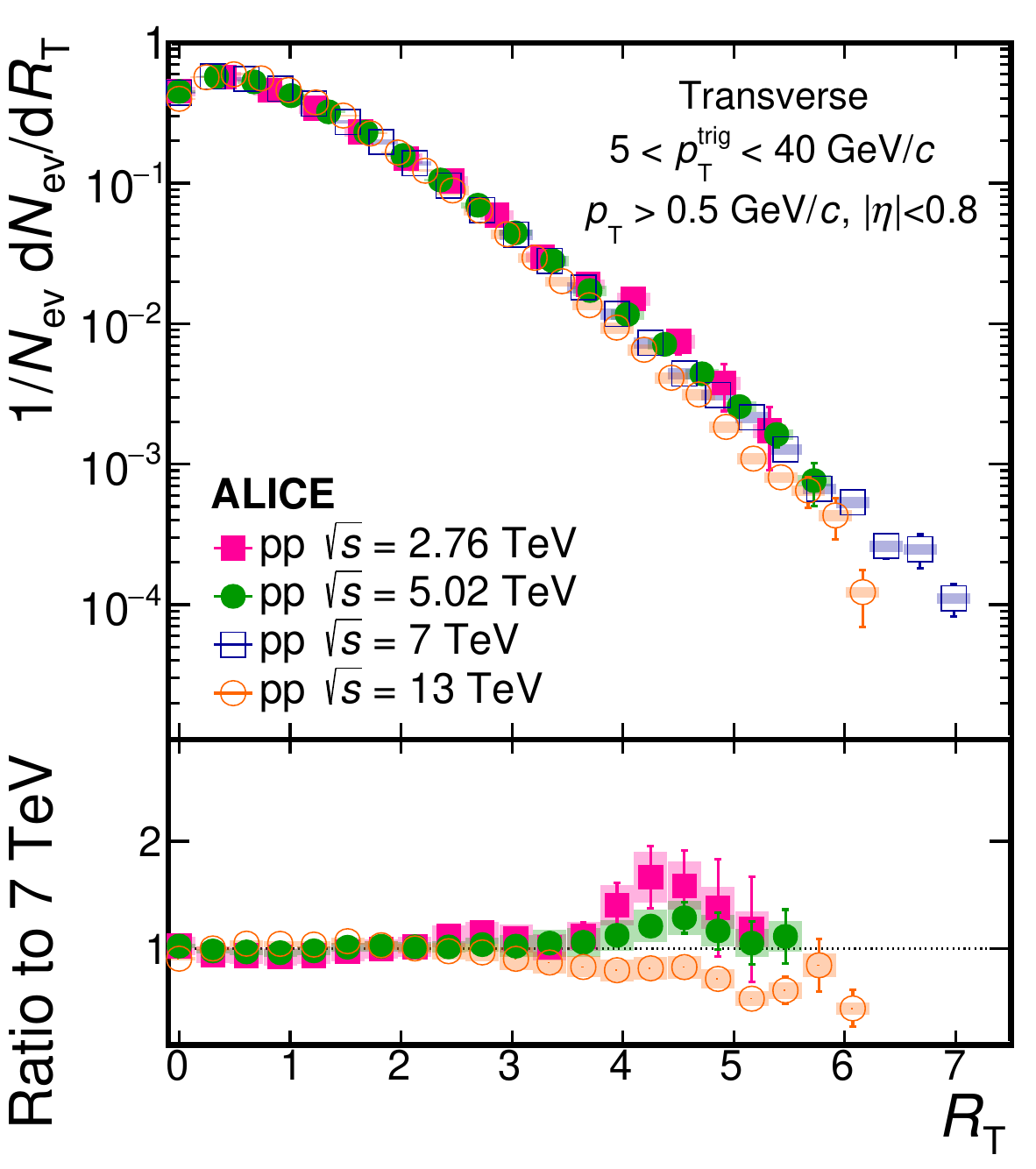}}
(a) 
\\
\end{minipage}
\hfill
\begin{minipage}[h]{0.32\textwidth} 
\center{\includegraphics[width=1.0\linewidth]{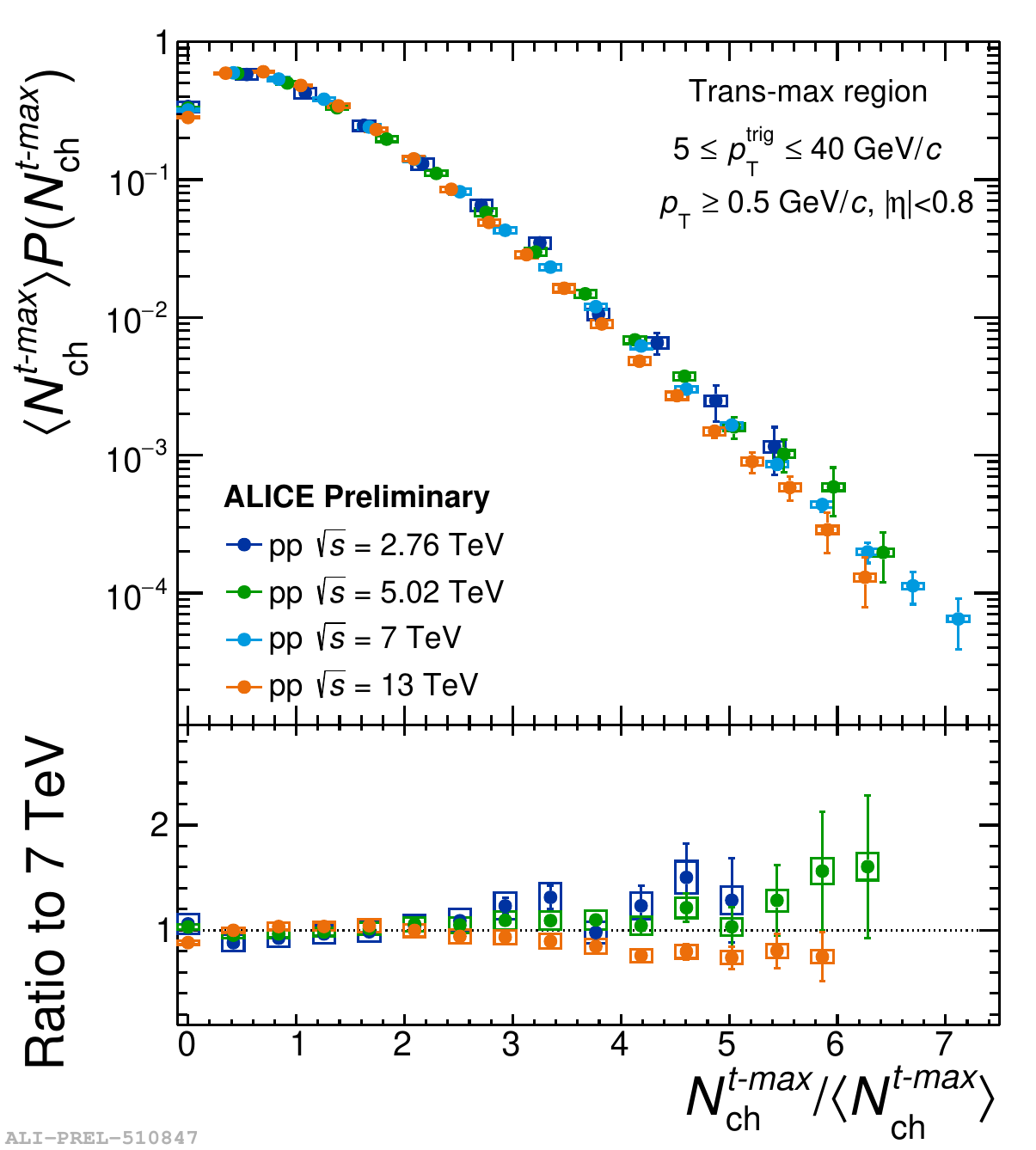}}
(b)
\\
\end{minipage}
\hfill
\begin{minipage}[h]{0.32\textwidth} 
\center{\includegraphics[width=1.0\linewidth]{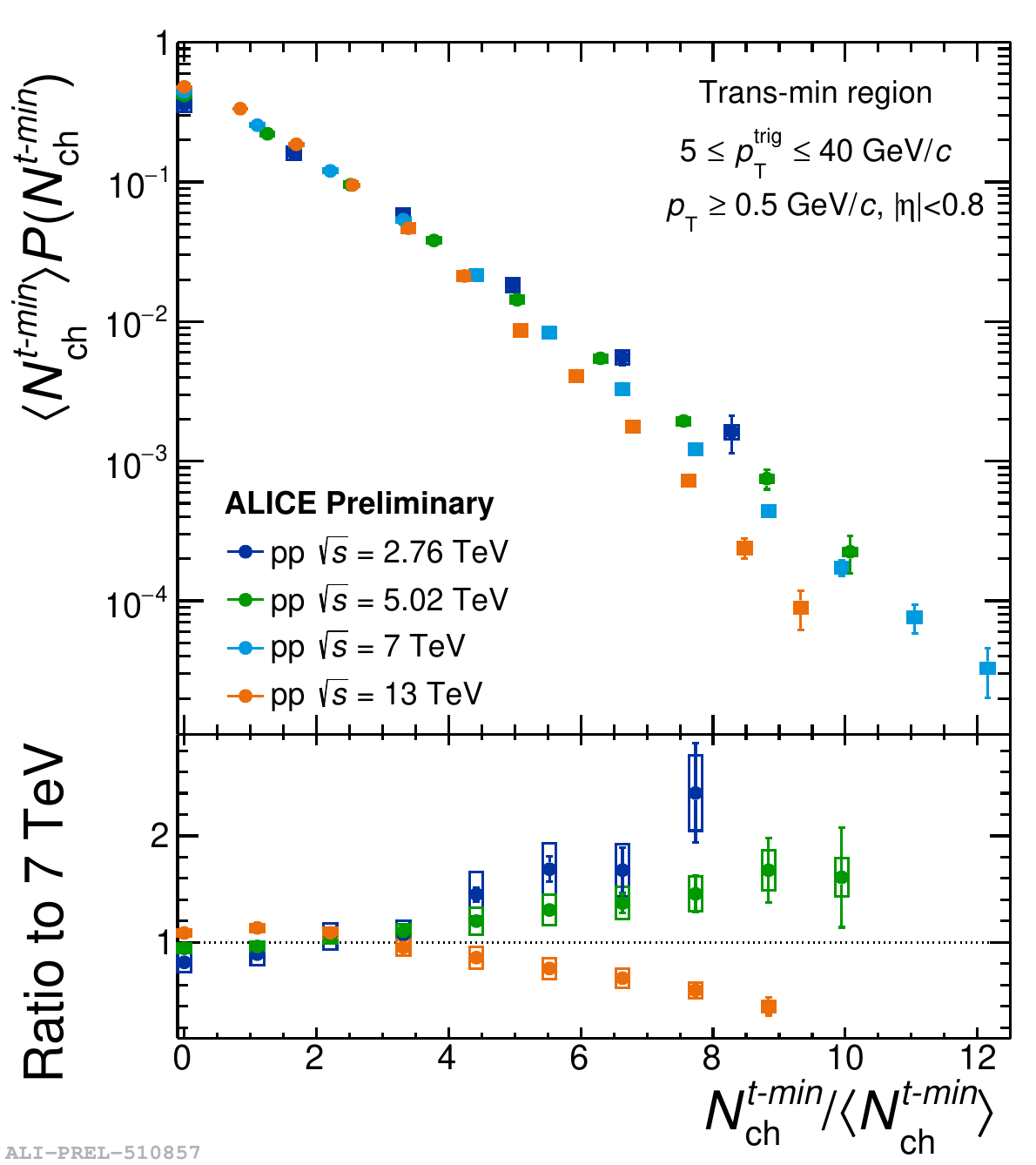}}
(c)
\\
\end{minipage}
\caption{
Top: 
The ALICE KNO primary charged-particle  multiplicity distributions   
as a function of the scaled multiplicity  
for \( pp\)  collisions at the different centre-of-mass energies
\(\sqrt{s} = 2.36,\ 5.02,\ 7\) and  \(13\)~\TeV\ for events in the kinematic range
\( 5 \leq p_{\mathrm{T}}^{\mathrm{trig}} \leq 40\)~\GeV/c,  
\( p_{\mathrm{T}} > 0.5\)~\GeV/c,   and \( \mid\eta\mid <0.8\) 
in 
(a)
the overall transverse region,  
(b)
the tran-max region 
(the sub-transverse region with the largest multiplicity),  
and 
(c)
the trans-min region
(the sub-transverse region with the smallest multiplicity).  
Bottom: 
The KNO multiplicity distributions are normalised to  those at  \( \sqrt{s} = 7\)~\TeV. 
The ratio is calculated using a linear interpolation between adjacent points. 
The boxes and the error bars represent the systematic and statistical uncertainties, respectively.
The sample at \( \sqrt{s} = 13\)~\TeV\  is smaller than that used at  \( \sqrt{s} = 7\)~\TeV.
Taken from Refs.~\cite{Fan:2022bbp,ALICE:2023csm}.
}
\label{fig_KNO_ALICE_4}
\end{figure*}

The study of the  KNO scaling was done by ALICE using UE events \cite{ALICE:2023csm, Fan:2022bbp}.
Figure~\ref{fig_KNO_ALICE_4}  shows the  ALICE results for  the overall transverse \cite{ALICE:2023csm},  
the trans-max and trans-min \cite{Fan:2022bbp} UE regions
for charged-particle multiplicity distributions in KNO variables  for \(pp\) collisions at  
\( \sqrt{s}=2.76,\ 5.02,\  7\) and \(13\)~\TeV\ for events in the kinematic range 
\( 5 \leq p_{\mathrm{T}}^{\mathrm{trig}} \leq 40\)~\GeV/c, \( p_{\mathrm{T}} > 0.5\)~\GeV/c, 
and \( \mid\eta\mid < 0.8 \). 
The traditional UE analysis focuses on the study of particles in three topological
regions depending on their azimuthal angle relative to the leading particle, 
\( \mid \Delta \phi \mid = \mid \phi - \phi^{\mathrm{trig}} \mid \),
which is the one with the highest transverse momentum, \( p_{\mathrm{T}}^{\mathrm{trig}}\), in the event. 
%
%
%
The transverse region,  with \(\mid\Delta\phi\mid >2\pi/3\)~rad,  is dominated by the UE dynamics,
but it also includes contributions from initial- and final-state radiation (ISR-FSR) 
\cite{Bencedi:2021tst,Ortiz:2021gcr}.
The trans-max  and trans-min  regions of UE refer to the sub-transverse regions  with the largest and smallest 
charged-particle multiplicity,   which have  enhanced sensitivity to ISR-FSR and UE, respectively  
\cite{Bencedi:2021tst, CDF:2015txs}. 
The relative transverse activity classifier, 
\(R_{\mathrm{T}}=N_{\mathrm{ch}}^{\mathrm{T}}/\langle N_{\mathrm{ch}}^{\mathrm{T}}\rangle\), 
is the ratio of the primary charged-particle multiplicity in the transverse region, 
\(N_{\mathrm{ch}}^{\mathrm{T}} \),   obtained event-by-event to the average value,  \(\langle N_{\mathrm{ch}}^{\mathrm{T}}\rangle\)
\cite{ALICE:2019mmy,Martin:2016igp}.
%

Figure~\ref{fig_KNO_ALICE_4}(a)  shows the \(R_{\mathrm{T}}\) distributions for overall transverse region.
%
In the bottom panel, the KNO multiplicity distributions are normalised to  those  
at  \( \sqrt{s} = 7\)~\TeV\  are reported. 
For low \(R_{\mathrm{T}}\), smaller than \(4\), 
distributions are found to be approximately, within \(\pm 20\)\%,  
collision energy independent, which indicates a KNO scaling 
\cite{Ortiz:2017jaz}.
For \(R_{\mathrm{T}}\) values higher than \(4\),  a large deviation of the ratios from unity is seen.
%
A similar effect is observed in \textsc{PYTHIA\,8}  \cite{Ortiz:2017jaz,Ortiz:2021gcr}.
From an analysis aimed at measuring the MPI, it was observed that for 
\( N_{\mathrm{ch}}/\langle N_{\mathrm{ch}}\rangle > 3\) -- \(4\),  the number of MPI as a function of 
\( N_{\mathrm{ch}}/\langle N_{\mathrm{ch}}\rangle\) deviates from the linear trend suggesting 
the presence of high-multiplicity jets 
\cite{ALICE:2013tla,Ortiz:2021peu}.
The presence of high-multiplicity jets may also explain the breaking of KNO scaling properties 
observed at high \(R_{\mathrm{T}}\) in 
Fig.~\ref{fig_KNO_ALICE_4}(a).  
To increase the sensitivity of \(R_{\mathrm{T}}\) to MPI, it has been recently proposed to build the so-called 
\(R_{\mathrm{T}}^{\mathrm{min}}\) that is based on the charged multiplicity in the less active side of 
the transverse region 
\cite{Bencedi:2021tst}.

More detailed information was presented in Figs.~\ref{fig_KNO_ALICE_4}(b) and (c) for the trans-max and trans-min regions, 
for UE events study, 
respectively. 
Figure~\ref{fig_KNO_ALICE_4}(b) is similar to distributions in Fig.~\ref{fig_KNO_ALICE_4}(a)
and  shows that in the trans-max region, also within \(\pm 20\)\%,  
the KNO scaling is observed in a wider range, up to \(4\),  of scaled multiplicity
\(R_{\mathrm{ch}}^{\mathrm{t\mbox{-}max}} = N_{\mathrm{ch}}^{\mathrm{t\mbox{-}max}}/
\langle N_{\mathrm{ch}}^{\mathrm{t\mbox{-}max}}\rangle\),  
relative to the results reported in \cite{Ortiz:2021gcr},  
while for higher  values  \( R_{\mathrm{ch}}^{\mathrm{t\mbox{-}max}}  > 4\)   the scaling is broken. 
It is worth noticing that for trans-max,  both contributions are considered:  UE and ISR-FSR. 
%
In Fig.~\ref{fig_KNO_ALICE_4}(c)  for the trans-min region,  where the effect of ISR-FSR is suppressed, 
the KNO scaling also holds up to \(4\) for  
\(R_{\mathrm{ch}}^{\mathrm{t\mbox{-}min}} = N_{\mathrm{ch}}^{\mathrm{t\mbox{-}min}}/
\langle N_{\mathrm{ch}}^{\mathrm{t\mbox{-}min}}\rangle\), 
and then for \( R_{\mathrm{ch}}^{\mathrm{t\mbox{-}min}} > 4\)  
the KNO scaling  is still broken but  for higher values  reach is achieved,  
especially for scaled multiplicity values grater than \(6\),  a larger violation is observed. 
Events with high-multiplicity jets can contribute to the large violation of the scaling properties. 
%

\begin{figure*}[t!]
\centering
\begin{minipage}[h]{0.45\textwidth} 
\center{\includegraphics[width=1.0\linewidth]{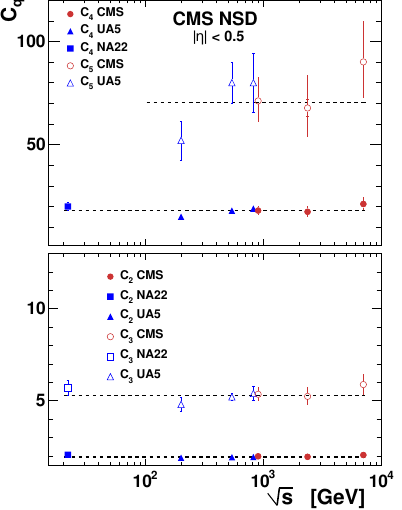}}
(a)
\\
\end{minipage}
\hspace{2mm}
\begin{minipage}[h]{0.46\textwidth} 
\center{\includegraphics[width=1.0\linewidth,height=0.41\textheight]{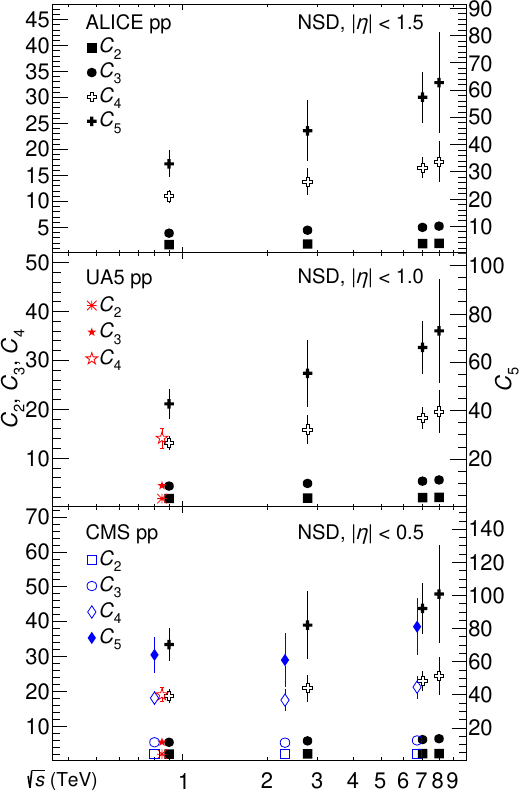}}
(b) 
\vspace{1mm}
\\
\end{minipage}
\caption{
The normalised moments \(C_{\mathrm{q}}\) of the primary charged-particle multiplicity distributions 
measurement by the  CMS for events with CM energies  at \(\sqrt{s} = 0.9,\ 2.36\) and \(7\)~\TeV\
for pseudorapidity regions 
(a)
\( \mid\eta\mid < 0.5\).
The 
\(C_{\mathrm{2}}\)
and
\(C_{\mathrm{3}}\) 
results  are shown in  the  bottom  panels,
and
\(C_{\mathrm{4}}\)
and
\(C_{\mathrm{5}}\) 
results  are shown in  the   top panel.
The results of  lower-energy experiments  NA22  \cite{EHSNA22:1987syd}  and 
UA5 \cite{UA5:1985hzd,UA5:1988gup} are included. 
Fits of the  \( \log{\sqrt{s}}\)   dependence of the normalised moments  \( C_{\mathrm{q}} \) 
of the multiplicity distribution for   \( \mid\eta\mid < 0.5\) (assuming no dependence) are shown.
The uncertainties represent the sum in quadrature of the statistical and systematic contributions.
Taken from Ref.~\cite{CMS:2010qvf}.
(b)
The ALICE  CM energy dependence of the  moments   \(C_{\mathrm{q}}\) 
(\(q = 2\) to \(4\), left-hand  scale, and  \(q = 5\), right-hand  scale) 
of the multiplicity distributions for NSD events in three different pseudorapidity intervals: 
(top) 		\( \mid\eta\mid < 1.5\),
(middle) 	\( \mid\eta\mid < 1.0\)
and 
(bottom) 	\( \mid\eta\mid < 0.5\)
\cite{ALICE:2015olq}.
The ALICE data (black) are compared to 
UA5  \cite{UA5:1988gup} (red) for \( \mid\eta\mid < 0.5\) and \( \mid\eta\mid < 1.0\) 
at \(\sqrt{s} = 0.9\)~\TeV\  
and CMS  \cite{CMS:2010qvf} (blue) at \(\sqrt{s} = 0.9\) and \(7\)~\TeV\  for \( \mid\eta\mid < 0.5\). 
The error bars represent the combined statistical and systematic uncertainties. 
The data at \(\sqrt{s} = 0.9\) and \(7\)~\TeV\  are slightly displaced horizontally for visibility.
Taken from Ref.~\cite{ALICE:2015olq}.
}
\label{fig_Cq_CMS}
\end{figure*}

Multiplicity distributions may be characterised  by their normalised  \(C_{\mathrm{q}}\)-moments 
where  \(q\)  is a positive integer studied  for  the values  \(2,\ 3,\ 4\) and \(5\).
%
The results obtained by different experiments  for  the \(C_{\mathrm{q}}\)-moment  dependence  
on \(\sqrt{s}\)  are shown in Fig.~\ref{fig_Cq_CMS}.
%
The CMS results  \cite{CMS:2010qvf}, which  are presented in Fig.~\ref{fig_Cq_CMS}(a),
show that the KNO scaling holds  for the central pseudorapidity region with \( \mid\eta\mid < 0.5 \)
and for the energy region from \(\sqrt{s} = 0.9\) to \(7\) \TeV.
The CMS results are complemented  by measurements at lower energies   in experiments 
NA22  \cite{EHSNA22:1987syd} and  UA5 \cite{UA5:1985hzd,UA5:1988gup}.
For  \( \mid\eta\mid < 0.5\)   the values of  \(C_{\mathrm{q}}\)-moments demonstrate  
in Fig.~\ref{fig_Cq_CMS}(a)
independence of energy and  the shape of  the KNO function is  similar for 
\(\sqrt{s} = 0.9\) and \(7\) \TeV, 
as can be seen in Fig.~\ref{fig_KNO_CMS}(b). 
Another energy dependence  the CMS results  \cite{CMS:2010qvf}  provide
for  the inclusive region with \( \mid\eta\mid < 2.4 \) in  Fig.~\ref{fig_Cq_ATLAS_1}(a), 
where  \(C_{\mathrm{q}}\)-moments  demonstrate the linear increase with energy. 
%
The ALICE results for \(C_{\mathrm{q}}\)-moments are published in Ref.~\cite{ALICE:2015olq}
for  three pseudorapidity intervals  \( \mid\eta\mid < 0.5 \),   \( \mid\eta\mid < 1.0 \),  
\( \mid\eta\mid < 1.5\) for the  energy from \(\sqrt{s} = 0.9\) to \(8\) \TeV\
and shown in Fig.~\ref{fig_Cq_CMS}(b).
In this case,  there are KNO scaling violations  for 
more inclusive pseudorapidity regions \( \mid\eta\mid < 1.0 \) (middle panel) 
and  \( \mid\eta\mid < 1.5\) (top panel) because \(C_{\mathrm{q}}\)-moments  
linear increase with  \( \log{\sqrt{s}}\):
\(C_{\mathrm{2}}\) remains constant over the energy range, 
\(C_{\mathrm{3}}\) shows a small increase with increasing energy for  the two largest \(\eta\) intervals, 
\(C_{\mathrm{4}}\) and  \(C_{\mathrm{5}}\) show an increase with increasing energy, 
which becomes stronger for larger \(\eta\)  intervals. 
Figure~\ref{fig_Cq_CMS}(b) show for the central interval  \(\mid\eta\mid <0.5\) 
(bottom panel)  that the KNO scaling holds for the energy region from \(\sqrt{s} = 0.9\) to \(8\) \TeV\ 
because the \(C_{\mathrm{q}}\)-moments energy distributions  
can be described by constant in the error bars.
These ALICE data  \cite{ALICE:2015olq} are consistent  with  the CMS \cite{CMS:2010qvf} and the UA5  
\(p\bar{p}\)  measurements at  \(0.9\)~\TeV\ \cite{UA5:1985hzd,UA5:1988gup} results. 
The energy dependence of the reduced moments  \(C_{\mathrm{q}}\)   shown in Fig.~\ref{fig_Cq_CMS}(b)
indicates a slight increase, which is not significant given the size of systematic uncertainties. 
Systematic uncertainties are assumed to be uncorrelated between energies. 

%
The results of KNO scaling research according to the data  from the ALICE, ATLAS and CMS experiments 
have been  analysed.
The shape evolution of the multiplicity distributions with a collision energy at ATLAS 
is studied in terms of KNO scaling variables  at \(\sqrt{s}\) from \(0.9\) to \(13\) \TeV\ 
with  the inclusive pseoudorapidity region \(\mid\eta\mid <2.5 \).
The KNO scaling and \(C_{\mathrm{q}}\)-moments were studied by the CMS 
at \(\sqrt{s}\) from \(0.9\) to \(7\) \TeV\  in central pseudorapidity   \( \mid\eta\mid <0.5\) region
and more inclusive  \( \mid\eta\mid < 2.4 \) regions \cite{CMS:2010qvf},
and ALICE  at \(\sqrt{s}\) from \(0.9\)  to \(13\) \TeV\ in three pseudorapidity regions 
 \(\mid\eta\mid < 0.5\),   \(\mid\eta\mid < 0.8\),  \(\mid\eta\mid <1.0\)   and  \(\mid\eta\mid <1.5\)
\cite{ALICE:2010cin,ALICE:2015olq,ALICE:2022xip,Fan:2022bbp}.
The charged-particle multiplicity distributions on the KNO scale for all experiments
have the similar shape and   decrease with increasing collision energy. 
For the ALICE and CMS experiments,  the KNO scaling  is  violated for energies from  \(0.9\) to \(13\) \TeV\
if taking into account more inclusive pseoudorapidity regions. 
The KNO scaling holds  for  the central pseudorapidity region with \(\mid\eta\mid <0.5 \)
and for the energy region from \(\sqrt{s}=0.9\) to \(8\) \TeV\ on the ALICE and CMS results.
The ATLAS data  demonstrate  a tendency for  the KNO scaling  to be independent of energy  
for the highest energies,  and the KNO scaling holds for a scaled multiplicity greater than \(1\). 

\section{Conclusions}
\label{summary}

ATLAS  studied MB events in \(pp\)  interactions  at  
the CM energies    \(\sqrt{s} = 0.9\), \( 2.36\), \(7\), \(8\) and \(13\)~\TeV\   
for the absolute pseudorapidity region less than \(2.5\)  in five separate 
PS regions
\(n_{\mathrm{ch}} \ge 2\),  \(p_{\mathrm{T}} > 100\)~\MeV\
and  
\(n_{\mathrm{ch}} \ge 1,\ 6,\ 20,\ 50\),  \(p_{\mathrm{T}} > 500\)~\MeV\
recorded in 2010 -- 2015. 
The data were taken in the special  configuration of the 
LHC with low beam currents and reduced beam focusing, 
producing a low mean number of interactions per 
bunch-crossing in the range \(0.003\) -- \(0.007\).

%
The charged-particle multiplicity 
depends
on pseudorapidity,  charged-particle multiplicity, 
and  transverse momentum, 
as  well as  the dependence of the mean transverse momentum on multiplicity, 
were presented for 
the
study 
of 
the soft-QCD phenomena. 
The measured distributions are presented as inclusive-inelastic distributions within a
given 
PS
region with minimal model-dependent corrections 
to facilitate the comparison with models.
%
Variables are tuned in event generators using these MB measurements
because there is 
variability in modelling since non-perturbative QCD is used.

The results are compared to the predictions of more than ten 
MC models tuned to a wide range of measurements.
Then variables  in  the MC event generators were tuned using  the MB measurements of 
the LHC and Tevatron experiments
because there was 
variability in modelling since non-perturbative QCD was used.

This review  reported that  the multiplicity distribution is not described perfectly
by any of the models; 
there are large discrepancies, 
especially at large multiplicities.
Having observed similar discrepancies at all measured energies, we conclude that
for every collision energy, model parameters usually need to be re-tuned in every
MC generator.
%
%
Reasonable agreement 
between the tunes used in   the MC models  
and  the data  was presented. 
The  models 
\textsc{EPOS} \textsc{LHC}, 
\textsc{PHOJET}, 
\textsc{QGSJET-II},
\textsc{Pythia\,6}
and 
\textsc{Pythia\,8} 
show big troubles in describing the whole
spectrum in 
the
data, 
but the best agreement is achieved with 
\textsc{EPOS}.
A new ATLAS \textsc{Pythia\,8}  \textsc{A3} tune was presented 
for result predictions at Run~3 of the LHC.

The comparisons of the charged-particle multiplicity
and   the average transverse momentum 
distributions  on  the basis of  the scaled multiplicity
using the results of the LHC experiments 
were presented.
The charged-particle multiplicity distributions on the KNO scale 
have 
a
similar shape  and decrease with increasing energy. 
The KNO scaling  was studied  using 
the results of the LHC  experiments. 
A test of the KNO scaling between \(0.9\)  and \(13\)~\TeV\  
confirms that the KNO scaling violation increases with decreasing collision energy.
The KNO distributions tend to be independent of energy for the highest energies.
The mean transverse momentum on the KNO scale has the same shape and increases with increasing energy. 


\section*{Acknowledgements}
%
The author thanks the ATLAS collaboration for the excellent experimental results that were used in this review.
Special  thanks go to Edward K.~Sarkisyan-Grinbaum  and   Stanislav~Tokar  for their  very productive discussions. 
The author is grateful to Pavel~Tsiareshka for 
technical support. 
%
\printbibliography
\end{document}